\documentclass[twoside, 11pt]{scrbook} 

\usepackage[utf8]{inputenc} 

\usepackage[margin=3.5cm]{geometry} 
\usepackage{graphicx} 
\usepackage{setspace}
\usepackage{appendix}
\usepackage{titletoc}

\usepackage{booktabs} 
\usepackage{array} 
\usepackage{paralist} 
\usepackage{verbatim} 
\usepackage{subfig} 
\usepackage{amsmath}
\usepackage{amssymb}
\usepackage{graphicx}
\usepackage{hyperref}
\usepackage{todonotes}
\usepackage{xy}
\usepackage[cc]{titlepic}

\usepackage{chngcntr}
\usepackage{etoolbox}
\usepackage{lipsum}

\AtBeginEnvironment{subappendices}{%
\chapter*{Appendix}
\addcontentsline{toc}{chapter}{Appendices}}

\newcommand{\bra}[1]{\langle#1 \vert}
\newcommand{\ket}[1]{ \vert #1\rangle}
\newcommand{\braket}[2]{\langle#1 \vert #2\rangle}
\newcommand{\ketbra}[2]{#1\rangle \!\ \langle#2|}
\newcommand{\avg}[1]{\langle #1 \rangle}
\newcommand{\tr}{\mbox{Tr}}
\newcommand{\re}{\text{Re}}
\newcommand{\im}{\text{Im}}

\newcommand{\uu}{\uparrow}
\newcommand{\dd}{\downarrow}

\newenvironment{psmallmatrix}
  {\left(\begin{smallmatrix}}
  {\end{smallmatrix}\right)}

\usepackage{fancyhdr} 
\usepackage{sectsty}
\allsectionsfont{\sffamily\mdseries\upshape} 

\usepackage[nottoc,notlof,notlot]{tocbibind} 
\usepackage[titles,subfigure]{tocloft} 

\begin{titlepage}
\title{\textbf{Power of photonic states: from computation to cosmology}}

\author{Nana Liu\\ Merton College, Oxford\\ Hilary term 2016}

\titlepic{\includegraphics[width=7cm]{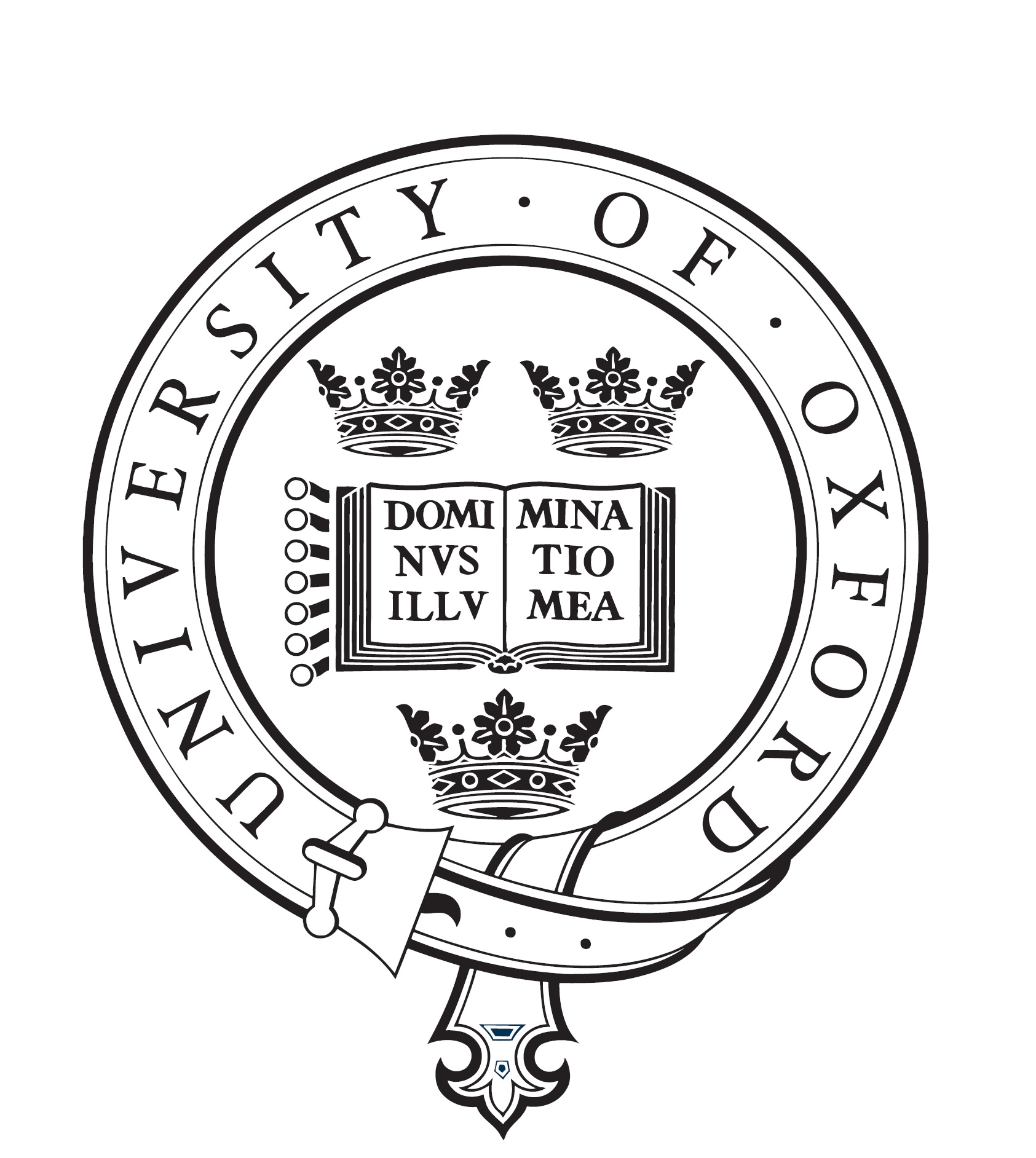}}
\date{\textit{A thesis submitted for the degree of Doctor of Philosophy} \\ in Atomic and Laser Physics \vspace{5mm}} 
\titlehead{A thesis submitted for the degree of \\ DPhil in Atomic and Laser Physics}
\end{titlepage}

\begin{document}
\maketitle

\frontmatter
\clearpage
\thispagestyle{plain}
\par\vspace*{0.35\textheight}{\centering To Mum...\\
\par}
\begin{figure}[ht!] 
\centering
\includegraphics[scale=0.5]{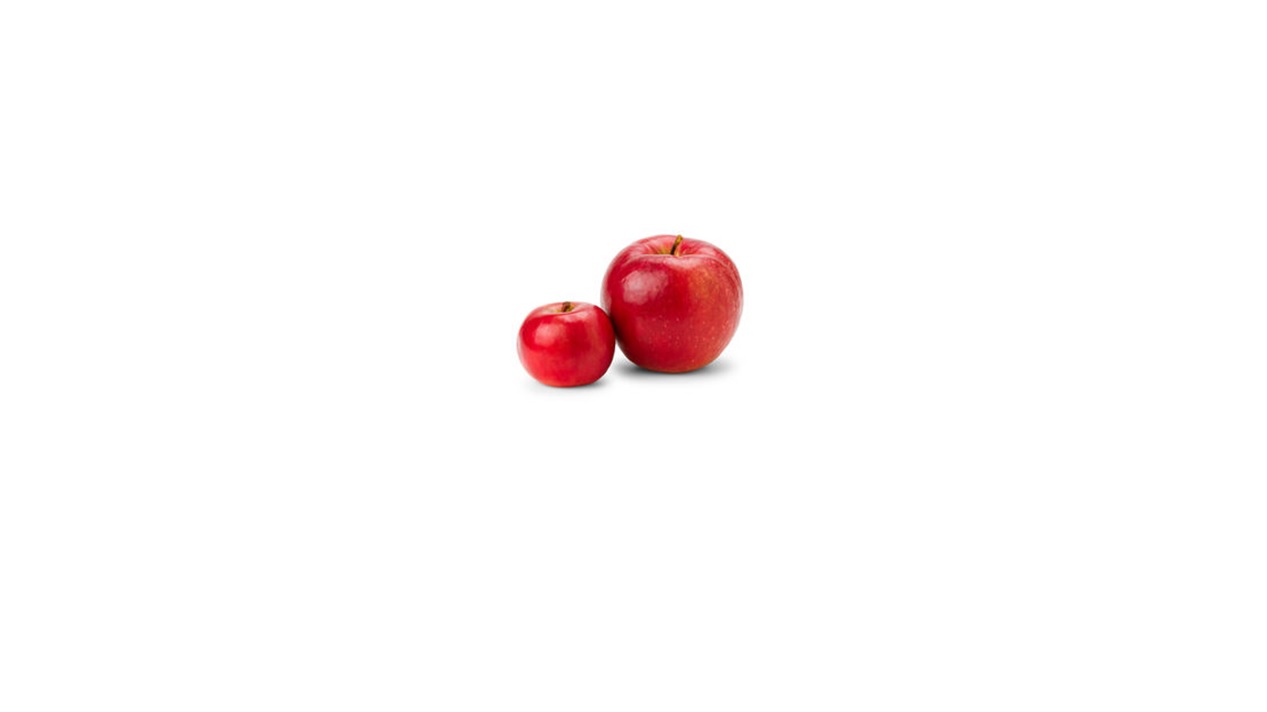}
\end{figure}

\setcounter{tocdepth}{1}

\chapter{}
\noindent \textbf{Supervisor}: \\

Prof. Vlatko Vedral \\

\noindent \textbf{Thesis examiners}: \\

Prof. Artur Ekert \\

Prof. Caslav Brukner
\chapter{Abstract}
This thesis is an exploration of the power of photonic resources, as viewed from several different but related perspectives. They range from quantum computation, precision parameter estimation to the thermodynamics of relativistic quantum systems, as applied to cosmology in particular. The use of photonic states allows us to address several important questions about the resources required in quantum mechanical processes. 

In chapter~\ref{chap:qumode}, we propose a new quantum computational model, called the `power of one qumode', that relies mainly on a single-mode photonic squeezed state. In particular, we show the amount of squeezing can quantitatively relate the resource requirements of factoring to the problem of finding the trace of large unitary matrices, a result with consequences for understanding how powerful quantum computation really is. Furthermore, we can connect squeezing to other known resources like precision, energy, qudit dimensionality and qubit number, which is a useful stepping stone to finding the resources that enable quantum computation. 

In chapter~\ref{chap:tomo}, we exploit the quantum mechanical properties of photonic states for use in precision parameter estimation of general linear optical processes, which is useful for a diverse number of applications, from characterising an unknown process in a photonic quantum computer to biological imaging. We introduce a formalism that quantifies this improvement in precision. We also provide conditions under which one can easily check for photonic states that are optimal to use in this context, which is a potentially important result for future experimental efforts. 

In chapter~\ref{chap:cosmo}, we explore the connection between two-mode squeezed states, commonly used in quantum optics, and relativistic quantum processes, in particular in cosmology. Using this connection, we apply recently developed tools from the thermodynamics of quantum systems perturbed far from equilibrium to address an old question of entropy production in cosmology from a surprising new angle. 
\chapter{Acknowledgements}
I am an incredibly lucky person to have spent these enjoyable and formative years in Oxford and around the world with so many amazing people, from whom I have learnt so much. I am grateful to all who have brightened my journey, especially those I mention here.  

First of all, I very warmly thank my supervisor Vlatko Vedral, for welcoming me into his wonderful group, for stimulating discussions and for his encouragement during my three years in his group. This has been amongst the most enjoyable three years of my life. I would also like to thank my former supervisor John March-Russell for his support during my first year in Oxford and for also letting me know that my changing fields once is a very good sign, as he himself changed fields and supervisors twice during his doctorate. 

I am greatly indebted to my collaborator Kavan Modi, who has also been an excellent mentor to me throughout my DPhil. Kavan has also hosted my multiple productive and pleasurable extended visits to his group in Melbourne, for which I am very thankful. In addition to being a solid and perceptive researcher, Kavan also takes good care of younger researchers by giving them practical guidance on many different aspects of academia. I am also very grateful to my collaborator Hugo Cable, not only for our very enjoyable collaboration and the thoroughness of his approach, but also for giving me many practical pieces of advice. I especially thank Hugo for the care he has taken in helping me edit the second chapter of this thesis. I would never have imagined that I would be lucky enough to one day work with one of my old `school' friends Jayne Thompson, when we became good friends almost seven years ago in Melbourne. Jayne has always been a kind and thoughtful sister to me and I am grateful not only for our fruitful scientific discussions during our collaboration which made this thesis possible, but especially for her warm friendship and advice throughout the years. I am also very thankful for my collaborator Mile Gu, for his insight and knowledge, excellent advice on numerous topics, our shared love of dinosaurs and of course his kind friendship. My sincere gratitude of course also goes to David Bruschi, for being such a fun and encouraging collaborator, as well as giving me innumerable and important advice on various aspects of academia. I thank David for our engaging scientific discussions and his friendship throughout my DPhil. I also won't forget our desert hike in Palestine while discussing the research forming the last chapter of this thesis. 

I also warmly thank all my other collaborators for their input into making this thesis possible. I thank Ivette Fuentes (and her group members) for inviting me to her former group in Nottingham so I can learn the `state of the art' of relativistic quantum information, an area which she helped establish. There was never a dull moment in Ivette's group. I also thank John Goold for introducing me to out-of-equilibrium thermodynamics. I thank Christian Weedbrook for his advice on several important issues and also to Seth Lloyd, for his great insight and humour. 

The research done in this thesis is also made possible by my wonderful office-mates, who has made my day-to-day life in Oxford so much more bright and colourful in the grey and damp weather. Of those I have spent the most time with, I want to thank especially Andy Garner, Benjamin Yadin, Cormac Browne, Davide Girolami, Fabio Anza, Felix Binder, Felix Pollock, Oscar Dahlsten, Tristan Farrow, Thomas Elliot and Mihai Vidrighin (our experimentalist neighbour) for our very stimulating physics discussions, our friendship and the constant sharing of jokes. I also thank all the newer members of our group and our constant stream of wonderful visitors for making our office even merrier. I also have the fortune of having Felix Pollock as my cheery office-mate in Melbourne and I look forward to every visit, where many parts of this thesis were written. Many thanks to Ben and Cormac for helping with the editing of this thesis. Special thanks also to Ben for our almost daily discussions from which I have learnt so much and for suggesting that I have access to over 120,000 symbols in Unicode when I complained about running out of Greek and Roman letters in one of my papers. 

I extend a special thank you to Professor Bei-Lok Hu of the University of Maryland, for his great kindness and warm encouragement, especially during a particularly slow period in my DPhil. `Hu Lao Shi' (Teacher Hu) has really made a profound difference during that time and I remain extremely grateful. 

There are many groups around the world I had the great pleasure of visiting. Of those I have not already mentioned, I would like to thank the Oxford Quantum Optics group, Gerardo Adesso and his group in Nottingham,  Dan Browne, Sougato Bose, Alessio Serafini and their groups in UCL, Mauro Paternostro and his group in Belfast, Sandu Popescu, Tony Short and their groups in Bristol, Guihua Zeng and his group in Shanghai, Qiongyi He and her group in Beijing, Lloyd Hollenberg and his group in Melbourne, Nick Menicucci and his group in Melbourne, Stephen Bartlett and his group in Sydney, Valerio Scarani and his group in Singapore, Joe Fitzsimons and his group in Singapore. 

A very special thank you also to the very kind and late Professor Jacob Bekenstein for inviting me to the University of Jerusalem. I spent two very happy and productive weeks there. 

I am also very greatly indebted to Merton college for their ongoing support throughout my time in Oxford. I am also very grateful to the porters and dinner staff who have been continuously kind and helpful during all my years in Oxford. 

I am also thankful to the Clarendon Fund for providing me with the Clarendon scholarship, which has supported me during my time in Oxford. 

My friends outside the office also made a great difference. I thank especially Tianyi Zhang, who has been my constant and cherished friend for most of my years in Oxford. I have learnt much not only from her excellent attitude and amazing biochemistry, but especially her steady and clear reasoning in everything from cooking, drawing to protein-folding. To my adventurous friend Jess Thorn for being my steady friend from the time we lived together, for her delicious vegan cooking and her amazing stories of field work in Nepal and Ghana. I also want to thank my very good friend Chiara Marletto for all our endlessly engaging conversations on everything from the nature of quantum mechanics to the nature of curiosity, and of course our constant laughter and our cooking together. Chiara is like my childhood friend. I have learnt much from Chiara on how to be a better listener. I am grateful also to my astrophysicist friend Kiz Natt for her great kindness and warm friendship. I am also thankful to Carolyn Lloyd-Davies and her wonderful family for hosting me during Easter and Christimas celebrations, making me feel at home while being at the other side of the world. A great thank you also to my good friends Emily Cliff and Ariell Ahearn for our little club and sharing cooking tips. I also thank my friends in the badminton society and the Kodaly choir at Merton for making fitness training and music even more fun. 

The UK is one of the farthest places from Australia and I am incredibly grateful to my family and friends outside Oxford for their steady encouragement and support. In particular, I want to thank my excellent friend Newton Langford from the Rotary club of Glenferrie, for his steady support and friendship over the last ten years. Newton has also given me much good advice on many important points and I always carry away some key lessons from our conversations. I thank also Debbie Leung for her great kindness and her wonderful support. I thank Robert Winspear for being our long-standing family friend and for sponsoring me to the National Youth Science Forum more than ten years ago and for his practically-minded advice. I am also grateful for the support of Sir Roderick Carnegie, for our chats and his solid advice on several important points. 

I extend also a very special thank you to two teachers in my early years. First to Dr. Antoinette Tordesillas of the Mathematics and Statistic Department at the University of Melbourne, for offering me the chance to conduct research in her group for many years and for having faith in me, although I was only in high school. I cannot forget the irreplaceable Mr. Rennie at MacRobertson Girls' High School, for everything he did to support my love of doing physics. 

I also thank my good friends Michelle Roche and Justin Matthys for our friendship, from which I have learnt a great deal. I always look forward to our reunions when I visit Melbourne. 

I would also like to thank my very curious and innovation-savvy grandfather. Although spending most of his career in chemistry and medicine, he is vitally interested in quantum technologies and chats with me about it every Saturday afternoon. My warmest thanks also to Bobby and Kuku for being such great companions. I also want to give a very special thank you to my stepdad Michael Weber for all your support during both the good times and also those slightly trickier times. 

Lastly to my beautiful mother Melody. I have no words to express my love and gratitude. Nothing would have been possible without you.
\tableofcontents
\chapter{Preface and overview}
In this thesis, we explore the use of quantum photonic states as resources in different but related quantum mechanical processes, from quantum computation, precision multi-parameter estimation to relativistic quantum processes, which includes cosmological particle creation.

Quantum photonic systems are finding an ever increasing diversity of applications, ranging from quantum cryptography, quantum imaging, quantum lithography to quantum computation and even cosmology, especially with the recent successful detection of gravitational waves.  The experimental manipulation of photonic systems has also been steadily improving in recent years, including a demonstration of all-photonic quantum repeaters and photonic small-scale quantum algorithms. This is, therefore, a good time to investigate the diverse applications of photonic states and the ways in which they serve as useful resources. 

There are three main questions one might ask when considering the resources required for a particular task, whether it is using cattle to plough a field, performing arithmetic on an abacus, using a steam engine for transportation or using photonic states for quantum computation, imaging and using spacetime expansion to drive particle creation in the early universe. One might ask:\\
1) What are the `minimal' resources for this task? \\
2) Can these resources be replaced by other resources that are more accessible? \\
3) What is the `waste' resulting from this task and how does this depend on one's resources? 

Chapters~\ref{chap:qumode} and~\ref{chap:tomo} explores the first two questions from the point of view of quantum computation and quantum precision multi-parameter estimation, using photonic states. Chapter~\ref{chap:cosmo} explores the last question from the angle of cosmological particle creation (and similarly other relativistic quantum scenarios), which can be interpreted as a photonic process, where `waste' is quantified by an entropy. In what follows, we provide a brief overview of the work in each chapter of this thesis (see also Fig. ~\ref{prefacepic}). 
\begin{figure}[ht!] 
\centering
\includegraphics[scale=0.5]{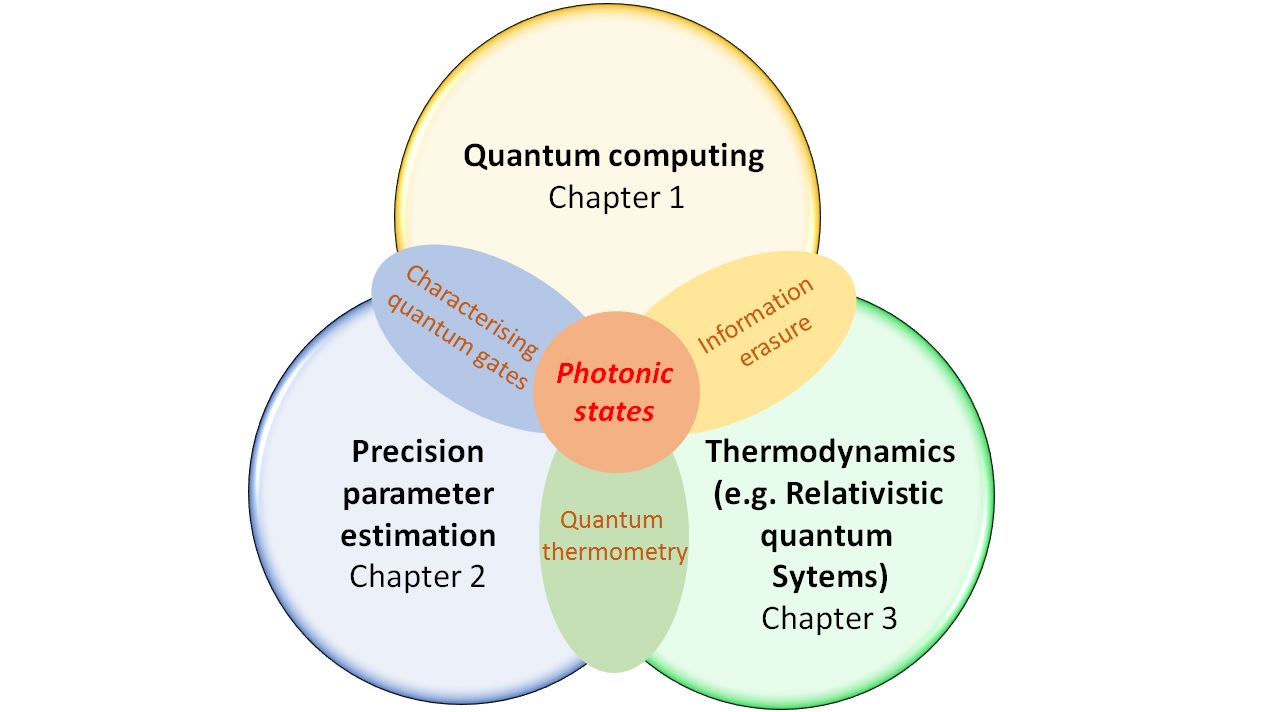}
\label{prefacepic}
\caption[\textit{Pictorial outline of thesis.}]{
\label{prefacepic}\textit{Basic outline of using photonic states as quantum mechanical resources}. In this thesis, photonic states are used in the application to quantum computing (chapter~\ref{chap:qumode}), precision parameter estimation (chapter~\ref{chap:tomo}) and the thermodynamics of relativistic quantum systems, in particular in cosmology (chapter~\ref{chap:cosmo}). Multiple connections exist between these different fields, some of which are mentioned above.}
\end{figure}
\section{Overview of thesis}
Quantum computing is a very notable example of how quantum resources can provide an enormous advantage over classical ones. It offers many useful applications like factoring and finding the trace of large matrices, which on a quantum computer can be solved in exponentially less time than on any classical computer. Yet, what resources a quantum computer actually exploits still remains to be fully understood. In our first chapter, \textbf{Power of one qumode}, we introduce a new computational model called the `power of one qumode' which relies on one pure continuous variable mode (qumode). This is a generalisation of the important computational model known as DQC1 (deterministic quantum computing with one quantum bit). Using this model we can show how some of the seemingly different resources of quantum computation that have been proposed in the past (including precision, energy, qudit dimensionality
and qubit number) can be recast in the form of squeezing – a long-standing notion of non-classicality commonly used in quantum optics. Furthermore, our framework demonstrates that the amount of squeezing used in our model allows us to quantitatively compare the complexity of two important quantum algorithms, specifically factoring and finding the trace of large matrices, which were not easily comparable before. The main results in this chapter are presented in the paper\\

\noindent \textbf{Power of one qumode for quantum computation} \cite{nanaqumode}

\noindent Authors: \textbf{Nana Liu}, Jayne Thompson, Christian Weedbrook, Seth Lloyd, Vlatko Vedral, Mile Gu, Kavan Modi.

\noindent Published in \textit{Physical Review A}, Vol 93, No.5, 052304, 2016. \\

Precision estimation of unknown parameters, useful in the characterisation of quantum gates in quantum computing, among other important applications, is another key area where quantum resources have shown to be more powerful than classical resources. In our second chapter, \textbf{Photonic multi-parameter estimation}, by using two-mode photonic states, we develop a formalism that allows us to study the efficiency of estimation of parameters of a unitary matrix describing general linear optical processes. Photonic states are vital in all applications of imaging and linear optical quantum computing, yet the problem of general multi-parameter estimation in the photonic context has not been previously studied using the same tools as single-parameter precision estimation. The main results in this chapter are presented in the forthcoming paper\\

\noindent \textbf{Quantum-enhanced multi-parameter estimation for unitary photonic systems} \cite{liu2016quantum}

\noindent Authors: \textbf{Nana Liu}, Hugo Cable. \\

In our last chapter, \textbf{Thermodynamics of a squeezed state in cosmology and other relativistic scenarios}, we present the connection between photonic squeezed states and relativistic quantum scenarios, in particular,  cosmological particle creation. By studying the thermodynamics of the creation of squeezed states using the recently developed concepts from the thermodynamics of fast processes (out-of-equilibrium) and quantum systems, we apply these results to investigate the relationship between entropy production and particle creation in an expanding universe. This allows us to gain new insight into some old questions in cosmological particle creation. This formalism also allows us to link for the first time quantum field theory in curved spacetime and concepts in out-of-equilibrium thermodynamics. The main results in this chapter are presented in the paper\\

\noindent \textbf{Quantum thermodynamics for a model of an expanding universe} \cite{nanacosmo}

\noindent Authors: \textbf{Nana Liu}, John Goold, Ivette Fuentes, Vlatko Vedral, Kavan Modi, David Edward Bruschi

\noindent Published in \textit{Classical and Quantum Gravity}, Vol 33, No. 3, 035003, 2016.
\section{How to read this thesis}
The content in this thesis involves three different areas: quantum computation, precision parameter estimation and thermodynamics of relativistic quantum systems. Although the central themes in these areas are related, the key concepts, results and notation are different enough that each chapter on each of these topics is self-contained and can be read independently, with its own introduction and motivation.

Chapter~\ref{chap:qumode} assumes a basic knowledge of the quantum circuit model and chapter~\ref{chap:cosmo} assumes some familiarity with terms used in equilibrium thermodynamics and elementary general relativity. We also work in natural units $\hbar=1=c=k_B=G$ throughout this thesis. 

To aid reading, a contents page appears before the beginning of each chapter and appendices appear after each chapter. Although the notation in each chapter is generally self-contained, there are overlaps in notation across the chapters. Although every effort is made for these to be as clear and as consistent as possible, there are a few places where notation in one chapter might differ slightly from another, but this should be clear from context. For example, the usual notation for the computational basis (or eigenvectors of the Pauli $\sigma_z$ matrix), denoted by $\ket{0}$, $\ket{1}$, are used everywhere except in chapter~\ref{chap:tomo}, where they are denoted by $\ket{\uu}$, $\ket{\dd}$. This is to prevent confusion with single-mode number states in chapter~\ref{chap:tomo}. 
\section{List of abbreviations used in this thesis}
\textbf{BQP} (bounded error quantum polynomial time); \textbf{CV} (continuous variable); \textbf{HV}, \textbf{DA} and \textbf{RL} (horizontal/vertical, diagonal/anti-diagonal and right/left circular polarisation); \textbf{DQC1} (deterministic quantum computing with one quantum bit); \textbf{gcd} (greatest common denominator); \textbf{GHZ} (Greenberger-Horne-Zeilinger state); \textbf{HB} (Holland-Burnett state); \textbf{MLE} (maximum likelihood estimation); \textbf{NP} (nondeterministic polynomial time); \textbf{P} (polynomial time); \textbf{QMA} (Quantum-Merlin-Arthur complexity class); \textbf{SLD} (standard logarithmic derivative).

\cleardoublepage
\phantomsection
\addcontentsline{toc}{chapter}{\listfigurename}
\listoffigures
\cleardoublepage
\phantomsection
\addcontentsline{toc}{chapter}{\listtablename}
\listoftables

\mainmatter

\newpage\null\newpage
\startlist{toc}
\printlist{toc}{}{\section*{\textbf{Power of one qumode}\\
Chapter contents}}

\chapter{Power of one qumode}
\label{chap:qumode}
\section{Introduction and motivation}
Quantum computing is a rapidly growing discipline that has attracted significant attention due to the discovery of quantum algorithms that are exponentially faster than the best-known classical ones \cite{dj, shor, grover, harrowlloyd}. One of the most notable examples is Shor's factoring algorithm \cite{shor}, which has been a strong driver for the quantum computing revolution. However, the essential resources that empower quantum computation remain elusive. Knowing what these resources are will have both great theoretical and practical consequences. This knowledge will motivate designs that take optimal advantage of such resources. In addition, it may further illuminate the quantum-classical boundary.

A computation is a physical process and constraints on physical resources during this process can limit the power of a computation. Given the enormous number and difficulty of computational problems of interest (e.g. simulating the human brain, weather forecast, finding the shortest commercial route through different cities, RSA code-breaking), finding the methods to minimize computational resources become essential. There are three main avenues where improvement can arise. Loosely speaking, they can be considered to occur on the \textit{software} and \textit{hardware} levels. 
The software level involves better algorithmic design. This involves no change on the computational model itself, but simply how it is used. For example, the same computer can be taught how to play chess as well as word processing, by using different algorithms. 

There are two kinds of changes that can occur on the hardware level, which is the physical structure that forms the computer and contains the physical degrees of freedom in which information is stored. One kind of change involves better engineering of existing technology. For example, by keeping traditional electronic circuitry, one can increase the density of the circuits on a chip or improve ventilation. 

There is a second kind of modification which is much more exciting and ground-breaking, which is to take advantage of new physical degrees of freedom in which information is stored that obey different physical \textit{laws}. This is the revolution of quantum computing, where information is stored in quantum states and the computation is performed via quantum mechanical processes. With changes on this hardware level, new algorithms must be invented that take advantage of these new ways of information processing. Quantum computers consume space and time resources like classical computers. However, there are also quantum mechanical resources that are not available to classical computers. 

The search for resources in quantum computation generally gravitates towards finding a single quantity. Apart from the simplicity of this picture, this enthusiasm is justified on two main accounts. The first is the tremendous success of quantum teleportation protocols in quantum cryptography and its necessary requirement of entanglement as a resource. This same simplicity is hoped for in quantum computation. Secondly, in pure state quantum computation, it is known that entanglement is a necessary resource to achieve a computational speed-up \cite{jozsalinden}. However, the picture becomes messier when we leave pure state quantum computation. Entanglement is no longer a crucial resource for mixed-state quantum computation, as we soon see. Here it is unclear if a single entity can quantify the computational resource in these models and there is no a priori reason why any single quantity should exist.

One possibility in what makes a single resource difficult to identify is perhaps the chameleon nature of quantum resources. The same underlying physics appears in different guises under different settings. For instance, entanglement in a many-body system with two-level constituents can be eliminated by simply rewriting it in terms of a single multi-level system. This simple observation \cite{ekertjozsa} can help explain why, for example, Grover's algorithm requires more precision resources when performed in the absence of entanglement \cite{meyer, lloyd1999, jozsa1997}. Entanglement is, in a sense, \textit{swapped} for precision. This illustrates why it is essential to better understand how different resources are related to each other. Without this knowledge, it may prove meaningless to test any single resource on every algorithm, where it might have transformed into other forms. 

By introducing a suitable model, we contribute to the literature on quantum resources by making more explicit the relationships between some common resources of quantum computation as well as using these resources to enhance our understanding of two key quantum algorithms. 

We begin by looking at one notable example of mixed-state computation called the \emph{deterministic quantum computation with one quantum bit} (DQC1) model \cite{knill1998power}. It is a powerful, though not a universal model of quantum computation~\footnote{A universal quantum computer is a model that is in principle capable of any computation allowed by quantum mechanics.}. This model contains little entanglement and purity \cite{white, datta2007}. Yet it can solve certain computational problems exponentially faster than the best-known classical algorithms by using only a \textit{single} pure control qubit and a highly mixed target state. One way to approach a better understanding of resources in computation is to compare this `basic model' DQC1 to other important problems like factoring. However, it is unclear how to compare the resources needed for DQC1 and factoring on an equal footing. One would require the same model to efficiently solve both problems. Although suggestions have been made that factoring requires more resources than DQC1 \cite{parkerplenio}, a direct quantitative relation between the two is still lacking. 

To address this challenge, we propose a continuous-variable (CV) extension of DQC1 by replacing the pure qubit with a CV mode, or \textit{qumode}. Qumodes can be used either directly in quantum computation or as a way of encoding qubits \cite{GKP, terhal, schoelkopf}. We call this new model the \emph{power of one qumode}. We demonstrate that our model is capable of reproducing DQC1 and factoring in polynomial time. This enables us to pinpoint a single resource in our model to compare factoring and DQC1 on the same level. We identify this to be the squeezing of the qumode. Squeezed states are also useful resources in other contexts, like gaining a quantum advantage in metrology \cite{caves, monras, pinel} and in CV quantum computation \cite{lloydCV, mile2009}. 

By inputting a squeezed state as the pure qumode in our model, we can perform both the hardest problem in DQC1 and phase estimation, which includes factoring.  As an application, we show that to factor an integer efficiently in time, we need a squeezing that grows exponentially with the number of bits to encode this integer. However, DQC1 can even be recovered with no squeezing. This shows how squeezing can serve as a unifying role between these two problems. 

This model also enables us to provide an example of a resource trade-off relation that is important for understanding how one resource can be interpreted in terms of others. We relate the squeezing to the degree of precision in phase estimation and the total computation time.  Our model also gives a wonderful opportunity to compare squeezing to other resources in addition to inverse precision, like energy, qubit number and qudit dimensionality. For example, it can be used to provide a clear illustration of how the key resources are changed if a qumode is replaced by qubits, without modifying the computational power of the model. This can be used to show a consistent way of seeing squeezing as a computational resource in terms of the equivalent number of pure qubits and qudit dimensionality. 

We note that the term `squeezing' in the quantum optical literature could refer to either the squeezing parameter $r$ or the squeezing factor $s_0=\exp(r/2)$. We use the term `squeezing' to refer exclusively to the squeezing factor unless otherwise stated.  For quantifying resources in the context of computational complexity, it is important to make a distinction between these two definitions since they are exponentially separated. We will motivate our use of the squeezing factor over the squeezing parameter by showing how it can be interpreted as inverse precision, which is a known resource in computational complexity \cite{algorithms}.

Before moving on, we remark that our architecture is an example of a hybrid computer: it jointly uses both discrete and CV systems. A similar hybrid model using a pure target state was given by Lloyd \cite{lloydhybrid} to find eigenvectors and eigenvalues.  Hybrid models for computing are interesting in their own right for providing an alternative avenue to quantum computing that bypasses some of the key obstacles to fully CV computation using linear optics or fully discrete-variable models \cite{lloydhybrid, hybridbook}. This creates an important best-of-both-worlds approach to quantum computing.\\

\textbf{Chapter outline} \\

We begin with a brief overview of how resources are discussed in quantum computation. Then we move to describing the relevant quantum algorithms like DQC1 and phase estimation, which includes factoring. We assume familiarity with quantum circuit diagrams. We provide the necessary notation for continuous variables in the last part of this introductory section. 

In section 2, we introduce our power of one qumode model. In subsequent sections 3 and 4, we demonstrate the squeezing resources required for DQC1 and factoring. This is followed in section 5 by a comparison of the squeezing resource with precision, energy, the equivalent qudit dimensionality and number of qubits. We show in what ways they are interchangeable and ways in which they are not. We end with our discussion on the key contributions of this work and avenues for future research.
\begin{table}[]
\centering

\label{my-label}
\begin{tabular}{|c|l|l|}
\hline
                   & \textbf{Classical}                                                        & \textbf{Quantum}                                                                                \\ \hline
\textbf{Resources} & \begin{tabular}[c]{@{}l@{}}Time \cite{cook71, papbook}\\ Space \cite{papbook}\\ Energy \cite{landauer}\\ Precision \cite{algorithms}\end{tabular} & \begin{tabular}[c]{@{}l@{}} Time, space, energy, precision \cite{bernvaz,bennett, meyer, lloyd1999}\\ Entanglement \cite{ekertjozsa}\\ Discord \cite{vlatkodiscord, zurekdiscord} \\ Quantum coherence \cite{coherenceref}\\ Negativity of Wigner function \cite{hakop} \end{tabular} \\ \hline
\end{tabular}
\caption[\textit{Resources in classical and quantum computation.}]{\label{table}\textit{Examples of computational resources accessible in the classical and quantum regimes.} We take an advantage of a squeezed state (which is a quantum resource) in our `power of one qumode' model. However, we later show its relationship to classically accessible resources like precision and energy.}
\end{table}
\subsection{Resources and computational complexity}
Understanding the resources required for a computation is not only useful on a practical level, it can also be used as a means to classify the difficulty of computational problems. This is the subject of computational complexity theory, which we very briefly describe. 

The most prominent examples of resources for computers are time and space. Spatial constraints limit the memory available for a computation and time limitations can mean a problem may practically never be solved when spatial resources are also bounded. The importance of space and time as resources come from a powerful result in classical computer science: the discovery of the Turing machine~\footnote{In fact, the \textit{probabilistic} Turing machine is a more powerful version of the original Turing machine. The basic idea still remains the same, except allowing for probabilistic outcomes of a computation.} \cite{turing}. This is a universal computer that all other classical computers are believed to be reducible to and it utilizes only spatial and temporal resources. 

Unlike in classical computation where space and time resources are known to be necessary, it is yet unknown which are the truly indispensable resources (if any) that are only available to quantum computers and which are responsible for their computational advantage. Quantum mechanics introduces resources that are not available to classical computers, which include quantum entanglement, discord, quantum coherence and negativity of the Wigner function. In addition to these purely quantum mechanical resources, resources like energy and precision may also be involved (see Table~\ref{table} for references).  

The study of computational complexity is in classifying the difficulty of a computational problem in terms of the minimal cost in resources required for the computation. A \textit{complexity class} is a set of computational problems that are grouped together based on similar resource requirements. For example, suppose the size of a problem we want to solve can be encoded in $n$ bits (e.g. factoring the number $N$ where $n=\log_2 N$) and we concentrate on time resources. Algorithms in this case can fall into two classes: those needing time resources that increase polynomially with $n$ (polynomial time) or those that increase faster than any polynomial in $n$ (collectively called exponential time)~\footnote{This classification based on `easy versus hard' is inspired by two observations, both stemming from classical computer science. The first observation is that there are very few examples of algorithms for classical computers which are not obviously easy or hard, which makes this grouping a natural choice. The second observation comes from the most important hypothesis in classical computer science known as the `strong Church-Turing thesis' \cite{church}. This hypothesis claims that all classical computers can be simulated using a probabilistic Turing machine up to a polynomial scaling difference in the number of elementary operations used. This result strengthens the usefulness of any classification where models differing by a polynomial scaling in resources are considered equivalent.}. These are known respectively as easy/efficient and hard/inefficient algorithms. Two well-known complexity classes are P and NP~\footnote{Although there is no rigorous proof that P $\neq$ NP, there is a popular belief that this is true, from a recent poll taken of computer scientists \cite{PNP}.}.  Problems belonging to P (or `polynomial') are those problems which are both easy to solve and easy to check the solution once given. NP (`nondeterministic polynomial') problems are those problems that are hard to solve but easy to check the solution once given, like the travelling salesman problem \cite{pap}~\footnote{This is also known as an NP-complete problem \cite{garey}, meaning that if this problem is solved in some time $t$, then every other problem in NP can be solved in time polynomial in $t$.}. Factoring is strongly believed to be in NP and outside P, though a rigorous proof is still lacking. 

The introduction of quantum computers, which demand different resources, require the formation of new complexity classes. The class BQP (bounded error quantum polynomial time) denote the class of problems that are solvable by a quantum computer in polynomial time (allowing for error at most $1/3$). It is known that quantum computers are at least as powerful as classical computers and P $\subset$ BQP. The class BQP also contains some problems believed to be in NP, like factoring. However, BQP problems do not contain the most difficult problems in NP (called NP-complete). Important models of quantum computing are also given their own complexity classes. For example, the DQC1 complexity class refers to all problems that can be solved in polynomial time using the DQC1 model. Understanding how these complexity classes relate to each other will provide more information on what resources really lie behind the power of quantum computers. 

\begin{figure}[ht!] 
\centering
\includegraphics[scale=0.5]{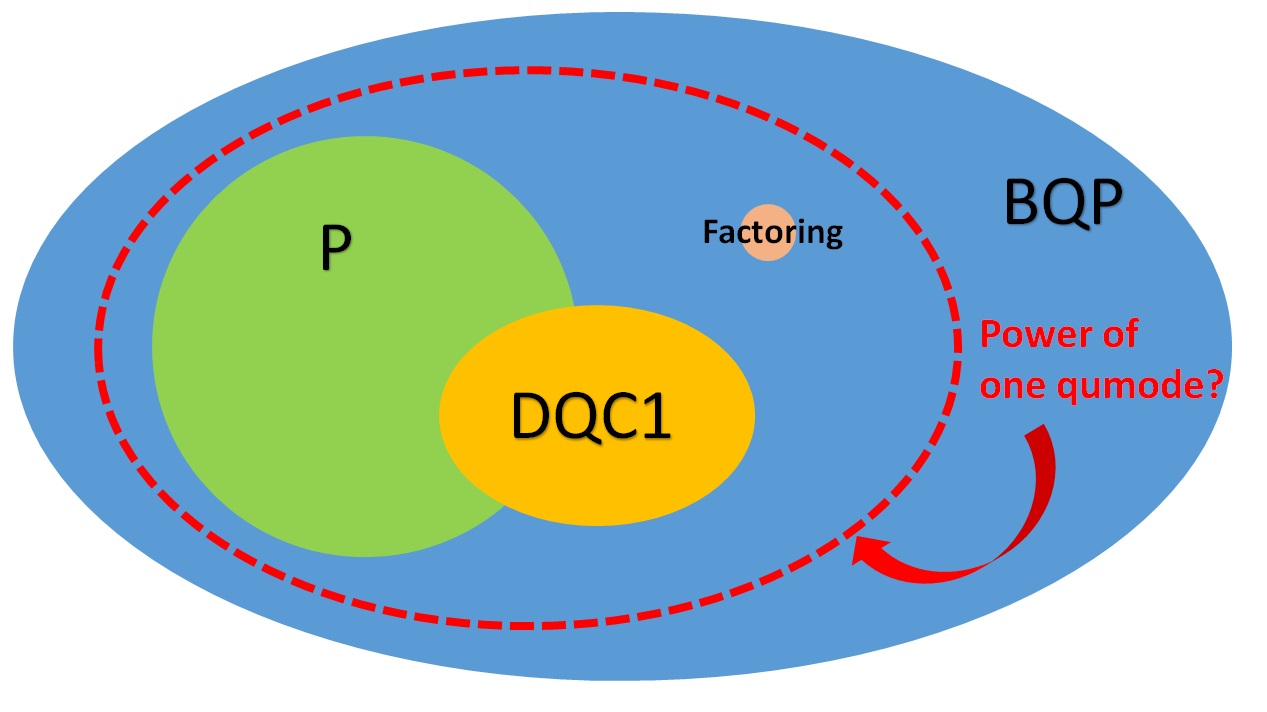}
\caption[\textit{Computational classes}.]{\label{figcomp}\textit{Computational classes.} This is a schematic diagram showing the relationship between computational complexity classes P, DQC1 and BQP as currently conjectured at this time. NP is not shown. Some problems believed to be in NP, like factoring, lie within BQP and believed to be outside P. NP-complete problems lie outside BQP. The `power of one qumode' model can solve both factoring and problems in DQC1 and may solve other problems in BQP. It is not yet known what are the most difficult problems this new model can solve. }
\end{figure}

However, much remains unknown. For example, there is not even a solid proof of whether DQC1 and BQP are actually equivalent classes (i.e. DQC1$=$BQP), even though DQC1 relies on apparently very few resources. Therefore, it is extremely important to find the boundary of DQC1 within BQP. One way to begin is to find the relationship between DQC1 and a particular BQP problem, like factoring. We introduce a new model, `power of one qumode', that contains both DQC1 and factoring. We show later that this model suggests factoring is more difficult than DQC1, which supports DQC1 $\neq$ BQP (see Fig.~\ref{figcomp}). We refer to the `power of one qumode' as both a model as well as denoting the class of problems that can be solved in polynomial time using this model. 
\subsection{Quantum algorithms}
Despite the great number of quantum algorithms now available~\footnote{See Quantum Algorithm Zoo for a comprehensive listing of and references to over 50 quantum algorithms to date: \textit{http://math.nist.gov/quantum/zoo/}.}, they are still based on a very small number of primitives that date back to the beginning of quantum computing. These algorithms fall into three basic classes: those based on the quantum Fourier transform, quantum search algorithms and quantum simulation. 

The quantum Fourier transform is used as an essential element in many of the most well-known algorithms, including Shor's factoring algorithm, the quantum phase estimation algorithm, Deutsch-Josza algorithm and the verification of certain quantum algorithms (like verifying Quantum-Merlin-Arthur or QMA). It is also a feature used in the DQC1 model and other non-universal computing models like boson sampling \cite{lund2014boson}. A common characteristic of algorithms using quantum Fourier transforms is the exponential speed-up over the best known classical algorithms it attains.

Quantum search algorithms, like Grover's algorithm, generally achieve only a quadratic speed-up compared to its known classical counterparts, but so far it enjoys algorithms of a greater range of practical applicability. Quantum simulations enjoy many practical applications and furthermore do not require a universal quantum computer. 

Our purpose is to compare DQC1 and factoring on an equal footing, both of which rely on the quantum Fourier transform. We focus on briefly describing these two algorithms only and refer the interested reader to a good recent overview of other algorithms in \cite{mon} and references therein. 

A quantum Fourier transform is an analogue of the classical discrete Fourier transform and is a particular linear transformation on quantum degrees of freedom. It is also a unitary transformation, which makes it possible for a quantum computer to implement. If we have $m$ qubits, then this has $2^m$ degrees of freedom which can be labelled by an integer $j$ in the range $0\leq j \leq 2^m-1$. If we encode these degrees of freedom into a state $\ket{j}$, then the quantum Fourier transform $\mathcal{F}$ takes $\ket{j}$ into a linear superposition 
\begin{align} \label{eq:fouriertransformdefinition}
\mathcal{F} \ket{j}=\frac{1}{2^m} \sum_{k=0}^{2^m-1} e^{2\pi i jk/2^m}\ket{k}.
\end{align}
The simplest quantum Fourier transform, acting on a single qubit ($m=1$), can be described by the Hadamard matrix $h$ where 
\begin{align}
h \equiv \frac{1}{\sqrt{2}}\begin{pmatrix}
1 & 1 \\
1 & -1 
\end{pmatrix}.
\end{align}
For example, if the qubit is encoded in photonic degrees of freedom, this transformation can be achieved by simply using a 50:50 beam-splitter. The quantum Fourier transform applied even on a single qubit is surprisingly powerful. In fact, it has been shown that every quantum transformation can be approximated by using Hadamard gates and purely classical gates (Toffoli gate) \cite{shi}. We use the Hadamard gate in DQC1. 

A more general quantum Fourier transform applied to more qubits is used in the quantum phase estimation algorithm, which efficiently finds the eigenvalue of a given unitary matrix. Most of the early quantum algorithms are based on this, including the factoring algorithm, which we briefly describe later. 
\subsubsection{Deterministic quantum computing with one quantum bit (DQC1)} 
DQC1 was a model introduced by Knill and Laflamme that confronted the belief that entanglement and purity
are the most important resources behind the computational advantage that quantum systems display. This model has only a single 
pure qubit (as the control qubit) and maximally mixed states (as the target state) as input and uses very little entanglement. Yet, despite missing these hallmarks of quantum systems, it still provides an exponential speed-up in certain computations over the best classical algorithms. 

\begin{figure}[ht!]
\centering
\includegraphics[scale=0.3]{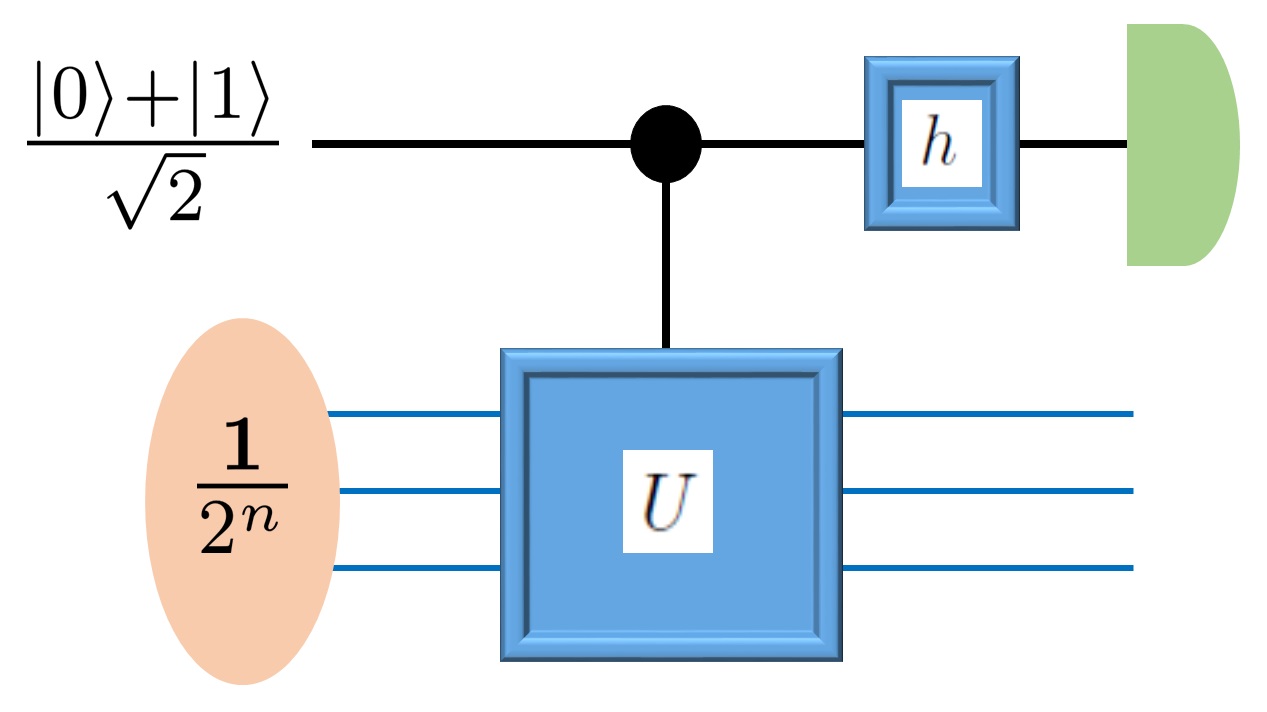}
\caption[\textit{DQC1 circuit}.]{\label{fig1}\textit{DQC1 circuit.} The control state is $\ket{+}=(\ket{0}+\ket{1})/\sqrt{2}$ and the target state is $n=\log_2N$ qubits in a maximally mixed state. Here $U$ encodes an $N \times N$ matrix and one can measure the final average spin of the control state to recover
the normalised trace of the matrix represented by $U$.}
\end{figure}

The most difficult problem DQC1 can solve, called DQC1-complete, is estimating the normalised trace of a unitary matrix $U$ \cite{shorjordan, shepherd}. This problem turns out to be important for a diverse set of applications, such as in quantum metrology \cite{hugoDQC1}, calculating the fidelity decay in quantum chaos \cite{fidelitydecay}, quadratically signed weight numerators \cite{knill2001} and estimating the Jones polynomial \cite{shorjordan}. In DQC1, the time required to find the normalised trace of $U$ is independent of the size of $U$. The DQC1 model can be represented by the circuit diagram in Fig.~\ref{fig1}. 
We begin with a single pure control qubit in the state $\ket{+}$ and a target state of $n=\log_2 N$ qubits in a fully-mixed state $\bf{1}$$/N$ where $\bf{1}$ is the identity matrix. The total initial state is $\rho_i=\ket{+}\bra{+} \otimes (\bf{1}$$/N)$. Here $\ket{+}=(\ket{0}+\ket{1})/\sqrt{2}$ where $\ket{0} \equiv (1 \, \, \, 0)^T$, $\ket{1} \equiv (0 \, \, \, 1)^T$ are the $+/-$ eigenstates of the Pauli matrix $\sigma_z$ respectively. Let the interaction between the control qubit and the target state be a control-unitary gate $\Gamma_U$, represented by
\begin{align}
\Gamma_U=\ket{0}\bra{0} \otimes \mathbf{1}+\ket{1}\bra{1} \otimes U \equiv \begin{pmatrix}
\bf{1} & 0 \\
0 & U \end{pmatrix},
\end{align}
where $U$ acts on the target qubits. The initial state can be represented in matrix form as
\begin{align}
\rho_i=\ket{+}\bra{+} \otimes \frac{\bf{1}}{N} \equiv \frac{1}{2N} \begin{pmatrix}
\bf{1} & \bf{1} \\
\bf{1} & \bf{1} \end{pmatrix}.
\end{align}
After the application of the control-unitary gate, the state becomes
\begin{align}
\rho' \equiv \Gamma_U\rho_i \Gamma_U^{\dagger}= \frac{1}{2N}\begin{pmatrix}
\bf{1} & U^{\dagger} \\
U & \bf{1} \end{pmatrix},
\end{align}
where we used the unitarity of $U$. A Hadamard gate is then applied to the control qubit. After measuring the control qubit in the computational basis,
final state of the control qubit is (after tracing out the target states)
\begin{align}
\rho_f &= \tr((h \otimes \mathbf{1})\rho'(h \otimes \mathbf{1})) \nonumber \\
           &=\frac{1}{2N} \begin{pmatrix}
\tr(\mathbf{1}+(U+U^{\dagger})/2) & \tr(U-U^{\dagger})/2 \\
\tr(-U+U^{\dagger})/2 & \tr(\mathbf{1}-(U+U^{\dagger}))/2 \end{pmatrix}\nonumber \\
           &=\frac{1}{2}\begin{pmatrix}
1+\text{Re}\left(\tr(U)\right)/N & i\text{Im}(\tr(U))/N \\
 -i\text{Im}(\tr(U))/N& 1-\text{Re}(\tr(U))/N \end{pmatrix},\nonumber \\         
\end{align}
where $\tr(\mathbf{1})=N$. If we measure the probability distribution of the measurement outcome of the control qubit in the computational basis (i.e. by measuring
with $\sigma_z$), we find that the  probability of getting the state $\ket{0}\bra{0}$ is $P(\ket{0}\bra{0})=(1+\text{Re}(\tr(U))/N)/2$ (reading off the top left corner of the above matrix).
The probability in finding $\rho_f$ in state $\ket{1}\bra{1}$ is $P(\ket{1}\bra{1})=(1-\text{Re}(\tr(U))/N)/2$. This means the expectation value of $\sigma_z$ is
\begin{align} \label{eq:sigmaz}
\avg{\sigma_z}=\tr(\rho_f \sigma_z)=\frac{\text{Re}(\tr(U))}{N}.
\end{align}
If we change our measurement basis to the eigenstates of $\sigma_y$, then we can measure $\avg{\sigma_y}=\tr(\rho_f \sigma_y)=-\text{Im}(\tr(U))/N$ to recover the imaginary part of $\tr(U)/N$ also. Thus we can recover all of $\tr(U)$ by measurements of $\sigma_x$ and $\sigma_y$. To estimate $\tr(U)/N$ to error $\delta$, that is, $\tr(U)/N \pm \delta$, we need to run the computation $T_{\text{DQC1}} \sim1/[\text{min}\{\text{Re}(\delta), \text{Im}(\delta)\}]^2$ times \cite{datta2005}. Since $\delta$ is independent of the size of $U$, this computation is efficient and DQC1 has an exponential advantage over the best-known classical algorithms \cite{animeshphd}. 

We also note that since $\tr(U)$ is really a sum of the eigenvalues of $U$, the basic structure of DQC1 may be expected to be helpful in devising a protocol that can find the individual eigenvalues of $U$~\footnote{This we later achieve using the `power of one qumode' model, which can be considered a CV analogue of DQC1.}. This is the quantum phase estimation problem to be described next.
\subsubsection{Quantum phase estimation}
A protocol that finds an eigenvalue to a precision of $1/2^n$ (the $n^{\text{th}}$ binary digit) to a high probability with number of required measurements or time $T_{\text{phase}} \sim \mathcal{O}(\text{poly}(n))$ is said to solve the phase estimation problem. It turns out, perhaps not too surprisingly, that the phase estimation protocol relies on similar basic elements as DQC1, namely the quantum Fourier transform, the control-unitary gate and an initial state constructed from a superposition of states $\ket{0}$ and $\ket{1}$. Suppose we are given a particular eigenvector $\ket{u}$ of $U$ where $U \ket{u}=\exp(2 \pi i \phi) \ket{u}$. Let $\phi$ be expanded exactly as $\phi=\phi_1/2+\phi_2/2^2+...\phi_m/2^m$ where $\phi_1, \phi_2,...,\phi_m=0,1$ are known as the binary digits of $\phi$. The phase estimation protocol, to find $\phi$, runs as follows (see Fig~\ref{fig2}):

\begin{figure}[ht!] 
\centering
\includegraphics[scale=0.5]{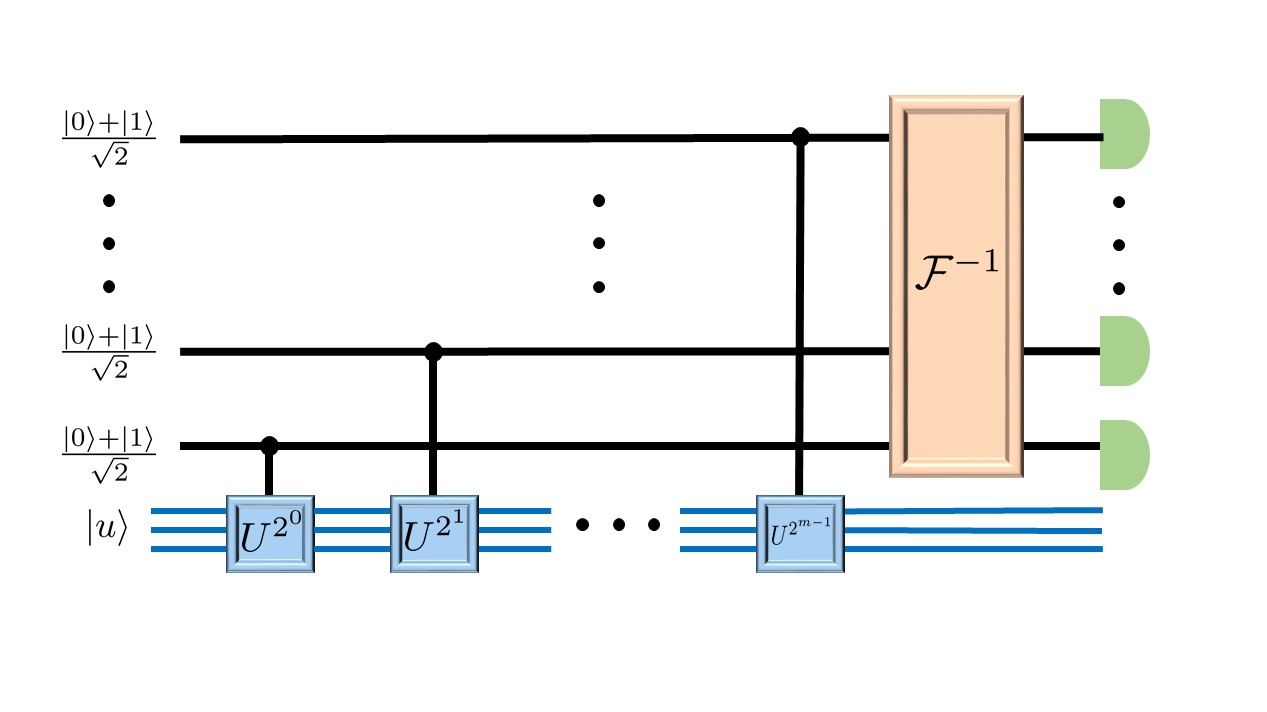}
\caption[\textit{Phase estimation protocol}.]{\label{fig2}\textit{Phase estimation protocol.} One inputs $m$ control pure qubits in state $(\ket{0}+\ket{1})/\sqrt{2}$ and a pure target register state $\ket{u}$ into the above quantum circuit. The circuit contains $m$ controlled unitary operations (of form $U^{2^j}$ for integer $0 \leq j \leq m-1$), an inverse quantum Fourier transform $\mathcal{F}^{-1}$ and final measurements of the control qubits in the computational basis. From these measurement results, it is possible to retrieve an eigenvalue of $U$ (with eigenstate $\ket{u}$ to precision $1/2^m$).}
\end{figure}
(i) Begin with $m$ control qubits each in a superposition $(\ket{0}+\ket{1})/\sqrt{2}$ and let $\ket{u}$ be the target state. Then the total initial state is $((\ket{0}+\ket{1})^{\otimes m}/2^{m/2}) \otimes \ket{u}$. \\

(ii) Apply control-unitary gate $\ket{0}\bra{0} \otimes \mathbf{1}+\ket{1}\bra{1} \otimes U^{2^{\mu-1}}$ to the $\mu^{\text{th}}$ qubit for every integer $\mu \in [1, m]$. This transforms the initial
state into $(\ket{0}+\ket{1})\otimes (\ket{0}+\exp(2\pi i (2 \phi))\ket{1}) \otimes...\otimes (\ket{0}+\exp(2\pi i (2^{m-1} \phi))\ket{1}) \otimes \ket{u}/2^{m/2}$. We can map the $m$ control qubits into an $2^m$ qudit. This state can be rewritten in the form
\begin{align} \label{eq:binarystate}
\frac{1}{2^{m/2}}\sum_{\mu=0}^{2^{m}-1} e^{2\pi i \mu \phi} \ket{\mu} \otimes \ket{u}.
\end{align}
We can re-express this state in a form using $\ket{u}$ and $m$ control qubits by expanding in binary digits (i.e. if $\mu=\mu 2^{m-1}+\mu_2 2^{m-2}+...+\mu_m 2^0$ where $\mu_1,...,\mu_m=0,1$, then $\ket{\mu}$ can be rewritten as the $m$ qubit state $\ket{\mu_1,\mu_2,...,\mu_m}$). \\

(iii) A vital next step is to take the inverse quantum Fourier transform of the control qubits in Eq.~\eqref{eq:binarystate} to recover $\phi$ (that gives us the eigenvalue of $U$). By inspection using Eq.~\eqref{eq:fouriertransformdefinition}, we arrive at
\begin{align}
\ket{2^m \phi}=\mathcal{F}^{-1}\frac{1}{2^{m/2}}\sum_{\mu=0}^{2^m-1} e^{2\pi i \mu \phi} \ket{\mu}.
\end{align}
Here $\ket{2^m \phi}$ can be rewritten (in terms of its binary representation) as $\ket{\phi_1 \phi_2...\phi_m}$. This means that, by measuring the output of the control qubits (to see if it is in state $\ket{0}$ or $\ket{1}$), we can directly read off the binary digits of $\phi$. \\

In this case the number of required measurements (or time) $T_{\text{phase}}$ to recover $\phi$ to accuracy $1/2^m$ is exactly $T_{\text{phase}}=m$. Since $m$ is the number of qubits needed to encode $\phi$ (in binary digits) and $T_{\text{phase}}$ is polynomial in $m$, this protocol is efficient in time. In the more general case where $\phi$ cannot be expanded exactly in $m$ binary digits as above, we can instead bound the probability of correctly finding $\phi$ to precision $1/2^m$ to be larger than some acceptable tolerance $1-\epsilon$. In this case, the number of required measurements is $T_{\text{phase}}=m+\log_2(2+1/(2\epsilon))$ \cite{cleve}, which is also polynomial in $m$ for given $\epsilon$. 

We can also make a rough comparison to how DQC1 can be used to find eigenvalues. We know with DQC1 it is possible to find $\tr(U)/N$ with DQC1 to some precision $\delta$ in time $T_{\text{DQC1}} \sim1/[\text{min}\{\text{Re}(\delta), \text{Im}(\delta)\}]^2$. Then let us consider a fully-degenerate matrix $U$ with eigenvalue $\exp(i\phi)$. Thus $\tr(U)/N=\exp(i\phi)$. To know $\phi$ to precision $1/2^m$ is equivalent to $\text{Re}(\delta), \text{Im}(\delta) \leq 1/2^m$. Thus DQC1 requires $T_{\text{DQC1}} \geq 4^m$, which is exponential in $m$. This is a first quantitative indication that DQC1 is less powerful than phase estimation. We explore the computational differences between DQC1 and phase estimation in more depth later in the chapter. 
\subsubsection{Factoring}
Factoring is the problem of reducing the integer $N$ to its prime factors. While it is easy to check the solution just by multiplying the given
factors, it requires an exponentially larger number of trials to find unknown factors as $N$ becomes larger. It turns out that the hard part of
factoring can be reduced to the phase estimation problem, based on two main results. 

The first result is the reduction of factoring to another classical hard problem called order-finding. This reduction says that for some number $x$ in range 
$1<x < N$, at least one of $\gcd(x\pm 1, N)$~\footnote{$\gcd(x,y)$ is the greatest common denominator of $x$ and $y$. Given $x$ and $y$, this can be found using 
the Euclidean algorithm. This is a classically easy problem that can be delegated to a classical computer.} is guaranteed to be a non-trivial factor of $N$, \textit{provided}
$x$ satisfies $x^2=1 \mod N$ and $x \neq \pm 1 \mod N$~\footnote{This is guaranteed to be true expect when $N$ is even or $N=a^b$ where $a \geq 1$, $b \geq 2$ are integers.}. Thus, it is sufficient if we find just one such $x$. Suppose we rewrite $x=q^l$ where $1<q<N-1$ and $l$ is another integer. Then $x$ is found if the smallest integer $r \leq N$ obeying condition $q^r \equiv 1 \mod N$ is an even number. $r$ is called the order of $q$ and solving for this $r$ is the order-finding problem~\footnote{Once $r$ is found, it's an easy problem to check if it is an even number. In fact, it turns out for large $N$, almost all $r$ found will be even \cite{nandc}.}. 

The second result is solving the order-finding problem using phase estimation. We begin by encoding 
$1 \mod N$ into a quantum state $\ket{1 \mod N}$, then applying an operation $V$ where
$V \ket{1 \mod N}=\ket{q \mod N}$. The repeated application of $V$ generates a cyclic sequence $\ket{1 \mod N} \rightarrow \ket{q \mod N} \rightarrow \ket{q^2 \mod N} \rightarrow ... \rightarrow \ket{q^r \mod N}=\ket{1 \mod N}$ whose number of independent elements is $r$. Then it can be shown that one set of eigenstates $\ket{v_m}$ of $V$ can be formed by a linear superposition of the states in this sequence. The eigenstates can be written 
$\ket{v_m}=(1/\sqrt{r}) \sum^{r-1}_{k=0} \exp(-(2 \pi i m k/r)) \ket{q^k \mod N}$ where $m$ is an integer in the range $0 \leq m \leq r-1$ and $q$ is coprime to $N$. Thus $V \ket{v_m}=\exp(2 \pi i m/r) \ket{v_m}$, where $\exp(2 \pi i m/r)$ are the eigenvalues of $V$, which can be solved efficiently using the phase estimation protocol we just described. However, the protocol must be modified a little, since it requires the use of the eigenvector in the target register that already assumes knowledge of $r$. Instead we can use state $\ket{1 \mod N}=(1/\sqrt{r})\sum_{m=0}^{r-1}\ket{v_m}$ in the target register, which requires no prior knowledge of $r$. Since $r$ is an integer, we want sufficient precision to find the fraction $m/r$ \textit{exactly}. This is possible if the precision of phase estimation is on the order $1/N^2$ and $m/r$ is retrieved by using a classically easy algorithm, called the continued fractions algorithm \cite{nandc}. From $m/r$, $r$ can be precisely found if $m$ and $r$ are coprime, which can be shown to occur with probability $\mathcal{O}(\ln(\ln(N)))$ \cite{nandc}. 

For our `power of one qumode' model that we later introduce, instead of using the pure state $\ket{1}$ in the target register, we use a fully-mixed target state $\bf{1}$$/N$. This is based on a similar trick as using $\ket{1}$ but now the mixed state is a classical sum of the eigenstates instead of a quantum superposition of the eigenstates of $V$. We see this in more detail when we show a factoring algorithm using the power of one qumode model. 
\subsection{Continuous variable states}
\label{sec:CVstates}
The most familiar state in quantum information processing is the qubit, which we know is formed from the basis states $\{\ket{Q}\}$ where $Q$ can take only two values. In the computational basis, typically $Q=0,1$. In general, $Q$ can take on more than two values. If it takes on $D$ values, the state formed from $\{ \ket{Q}\}$ is known as a qudit of dimension $D$. 

$Q$ can even be a continuous variable. In this case, the state formed using the basis set $\{\ket{Q}\}$ is known as a continuous variable or CV state, where the basis states obey the orthogonality conditions $\braket{Q}{Q'}=\delta(Q-Q')$ for $Q,Q' \in \Re$. In the context of quantum computation, CV states can be used in three ways: as a way of encoding qudits in qubit quantum computation, in fully-CV computation~\footnote{These are quantum versions of classical analog computing, which includes the slide rule as well as the Antikythera mechanism, dating from as early as 205 BC, used for predicting eclipses and planetary motion \cite{freeth2006decoding}.} or in discrete variable-CV hybrid computing. CV states in quantum computation enjoy many advantages, including fast methods of state characterisation using homodyne/heterodyne detection, easier generation and control of entangled states and the simple implementation of quantum Fourier transforms, which lie at the heart of many quantum algorithms. 

The observables of CV states also form a continuous spectrum. Two examples are the position and momentum observables.  They can be found using the position $\hat{x}$ and momentum $\hat{p}$ operators, which obey the commutation relation $[\hat{x},\hat{p}]=i$. The eigenstates of the position observable $\{\ket{x}\}$ and the momentum operators $\{\ket{p}\}$ each form a basis set for a CV state and they satisfy $\braket{p}{x}=(1/\sqrt{2\pi})\exp(-ixp)$. 

We can now define the quantum Fourier transform $\mathcal{F}$ on CV states 
\begin{align}
\mathcal{F}\ket{x}&=\frac{1}{\sqrt{2\pi}}\int^{\infty}_{-\infty} dz e^{ixz}\ket{x}=\ket{\zeta},
\end{align}
where $\ket{\zeta}$ is the momentum eigenstate with value $x$. Thus the quantum Fourier transform changes a position eigenstate into a momentum eigenstate. 

A CV state can also be expanded in terms of the number or Fock basis $\{\ket{n}\}$ where $n \geq 0$ is an integer. They are eigenstates of the number operator $\hat{n}=a^{\dagger}a$, where $a$ and $a^{\dagger}$ are respectively the annihilation and creation operators. These operators can also be written in terms of the position and momentum operators as $a=(\hat{x}+i\hat{p})/\sqrt{2}$ and $a^{\dagger}=(\hat{x}-i\hat{p})/\sqrt{2}$. 

An important class of CV states are known as Gaussian states, which are not only experimentally accessible but also have simple mathematical properties that enable easy analytics~\footnote{A very elegant formalism dedicated to Gaussian states is the covariance matrix formalism. We refer the interested reader to an excellent introduction to key mathematical methods for Gaussian states \cite{gerardoreview}.}. These are defined as states saturating the Heisenberg uncertainty relation $\Delta x \Delta p=1/2$, where $\Delta x \equiv \sqrt{\avg{\hat{x}^2}-(\avg{\hat{x}})^2}$, $\Delta p \equiv \sqrt{\avg{\hat{p}^2}-(\avg{\hat{p}})^2}$ where $\avg{.}$ denotes the expectation value. Thus, they are often viewed as `almost classical' states. They are termed `Gaussian' states because their probability profiles are Gaussian when they are measured in position or momentum. This also means that Gaussian states can be characterised using only two numbers: the expectation value of position/momentum and the standard deviation of position/momentum. Below we list some brief facts about some Gaussian states we will encounter in this chapter: vacuum state, coherent state, single-mode squeezed state. Another important Gaussian state is the two-mode squeezed state, which we will encounter in the last chapter. 

\textit{Vacuum state $\ket{0}$.} The vacuum state is defined in terms of position eigenstates as $\ket{0}=(1/\pi^{1/4})\int \exp(-x^2/2)\ket{x}$, where its position and momentum expectation values are both zero. It is also the state that is annihilated by the annhilation operator, i.e. $a\ket{0}=0$. 

\textit{Coherent state $\ket{\alpha}$.} A coherent state is defined as an eigenstate of the annihilation operator $a\ket{\alpha}=\alpha \ket{\alpha}$ where $\alpha$ is in general a complex number. The special case of $\alpha=0$ is the vacuum state. It can be expanded in terms of the position eigenstates as \sloppy $\ket{\alpha}=\int dx \ket{x}(1/(\pi x_0^2))^{1/4} \exp(-(1/(2x_0^2))(x-\text{Re}(\alpha))^2)\exp(i\text{Im}(\alpha)x/x_0)\exp(-i\text{Re}(\alpha)\text{Im}(\alpha)/2)$. Here $x_0\equiv 1/\sqrt{m \omega}$ and $m, \omega$ are the mass and frequency scales of the corresponding quantum harmonic oscillator. The real and imaginary parts of $\alpha$ are $\text{Re}(\alpha)=\sqrt{m \omega/2} \langle x\rangle$ and $\text{Im}(\alpha)=(1/\sqrt{2m\omega})\langle p \rangle$. Coherent states are also characterised by the equality $\Delta x=\Delta p$. We also note that $\ket{\alpha}=D(\alpha)\ket{0}$ where $D(\alpha)$ is the displacement operator acting on the vacuum state with $D(\alpha)=\exp(\alpha a^{\dagger}-\alpha^*a)$. This displacement operation can be considered as inducing a `translation' in the annihilation and creation operators, i.e. $D(\alpha)^{\dagger} aD(\alpha)=a+\alpha$ and $D(\alpha)^{\dagger} a^{\dagger}D(\alpha)=a+\alpha^*$. Using this we can derive the expected particle number $\avg{a^{\dagger}a}=\langle \hat{n} \rangle=|\alpha|^2$.

\textit{Single-mode squeezed state $\ket{s}$.} A single-mode squeezed state is a single-mode Gaussian state where $\Delta x \neq \Delta p$. It results from applying the squeezing operator~\footnote{Here we consider the simplest case where $s$ is a real number. This will not affect our general conclusions in the chapter.} $S(s)=\exp(i \ln(s)(a^2-a^{\dagger 2}))$ to the  vacuum state, where $S(s)$ rescales the position and momentum operators like $S^{\dagger}(s) \hat{x} S(s)=s \hat{x}$ and $S^{\dagger}(s) \hat{p} S(s)=\hat{p}/s$. This means that with higher $s$, the momentum becomes more sharply defined with smaller $\Delta p$ than the vacuum. To preserve the volume element in $x$ and $p$ from $\Delta x \Delta p=1/2$, the position becomes less sharply defined with $\Delta x$ larger than the vacuum. This is called squeezing in the momentum quadrature~\footnote{Squeezing along any direction in the position-momentum phase space can be experimentally accomplished in quantum optics \cite{knightgerry}.}. Here $s=s_0 x_0$ where we call $s_0$ the squeezing factor. Another way to parameterise the amount of squeezing is the squeezing parameter $r=2 \ln(s)$. The squeezed state can be expanded in terms of the position eigenstates as $\ket{s}=\int dx \ket{x} (1 / (\sqrt{s}\pi^{1/4})) \text{exp}(-x^2/(2s^2))$. The squeezing operator acts on annihilation and creation operators like $S^{\dagger}(s)aS(s)=a\cosh(2 \ln(s))-a^{\dagger}\sinh(2 \ln(s))$ and $S^{\dagger}(s)a^{\dagger}S(s)=a^{\dagger}\cosh(2 \ln(s))-a\sinh(2 \ln(s))$. From this one can derive the expected particle excitation as $\avg{\hat{n}}=\sinh^2(2 \ln(s))$. 
\section{Power of one qumode}
\begin{figure}[ht!]
\centering
\includegraphics[scale=0.3]{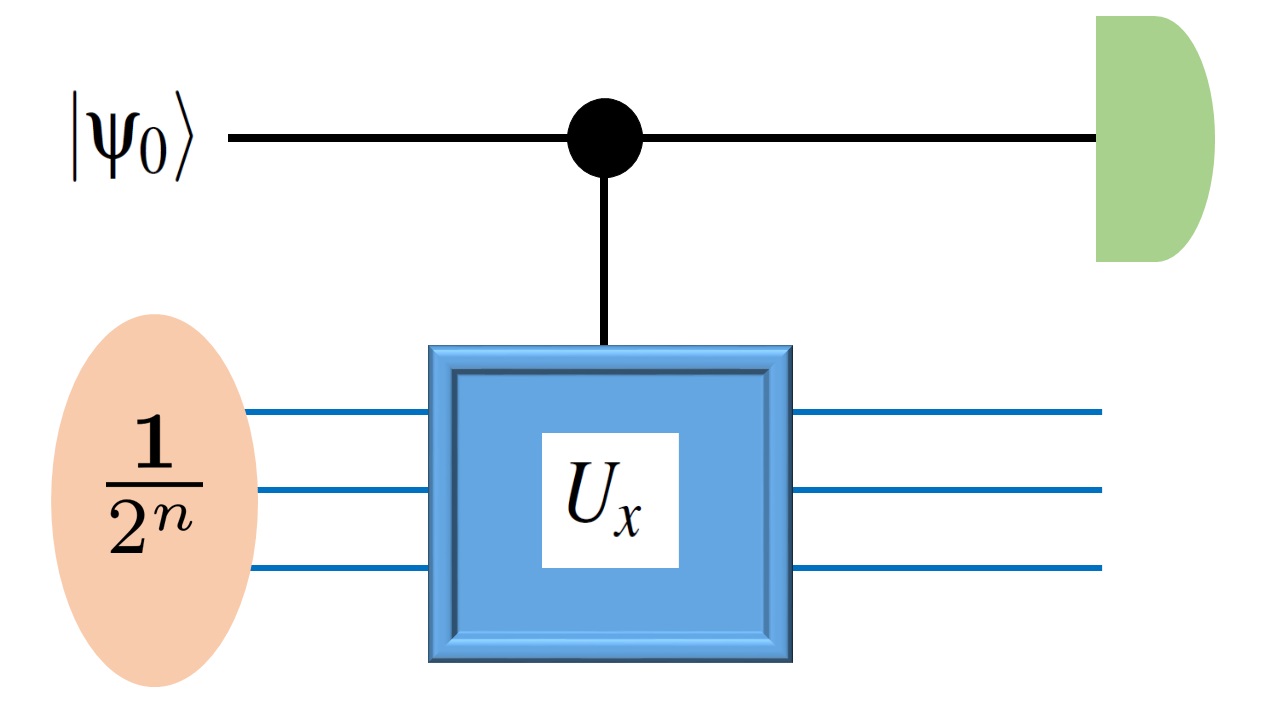}
\caption[\textit{Power of one qumode model}.]{\label{fig3}\textit{Power of one qumode model.} We can use a squeezed state $\ket{\psi_0}$ as the control state. The target state consists of $n=\log_2 N$ qubits in a maximally mixed state, like in DQC1. Here $U_x \equiv \exp(i x H \tau/x_0 )$ where $x_0$ is a constant and $\tau$ is the gate running time. Its relationship to the unitary $U$ in DQC1 is $U_x=U^{x \tau/x_0}$. We make final measurements of the control state in the momentum basis. The momentum measurements in this model can be used to recover the normalised trace of an $N \times N$ matrix $U$ and also to factor the integer $N$.}
\end{figure}

In this chapter we generalise DQC1 by replacing the pure control qubit with a pure CV state (qumode), while keeping the target register the same. The total input state in our model is thus a hybrid state of discrete-variable states and a CV state. See Fig.~\ref{fig3} for the circuit diagram of our model. We first show how our model can perform the quantum phase estimation algorithm \cite{cleve}. We use this to efficiently compute a DQC1-complete problem, thus showing that this model contains DQC1. Next, we show that our model can perform Shor's factoring algorithm, which is based on the phase estimation algorithm. 

The aim in the phase estimation problem is to find the eigenvalues of a Hamiltonian, $H \ket{u_j}=\phi_j \ket{u_j}$. The complete set of eigenvalues of $H$ is given by $\{ \phi_j \}$, thus $H=\sum_j \phi_j \ket{u_j}\bra{u_j}$. We encode the Hamiltonian $H$ into a unitary transformation, $C_U$, that acts on the hybrid input state. We call $C_U$ the hybrid control gate and is defined as $C_U=\exp(i \, \hat{x} \otimes H \tau/x_0)$, where the position operator $\hat{x}$ acts on the qumode~\footnote{It is also possible to define a control gate controlled on the particle number operator instead of $\hat{x}$. However, analytical solutions in this case are not straightforward and for our purposes it suffices to look at our current hybrid control gate.} and $\tau$ is the running time of the hybrid gate. Here $x_0 \equiv 1/\sqrt{m \omega}$, where $m, \omega$ are the mass and frequencies of the harmonic oscillator corresponding to the qumode. Like the control gate $\Gamma_U$ in DQC1, the hybrid control gate can also be reduced into elementary operations (see Appendix~\ref{sec:elementarygates}). If the qumode is in a position eigenstate $\ket{x}$ and $\ket{u_j}$ is a state of target register qubits, the action of the hybrid control gate is
\begin{gather}
C_U \ket{x}\otimes \ket{u_j}=\ket{x} \otimes U_x \ket{u_j}
= \ket{x} \otimes e^{i \phi_j x \tau/x_0} \ket{u_j},
\end{gather}
where $x$ is the eigenvalue of $\hat{x}$ and $U_x \equiv \exp(i x H \tau/x_0 )$. In our model, we apply $C_U$ to a maximally mixed state of $n$ qubits and a qumode state $\ket{\psi_0} = \int G(x) \ket{x} dx$. $G(x)$ is the wave-function of the initial qumode in the position basis. After implementing this gate, the target register is discarded, and the qumode is in the state
\begin{gather}
\rho_f=\frac{1}{2^n} \iint G(x) G^*(x') \tr[e^{i(x-x')H \tau/x_0}] \ket{x}\!\bra{x'} \text{d}x \, \text{d}x'.
\end{gather}
Next, we measure this state in the basis of the momentum operator $\hat{p}$, i.e. $\braket{p} {\rho_f \vert p}$. This measurement yields the momentum probability distribution 
\begin{align} \label{eq:momentum1}
\mathcal{P}(p) &=\frac{1}{2^n} \sum_m \iint G(x)G^*(x') e^{i(x-x')\phi_m \tau/x_0} \braket{p}{x}\! \braket{x'}{p} \text{d}x \, \text{d}x' \nonumber \\
        &=\frac{1}{2^n} \sum_m \mathcal{G}(\phi_m \tau/x_0-p) \mathcal{G}^*(p-\phi_m \tau/x_0),
\end{align}
where we used $\braket{p}{x}=(1/\sqrt{2\pi})\exp(-ixp)$ and the Fourier transform of $G(x)$ is denoted by $\mathcal{G}(p)=(1/\sqrt{2\pi})\int^{\infty}_{-\infty} \exp(ixp)G(x)\text{d}x$. 
If we choose our wavefunction $G(x)$ carefully, we can employ our model to recover the eigenvalues of $H$. Suppose we initialized the control mode in a coherent state $\ket{\alpha}$, chosen for its experimental accessibility \cite{knightgerry}. If we measure the probability distribution of $p_{\text{E}} \equiv px_0/\tau$ where $x_0$ and $\tau$ are known inputs and $p_{\text{E}}$ has dimensions of energy, we find (see Appendix~\ref{sec:coherent} for a derivation)
\begin{gather}\label{eq:coherentp}
\mathcal{P}(p_{\text{E}})=\frac{\tau}{\sqrt{\pi} 2^n} \sum_{m=1}^{2^n} e^{-\tau^2 \left[p_{\text{E}}-\left(\phi_m+\frac{\text{Im}(\alpha)}{\tau}\right)\right]^2},
\end{gather}
where $\text{Im}(\alpha)$ is the imaginary component of $\alpha$~\footnote{This is equivalent to the initial expectation value of momentum of the coherent state. Please see section ~\ref{sec:CVstates}.}. We can see that the probability distribution is a sum of Gaussian distributions. It has individual peaks centred at each shifted eigenvalue $\phi_j+\text{Im}(\alpha)$ with an individual spread given by the inverse of $\tau$. By sampling this probability distribution we can infer the position of the peaks to any finite precision. Thus it is possible to perform phase estimation to arbitrary accuracy just by increasing $\tau$ alone. However, to estimate eigenvalues to a precision better than a polynomial in $n=\log_2 N$, we require $\tau$ greater than polynomial in $n=\log_2 N$. Thus the coherent state no longer suffices for Shor's factoring algorithm, which requires high precision phase estimation. In such cases, we require a further resource that we identify to be the squeezing factor.

A finite squeezed state is defined by $G(x) = (1 / (\sqrt{s}\pi^{\frac{1}{4}})) \text{exp}(-x^2/(2s^2))$ where $s\equiv s_0 x_0$ and $s_0$ parameterises the amount of squeezing in the momentum direction~\footnote{Here $s_0$ is a real number in the range $s_0 \in [1, \infty)$.}. We call $s_0$ the squeezing factor. Its wavefunction in $x$ has a Gaussian profile with standard deviation $1/s_0$. By inputting a squeezed state into our model, the probability distribution in $p_{\text{E}}$ becomes
\begin{gather} \label{eq:pp}
\mathcal{P}(p_{\text{E}})=\frac{s_0 \tau}{2^n \sqrt{\pi}} \sum_{m=1}^{2^n} e^{-(s_0 \tau)^2(p_{\text{E}}-\phi_m)^2}.
\end{gather}
Comparing this to Eq.~\eqref{eq:coherentp} we see the coherent state plays the same role as an unsqueezed state (i.e. $s_0=1$). The method for retrieving the eigenvalues is now identical to that of the coherent state, except now we can take advantage of a large squeezing factor instead of non-polynomial gate running time. 

We can see the relationship between the squeezing factor and gate running time more explicitly. Let $T_{\text{bound}}$ be the upper bound on the total number of momentum measurements we are willing to make for phase estimation. If we need to recover any eigenvalue of the Hamiltonian to accuracy $\Delta_{\text{E}}$, the following `time-energy' condition is satisfied (see Appendix~\ref{sec:phaseestimate} for a derivation)
\begin{gather}\label{eq:tbound}
T_{\text{bound}}\tau s_0 \Delta_{\text{E}} \gtrsim 1,
\end{gather}
where $\Delta_{\text{E}}$ can be a function of the size of the Hamiltonian. In an efficient protocol the maximum total gate running time $T_{\text{bound}} \tau$ is bounded by a polynomial in $n$. When the inverse of $\Delta_{\text{E}}$ is also a polynomial in $n$, efficient phase estimation is still possible for a squeezing factor polynomial in $n$. For example, this is useful for the verification of problems in the QMA complexity class, which includes the local Hamiltonian problem \cite{localhamiltonian}. For an exponentially greater precision in phase estimation, however, an exponentially higher squeezing factor is needed. We see from Eq.~\eqref{eq:tbound} that the squeezing factor serves as a rescaling of the energy `uncertainty' $\Delta_{\text{E}}$. Similarly to phase estimation, increased squeezing can also retrieve the corresponding eigenvectors to greater precision~\footnote{See Appendix~\ref{sec:eigenvectors} for our algorithm on retrieving eigenvectors. Also, see \cite{abramslloyd} for another algorithm on eigenvector retrieval.}. 

We can see the relationship between the squeezing factor and inverse precision from Eq.~\eqref{eq:tbound} by considering when the maximum total gate running time resource is constrained. When the time resource is constant, the minimum squeezing factor required for efficient phase estimation is the inverse precision, i.e. $s_0 \sim 1/\Delta_E$.

This relationship can be seen more intuitively by considering a problem whose solution is given by the central position $x_0$ of a squeezed state with squeezing factor $s_0$. From the central limit theorem, it requires $t \sim 1/(s_0^2 \eta^2)$ measurements of the position $x$ to get within precision $\eta=|x-x_0|$ of the centre. Thus for a fixed number of measurements (or time), the squeezing factor scales as the inverse of precision $s_0 \sim 1/\eta$.

Another way we can see $s_0$ as the inverse precision is to consider when we are trying to resolve the distance between two adjacent Gaussian peaks $\Delta \phi$. We see later that factoring in our model is essentially this problem with $\Delta \phi \sim 1/N=1/2^n$, where $N$ is the number to be factored. Each Gaussian has standard deviation $1/s_0$. If the distance between the centres of these peaks is closer than this length scale, it becomes difficult to resolve the two peaks. Thus, $1/s_0$ is the maximum resolution for $\Delta \phi$, which is another precision scale. This fact is used when we later examine both the qubit and qudit encodings in our model.
\section{Recovering DQC1}
We begin with an observation that the average of $\exp(ip_{\text{E}})$ can reproduce the normalised trace of $U \equiv \exp(iH)$ in the following way
\begin{gather}
\int e^{ip_{\text{E}}} \mathcal{P}(p_{\text{E}}) \text{d}p_{\text{E}}=e^{-\frac{1}{4 s_0^2}} \frac{\tr(U_{\tau})}{2^n},
\end{gather}
where $\mathcal{P}(p_{\text{E}})$ is given by Eq.~\eqref{eq:pp} and $U_{\tau}\equiv \exp(iH \tau)$. For an $N \times N$ matrix $U_{\tau}$, we use $n=\log_2 N$. If we wish to recover the normalised trace of $U$ to within an error $\delta$ (i.e. $\tr(U)/2^n \pm \delta$), we require $\tau=1$ and $T_{\text{DQC1}}$ measurements of momentum~\footnote{Note that the number of momentum measurements and $p_{\text{E}}$ measurements needed are equivalent.} in our model. This is equivalent to running our hybrid gate once per momentum measurement and then averaging the corresponding values $\{ \exp(ip_{\text{E}}) \}$. 

This computation of the normalised trace is as efficient as DQC1 if $T_{\text{DQC1}}$ is independent of $N=2^n$. By employing the central limit theorem we find (see Appendix~\ref{sec:DQC1measure} for a derivation) 
\begin{gather}\label{eq:tF}
T_{\text{DQC1}} \lesssim \frac{F(s_0)}{[\text{min}\{\text{Re}(\delta), \text{Im}(\delta)\}]^2},
\end{gather}
where $F(s_0)=\text{sinh}(1/(2s_0^2))+\exp(-1/(2 s_0^2))$ and $F(s_0) \rightarrow 1$ very quickly with increasing $s_0$. The $F(s_0)$ overhead is analogous to the case in DQC1 when using a slightly mixed state probe state instead of the pure state {$\ket{+}\!\bra{+}$} \cite{animeshphd}. In that scenario, the degree of mixedness does not affect the result that the computation is efficient. The amount of squeezing in our model thus corresponds to the degree of mixedness in the input state of DQC1. Higher squeezing in our model corresponds to greater purity in the control state of the DQC1 model.

We observe from Eq.~\eqref{eq:tF} that $T_{\text{DQC1}}$ is upper-bounded by a quantity dependent only on the squeezing and not on the size of the matrix. In fact, even when $s_0=1$ (equivalent to a coherent state input) our qumode model is sufficient to efficiently compute the normalised trace of $U$, thus reproducing DQC1. This can also be viewed as a consequence of $\Delta_{\text{E}}$ being independent of $N=2^n$ in Eq.~\eqref{eq:tbound}. 
\section{Factoring using power of one qumode}
Factoring is the problem of finding a non-trivial multiplicative factor of an integer $N$. We saw earlier that the classically hard part of factoring can be reduced to a phase estimation problem, where the quantum advantage in phase estimation can be exploited. We show how the corresponding phase estimation problem can be solved in our model and how much squeezing resource is required. 

We recall that factoring can be reduced to finding the order $r$ of a random integer $q$ coprime to $N$ in the range $1 < q < N$. The order $r$ is an integer $r \leq N$ satisfying $q^r \equiv 1 \mod N$. This integer $r$ can be found from the eigenvalues of a suitably chosen Hamiltonian $H_q$, which is a phase estimation problem.

Here we begin with a squeezed control state and a target state of $n=\log_2 N$ qubits in a maximally mixed state. Let our hybrid control gate be $C_{U_q}=\exp(i \hat{x} \otimes H_q \tau/x_0)$. Next we choose a suitable Hamiltonian $H_q$ whose eigenvalues contain the order $r$. By suitable, we mean choosing a Hamiltonian such that the unitary $U_q=\exp(iH_q)$ acts on a qubit state $\ket{l \text{mod} N}$ like $U_q\ket{l \text{mod} N}=\ket{lq \text{mod} N}$, where $l$ is an integer $1 \leq l \leq N-1$. In our introduction to factoring, we only considered when $l=q^k$ for an integer $k \leq r$. In this case, $\exp(iH_q r)\ket{l \mod N}=\ket{q^k q^r \mod N}=\ket{q^k \mod N}=\ket{l \mod N}$. Therefore $\ket{q^k \mod N}$ is an eigenstate of $U_q$ with eigenvalue $1$. This means the eigenvalues of $H_q$ are $2 \pi m/r$ where $m$ is an integer $1 \leq m<r$. 

However, when we begin with $n$ qubits in a fully-mixed state $\mathbf{1}/2^n=\mathbf{1}/N$, we also require cases where $l \neq q^k$. This is because $\mathbf{1}/N$ is a classical mixture using \textit{all} the eigenvectors of $U_q$, which can comprise of $\{\ket{l \mod N}\}$ for all integers $1 \leq l \leq N-1$. As an example, we look at the simple case of $N=15$ and $q=7$. Applying $U_q$ to each $\ket{l \mod N}$ we arrive at the following set of series~\footnote{This example is similar to that presented in Parker and Plenio \cite{parkerplenio}.}
\begin{align}
d=1: \,\,\, & \ket{1 \mod 15} \rightarrow \ket{7 \mod 15} \rightarrow \ket{4 \mod 15} \rightarrow \ket{13 \mod 15} \rightarrow \ket{1 \mod 15} \nonumber \\
d=2: \,\,\, & \ket{2 \mod 15} \rightarrow \ket{14 \mod 15} \rightarrow \ket{8 \mod 15} \rightarrow \ket{11 \mod 15} \rightarrow \ket{2 \mod 15} \nonumber \\
d=3: \,\,\, & \ket{3 \mod 15} \rightarrow \ket{6 \mod 15} \rightarrow \ket{12 \mod 15} \rightarrow \ket{9 \mod 15} \rightarrow \ket{3 \mod 15} \nonumber \\
d=4: \,\,\, & \ket{5 \mod 15} \rightarrow \ket{5 \mod 15} \nonumber \\
d=5: \,\,\, & \ket{10 \mod 15} \rightarrow \ket{10 \mod 15}. 
\end{align}
In this case, we have five series instead of just one when $l=q^k$. For each of these series, there is a different `order' (i.e. size of each cyclic series). For example, in the first three series, the size is $4$, whereas the size in the last two series is $1$. Thus we can define a generalised `order' $r_d$ corresponding to $l=l_d$ and $q$, where $d$ labels the series. In the above example, we can use $d=1,2,...5$ to label the five series, thus $l_1=1$ and $l_4=5$.  The ``order"  $r_d$ can be formally defined as an integer $r_d \leq r$ satisfying $l_d q^{r_d} \mod N=l_d \mod N$. This means $\{\ket{l_d \mod N}\}$ are all eigenstates of $U_q$ with eigenvalue $1$. Thus for general $l_d$, the eigenvalues of $H_q$ can be written as $2 \pi m_d/r_d$ where $m_d$ is an integer $1 \leq m_d <r_d$. 

These eigenvalues can provide $m_d/r_d$ but do not give $r$ directly. However, we can always rewrite $m_d/r_d$ in the form $m/r$ since $r_d$ is a factor of $r$. In general, there will be a single fraction $m/r$ corresponding to many possible $m_d$ and $r_d$. If we call this multiplicity $c_m$ for a given $m/r$, then following Eq.~\eqref{eq:pp} we can write the $p_{\text{E}}$ probability distribution as measured by the final control state as
\begin{align}\label{eq:PDFfactor}
\mathcal{P}(p_{\text{E}}) &=\frac{s_0 \tau}{\sqrt{\pi} N} \sum_{d} \sum_{m_d=0}^{r_d-1} e^{-(2 \pi s_0 \tau)^2 \left(\frac{p_{\text{E}}}{2 \pi}-\frac{m_d}{r_d}\right)^2} \nonumber \\
       &=\frac{s_0 \tau}{\sqrt{\pi} N} \sum_{m=0}^{r-1} c_m e^{-(2 \pi s_0 \tau)^2 \left(\frac{p_{\text{E}}}{2 \pi}-\frac{m}{r}\right)^2}.
\end{align}
This probability distribution is a sum of Gaussian functions with amplitudes $c_m$ and centered on $m/r$. To recover the order $r$ from the above probability distribution, it is sufficient to satisfy two conditions. The first condition is to be able to recover the fractions $m/r$ to within the interval $[m/r-1/(2N^2), m/r+1/(2N^2)]$. This ensures that $m/r$ is recovered exactly by using the continued fractions algorithm~\footnote{See \cite{nandc} for an explicit demonstration.}. Thus the larger the number we wish to factor, the more squeezing we need to improve the precision of the phase estimation. The second requirement is for $m$ and $r$ to be coprime, which enables us to find $r$. This requirement is satisfied with probability less than $\mathcal{O}(\ln[\ln(N)])$. 

Subject to the above two conditions, we can compute the probability that a correct $r$ is found using the momentum probability distribution in our model. We derive in Appendix~\ref{sec:factoringderive} the number of runs $T_{\text{factor}}<\mathcal{O}(\ln[\ln(N)])/\text{erf}(\pi s_0 \tau/N^2)$ needed to factor $N$, which is inversely related to the probability of finding a correct $r$. In the large $N$ limit, to achieve the same efficiency as Shor's algorithm using qubits, which is $T_{\text{factor}} \sim \mathcal{O}(\ln[\ln(N)])=\mathcal{O}(\ln[\ln(N)]) T_{\text{bound}}$~\footnote{Note that $T_{\text{bound}}$ in this case corresponds to the number of momentum measurements needed to find the correct eigenvalue of the Hamiltonian. From the eigenvalue, one still needs an extra classically efficient step to find the $r$, so $T_{\text{factor}}>T_{\text{bound}}$.} it is thus sufficient to choose 
\begin{align} \label{eq:ssufficient}
s_0 \tau \sim N^2,
\end{align}
which is exponential in the number of target qubits $n=\log_2 N$. 
This can also be derived from Eq.~\eqref{eq:tbound} using $\Delta_{\text{E}}=2\pi/(2N^2)$, where $T_{\text{bound}} \sim 1$. If we let $s_0=1$ for the coherent state, this requires total computing time to scale exponentially with the size of the problem (i.e. $\log_2 N$). Thus to ensure polynomial total computing time, we can choose instead $\tau \sim 1$ and $s_0 \sim N^2$. 

For experimental purposes, however, factoring large numbers is still beyond the reach of current experimental capability. To factor $N=21$ in our model for example, a squeezing of $d \approx 26$ dB is required, whereas the current experimental record stands near half the number of decibels needed \cite{andersen201530}.
\section{Comparison of resources}
We saw that the squeezing factor can be interpreted as an inverse precision. They can be considered equivalent resources since the two quantities are also polynomially related. There are also other quantities polynomially related to the squeezing factor like energy, qubit number and the dimensionality of the qudit that can be encoded in our squeezed state. We discuss their relationship to the squeezing factor and in what ways they can and cannot also be considered resources.

\textbf{Energy}. We know energy to be a familiar resource when it takes the form of thermodynamical work. Energy is also a resource in computational tasks when there is the erasure of information. Since we learn from Landauer's principle~\footnote{The basic idea behind Landauer's principle is that a minimal energy cost of $k_B T \ln 2$ is incurred for erasure of one bit of information \cite{lan}.} that energy dissipation occurs in these cases, such computers must consume energy to remain functional. Energy can also be considered a resource if it is required in the initial preparation of the necessary input states. In a quantum optical setting, for example, energy is required for preparing a squeezed state resource. The minimum energy $E_{\text{min}}$ required is that needed to create the number of particle excitations $\avg{\hat{n}}$ corresponding to a certain amount of squeezing since $E_{\text{min}} \propto \avg{\hat{n}}$. The number of particle excitations is itself regularly considered as the primary resource in the context of quantum metrology. For our squeezed state $\langle \hat{n} \rangle =\text{sinh}^2(\ln (2 s_0))$, where for a large squeezing factor $\avg{\hat{n}} \propto s_0^2$. Thus, energy and the squeezing factor are polynomially related.

\begin{figure}[ht!]
\centering
\includegraphics[scale=0.5]{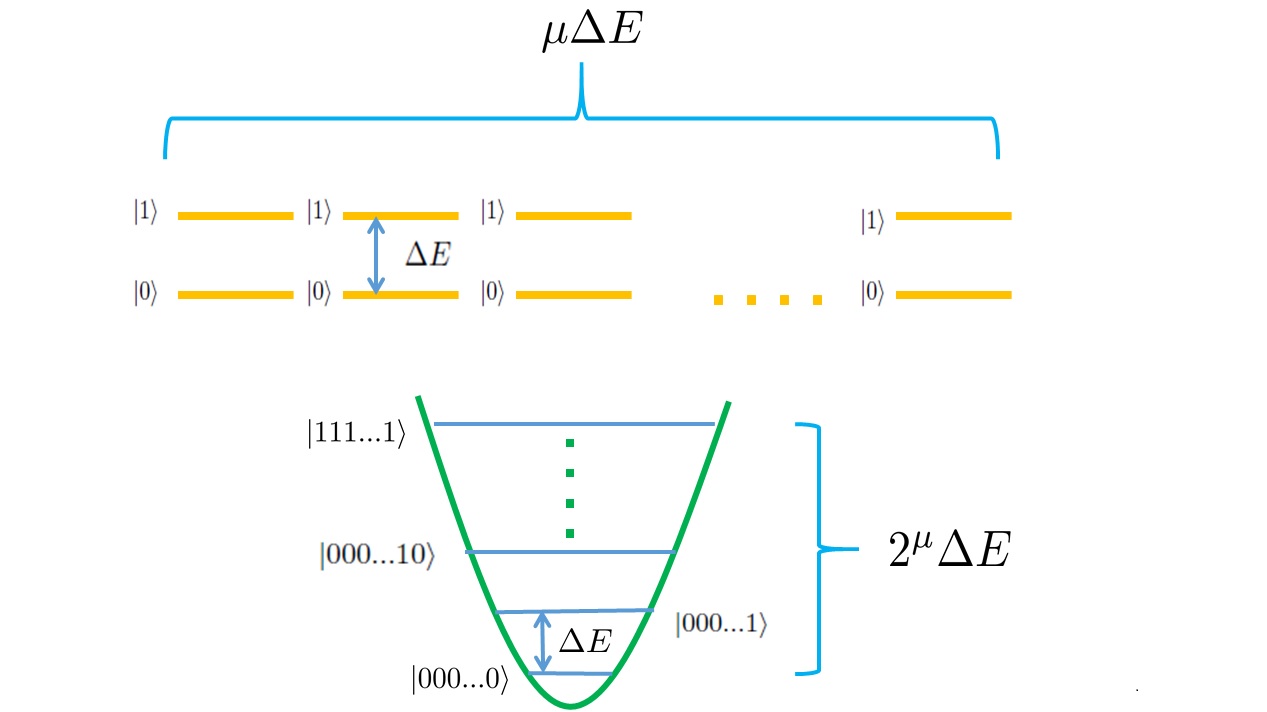}
\caption[\textit{Energy cost in discrete and continuous variables}.]{\label{fig4}\textit{Energy cost in discrete and continuous variables}. This is an illustration of the difference in energy cost in enabling the computation of $\ket{0}^{\otimes \mu} \rightarrow \ket{1}^{\otimes \mu}$ using qubits (2-level systems) versus a harmonic oscillator mode. Using $\mu$ two level systems, the total excitation energy required is $\mu \Delta E$. To perform the same computation using a harmonic oscillator mode with energy level gap $\Delta E$, the excitation energy for the computation is $2^{\mu} \Delta E$.}
\end{figure}

This interpretation of the squeezing factor as an energy can help us understand why $s_0$ of the order $\mathcal{O}(\exp(n))$ may be necessary for factoring. We can consider performing factoring in our model as swapping $m=\log_2 N$ pure control qubits in the qubit factoring protocol with a single qumode. A simple example to illustrate this phenomenon is to consider a simple computation $\ket{0}^{\otimes \mu} \rightarrow \ket{1}^{\otimes \mu}$. Suppose the computation is performed using $\mu$ qubits encoded in $\mu$ two-level atoms. Let the energy gap between the ground ($\ket{0}$) and the first excited state ($\ket{1}$) be $\Delta E$. Then a total energy of $\mu \Delta E$ is required for the computation. If we use a single CV mode instead, for instance, a harmonic oscillator with $2^{\mu}$ energy levels, the total energy required to perform this computation is $2^{\mu} \Delta E$, which has the exponential scaling in $\log_2 N$ we observe in our model. See Fig.~\ref{fig4}.
However, there are also two reasons why it is not ideal to consider energy as a resource.  Firstly, having no energy does not guarantee that the computational power of a high squeezing factor cannot be achieved. An example is spin-squeezing in the case of energy-degenerate spin states. Secondly, having large amounts of available energy also does not guarantee more efficient computation. If we instead use a coherent state with high coherence $\alpha$ and hence large energy (since $\langle \hat{n} \rangle=|\alpha|^2$), we still cannot factor in polynomial time.

\textbf{Qudit dimensionality.} There is a known way of encoding a qudit, or a discrete variable quantum state with $D$ dimensions~\footnote{D$=$2 is equivalent to a qubit.} into a CV mode, called the GKP (Gottesman-Kitaev-Preskill) encoding \cite{GKP}, which we can now use for illustration. This can work for CV states whose probability distribution (in momentum, for example) can be described as a sum of Gaussian functions each with width $w$ and separated by a distance $\Delta \phi$. Since the precision associated with each peak is on the order $w$, we can fit $\Delta \phi/w$ distinguishable copies of these distributions together, where each profile is translated along the momentum axis by a unit $\Delta \phi/w$. If we represent each degree of freedom by one such distribution, then there are $D=\Delta \phi/w$ degrees of freedom available to this CV state just by displacement in momentum. These $D$ degrees of freedom can be mapped onto a qudit of dimensionality $D$.

Given an encoding like GKP, we can write $D \sim s_0 \Delta \phi$ since in our case $w=1/s_0$. Thus here $s_0$ is interpreted as the inverse precision $1/w$. Since $\Delta \phi$ is the distance between adjacent Gaussian peaks in our probability distribution $P(p_E)$, to accomplish factoring, we require $s_0=2^{2n}=N^2$ and $\Delta \phi=1/N$, so $D=N$. For DQC1, $s_0=1$ and $D=2$ (since we only need a single qubit). Thus $D$ and $s_0$ are also polynomially related.

\textbf{Qubit number.} A qudit of dimension $D$ is equivalent to $m=\log_2 D$ pure qubits, where $D$ is polynomially related to $s_0$. Thus for factoring, the required number of control qubits scales as $m \sim \mathcal{O}(\text{poly}(n))$ compared to $m=1$ for DQC1, where $n=\log_2 N$ is the number of target register qubits. Here we see that the number of qubits for the two problems are not exponentially separated. There is an  important result of Shor and Jordan \cite{shorjordan}, which compares the computational power of DQC1 with an $n$-qubit target register and a model that is an $m$-control qubit extension of DQC1. Their result claims that if $m$ is logarithmically related to $n$, then this model still has the same computational power as DQC1. On the other hand, if $m$ is polynomially related to $n$, then this model is computationally harder than DQC1. If we use $n=\log_2 N$, then the Shor and Jordan result make clear that the number of control pure qubits $m$ in these two different models are not separated exponentially, even though one model has higher computational power. However, like the time resource in these two models, $D=2^m$ in these two models are exponentially separated, which suggests that $D$ is preferred over $m$ as a good quantifier for a computational resource.

That the required number of control qubits scales as $m \sim \mathcal{O}(\text{poly}(n))$ is not too surprising if we observe a similarity between our model and standard quantum phase estimation. Our model has more in common with standard quantum phase estimation than DQC1, even though it is a hybrid generalisation of DQC1. We can see that by taking the average of momentum measurements in our model, we obtain the average of the eigenvalues of the Hamiltonian. The momentum average, however, does not give the normalised trace of the unitary matrix $U$ as may be expected from DQC1. This can be understood by taking a discretized version of our model, where one uses instead $\ket{x}$ for $x=0, 1, 2, ...,N$. Then the circuit reduces to the standard phase estimation circuit, which requires the $m=\log_2 N$ pure control qubits which we traded for a single qumode. From this, we can also see that our model using an infinite squeezing factor is an analogue of the standard phase estimation using an infinite number of qubits, which in both models allow us to attain infinite precision in phase estimation.

We add that this comparison with standard quantum phase estimation further strengthens our claim that $s_0 \sim N^2$ may be necessary for factoring the number $N$. Suppose if we instead only need an exponentially smaller squeezing factor for factoring in a new algorithm. This would imply that the qubit analogue of this new algorithm is exponentially more powerful than standard quantum phase estimation.

While qumodes like squeezed states can be used as a way of encoding qudits and qubits \cite{GKP, terhal, schoelkopf}, the squeezing factor is still a resource that should be considered in its own right. Its emphasis over qudits is important for practical consideration. The practical advantages of considering qumode resources, in general, are that CVs typically use affordable off-the-shelf components and widely leveraged quantum optics techniques. They also have higher detection efficiencies at room temperature and can be fully integrated into current fiber optics networks \cite{braunstein2005, christian2012}.

We note that the replacement of a squeezed state by qubits is still heuristic proof and should be considered as a very suggestive, though not solid, demonstration. It is still necessary to use $s_0$ as a quantifier of hardness in our particular model. However, these comparisons serve supporting roles in justifying the use of the squeezing factor as well as demonstrating the versatility in our model to allow these comparisons to be made. 
\section{Discussion and further work}
We have approached the question of what quantum resource can quantify the hardness of computational problems using the important and so far unresolved example of DQC1 and factoring. We have found how one such resource, which we identify as squeezing, can be used to quantitatively compare the hardness of DQC1 and factoring. We show explicitly how squeezing can also be interpreted in terms of other resources, like precision, energy, qudit dimensionality and qubit number. This highlights in a clear way how computational power may be dependent upon many different, but in many ways equivalent resources.

The chief contributions in this chapter are the following: 1) The discovery that squeezing, an exclusively continuous variable resource, can serve a unifying role between DQC1 and phase estimation. This is particularly important since DQC1 and factoring are key algorithms. Understanding how they can be compared brings us a step closer to understanding how resources can quantify quantum computational power. 2) We provide a detailed discussion of the resource trade-off between running time, interaction strength and required squeezing factor and precision in the qumode model. In so doing, we gain a better understanding of which resources play a more prominent role under which circumstances, while still maintaining the same computational power. 3) We also provide an interpretation of squeezing in terms of precision and energy. We discuss when they can and cannot be considered equivalent resources to squeezing. 4) We can also interpret our result in terms of the number of qubits and qudit dimensionality. This encourages further dialogue on how resources can `interchange' to others while the computational power in the model as a whole does not change. This becomes more important as hybrid and continuous variable computing becomes further developed and they need to be compared to their purely discrete variable counterparts. 5) Hybrid quantum information is an exciting emerging field that attempts to make the most of both discrete and continuous variable resources in a single device \cite{andersen}. Here we have provided the first protocols in hybrid quantum computing for both factoring and a DQC1-complete problem. 

There are five current main directions of further research that this work motivates: 1) The first direction is the experimental implementation of our model. We are undergoing discussions with an experimental group on using their superconducting qubits to implement a DQC1 protocol (which would be a first using superconductors), which is within current experimental possibilities;  2) The second direction is considering other resources that are present in our model, like coherence, entanglement or discord; 3) A third direction is considering extensions of our model to better understand the computational power of a single qumode and the role of mixedness of state present in the model. For example, instead of a maximally mixed target state, to use a target state with a variable mixedness, which can be associated with an effective temperature. An explicit example is to compare a target state with infinite effective temperature (or maximally mixed state) with a pure target state (zero effective temperature) and to compare the computational power of these two models. Using a pure target state it may be possible implement protocols of efficiently solving linear equations, which is a BQP-complete problem \cite{harrowlloyd}; 4) It is also interesting to explore ways of adapting the `power of one qumode', where the maximally-mixed state is replaced by an unknown state, as a probe of many-body systems; 5) More speculatively, exploring further the `time-energy' relationship in Eq.~\eqref{eq:tbound} in terms of a `time-energy' uncertainty principle like the Mandelstam-Tamm inequality \cite{mandelstam}. 
\begin{subappendices}
\section{Reducing the hybrid control gate to elementary operations}
\label{sec:elementarygates}
We note that in DQC1, there is a method of reducing the control gate $\Gamma_U=\ket{0}\bra{0} \otimes \mathbf{1}+\ket{1}\bra{1} \otimes U$ in terms of elementary (e.g. one or two-qubit) circuits \cite{animeshphd}. The analogous gate in the power of one qumode model is the hybrid control gate $C_U=\exp(i \hat{x} \otimes H \tau/x_0)$, where we now set $\tau=x_0$ for convenience. We demonstrate how this gate can also be reduced to elementary operations to further clarify the relationship between DQC1 and the power of one qumode model.

We first write down the DQC1 set-up. The DQC1 set-up begins with a polynomial sequence of elementary (e.g. one or two qubit) gates $\{u_k=\exp(ih_k)\}$. We define the product of these gates to be $\prod_k u_k \equiv U=\exp(iH)$. The next step is to implement a control-unitary on each $u_k$, so our collection of elementary gates is now transformed into the set  $\{\lambda_u \equiv \ket{0}\bra{0} \otimes \mathbf{1}+\ket{1}\bra{1} \otimes u_k\}$. The product of these gates will recover the controlled-unitary operation $\Gamma_U=\ket{0}\bra{0}\otimes \mathbf{1}+\ket{1}\bra{1}\otimes U$ appearing in the description of DQC1, since
\begin{align}
\prod_k \lambda_u &=\prod_k \ket{0}\bra{0} \otimes \mathbf{1}+\ket{1}\bra{1} \otimes u_k \nonumber\\
&=\ket{0}\bra{0}\otimes \mathbf{1}+\ket{1}\bra{1} \otimes \prod_k u_k \nonumber \\
&=\ket{0}\bra{0}\otimes \mathbf{1}+\ket{1}\bra{1}\otimes U=\Gamma_U.
\end{align}
The analogus requirement for the power of one qumode model is to begin from a polynomial sequence of elementary gates which can form the hybrid control-unitary operation $C_U=\exp(i\hat{x}\otimes H)$. We show how this can be achieved.

Let us begin with the same set of elementary gates $\{u_k=\exp(ih_k)\}$. Instead of implementing the usual control-unitary on each $u_k$, we implement a \textit{hybrid} control unitary on each $u_k$. This means our set of elementary gates is modified into the new set $\{c_u \equiv \exp(i \hat{x} \otimes h_k)\}$. We can take the product of these operations and recover $C_U$ in the following way
\begin{align}
\prod_k c_u &=\prod_k \exp(i \hat{x} \otimes h_k)=\prod_k \int dx \ket{x}\bra{x} \otimes e^{ixh_k} \nonumber \\
&=\int dx \ket{x}\bra{x} \otimes \prod_k e^{ixh_k} \nonumber \\
&=\int dx \ket{x}\bra{x} \otimes e^{ixH}=e^{i\hat{x}\otimes H}=C_U,
\end{align}
where $x$ is a number and we used 
\begin{align}
\prod_k e^{ixh_k}\equiv e^{ixH},
\end{align}
which must be satisfied for all $x$. This condition, combined with the definition that $\prod_k u_k=\prod_k \exp(ih_k)=\exp(iH)=U$, implies that $[h_k, h_{k'}]=0$ for all $k,k'$ in the product $\prod_k$~\footnote{This also implies $H=\sum_k h_k$.}. Equivalently, this means $\{u_k\}$ must be a commuting set of operators.

We can show that such a set $\{u_k\}$ where $U=\exp(iH)=\prod_k u_k$ exists for the factoring problem. We know that factoring the number $N$ is equivalent to finding the order $r$ of a random integer $q$ where $1<q<N$, which requires $U \ket{1 \mod N}=\exp(iH) \ket{1 \mod N}=\ket{q \mod N}$. Since $q$ is an integer, we can make a binary decomposition $q-1=2^0b_0+2^1b_1+2^2b_2+...+2^{f}b_f$ where $f$ is an integer and $b_j=0,1$. Then if choose $u_k$ to be an elementary operation defined by $u_k \ket{1 \mod N}=\ket{(1+2^kb_k) \mod N}$, we can see that all operators in $\{u_k\}$ commute and $\prod_{k=0}^f u_k \ket{1 \mod N}=\ket{q \mod N}=U \ket{1 \mod N}$. 
\section{Coherent state in power of one qumode model}
\label{sec:coherent}
Suppose we begin with a coherent state $\ket{\alpha}$ in our model. The coherent state can be written in the position basis as
\begin{gather}
\ket{\alpha}=\int \braket{x}{\alpha} \ket{x} dx,
\end{gather}
whose position wavefunction is  
\begin{gather}
\braket{x}{\alpha}=\left(\frac{1}{\pi x_0^2}\right)^{\frac{1}{4}} e^{-\frac{1}{2x_0^2}(x-\text{Re}(\alpha))^2}e^{i\text{Im}(\alpha)x/x_0}e^{-\frac{i}{2}\text{Re}(\alpha)\text{Im}(\alpha)},
\end{gather}
where $x_0\equiv 1/\sqrt{m \omega}$ and $m, \omega$ are the mass and frequency scales of the corresponding quantum harmonic oscillator. The last term is a constant phase factor. We note that $\text{Re}(\alpha)=\sqrt{m \omega/2} \langle x_{\alpha}\rangle$ where $\langle x_{\alpha}\rangle$ is the expected value of the position operator. Likewise $\text{Im}(\alpha)=(1/\sqrt{2m\omega})\langle p_{\alpha}\rangle$ where $\langle p_{\alpha}\rangle$ is the average value of the momentum operator. 

By using $G(x) \equiv \braket{x}{\alpha}$ in Eq.~\eqref{eq:momentum1}, we find the momentum probability distribution of the final control state to be
\begin{align}
\mathcal{P}(p) &=\frac{1}{N} \sum_n \iint G(x)G^*(x') e^{i(x-x')\phi_n \tau/x_0} \braket{p}{x}\! \braket{x'}{p} \text{d}x \, \text{d}x' \nonumber \\
        &=\frac{x_0}{\sqrt{\pi} N} \sum_{n=0}^{N} e^{-x_0^2 \left[p-\frac{\tau}{x_0}\left(\phi_n+\frac{\text{Im}(\alpha)}{\tau}\right)\right]^2}.
\end{align}
If we measure variable $p_{\text{E}} \equiv p x_0/\tau$ (where inputs $x_0$ and $\tau$ are initially known), the probability distribution for $p_{\text{E}}$ is
\begin{gather}
\mathcal{P}(p_{\text{E}})=\frac{\tau}{\sqrt{\pi} N} \sum_{n=0}^{N} e^{-\tau^2 \left[p_{\text{E}}-\left(\phi_n+\frac{\text{Im}(\alpha)}{\tau}\right)\right]^2}.
\end{gather}
Thus the coherent state can be used for phase estimation, where the accuracy of the phase estimation improves with increasing running time of the hybrid gate. 
\section{Phase estimation using power of one qumode model}
\label{sec:phaseestimate}
Suppose we want to recover any eigenvalue of our Hamiltonian to accuracy $\Delta_{\text{E}}$. The total number of $p_{\text{E}}$ measurements required for an average of one success is
\begin{gather} \label{eq:tprob1}
T_{\text{measure}} \sim \frac{1}{P_{\Delta_{\text{E}}}},
\end{gather}
where $P_{\Delta_{\text{E}}}$ is the probability of retrieving the eigenvalues to within the interval $[\phi_j-\Delta_{\text{E}}, \phi_j+\Delta_{\text{E}}]$. Using Eq.~\eqref{eq:pp} we find
\begin{align}
P_{\Delta_{\text{E}}}&\equiv \mathcal{P} \left(p_{\text{E}};\left|p_{\text{E}}-\phi_n\right| \leq \Delta_{\text{E}}\right) \nonumber \\
                   & =\frac{s_0 \tau}{\sqrt{\pi}2^n} \sum_{l=1}^{2^n} \int^{\phi_l+\Delta_{\text{E}}}_{\phi_l-\Delta_{\text{E}}} \sum_{m=1}^{2^n}e^{-(s_0 \tau)^2 \left(p_{\text{E}}-\phi_m\right)^2} \, \text{d}p_{\text{E}} \nonumber \\
                  & \equiv P(l=m)+P(l \neq m),
\end{align}
where
\begin{align}
P(l=m) &=\frac{s_0 \tau}{\sqrt{\pi}2^n} \sum_{m=1}^{2^n} \int^{\phi_m+\Delta_{\text{E}}}_{\phi_m-\Delta_{\text{E}}} e^{-(s_0 \tau)^2 (p_{\text{E}}-\phi_m)^2}\text{d}p_{\text{E}} \nonumber \\
                  &=\text{erf} \left(s_0 \tau \Delta_{\text{E}}\right)
\end{align}
and $P(l \neq m)=(1/2^n)\sum_{l \neq m=1}^{2^n}\{\text{erf}\{s_0 \tau [(\phi_l-\phi_m)/r+\Delta_{\text{E}}]\}-\text{erf}\{s_0 \tau [(\phi_l-\phi_m)/r-\Delta_{\text{E}}]\}\}>0$. 
These two contributions to the total probability distribution $P_{\Delta_{\text{E}}}$ can be interpreted in the following way. $P(l=m)$ is the probability of finding $\phi_n$ to within $\Delta_{\text{E}}$ 
if the Gaussian peaks are very far apart. This occurs when the spread of each Gaussian is much smaller than the distance between neighbouring Gaussian peaks $1/(s_0 \tau) \ll \Delta \phi_{\text{min}}$ where $\Delta \phi_{\text{min}}$ is the minimum gap between adjacent eigenvalues. $P(l \neq m)$ captures the overlaps between the Gaussians. This overlap contribution vanishes for large $N$, so for simplicity we will neglect this term. This neglecting will not affect the overall validity of our result. We can now write
\begin{align}\label{eq:probDelta}
P_{\Delta_{\text{E}}}>P(l=m)=\text{erf} \left(s_0 \tau \Delta_{\text{E}}\right).
\end{align}
By demanding $T_{\text{measure}} < T_{\text{bound}}$, then using Eqs. ~\eqref{eq:tprob1} and ~\eqref{eq:probDelta}, we find it is sufficient to satisfy
\begin{gather}\label{eq:Tbound}
T_{\text{bound}} \text{erf}(\tau s_0 \Delta_{\text{E}}) \gtrsim 1.
\end{gather}
For large $\tau s_0 \Delta_{\text{E}}$, the above inequality is automatically satisfied. This assumes that $\tau s_0$ grows more quickly in $N$ than the inverse of the eigenvalue uncertainty $\Delta_{\text{E}}$ that we are willing to tolerate. More generally however, it is the time and squeezing resources
we want to minimise for a given precision, so $\tau s_0 \Delta_{\text{E}}$ is small. In this case, Eq.~\eqref{eq:Tbound} becomes
\begin{gather}
T_{\text{bound}}\tau s_0 \Delta_{\text{E}} \gtrsim 1.
\end{gather}
\section{Retrieving eigenvectors in the power of one qumode}
\label{sec:eigenvectors}
Here we provide a brief argument of how eigenvectors of the Hamiltonian $\{ \ket{\phi_j} \}$ can also be found using our model. The hybrid state $\rho_{\text{total}}$ after application of the hybrid gate is
\begin{align}
\rho_{\text{total}} =&\frac{1}{2^n} \iint G(x) G^*(x') e^{i(x-x')H \tau/x_0} \ket{x}\!\bra{x'} \text{d}x \, \text{d}x' \nonumber \\
                             =&\sum_m \frac{1}{2^n} \iint G(x) G^*(x') e^{i(x-x')\phi_m \tau/x_0} \nonumber \\
                             & \times \ket{\phi_m}\! \bra{\phi_m} \otimes \ket{x}\!\bra{x'} \text{d}x \, \text{d}x'.
\end{align}
After a momentum measurement we are in the following state of the target register
\begin{align}
&\bra{p} \rho_{\text{total}} \ket{p} \nonumber\\
&=\frac{1}{2^n} \sum_m \mathcal{G}(\phi_m \tau/x_0-p) \mathcal{G}^*(p-\phi_m \tau/x_0) \ket{\phi_m}\! \bra{\phi_m}.
\end{align}
For a squeezed state  $G(x) = (1 / (\sqrt{s}\pi^{\frac{1}{4}})) \text{exp}(-x^2/(2s^2))$ the final state of the target register becomes
\begin{align}
\bra{p} \rho_{\text{total}} \ket{p}=\frac{s}{2^n \sqrt{\pi}} \sum_m e^{-s^2(p-\phi_m \tau/x_0)^2} \ket{\phi_m}\! \bra{\phi_m}.
\end{align}
Approximate eigenvectors can thus be obtained by measurement of the target state. The probability of obtaining the eigenvectors of the Hamiltonian is distributed in the same way as for the eigenvalues. Eigenvector identification therefore also improves with an increase in the squeezing factor. 
\section{Number of measurements to retrieve $\tr(U)/N$ in power of one qumode model}
\label{sec:DQC1measure}
Here we derive the number of momentum measurements $T_{\text{DQC1}}$ in our model needed to recover the normalised trace of $U \equiv \exp{iH}$ to within error $\delta$. We show this is upper-bounded by a quantity independent of the size of $U$. 

Let us begin by introducing a new random variable $y \equiv \exp(ip_{\text{E}} x_0)$ where $p_{\text{E}}$ are the measurement outcomes from our model. The probability distribution function with respect to $y$ can be rewritten as
\begin{gather} \label{eq:py}
\mathcal{P}_y(y)=\int^{\infty}_{-\infty} \delta(y-e^{ip_{\text{E}}x_0}) \mathcal{P}(p_{\text{E}}) \text{d}p_{\text{E}},
\end{gather}
where $\mathcal{P}(p_{\text{E}})$ is given by Eq.~\eqref{eq:pp}. 
We find that the average of $y$ is related to the normalised trace of unitary matrix $U$ 
\begin{align}
 \int y \mathcal{P}_y(y) \text{d} y &=\int e^{ip_{\text{E}}x_0} \mathcal{P}(p_{\text{E}}) \text{d}p_{\text{E}} \nonumber \\
         &=e^{-\frac{1}{4 s_0^2}} \left[\frac{\tr(U_{\tau})}{2^n}\right].
\end{align}
We now let $\tau=1$ since $U_{\tau=1}=U$. 

To find the normalised trace of $U$ to error $\delta$ is equivalent to finding the average of $y$ to within $\epsilon$ where
\begin{gather}
\int y\mathcal{P}_y(y) \text{d} y \pm \epsilon=e^{-\frac{1}{4 s_0^2}}\left(\frac{\tr(U)}{2^n} \pm \delta \right).
\end{gather}
Therefore 
\begin{gather}\label{eq:epsilondelta}
\epsilon=e^{-\frac{1}{4 s_0^2}} \delta.
\end{gather}
For concreteness, we will first separately examine recovering the real part of the normalised trace of $U$ to within $\re(\delta)$ then the imaginary part of the trace to within $\im(\delta)$.\\

\textit{Real part of the normalised trace of $U$.---}
We define a new random variable $y_R\equiv \re(y)=\cos(p_{\text{E}}x_0)$ whose average is within $\re(\epsilon)$ of the real part of the normalised trace of $U$. The probability distribution with repect to $y_R$ is
\begin{gather}
\mathcal{P}_{y_R}(y_R)=\int^{\infty}_{-\infty} \delta(y_R-\cos(p_{\text{E}} x_0)) \mathcal{P}(p_{\text{E}}) \text{d}p_{\text{E}}.
\end{gather}
We can employ the central limit theorem~\footnote{Since we are selecting our random variable independently and from the same distribution which has finite mean and variance, it is valid to use the central limit theorem.} and Eq.~\eqref{eq:epsilondelta} to find the number $t_R$ of necessary $p_{\text{E}}$ measurements to be
\begin{gather}\label{eq:clt}
t_R \sim \frac{\Sigma^2_{R}}{\re(\epsilon)^2}=\frac{\Sigma^2_{R} e^{\frac{1}{2 s_0^2}}}{\re(\delta)^2},
\end{gather}
where $\Sigma^2_R$ is the variance of the probability distribution with respect to $y_R$. Using Eqs.~\eqref{eq:py} and ~\eqref{eq:pp} we can show
\begin{align} \label{eq:SigmaR}
\Sigma^2_R \equiv & \int y^2_R \mathcal{P}_{y_R}(y_R) \text{d} y_R-\left(\int y_R \mathcal{P}_{y_R}(y_R) \text{d} y_R\right)^2 \nonumber \\
                    = &\int \cos^2(p_{\text{E}}x_0) \mathcal{P}(p_{\text{E}}) \text{d}p_{\text{E}}-\left(\int \cos(p_{\text{E}}x_0) \mathcal{P}(p_{\text{E}}) \text{d}p_{\text{E}} \right)^2 \nonumber \\
                    = &e^{-\frac{1}{2 s_0^2}}\text{sinh}\left(\frac{1}{2 s_0^2}\right) \nonumber \\
                   &+e^{-\frac{1}{4 s_0^2}}\frac{1}{2^n} \sum_{m=1}^{2^n} \cos^2(\phi_m)-e^{-\frac{1}{2 s_0^2}}\left(\frac{1}{2^n} \sum_{m=1}^{2^n} \cos(\phi_m) \right)^2 \nonumber \\
                    \leq & e^{-\frac{1}{2 s_0^2}}\text{sinh}\left(\frac{1}{2 s_0^2}\right)+e^{-\frac{1}{4 s_0^2}}\frac{1}{2^n} \sum_{m=1}^{2^n} \cos^2(\phi_m) \nonumber \\
                    \leq & e^{-\frac{1}{2 s_0^2}}\left(\text{sinh}\left(\frac{1}{2 s_0^2}\right)+e^{-\frac{1}{2s_0^2}} \right).
\end{align}
We can now use Eqs.~\eqref{eq:clt} and ~\eqref{eq:SigmaR} to find an upper bound to the number of measurements
\begin{gather}
t_R \lesssim \frac{F(s_0)}{\re(\delta)^2},
\end{gather}
where
\begin{gather}
F(s)=\text{sinh}\left(\frac{1}{2s_0^2}\right)+e^{-\frac{1}{2 s_0^2}}.
\end{gather}
\textit{Imaginary part of the normalised trace of $U$.---}
To recover the imaginary part of the normalised trace of $U$ to within an error $\im(\delta)$, we average $y_I \equiv \im(y)=\sin(p_{\text{E}}x_0)$. The probability distribution with respect to $y_I$ is
\begin{gather}
\mathcal{P}_{y_I}(y_I)=\int^{\infty}_{-\infty} \delta(y_I-\sin(p_{\text{E}}x_0)) \mathcal{P}(p_{\text{E}}) \text{d}p_{\text{E}}.
\end{gather}
We can similarly use the central limit theorem in this case to find the necessary number of measurements $t_I$ 
\begin{gather}
t_I \sim \frac{\Sigma^2_{I}e^{\frac{1}{2 s_0^2}}}{\im(\delta)^2},
\end{gather}
where $\Sigma^2_I$ is the variance with respect to probability distribution $P_{y_I}(y_I)$. We can show
\begin{align}
\Sigma^2_I \equiv & \int y^2_I \mathcal{P}_{y_I}(y_I) \text{d} y_I-\left(\int y_I \mathcal{P}_{y_I}(y_I) \text{d} y_I\right)^2 \nonumber \\
               =& e^{-\frac{1}{2 s_0^2}}\text{sinh}\left(\frac{1}{2 s_0^2}\right) \nonumber \\
                   &+e^{-\frac{1}{4 s_0^2}}\frac{1}{2^n} \sum_{m=1}^{2^n} \sin^2(\phi_m)-e^{-\frac{1}{2 s_0^2}}\left(\frac{1}{2^n} \sum_{m=1}^{2^n} \sin(\phi_m) \right)^2 \nonumber \\
               \leq & e^{-\frac{1}{2 s_0^2}} F(s_0).
\end{align}
Thus
\begin{gather}
t_I \lesssim \frac{F(s_0)}{\im(\delta)^2}.
\end{gather}
This means the number of required measurements $t$ to recover the normalised trace of $U$ to within $\delta$ has the upper bound
\begin{gather}
T_{\text{DQC1}} =\text{max}(t_R,t_I) \lesssim \frac{F(s_0)}{[\text{min}\{ \text{Re}(\delta), \text{Im}(\delta)\}]^2}.
\end{gather}
\section{Number of measurements for factoring in the power of one qumode model}
\label{sec:factoringderive}
Here we give the derivation of the number of runs $T_{\text{factor}}$ needed to recover a non-trivial factor of $N$ given the momentum probability distribution (Eq.~\eqref{eq:PDFfactor})
\begin{gather}
\mathcal{P}(p_{\text{E}})=\frac{s_0 \tau}{\sqrt{\pi} N} \sum_{m=0}^{r-1} c_m e^{-(2 \pi s_0 \tau)^2 \left(\frac{p_{\text{E}}}{2 \pi}-\frac{m}{r}\right)^2}.
\end{gather}
We want to find the probability $P_{r}$ in which one can retrieve the correct value of the order $r$. The number of runs required on average to find a non-trivial factor of $N$ is inversely related to this probability
\begin{gather} \label{eq:timefactor}
T_{\text{factor}} \sim \frac{1}{P_{r}}.
\end{gather}
Here we derive a lower bound to $P_{r}$ (hence an upper bound to the number of runs) that satisfies the following two conditions. To recover $r$ it is sufficient to (i) know $m/r$ to an accuracy within $1/(2N^2)$ and (ii) to choose when $m$ and $r$ have no factors in common so their greatest common denominator is one (i.e. $\gcd(m,r)=1$). 

The first condition comes from the continued fractions algorithm \cite{nandc}, which can be used to exactly recover the rational number $m/r$ given some $\phi'$ when $\left|\phi'-m/r\right|\leq 1/(2r^2)$. Since $r \leq N$, a sufficient condition is $\left|\phi'-m/r\right|\leq 1/(2N^2)$. The second condition ensures we recover $r$ instead of a non-trivial factor of $r$. We will see how to satisfy the second condition later on.

To satisfy the first condition, we see that the probability of finding $m/r$ to within $1/(2 N^2)$ when measuring $p'_{\text{E}} \equiv p_{\text{E}}/(2 \pi)$ is
\begin{align}
P_{r}&\equiv \mathcal{P} \left(p'_{\text{E}};\left|p'_{\text{E}}-\frac{m}{r}\right| \leq \frac{1}{2N^2}\right) \nonumber \\
                   & =\frac{s_0 \tau}{\sqrt{\pi}N} \sum_{l=0}^{r-1} \int^{\frac{l}{r}+\frac{1}{2N^2}}_{\frac{l}{r}-\frac{1}{2N^2}} \sum_{m=0}^{r-1} c_m e^{-(2 \pi s_0 \tau)^2 \left(p'_{\text{E}}-\frac{m}{r}\right)^2} 2\pi \, \text{d}p'_{\text{E}} \nonumber \\
                  &>\frac{s_0 \tau}{\sqrt{\pi}N} \sum_{m=0}^{r-1} c_m \int^{\frac{m}{r}+\frac{\pi}{2N^2}}_{\frac{m}{r}-\frac{\pi}{N^2}} e^{-(2 \pi s_0 \tau)^2 (p'_{\text{E}}-\frac{m}{r})^2} 2 \pi \text{d}p'_{\text{E}} \nonumber \\
             &= \sum_{m=0}^{r-1} \frac{c_m}{N} \text{erf} \left(\frac{\pi s_0 \tau}{N^2}\right)=\sum_{m=0}^{r-1} \frac{c_m}{N} \text{erf} \left(\frac{\pi s_0 \tau}{N^2}\right).
\end{align}

Note that we do not require contributions to the probability from every $m$ in the summation. In order to successfully retrieve $r$ from the fraction $m/r$, we
need only consider the cases where $\gcd (m,r)=1$. Euler's totient function $\Phi(r)$ represents the number of cases where $m$ and $r$ are coprime with $m<r$. It can be
shown that $\Phi(r) >r/\{e^{\gamma} \ln[\ln(r)]\}$ where $\gamma$ is Euler's number \cite{shor}. In the cases where $\gcd (m,r)=1$, the amplitude
$|c_m| \equiv M$, where $M$ is the number of cases where $r_d=r$. It is also possible to show that when $N=v_1v_2$ (where $v_1$ and $v_2$ are prime numbers),
$M> (v_1-1)(v_2-1)$ \cite{parkerplenio}. 

Then the probability of retrieving the correct $r$ from the probability distribution is at least
\begin{align}\label{eq:prbound}
P_{r} &>\sum_{m=0}^{r-1} \frac{c_m}{N} \text{erf} \left(\frac{\pi s_0 \tau}{2^{2n}}\right)>\frac{M \Phi(r)}{N} \text{erf}\left(\frac{\pi s}{N^2}\right) \nonumber \\
                    &> \frac{(v_1-1)(v_2-1) r}{e^{\gamma} N \ln[\ln(r)]} \text{erf}\left(\frac{\pi s_0 \tau}{N^2}\right) \nonumber \\
                    &>  \frac{(v_1-1)(v_2-1)}{e^{\gamma} N\ln[\ln(r)]} \text{erf}\left(\frac{\pi s_0 \tau}{N^2}\right).
\end{align}
From Eqs.~\eqref{eq:timefactor} and ~\eqref{eq:prbound} we now have an upper bound to the time steps required
\begin{gather}\label{eq:tfactorize}
T_{\text{factor}} <\frac{1}{P_r}=\frac{e^{\gamma} N \ln[\ln(N)]}{(v_1-1)(v_2-1) \text{erf}\left(\frac{\pi s_0 \tau}{N^2}\right)}.
\end{gather}
The large $N$ limit (where $v_1,v_2 \gg 1$) gives our result 
\begin{gather}
T_{\text{factor}}<\frac{e^{\gamma}  \ln[\ln(N)]}{\text{erf}\left(\frac{\pi s_0 \tau}{N^2}\right)}.
\end{gather}
\end{subappendices}
\newpage\null\newpage

\startlist{toc}
\printlist{toc}{}{\section*{\textbf{Photonic multi-parameter estimation}\\
Chapter contents}}
\chapter{Photonic multi-parameter estimation}
\label{chap:tomo}
\section{Introduction and motivation}
Advancements in precision measurement are playing an ever more important role in technological development. From biological imaging \cite{bioimaging}, magnetometry \cite{magneto}, precision clocks \cite{quantumclock, bloom2014optical}, lithography \cite{quantumlitho1, quantumlitho2}, thermometry \cite{luis, robthermo}, to the recent detection of gravitational waves \cite{LIGO}, there is ever increasing demand for higher precision in parameter estimation schemes. However, precision estimation of a single parameter using classical resources is always bounded by the shot noise limit. Quantum resources have been shown to exceed this bound and this has allowed the development of quantum metrology. This enhanced precision can also be carried over to the simultaneous estimation of multiple parameters. In particular, this can be applied to the problem of estimating a unitary matrix describing general linear optical processes. 

Improving the precision in characterising general linear optical processes with limited resources has twofold benefits. The first is the enormous number of important potential applications such a study would advance, including the characterisation of quantum gates in quantum computing, measuring the birefringence in fibre optics and imaging of sensitive specimens. The central focus here is on the development of experimentally accessible protocols that can at the same time provide improved precision.

Another motivation comes from a need to theoretically understand differences between single-parameter and multi-parameter estimation in the resources they require for attaining optimal precision. This will have an impact on how we understand the role of quantum processes and quantum resources providing an advantage over their classical counterparts. This requires the development of appropriate theoretical tools, which remains to be further explored for photonic systems. 

In our study, we take both considerations into account. We develop a suitable theoretical machinery for multi-parameter estimation using photonic systems that is at the same time adapted to an experimentally accessible protocol. 

One primary resource restriction in parameter estimation is average particle number $\langle N\rangle $. High $\langle N\rangle$ can either harm sensitive samples during imaging or do not easily enable the optimal resource state to be created~\footnote{For example, in single-parameter estimation, the most theoretically useful state to use are $N00N$ states, provided there are no losses. However, they are very challenging to generate with high $N$ \cite{cable2007efficient}.}. In single-parameter estimation in metrology, measurements using classical resources are subject to the shot noise limit~\footnote{Also known as the standard quantum limit.}, where the mean square error of estimates of the parameter scale as $1/N$. This results from the central limit theorem in classical statistics. However, just as was discovered for quantum computing, this limit does not take into account the full capability of quantum mechanical states and processes. Caves \cite{caves} discovered that squeezed photonic states can be used to achieve precision improved beyond the shot noise limit~\footnote{The basic idea behind this protocol has proved to be integral part of the LIGO experiment that led to the recent long-awaited success of the detection of gravitational waves \cite{LIGO}.}. The best precision scaling that quantum mechanics can allow is related to the limits imposed by the Heisenberg uncertainty principle \cite{carlos}. This precision scaling is a quadratic scaling~\footnote{For states with number fluctuations, it has been argued that $1/\langle N^2\rangle$ is better at capturing the precision improvement \cite{hofmann2009}. We confine ourselves in this chapter to states with definite particle number.} $1/\langle N\rangle^2$, also known as Heisenberg scaling~\footnote{One of the earliest origins of Heisenberg scaling appears in \cite{heitler}.}. In fact, the states which can uniquely attain this optimal precision for single-parameter estimation are $N00N$ states. 

A traditional technique for estimating general linear optical processes is called process tomography. The communities studying process tomography and quantum metrology have largely evolved separately. In process tomography, single-photon probes are used and measurement precision is limited to the shot noise limit. A recent work \cite{hugotomo} was the first to experimentally demonstrate a protocol extending traditional process tomography using multi-photon probes (a Holland-Burnett state) to attain a Heisenberg scaling, just like for protocols in quantum metrology. Although a true quantum enhancement for precision was demonstrated, it was not known whether the scheme is optimal and what other input states could be used to show the same scaling. These are some of the points we address in this work. 

A limitation of the analysis in \cite{hugotomo} was a reliance on a quantity traditional to process tomography, called process fidelity. However, a primary theoretical tool already exists from the metrology community that is especially suitable for quantifying precision. This is known as Fisher information and it can be used to bound the highest precision possible with given $N$. In photonic systems, this tool has already been exploited for some very specific cases of multi-parameter estimation where substantial simplifications occur, such as when the parameters can be encoded in simultaneous, multiple single-parameter estimation schemes \cite{peterh, yue} or some other more specific unitaries of importance in experiments \cite{mihai, philip}. However, for \textit{general} linear optical evolution, questions surrounding quantum advantages for precision have not previously been addressed for photonic systems. We approach these questions by studying carefully the simplest case of $N$-particle two-mode photonic states. This is equivalent to estimating the three unknown independent parameters of an $SU(2)$ matrix. 

Although Heisenberg scaling for general $SU(2)$ estimation in photonic systems have not been studied except in \cite{hugotomo}, there has been research on spin systems in related contexts. Analyses on parameter estimation for these systems have mostly concentrated on using process fidelity measures instead of the Fisher information formalism, for historical reasons. One of the first indications that Heisenberg scaling appears in the general $SU(2)$ estimation can be found in \cite{childspreskill}, where entanglement of a spin-$1/2$ probe state is used for estimating the direction of a magnetic field. Following soon after this work, in \cite{peres}, a fidelity measure is used to numerically find $N$-particle spin-$1/2$ states which have Heisenberg scaling, but without any proofs of optimality. A succession of theoretical studies \cite{bagan1, bagan2, chiri, hayashi}, also using fidelity measures, demonstrated the role of entanglement in states which help achieve Heisenberg scaling, especially in the context of quantum protocols for aligning quantum references frames. Although optimal states and measurements are derived, the theoretical results are not motivated by states and measurements that are experimentally accessible. Furthermore, no comparisons were made between states useful for single and multi-parameter parameter estimation problems. 

One of the first theoretical studies \cite{fujiwara} focusing on $SU(2)$ estimation using the Fisher information formalism adopts the viewpoint of information geometry. However, this work was more mathematically motivated and neither a protocol nor mention of Heisenberg scaling was given. Matsumoto \cite{matsumoto} followed with a more general formalism, including the conditions under which the Fisher information matrix can attain its maximum value (i.e. can equal to the quantum Fisher information matrix) using projective measurements. His emphasis was also on mathematical aspects of the problem, and like \cite{fujiwara}, no protocols or Heisenberg scaling was mentioned. Later \cite{fujiwaralate} built upon \cite{fujiwara} for general $SU(d)$ \footnote{See \cite{kahn} for another work on $SU(d)$ estimation.}, but focussed on maximally-entangled input states instead of more general states. Ballester \cite{ballester2004, ballester2005} furthered these works with protocols that demonstrated the role of entanglement to realise Heisenberg scaling, validating previous findings using fidelity measures. Although optimality results were found, the states and measurements achieving the optimal bounds were also not experimentally motivated. A very recent work \cite{animesh} revived the study of $SU(2)$ estimation following practical considerations, in the context of estimating the direction of a magnetic field. However, here only one type of state was studied and there is no proof for optimality of the protocol presented. 

In our work, we extend quantum-enhanced $SU(2)$ estimation to $N$-particle photonic systems, in the absence of photon loss or decoherence. We employ the frequently-used Fisher information formalism to make easier proofs of optimal precision as well as to bridge the gap between the methodology commonly used by those studying process tomography and quantum metrology. With an emphasis on experimentally accessible strategies, we use the novel working protocol from \cite{hugotomo} and present an optimality proof adapted into our protocol from results in \cite{ballester2005}. We use this tool to show how far experimentally accessible states and measurements are from obtaining optimal precision. This enables future comparison of resources between the single-parameter estimation case and $SU(2)$ estimation. Our formalism also makes it much easier to check the quantum advantages expected of specific quantum states as well as giving simple-to-check conditions to test which states can provide optimal precision.  In addition, by presenting a mapping between two-mode photonic states and linear processes with spin states and processes, we can show the connection between optimal states in both contexts. This is the first time such a tool is employed in multi-parameter estimation, while in quantum metrology a related analogy was proposed in \cite{shajicaves}. The mapping developed in this chapter will be a useful tool for all future comparisons between photonic and spin state multi-parameter estimation schemes. 

One might wonder why multi-parameter estimation deserves a separate theoretical study. It is crucial to emphasise that general multi-parameter estimation cannot be regarded as simple generalisations of the common quantum metrology scheme for single-parameter estimation. For example, the optimal measurements of the different parameters may be incompatible, so they cannot be performed simultaneously to achieve optimal precision. One key observation made possible by our formalism is the fact that results of useful states and measurements for single-parameter estimation in metrology do not necessarily coincide with the multi-parameter cases. For example, popular states used together with photon-number-counting measurements in metrology, such as $N00N$ states and Holland-Burnett states, allow for measurement precision independent of the unknown parameter for commonly-considered set-ups for estimating a single phase parameter. However, we show this is not true for general $SU(2)$ estimation. As another example, for single-parameter estimation, Holland-Burnett states give sub-optimal precision \cite{HB1993} while $N00N$ states are optimal, while from an experimental standpoint, Holland-Burnett states are more tolerant to photon loss \cite{datta2011quantum}.  For the multi-parameter case, the comparison of the two states is more complicated even in the lossless case, since it depends on which unitaries are compared. \\

\textbf{Chapter outline}\\

In section ~\ref{sec:FIintro} we briefly review Fisher information, related tools and their role in single-parameter and multi-parameter estimation. We highlight some key states for single-parameter estimation in metrology. We later test these states in an $SU(2)$ estimation protocol using a multi-photon probe, that we introduce in section \ref{sec:protocol}. In section ~\ref{sec:mapping}, we provide an explicit mapping between a two-mode linear optical process and the corresponding process with spin-$1/2$ particles. This we use later in our proofs for optimal precision as well as more clearly highlighting the analogy between the two kinds of physical systems. The next step is to construct the Fisher information formalism appropriate for our protocol in section ~\ref{sec:FImatrix}. We then move on to section ~\ref{sec:QFI} to prove the theoretical optimal precision in our protocol (via the quantum Cramer-Rao bound) and general conditions to test which states can achieve this. In the application of the formalism we developed, we study two classes of popular states used in single-parameter estimation in metrology. In particular, we focus on $N00N$ states and Holland-Burnett states and compare their usefulness in $SU(2)$ estimation with photon number-counting measurements in section ~\ref{sec:applications}. We summarise our main results and highlight key directions for future work in section ~\ref{sec:discussion}. 
\subsection{Fisher information and parameter estimation}
\label{sec:FIintro}
In any scheme to estimate unknown parameters in a system, it is useful to bound the variance of those parameters as a way of characterising the precision of that scheme. The inverse of a quantity known as the Fisher information provides such a bound, known as the Cramer-Rao bound, which is what makes this quantity so crucial. There are three independent choices in one's estimation scheme, which are one's initial state, one's process (or interaction containing unknown parameters) and the measurement one uses. In any given process, the goal is to find the initial state and measurement maximising the Fisher information while being subject to some given constraints of one's resources. Here we take this resource to be the number of particles of our input state. We look at the Fisher information and related important quantities for both single-parameter and multi-parameter estimation. 
\subsubsection{Introducing single-parameter estimation}
\begin{figure}[!tbp]
  \centering
  \subfloat[Single-photon input state.]{\includegraphics[width=0.4\textwidth]{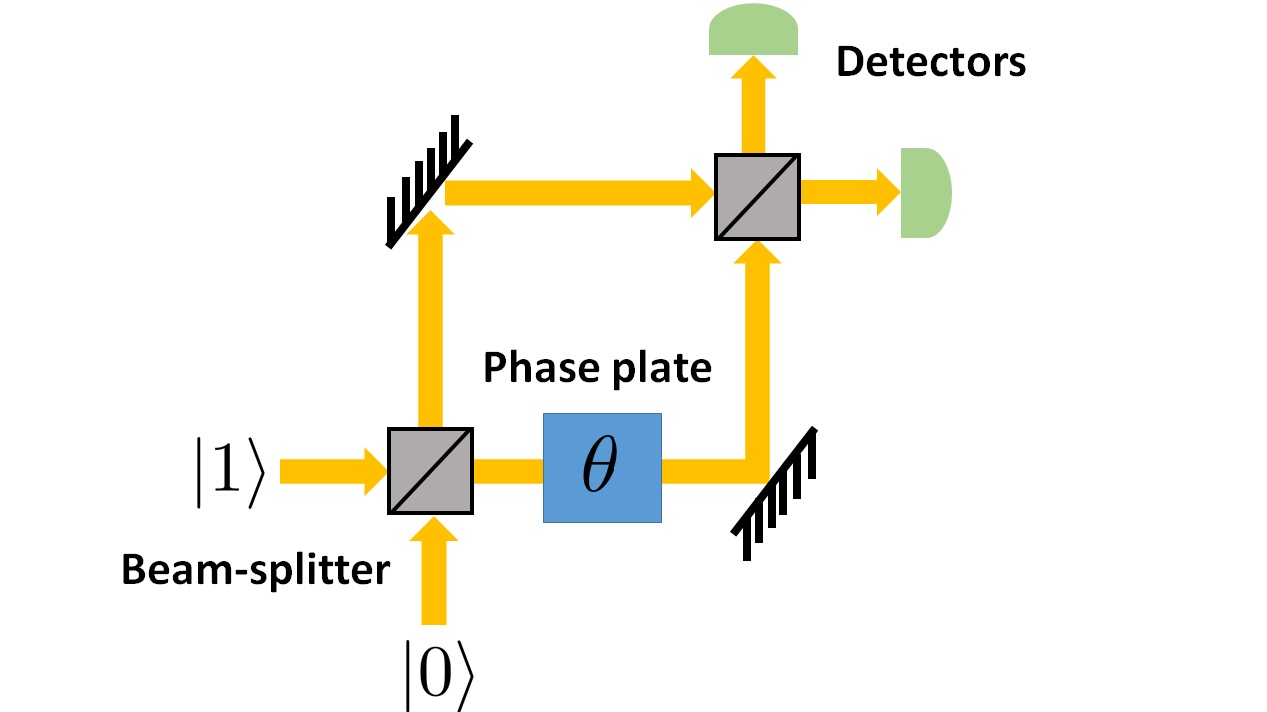}\label{fig:f1}}
  \hfill
  \subfloat[$N00N$ input state.]{\includegraphics[width=0.4\textwidth]{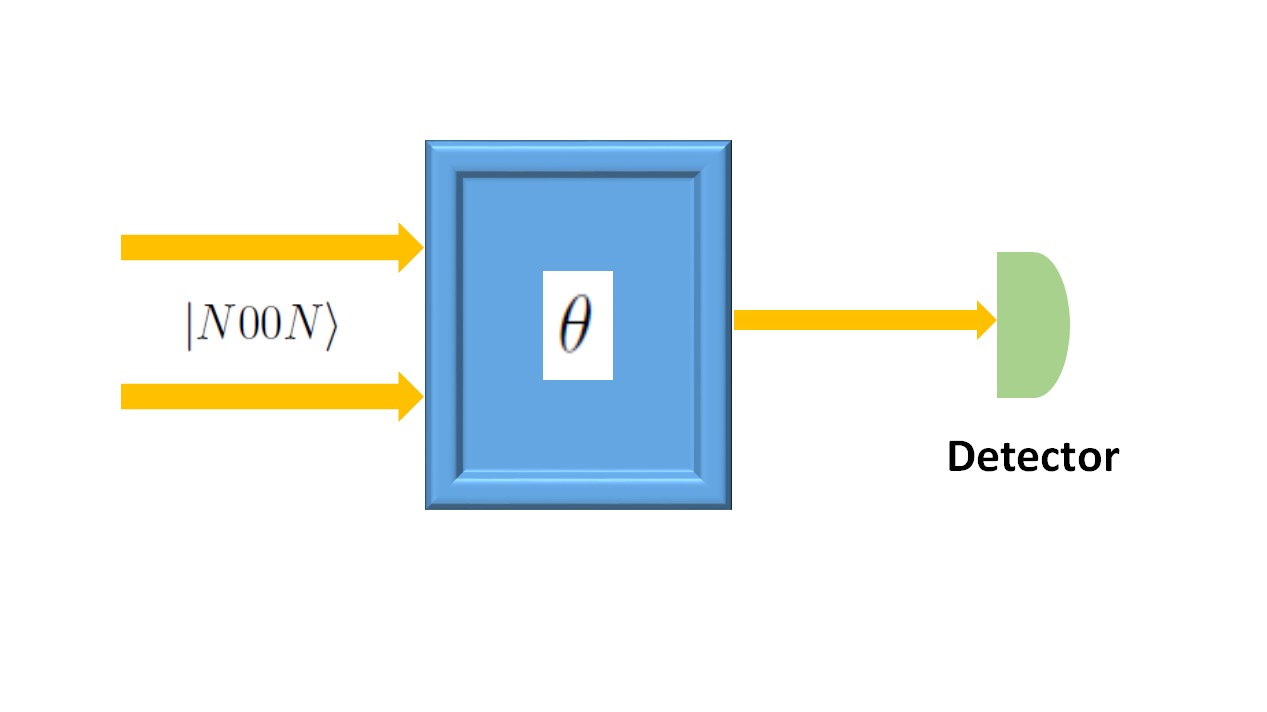}\label{fig:f2}}
  \caption[\textit{Interferometer with (a) $\ket{10}$ input state and (b) $N00N$ input state}.]{\textit{Interferometer with $\ket{10}$ input state and $N00N$ input state}.$(a)$ Schematic representation of Mach-Zehnder interferometer with a phase plate along one of its arms with unknown phase $\theta$ and input state $\ket{10}$ (i.e. state $\ket{1}$ in the left-incoming arm and state $\ket{0}$ from the bottom-incoming arm) and final detection in the $\ket{01},\ket{10}$ basis, and schematic representation of input state $(b)$ $N00N$ in single-parameter estimation scheme with final measurement in basis $\ket{N00N_{\pm}}$.}
\end{figure}

We begin with a simple single-parameter estimation protocol for illustration. Imagine we have a Mach-Zehnder interferometer with a single in-going photon. The interferometer consists first of a 50:50 beam-splitter, a phase plate with phase $\theta$ that we want to estimate and a second 50:50 beam-splitter at the end, see Fig.~\ref{fig:f1}.
We consider the state of the photon passing through the interferometer as consisting of two modes $a$ and $b$, which have corresponding creation operators $a^{\dagger}$ and $b^{\dagger}$. The creation operators $a^{\dagger}$, $b^{\dagger}$ creates a photon along the top and bottom arms of the interferometer respectively. Let our initial photon belong to mode $a$ and it is represented by $a^{\dagger}\ket{00}$ where $\ket{00}$ is the two-mode vacuum state. The 50:50 beam-splitter takes acts on the creation operators like $a^{\dagger} \rightarrow (a^{\dagger}+b^{\dagger})/\sqrt{2}$, $b^{\dagger} \rightarrow (a^{\dagger}-b^{\dagger})/\sqrt{2}$ and our phase plate acts only along the top arm so takes $a^{\dagger} \rightarrow \exp(i\theta) a^{\dagger}$, $b^{\dagger} \rightarrow b^{\dagger}$. Thus our single-photon state evolves as $\ket{10} \equiv a^{\dagger}\ket{00} \longrightarrow (1/2)((e^{i\theta}+1)\ket{10}+(e^{i\theta}-1)\ket{01}) \equiv \ket{\theta_f}$ into the final state $\ket{\theta_f}$. The probability that the photon in state $\ket{\theta_f}$ is detected in either mode $a$ or $b$ are $P(a)=|\braket{10}{\theta_f}|^2=(1/2)(1+\cos(\theta))$ and $P(b)=|\braket{01}{\theta_f}^2=(1/2)(1-\cos(\theta))=1-P(a)$~\footnote{We note that this probability distribution can equivalently be obtained without the second beam-splitter and making a measurement in basis $\{(\ket{10}\pm\ket{01})/\sqrt{2} \}$. This is used later in the $N00N$ state case.}. When we make $N$ number of measurements of photon number, it can be shown that the variance in $\theta$ is
\begin{align}
\delta^2 \theta=\frac{1}{N}.
\end{align}
That $\delta^2 \theta$ scales as $1/N$ is known as the shot noise limit. It is the best precision a single-particle quantum state can be used to achieve in parameter estimation. Notice that the DQC1 model studied in the first chapter of this thesis also exploits this scaling, where we can consider our photonic states $\ket{10}$, $\ket{01}$ as representing the two orthogonal states of a qubit. However, unlike in DQC1 which is restricted to a single control qubit unentangled with the other qubits, in single-parameter estimation we are free to use multiparticle states which are not separable. We can, in fact, use these states to go beyond the shot-noise limit. 

With $N$ particles as resource, instead of making repeated measurements using a single-particle state $N$ times, we can use a single $N$-particle state. One of the most popular states to consider in this context is known as a $N00N$ state, defined as 
\begin{align}\label{eq:noondef}
\ket{N00N} \equiv \frac{\ket{N0}+\ket{0N}}{\sqrt{2}},
\end{align}
which we can consider as a generalisation of the single-photon state $(\ket{10}+\ket{01})/\sqrt{2}$ after the first beam-splitter in Fig.~\ref{fig:f1}.
The $N00N$ state evolves in the interferometer (see Fig.~\ref{fig:f2}) as $(\ket{0N}+\ket{N0})(\sqrt{2})  \longrightarrow (\ket{0N}+e^{i N\theta} \ket{N0})(\sqrt{2})=\ket{\theta_f}$. Here, instead of using a second beam-splitter, we choose to make a measurement that projects the final state $\ket{\theta_f}$ onto the basis $\{\ket{N00N_+}, \ket{N00N_-}\}$ where $\ket{N00N_{\pm}}=(\ket{N0} \pm \ket{0N})/\sqrt{2}$. Then our probability distribution for measuring the final state in either $\ket{N00N_+}$, $\ket{N00N_-}$ is given by $P(\ket{N00N_+})=|\braket{N00N_+}{\theta_f}|^2=(1/2)(1+\cos(N\phi))$ and $P(\ket{N00N_-})=|\braket{N00N_-}{\theta_f}|^2=(1/2)(1-\cos(N\phi))=1-P(\ket{N00N_+})$. Now we find the variance in parameter $\theta$ is
\begin{align}
\delta^2 \theta=\frac{1}{N^2}.
\end{align}
There is a factor of $N$ enhancement here. This enhancement is known as the Heisenberg scaling. While $\delta \theta^2=1/N^2$ has been shown to be the optimal precision achievable using quantum states \cite{Bollinger1996} (also known as the Heisenberg limit), any scaling $\delta^2 \theta=\mathcal{O}(1/N^2)$ is called by the name of Heisenberg scaling and is a major improvement over the shot noise limit for large $N$. In fact $N00N$ states have been shown to uniquely give the Heisenberg limit \cite{durkin2007}. There is in a fact a theoretical framework to calculate the lower bound to $\delta^2\theta$ when given an input state, interaction and measurement. It can be proved that, when an estimator for $\theta$ is unbiased~\footnote{An unbiased estimator for a parameter is an estimator whose expectation value, derived from data fitting, is equivalent to the true value of this parameter.}, then a bound, called the Cramer-Rao bound 
\begin{align}
\delta^2\theta \geq \frac{1}{F_{\theta}},
\end{align}
always holds, where $F_{\theta}$ is called the Fisher information, to be defined later in this section. In fact, the equality can always be shown to hold in the asymptotic limit by using a standard statistical tool called maximum likelihood estimation (MLE)~\footnote{In this case, MLE is based on sampling independent and identically distributed outcomes from measuring the final state. Then a best-fit to a known form of the probability distribution (to be used in Fisher information) is achieved by maximizing the \textit{likelihood function} for the recorded measurement outcomes. The asymptotic limit refers to the number of data points, from which to infer the probability distribution, going to infinity. For more details, see \cite{andersen1970asymptotic, le1990maximum}.}.
\subsubsection{Fisher information}
The Fisher information is dependent only on $P(x;\theta)$, which is the probability that the final state $\rho_{\theta}$ in a single-parameter estimation protocol (with unknown parameter $\theta$) undergoes a projective measurement $\ket{x}\bra{x}$ with measurement outcome $x$. Suppose we restrict to only allowing projective measurements~\footnote{All the results we mention hold for general positive-operator valued measure or POVM, but we only concentrate on projective measurements in this chapter.}. Then the probability of $\rho_{\theta}$ having some measurement outcome $x$ is
\begin{align}
P(x;\theta)=\tr(\ket{x}\bra{x} \rho_{\theta}),
\end{align}
where the normalisation $\sum_x \ket{x}\bra{x}=\mathbf{1}$~\footnote{If $x$ is a continuous variable, like a measurement of position, an integral can be used instead of the discrete sum in the normalisation.}. Then the Fisher information $F_{\theta}$ for a given process with a single unknown parameter $\theta$, known initial state $\rho_0$ and projective measurements onto $\{\ket{x}\}$ is defined by
\begin{align}
F_{\theta} &=\sum_x \frac{1}{P(x;\theta)} \left(\frac{\partial P(x;\theta)}{\partial \theta}\right)^2.
\end{align}
Now $F_{\theta}$ is dependent on the measurement chosen, so it is not necessarily true that any Fisher information for a given initial state gives the true lowest bound. We thus want to optimise over all possible
measurements. We define $I_{\phi}\equiv \text{max} F_{\phi}$ as the maximum Fisher information with respect to all measurements and it is called the \textit{quantum} Fisher information. 
\subsubsection{Quantum Fisher information}
To find the maximum value of Fisher information with respect to any measurements (i.e. the quantum Fisher information), it is sufficient to find an upper bound to Fisher information that is independent of any measurement. Suppose we begin with a state $\rho_0$ that is passed through an interferometer with one of the arms subject to an unknown parameter $\theta$ (e.g. via a phase plate). Just before this state is measured as it passes through the interferometer, it evolves to state $\rho_{\theta}$. Suppose we choose to measure a quantity $x$ (e.g. particle number), using the projective measurement $\ket{x}\bra{x}$. The Fisher information depends only on $P(x;\theta)$ and its first derivative. We can also rewrite the Fisher information in terms of a Hermitian operator $\lambda_{\theta}$ called the standard logarithmic derivative (SLD) \footnote{For pure state inputs. the eigenvectors of the SLD can be used directly to find optimal measurements \cite{parisreview}.}, that is defined by the following equation
\begin{align} \label{eq:SLDdef}
\frac{\partial \rho_{\theta}}{\partial \theta}=\frac{\lambda_{\theta}\rho_{\theta}+\rho_{\theta}\lambda_{\theta}}{2}.
\end{align}
Using the Cauchy-Schwarz inequality, it can be shown that the Fisher information is upper-bounded
\begin{align} \label{eq:fineq2}
F(\theta) \leq \tr(\rho_{\theta} \lambda_{\theta}^2) \equiv I_{\theta},
\end{align}
where $\tr(\rho_{\theta} \lambda_{\theta}^2)$ is  independent of the measurement and is thus an ultimate limit to the possible precision in a single-parameter estimation scheme. For pure states, we can choose the SLD to be $\lambda_{\theta}=2 (\partial \rho_{\theta})(\partial \theta)$ which easily satisfies the definition in Eq.~\eqref{eq:SLDdef} since $\rho_{\theta}^2=\rho_{\theta}$. The Cramer-Rao bound using Fisher information can now be extended to
\begin{align}
\delta^2\theta \geq \frac{1}{F_{\theta}} \geq \frac{1}{I_{\theta}},
\end{align}
known as the \textit{quantum} Cramer-Rao bound. When this new lower bound is reached, the estimation scheme (the initial state and measurement) is called \textit{optimal}. The state that can achieve this is defined as the \textit{optimal} state. It can be shown that there is always a measurement that can saturate this inequality (i.e. the quantum Fisher information is equivalent to the Fisher information) \cite{parisreview, braunsteincaves}. However, showing there is a theoretical solution does not provide a recipe for constructing experimentally accessible measurements that can also achieve this equality. One way to test for experimentally accessible schemes is to directly find the Fisher information with respect to a given experimentally accessible measurement. This is the method we adopt later for $SU(2)$ estimation. 

Before moving on to the multi-parameter estimation setting, we first take a look at some photonic states that are useful in single-parameter estimation. 
\subsubsection{Interesting states in single-parameter estimation}
We restrict ourselves to discrete variable photonic states~\footnote{For example, squeezed states play a very prominent role in continuous variable metrology and can achieve Heisenberg scaling. This lies outside the current scope of our analysis.} and take as illustrative examples three classes of these states achieving a scaling $\mathcal{O}(1/N^2)$ in the absence of photon loss or decoherence~\footnote{In the presence of photon loss and decoherence, different optimal states must be found compared to the ideal case. For example, the presence of photon loss has been considered in \cite{dorner2009optimal, knysh2011scaling, lee2009optimization}.}. These are the $N00N$ states, Holland-Burnett (or sometimes double-Fock) states and the Yurke states. We want to address the question of how these important states fare in multi-parameter estimation schemes, so we first briefly summarise their features in single-parameter estimation schemes. \\

\textit{$N00N$ states}. 
$N$-particle $N00N$ states are an equal superposition between an $N$-particle state in two different modes (or two different arms of a Mach-Zehnder interferometer), see Eq.~\eqref{eq:noondef}. It is able to saturate the tightest known bound to precision, reaching the Heisenberg limit by making number-counting measurements through the use of photon-number resolving detectors \cite{hofmann2009}. Thus, these are optimal states for metrology with the Fisher information and quantum Fisher information coinciding at
\begin{align}
F=I=\frac{1}{N^2}.
\end{align}
Unlike some other states considered in single-parameter estimation (like Yurke states), the $N00N$ state also has the advantage of its Fisher information being independent of the phase that is being estimated, hence obviating the need for adaptive schemes. Despite these advantages, there are some major experimental shortcomings. For example, $N00N$ states with high $N$ are notoriously difficult to generate for photonic systems, with the current record standing at $N=5$ \cite{highnoon}. Furthermore, these states no longer become sensitive to the unknown parameter with the loss of even a single photon. Thus, with current technology,  these make interesting but impractical states for larger values of $N$. \\

\textit{Holland-Burnett states.} $N$-particle Holland-Burnett (HB) states $\ket{HB}$ are defined as states with $N/2$ particles in each mode or
\begin{align}
\ket{HB} \equiv \ket{\frac{N}{2}, \frac{N}{2}}.
\end{align}
It can be shown that the best precision attainable by these states can also be achieved (i.e. $F=I$) using number-counting measurements \cite{hofmann2009}. Although it can exhibit Heisenberg scaling, it is a sub-optimal state since it is a factor of 2 greater the Heisenberg limit
\begin{align}
F=I=\frac{2}{N(N+2)}.
\end{align}
The main advantage of these states is their experimental accessibility \cite{datta2011quantum, thomas2011}. In addition, there are also much more resistant to photon loss compared to $N00N$ states. \\

\textit{Yurke states.} $N$-particle Yurke states $\ket{\text{Yurke}}$ are defined as
\begin{align}
\ket{\text{Yurke}}=\frac{1}{\sqrt{2}}\left(\ket{\frac{N}{2}-1, \frac{N}{2}+1}+\ket{\frac{N}{2}+1, \frac{N}{2}-1}\right).
\end{align}
Yurke states can display Heisenberg scaling, but in the best case, attains a precision with a factor of 4 greater than the Heisenberg limit. In the Mach-Zehnder interferometer setting we have introduced, if we make final measurements in photon number difference between the two modes, the Fisher information is dependent on the unknown phase $\theta$ 
\begin{align}
F=\frac{\cos^2\theta+\sin^2\theta(N^2/4+N/2-1)}{(\sin\theta+\cos\theta\sqrt{N(N/2+1)/2})^2}.
\end{align}
Here the highest precision is attained at $\theta=0$, giving $F=4/(N(N+2))$. 

\subsubsection{Multi-parameter estimation}
The formalism for Fisher information and quantum Fisher information can be generalised to the multi-parameter estimation setting where both quantities become matrix quantities. There is also a corresponding quantity to capture precision with multiple unknown parameters as well as a generalised Cramer-Rao bound. Unlike in single-parameter estimation, it is not always possible to find measurements that saturate the generalised quantum Cramer-Rao bound. 
\subsubsection{Fisher information matrix}
For multi-parameter estimation, the relevant probability distributions now depend upon a family of parameters $\underline{\theta}\equiv\{\theta_{\alpha}\}$. Thus, we need to generalise the Fisher information to a matrix, called the Fisher information matrix. We define the Fisher information matrix $F_{\alpha \beta}$ as 
\begin{align} \label{eq:fisherdefinition}
F_{\alpha \beta} &=\sum_x  (\frac{\partial \log P(x; \underline{\theta})}{\partial \alpha} ) (\frac{\partial \log P(x; \underline{\theta})}{\partial \beta} ) P(x; \underline{\theta}) \nonumber \\
         &=\sum_x \frac{1}{P(x; \underline{\theta})} (\frac{\partial P(x; \underline{\theta})}{\partial \alpha} ) (\frac{\partial P(x; \underline{\theta})}{\partial \beta} ) 
\end{align}
where $P(x; \underline{\theta})$ is the probability of measuring observable $x$ satisfying $\sum_x P(x; \underline{\theta})=1$ and $\underline{\theta}$ denotes the set of parameters one recovers from the measurements from which to estimate the parameters of $U$. For convenience we use the notation $\partial/\partial \theta_{\alpha} \equiv \partial /\partial \alpha$ from now on. 

It can be shown that a generalised Cramer-Rao bound in the multi-parameter case also holds, which states that the covariance matrix of the parameters $\{ \theta_{\alpha} \}$, denoted $C_{\alpha \beta}$,  is lower-bounded by the corresponding inverse Fisher information matrix
\begin{equation}
C^{\theta}_{\alpha \beta} \geq F^{-1}_{\alpha \beta},
\end{equation}
where the covariance matrix is defined as $C^{\theta}_{\alpha \beta} \equiv \langle \theta_{\alpha}\theta_{\beta}\rangle-\langle \theta_{\alpha}\rangle \langle \theta_{\beta}\rangle$ and $\langle \theta_{\alpha}\rangle$ denotes an average over all measurements of $\theta_{\alpha}$. It is then natural to take the trace of the covariance matrix to capture the net precision in estimating multiple parameters to get
\begin{equation}\label{eq:traceineq}
\tr(C^{\theta}_{\alpha \beta}) \geq \tr(F^{-1}_{\alpha \beta}),
\end{equation}
where now $\tr(F^{-1}_{\alpha \beta})$ becomes the key quantity to study. This lower bound can always be saturated using the MLE, like in the single-parameter estimation case. The quantum Fisher information matrix $I_{\alpha \beta}$ plays a similar role to quantum Fisher information, but is now a matrix
\begin{align}
I_{\alpha \beta} \equiv \text{Re}[\tr(\rho_f \lambda_{\alpha} \lambda_{\beta})],
\end{align}
where $\rho_f$ is the final state, before the final measurement, dependent upon all the unknown parameters $\{\theta_{\alpha} \}$ and the symmetric logarithmic derivatives for pure states is defined in the same way in the single-parameter case as
\begin{align}
\lambda_{\alpha}=2\frac{\partial \rho_f}{\partial \alpha}. 
\end{align}
It has been shown that the inequality $F(\theta)\geq I(\theta)$ always holds in the single-parameter case. Similarly, there is a generalised matrix inequality \cite{parisreview} that is shown to hold
\begin{align}
F_{\alpha \beta} \geq I_{\alpha \beta}.
\end{align}
We can thus extend Eq.~\eqref{eq:traceineq} to a multi-parameter quantum Cramer-Rao bound
\begin{align}
\tr(C^{\theta}_{\alpha \beta}) \geq \tr(F^{-1}_{\alpha \beta}) \geq \tr(I^{-1}_{\alpha \beta}),
\end{align}
and the minimum value of $\tr(I^{-1}_{\alpha \beta})$ will quantify the highest precision multi-parameter estimation. However, unlike in single-parameter estimation where a measurement always exists that saturates the Cramer-Rao bound, a measurement does not always exist to saturate the multi-parameter Cramer-Rao bound. We examine this in section ~\ref{sec:QFI}.
\subsection{Protocol with multi-photon probe}
\label{sec:protocol}
In identifying unknown optical processes, the traditional approach has been the use of process tomography, which relies on single-photon probes. However, lessons from single-parameter estimation tell us that even in characterising a single unknown parameter, single-photon probes are only capable of achieving the shot noise precision. 

Luckily, we also know from single-parameter estimation that by using multi-photon probes, a much greater precision, when given the same number of photons, may be achieved, like reaching the Heisenberg scaling. This would be especially useful in the probing of highly light-sensitive samples. The first experiment to demonstrate such a multi-photon probe scheme was recently proposed \cite{hugotomo}, which served as our initial inspiration. However, a Heisenberg scaling was only shown for very small photon numbers and not using the language of Fisher information. This makes the result difficult to generalise and to interpret. Here we will make use of this multi-photon probe estimation protocol and provide the first theoretical justification for these results using the language familiar from single-parameter estimation. 

It is useful first to look at how our unknown unitary can be recovered in the single-photon probe scheme. In a general two-mode linear optical process, we can consider the two modes $a$ and $b$ to be the two different polarisation degrees of freedom (e.g. horizontal/vertical polarisations) and can be represented by $\ket{n_a,n_b}$, where $n_a$ and $n_b$ represents particle number in modes $a$ and $b$ respectively. We can describe our initial single-photon probe state as $\ket{\psi}=c_{a}\ket{1,0}+c_{b}\ket{0,1}$ where $c_a$, $c_b$ are complex numbers. Then a general two-mode linear optical process, characterised by an $SU(2)$ matrix $u$, takes $\ket{\psi} \rightarrow c'_{a}\ket{1,0}+c'_{b}\ket{0,1}$. This matrix can be represented by
\begin{align}
u=\begin{pmatrix}
a_1+i a_2 & a_3+i a_4 \\
-a_3+i a_4 & a_1-ia_2
\end{pmatrix},
\end{align}
where $a_j$ for $j=1,2,3,4$ are real numbers satisfying $a_1^2+a_2^2+a_3^2+a_4^2=1$, coming from the unitarity condition. 

Since there are three unknown real parameters, to identify them we need at least three different kinds of measurements. In this case, one can rotate the initial state and make the final measurement in three different polarisation bases. The first pair of polarisation degree of freedom is the horizontal and vertical polarisation (collectively called $HV$), represented by states $\ket{1,0}_{HV}$ and $\ket{0,1}_{HV}$ respectively. Then we also have the diagonal and antidiagonal polarisation ($DA$), defined respectively by $\ket{1,0}_{DA}=(\ket{1,0}+\ket{0,1})/\sqrt{2}$ and $\ket{0,1}_{DA}=(\ket{0,1}-\ket{1,0})/\sqrt{2}$. Finally we have the right and left circular polarisation ($RL$), represented by $\ket{1,0}_{RL}=(\ket{0,1}-i\ket{1,0})/\sqrt{2}$ and $\ket{0,1}_{RL}=(\ket{0,1}+i\ket{1,0})/\sqrt{2}$. It can be shown that starting from the probe state in the horizontal basis, the probability of finally measuring a single-photon in the horizontal basis is given by $p_{H}=a^2_1+a^2_2$. Likewise, in starting from a single-photon in the diagonal/right polarisation, the probability that this photon is still in the diagonal/right polarisations is given respectively by $p_D=a^2_1+a^2_4$ and $p_R=a^2_1+a^2_3$. From here all three unknown parameters can be found, but the precision in determining these probabilities is restricted by the shot noise limit. 

The multi-photon probe protocol is a straightforward generalisation of the single-photon probe method. A \textit{single run} of this protocol is defined as inputting three $N$-particle states (of the same form but in different polarisation bases) into an unknown unitary (which we call $U$) before making photon number measurements. For example, if the initial state in the $HV$ basis is $\ket{M, N-M}_{HV}$ (where $M$ is an integer $0\leq M\leq N$), then the other two states are $\ket{M, N-M}_{DA}$ and $\ket{M, N-M}_{RL}$. State $\ket{M, N-M}_{HV}$ passes through the unknown unitary, which we call $U$, before photon counting measurements are made in the $HV$ basis. The two other states $\ket{M, N-M}_{DA}$ and $\ket{M, N-M}_{RL}$ follow the same procedure, but with photon measurements made in basis $DA$ and $RL$ respectively. These measurement probabilities then contain enough information to recover the full unitary matrix. 

In this chapter, we use the Fisher information matrix and quantum Fisher information matrix to characterise the precision to which we can estimate the unknown unitary. We show later that both these quantities are additive with respect to initial product states (like their single-parameter estimation counterparts). This means that by adding the Fisher information matrix corresponding to the procedure using the $HV$ basis to that with respect to the $DA$ and $RL$ basis, we have total Fisher information matrix for a single run of this protocol (i.e. three identical protocols in three polarisation bases). The same is true for the quantum Fisher information matrix. 
\section{Mapping between photonic and spin states, unitaries and measurements}
\label{sec:mapping}
We establish a mapping between an $N$-particle two-mode linear optical process and an $N$ spin-$1/2$ particle process. This allows for the first time a clear analogy to be made between spin and photonic processes in multi-parameter estimation. In addition to making our derivation of the quantum Fisher information simpler in section \ref{sec:QFI}, this mapping also offers clearer insight on the correspondence between seemingly different physical processes and their analogous roles in precision estimation. We begin with the following one-to-one correspondence between an $N$-particle two-mode photonic state $\ket{M, N-M}$~\footnote{The two modes of the photonic state typically correspond to two polarisation or spatial degrees of freedom (e.g. horizontal/vertical polarisation).} and an $N$-particle symmetric spin-$1/2$ state $\ket{\xi_0}_{\text{spin}}$ (also known as Dicke states) \cite{dicke},
\begin{align}\label{eq:mapping}
\ket{M, N-M} \longrightarrow \frac{1}{\sqrt{\begin{psmallmatrix}
N \\
M
\end{psmallmatrix}}} \sum_j \Pi_j \left(\ket{\uu}^{\otimes M} \otimes \ket{\dd}^{\otimes (N-M)}\right) \equiv \ket{\xi_0}_{\text{spin}},
\end{align}
where the summation $\sum_j \Pi_j$ is over all the possible permutations of the product states and this is known as the symmetrisation of the spin state. For concreteness, we choose $\ket{\uu}$, $\ket{\dd}$ to be the spin-up and spin-down eigenstates of $\sigma_z$~\footnote{Here we choose two-mode photonic states in the horizontal/vertical polarisation to correspond to Dicke states written in the $\sigma_z$ basis. If we change into a different polarisation basis, we must make corresponding changes of basis in the Dicke states. For example, the diagonal/anti-diagonal polarisation basis would correspond to the $\sigma_x$ basis and the right/left circular polarisation basis would correspond to the $\sigma_y$ basis.} Another way to represent this correspondence is
\begin{align} \label{eq:mapping2}
&\frac{1}{\sqrt{\begin{psmallmatrix}
N \\
M
\end{psmallmatrix}}} \sum_j \Pi_j \frac{(a^{\dagger})^M}{\sqrt{M!}} \otimes \frac{(b^{\dagger})^{N-M}}{\sqrt{(N-M)!}} \ket{00}=\frac{(a^{\dagger})^M}{\sqrt{M!}} \otimes \frac{(b^{\dagger})^{N-M}}{\sqrt{(N-M)!}} \ket{00} \nonumber \\
&\longrightarrow \frac{1}{\sqrt{\begin{psmallmatrix}
N \\
M
\end{psmallmatrix}}} \sum_j \Pi_j \left((a^{\dagger}_{\uu})^{\otimes M} \otimes (a^{\dagger}_{\dd})^{\otimes (N-M)}\right)\ket{0...0},
\end{align}
where $a^{\dagger}$, $b^{\dagger}$ are the creation operators for the first and second photonic modes respectively. The creation operators corresponding to the up and down spin states are represented by $a^{\dagger}_{\uu}$ and $a^{\dagger}_{\dd}$, which satisfy the anticommutation relations $\{a^{\dagger}_{\uu},a_{\uu}\}=1=\{a^{\dagger}_{\dd},a_{\dd}\}$ and where all other anticommutation relations vanish. 

For example, in the single-particle case~\footnote{From Eq.~\eqref{eq:mapping2} we see that the mapping for $M$-particle excitations is $a^{\dagger M}/\sqrt{M!} \rightarrow a_{\uu}^{\dagger M}$ and $b^{\dagger M}/\sqrt{M!} \rightarrow a_{\dd}^{\dagger M}$.}, we have the correspondence $a^{\dagger} \ket{00} \longrightarrow a^{\dagger}_{\uu} \ket{0}=\ket{\uu}$ and $b^{\dagger} \ket{00}\longrightarrow a^{\dagger}_{\dd}\ket{0}=\ket{\dd}$. This is the mapping between the two photonic modes and the two spin degrees of freedom, as shown in Fig.~\ref{map}. We can use this mapping to show that the $N$-particle $N00N$ states $(\ket{N0}+\ket{0N})/\sqrt{2}$ map to $N$-particle GHZ~\footnote{Greenberger-Horne-Zeilinger.} states $(\ket{0}^{\otimes N}+\ket{1}^{\otimes N})/\sqrt{2}$. This can help explain why both $N00N$ states and GHZ states have been found to be optimal states in single-parameter estimation \cite{toth}, though appearing in different contexts. Another example is the correspondence between $N$-particle Holland-Burnett states $\ket{N/2, N/2}$ and symmetric Dicke states with $N/2$ excitations \cite{toth}.  
\begin{figure}[ht!]
\centering
\label{map}
\includegraphics[scale=0.5]{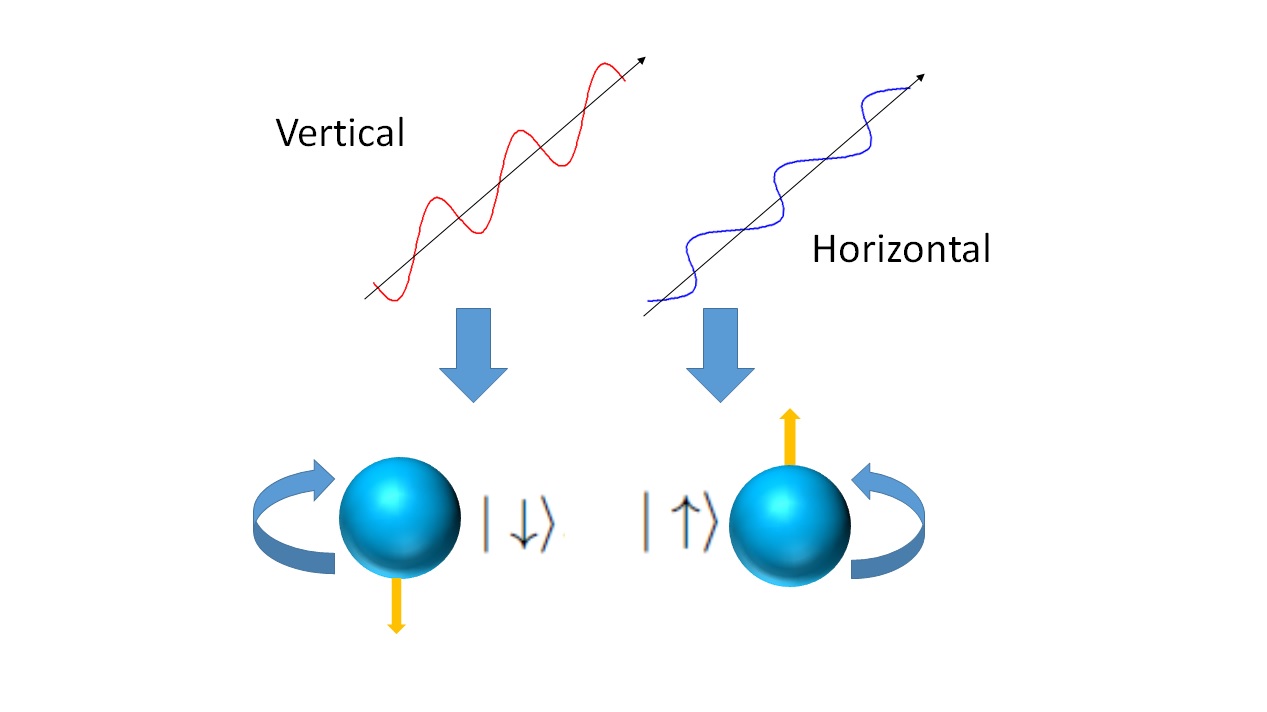}
\caption[\textit{Map between photonic and spin degrees of freedom}.]{\label{map}\textit{Map between photonic and spin degrees of freedom}. Map between photonic polarisation degrees of freedom and spin orientation for a spin-$1/2$ particle. Here using example of horizontal/vertical polarisation mapping to spin-up/spin-down along the $z$-axis.}
\end{figure}
Using this mapping, it turns out we can also describe a transformation of the two-mode photonic state under unitary operator $U$ in terms of the evolution of $N$ spin-$1/2$ particles, each transforming under the $2 \times 2$ represented by $\mathcal{M}$, where $\mathcal{M}$ and $U$ are related by 
\begin{align} \label{eq:spinphoton}
\begin{pmatrix}
\tilde{a}^{\dagger} \\
\tilde{b}^{\dagger}
\end{pmatrix} \equiv 
\begin{pmatrix}
U a^{\dagger} U^{\dagger} \\
U b^{\dagger} U^{\dagger}
\end{pmatrix}=
\begin{pmatrix}
\alpha & \beta \\
\gamma & \delta
\end{pmatrix} 
\begin{pmatrix}
a^{\dagger} \\
b^{\dagger}
\end{pmatrix} \equiv
\mathcal{M}^{T} \begin{pmatrix}
a^{\dagger} \\
b^{\dagger}
\end{pmatrix},
\end{align}
where $U$ is confined to a linear optical process. Here $\mathcal{M}$ can be interpreted as acting on a single-photon state. 

\begin{figure}[ht!]
\centering
\includegraphics[scale=0.5]{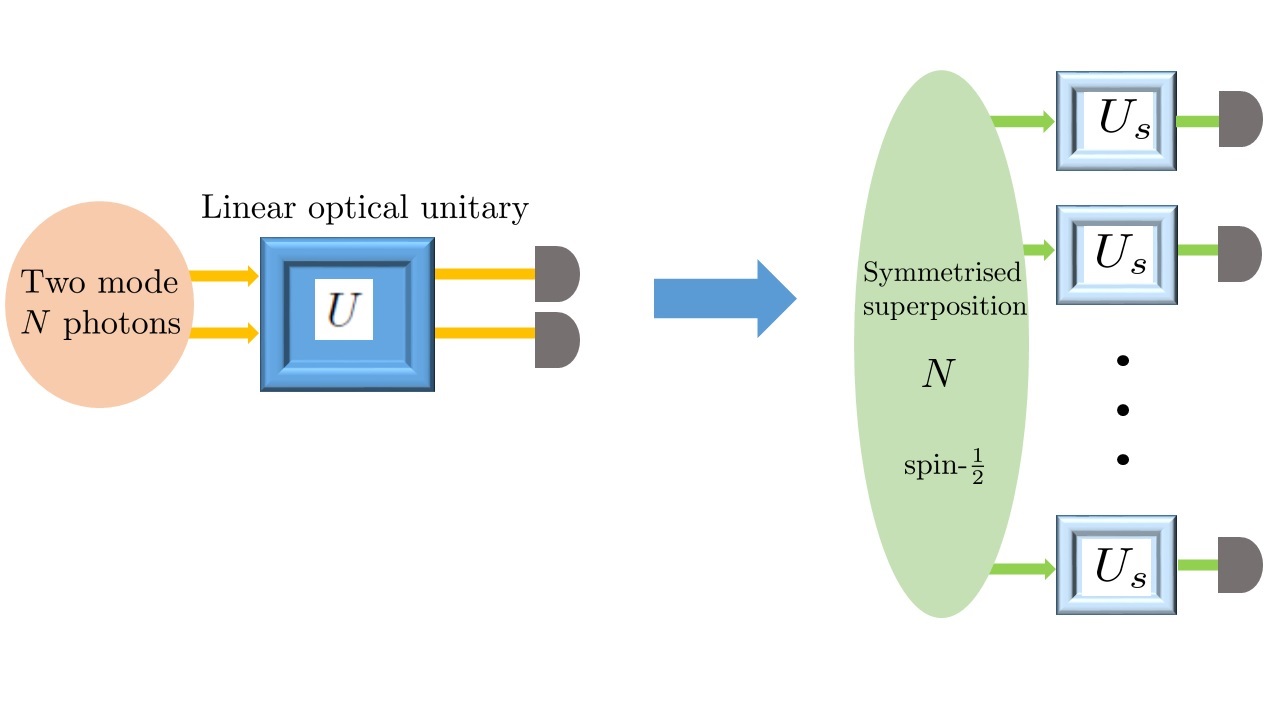}
\caption[\textit{Linear-optical process and evolution of spin degree of freedom}.]{\label{operatormap}\textit{Linear-optical process and evolution of spin degree of freedom}. Map between $N$-particle two-mode linear optical process captured by $U$ and the analogous process with $N$ spin-$1/2$ particles each undergoing evolution under $U_s$. The $N$-particle photons begin in two modes correspond to a symmetric superposition of $N$ spin-$1/2$ particles. Measurement of $R$ (or $N-R$) photons in the first (or second) mode correspond to measurement of $R$ spin-up (or spin-down) particles in the spin picture.}
\end{figure}

Since we have the corespondence between the creation operators of the photonic and spin states $a^{\dagger} \rightarrow a^{\dagger}_{\uu}$, $b^{\dagger} \rightarrow a^{\dagger}_{\dd}$, after unitary evolution we have the correspondence $U\ket{10}=\tilde{a}^{\dagger} \rightarrow U_s a^{\dagger}_{\uu} U_s^{\dagger} \equiv \tilde{a}^{\dagger}_{\uu}$ and $U\ket{01}=\tilde{b}^{\dagger} \rightarrow U_s a^{\dagger}_{\dd} U_s^{\dagger} \equiv \tilde{a}^{\dagger}_{\dd}$, where $U_s$ is the unitary operator acting on the spin degrees of freedom. To find the correspondence between $U_s$ and $U$, we note the relations $U a^{\dagger} U^{\dagger} \ket{00}=\alpha a^{\dagger} \ket{00}+\beta b^{\dagger} \ket{00} \rightarrow \alpha \ket{\uu}+\beta \ket{\dd}=U_s a^{\dagger}_{\uu} U_s^{\dagger} \ket{0}=U_s a^{\dagger}_{\uu} \ket{0}=U_s \ket{\uu}$ and $U b^{\dagger} U^{\dagger} \ket{00}=\gamma a^{\dagger} \ket{00}+\delta b^{\dagger} \ket{00}\rightarrow \gamma \ket{\uu}+\delta \ket{\dd}=U_s \ket{\dd}$. Thus we can write 
\begin{align}
\begin{pmatrix}
\tilde{a}^{\dagger}\\
\tilde{b}^{\dagger}
\end{pmatrix}\equiv
\begin{pmatrix}
U a^{\dagger} U^{\dagger} \\
U b^{\dagger} U^{\dagger}
\end{pmatrix} \longrightarrow 
\begin{pmatrix}
U_s a^{\dagger}_{\uu} U_s^{\dagger} \\
U_s a^{\dagger}_{\dd} U_s^{\dagger}
\end{pmatrix}=
\mathcal{M}^T 
\begin{pmatrix}
a^{\dagger}_{\uu} \\
a^{\dagger}_{\dd}
\end{pmatrix}.
\end{align}
We can find a matrix representation for $U_s$ by choosing a representation for the spin eigenstates
\begin{align}
\ket{\uu} \equiv
\begin{pmatrix}
1 \\
0
\end{pmatrix}, \, \, \, \, 
\ket{\dd} \equiv
\begin{pmatrix}
0 \\
1
\end{pmatrix}.
\end{align}
Inserting this representation into the relations derived earlier $U_s \ket{\uu}=\alpha \ket{\uu}+\beta \ket{\dd}$ and $U_s \ket{\dd}= \gamma \ket{\uu}+\delta \ket{\dd}$, we can see that a $2 \times 2$ matrix representation of $U_s$ when it acts on the spin states is equivalent to the matrix $\mathcal{M}$. We can now see correspondence between the evolution of the photonic two mode state and the evolution of the spin state
\begin{align}
U\ket{M, N-M} \longrightarrow U_s^{\otimes N} \ket{\xi_0}_{\text{spin}},
\end{align}
where a matrix representation of $U_s$ is equivalent to $\mathcal{M}$. This means that there is an equivalence between the protocol of $N$-particle photonic states in two modes undergoing a linear optical process represented by $U$ and an $N$-particle spin-$1/2$ particles with \textit{each} particle evolving under evolution represented by $U_s$ (see Fig.~\ref{operatormap} for a schematic representation of this correspondence). 

The final part of the protocol involves measurement of the final state. We note that a projective measurement $\ket{R,N-R}\bra{R,N-R}$ on the photonic state (where $R$ is an integer $0\leq R\leq N$) correspond in the spin picture to measurement of $R$ spin-up particles and $N-R$ spin-down particles, where there are $N!/(R!(N-R)!)$ equivalent such states with different permutations. 

Lastly, we note that it is also possible to map between multi-mode photonic states and spin states with higher spin values. For example, an $N$-particle $n+1$ mode photonic state maps to an $N$-particle symmetric spin-$n/2$ spin state where $n\geq 1$ is an integer. The mapping between these states is straightforward. For a spin-$n/2$ particle, there are $n+1$ possible spin values $-n/2, -n/2+1,...,n/2-1, n/2$. Each of the $n+1$ photonic modes correspond to one of these $n+1$ spin values. In the $n=1$ case we have seen that the first photonic mode corresponds to spin-up (or spin value $1/2$) and the second photonic mode corresponds to spin-down (or spin value $-1/2$.). In the case $n=2$, there is a mapping between $3$-mode photonics to superpositions of the spin-$1$ state, where the possible spin values are $-1,0,1$. Here for instance, the generalised $N00N$ states $(\ket{00N}+\ket{0N0}+\ket{N00})/\sqrt{3}$ correspond to the generalised GHZ state $(\ket{-1}^{\otimes N}+\ket{0}^{\otimes N}+\ket{1}^{\otimes N})/\sqrt{3}$. Another example is the generalised Holland-Burnett states $\ket{N/3, N/3, N/3}$, which correspond to the symmetrised sum $\propto \sum_j \Pi_j \left(\ket{-1}^{\otimes N/3}\ket{0}^{\otimes N/3} \ket{1}^{\otimes N/3}\right)$. This procedure follows straightforwardly for higher $n$ examples.  
\section{Fisher information matrix formalism for unitary-estimation}
\label{sec:FImatrix}
We develop in this section the tools to calculate the Fisher information matrix, with respect to photon number-counting measurements in our protocol introduced in section ~\ref{sec:protocol}. Photon number-counting measurements are chosen since they are known to be experimentally accessible \cite{numbercount}. We begin in section ~\ref{sec:prob} by computing the probability distributions needed to compute the Fisher information matrix, then develop a method calculate the Fisher information matrix with respect to using the three different polarisation bases in section ~\ref{sec:differentbases}. We also compute the trace of the inverse Fisher information matrix with respect to a new parameterisation in section ~\ref{sec:parameterisation}, which we later use in the multi-parameter Cramer-Rao bound. 
\subsection{Computing probabilities}
\label{sec:prob}
A basic step in our protocol is to input an $N$-particle two-mode photonic state $\ket{\Psi}=\sum_{M=0}^N c_M \ket{M, N-M}$ (where $c_M$ is a number) into an unknown linear optical unitary $U$ before detectors make photon number measurements in each mode. To compute the total Fisher information matrix for this protocol, we need to know the probability $P(\ket{\Psi}, M')$ that our state is finally measured by the photon number detector to be in state $\ket{M', N-M'}$, i.e.
\begin{align} \label{eq:prob1}
P(\ket{\Psi}, M')=|\bra{M', N-M'}U\ket{\Psi}|^2,
\end{align}
where $M$ is an integer $0\leq M \leq N$. Using the Schwinger representation~\footnote{This is a well-known mapping between an $N$-particle two-mode photonic state and a spin-$N/2$ particle, which is very distinct from our mapping in section ~\ref{sec:mapping}. See Appendix~\ref{sec:schwinger}.}, this transition probability for photonic states can be translated into the transition probability between spin-$j$ states with quantum numbers $m$, represented by $\ket{j,m}$, where $j=N/2$ and $m$ takes on values $-j, -j+1,...,j$. The probability distributions written in this way are identified with Wigner D-matrices, which have well-known forms. 

We can now consider $U$ as a $(2j+1) \times (2j+1)$ matrix acting on spin-$j$ states. This can be written in terms of the Euler angle decomposition
\begin{align}
U=e^{i \psi_1 J_z}e^{i \psi_2 J_y}e^{i \psi_3 J_z},
\end{align}
where $\psi_1$, $\psi_2$, $\psi_3$ are the Euler angles and $J_{x,y,z}$ are the total angular momentum operators for spin-$j$ states along the $x,y,z$ basis. These operators obey the commutation relations $[J_i,J_j]=i\epsilon_{ijk}J_k$, where $\epsilon_{ijk}$ is the Levi-Civita symbol. We can thus write Eq.~\eqref{eq:prob1} as
\begin{align}
P(\ket{\Psi}, M')&=|\sum_{M=0}^N c_M\bra{M', N-M'}U\ket{M, N-M}|^2 \nonumber \\
                           &=|\sum_{m=-j}^j c_m \bra{j=N/2, m'=M'-N/2}U\ket{j=N/2, m}|^2 \nonumber \\
                           &=|\sum_{m=-j}^j c_m e^{i(\psi_3-\psi_1)N/2}d^j_{m',m}(\psi_2)|^2,
\end{align}
where $d^j_{m',m}(\psi_2) \equiv \bra{j, m'} \exp(i\psi_2 J_y) \ket{j,m}$ is defined as the Wigner d-matrix \cite{sakurai}.
\subsection{Different measurement bases}
\label{sec:differentbases}
Our unitary-estimator protocol consists of three sets of measurements, each using a different polarisation basis, namely horizontal/vertical ($HV$), diagonal/anti-diagonal ($DA$) and right/left circular ($RL$), introduced in section ~\ref{sec:protocol}. To compute the Fisher information matrix with respect to the different polarisation bases, we observe that the functional form of the probability distributions does not change as the basis is changed. The only thing that changes is the value of the Euler angles and this change can be computed from the rotation matrix relating the two different bases. 

We want to relate the new probability distribution to the old probability distribution when there is a change in basis (e.g. from $HV$ to $DA$) in both the initial state and in the final measurement, from basis $B$ to $B'$. Suppose our initial state is written $\ket{N/2+m, N/2-m}$ and we define an operator $R_{B B'}$ as
\begin{equation}
R_{BB'}\ket{\frac{N}{2}+m, \frac{N}{2}-m}_B=\ket{\frac{N}{2}+m, \frac{N}{2}-m}_{B'}.
\end{equation}
Then the probability of going from state $\ket{\frac{N}{2}+m, \frac{N}{2}-m}_{B'}$ to a state $\ket{M, N-M}_{B'}$ can be calculated in terms of the old basis $B$, but with a change in the evolution operator
\begin{align}
P(M',m) &=\left|\bra{M', N-M'}_{B'} U \ket{\frac{N}{2}+m, \frac{N}{2}-m}_{B'}\right|^2 \nonumber \\
                &=\left|\bra{M', N-M'}_B U'\ket{\frac{N}{2}+m, \frac{N}{2}-m}_B\right|^2,
\end{align}
where $U' \equiv R_{BB'}^{\dagger} UR_{BB'}$. Since $U'$ is just another general rotaton, it also has its own Euler decomposition
\begin{equation}\label{eq:uprime}
U'=e^{i \psi'_1 J_z}e^{i \psi'_2 J_y}e^{i \psi'_3 J_z},
\end{equation}
where the original evolution operator has a representation in terms of the old Euler angles as $U=e^{i \psi_1 J_z}e^{i \psi_2 J_y}e^{i \psi_3 J_z}$. Therefore, by substituting the original Euler angles $\psi_j$ with the new Euler angles $\psi'_j$ (where $j=1,2,3$) in the original probability distribution computed in basis $B$, we have the new probability distribution with respect to basis $B'$. 

To compute angles $\{\psi'_1, \psi'_2, \psi'_3\}$ in terms of $\{\psi_1, \psi_2, \psi_3 \}$, we first introduce the matrix $\mathcal{M'}$, defined by 
\begin{align} \label{eq:mdef}
\mathcal{M'}^T \begin{pmatrix}
a^{\dagger} \\
b^{\dagger}
\end{pmatrix} \equiv \begin{pmatrix}
U' a^{\dagger} U'^{\dagger} \\
U' b^{\dagger} U'^{\dagger}
\end{pmatrix}=\begin{pmatrix}
e^{i\frac{\psi'_1+\psi'_3}{2}} \cos \left(\frac{\psi'_2}{2} \right) & -e^{i\frac{\psi'_3-\psi'_1}{2}} \sin \left(\frac{\psi'_2}{2} \right) \\
e^{-i\frac{\psi'_3-\psi'_1}{2}} \sin \left(\frac{\psi'_2}{2} \right) & e^{-i\frac{\psi'_1+\psi'_3}{2}} \cos \left(\frac{\psi'_2}{2} \right) 
\end{pmatrix} \begin{pmatrix}
a^{\dagger} \\
b^{\dagger}
\end{pmatrix},
\end{align}
where the last expression is derived using Eq.~\eqref{eq:uprime} (see Appendix~\ref{sec:schwinger}). From Eq.~\eqref{eq:mdef} we see that matrix $\mathcal{M'}$ also has an Euler decomposition. 
\begin{align}
\mathcal{M'}=e^{i \frac{\psi'_1}{2} \sigma_z}e^{i \frac{\psi'_2}{2} \sigma_y}e^{i \frac{\psi'_3}{2} \sigma_z} \equiv \begin{pmatrix}
\mathcal{M'}_{11} & \mathcal{M'}_{12}\\
\mathcal{M'}_{21} & \mathcal{M'}_{22}
\end{pmatrix} 
\end{align}
which has the same Euler angles as those appearing in $U'$. It is then straightforward to show 
\begin{align}\label{eq:psiprimes}
\psi_1' &=\tan^{-1}\left(\frac{\text{Im}(\mathcal{M'}_{11})}{\text{Re}(\mathcal{M'}_{11})}\right)-\tan^{-1}\left(\frac{\text{Im}(\mathcal{M'}_{21})}{\text{Re}(\mathcal{M'}_{21})}\right) \nonumber \\
\psi_2'&=2\cos^{-1}(|\mathcal{M'}_{11}|) \nonumber \\
\psi_3' &=\tan^{-1}\left(\frac{\text{Im}(\mathcal{M'}_{11})}{\text{Re}(\mathcal{M'}_{11})}\right)+\tan^{-1}\left(\frac{\text{Im}(\mathcal{M'}_{21})}{\text{Re}(\mathcal{M'}_{21})}\right),
\end{align}
This means we only need to find the two matrix elements $\mathcal{M'}_{11}$ and $\mathcal{M'}_{21}$ to find the new Euler angles. To find these matrix elements, it is sufficient to find how $a^{\dagger}$ transforms under
\begin{equation} \label{eq:aprime}
U' a^{\dagger} U'^{\dagger}=R^{\dagger} U (R a^{\dagger} R^{\dagger}) U^{\dagger} R.
\end{equation}
We first define a unitary matrix $r$ as 
\begin{align} \label{eq:R}
 \begin{pmatrix}
R a^{\dagger} R^{\dagger} \\
R b^{\dagger} R^{\dagger}
\end{pmatrix}=r
\begin{pmatrix}
a^{\dagger} \\
b^{\dagger}
\end{pmatrix}=\begin{pmatrix}
\alpha' & \beta' \\
-e^{i\zeta}\beta'^*&e^{i\zeta}\alpha'^*
\end{pmatrix} \begin{pmatrix}
a^{\dagger} \\
b^{\dagger}
\end{pmatrix},
\end{align}
where $|\alpha'|^2+|\beta'|^2=1$ and $\zeta$ is a real number. We can readily derive $\alpha'=1/\sqrt{2}=\beta'$ and $\zeta=0$ for basis change from $HV \rightarrow DA$, since $a^{\dagger}_{DA}\rightarrow (1/\sqrt{2})(a^{\dagger}_{HV}+b^{\dagger}_{HV})$ and  $b^{\dagger}_{DA}\rightarrow (1/\sqrt{2})(a^{\dagger}_{HV}-b^{\dagger}_{HV})$~\footnote{We can see this from the definition $\ket{1,0}_{DA}=a^{\dagger}_{DA} \ket{0,0}=Ra^{\dagger}_{HV} R^{\dagger} \ket{0,0}=(1/\sqrt{2})(\ket{1,0}_{HV}+\ket{0,1}_{HV})=(1/\sqrt{2})(a^{\dagger}_{HV}+b^{\dagger}_{HV}) \ket{0,0}$ and $\ket{0,1}_{DA}=b^{\dagger}_{DA}\ket{0,0}=(1/\sqrt{2})(b^{\dagger}_{HV}-a^{\dagger}_{HV})\ket{00}$.}. Likewise we can derive $\alpha'=-i/\sqrt{2}$, $\beta'=1/\sqrt{2}$ and $\zeta=-\pi/2$ for the basis change $HV \rightarrow RL$~\footnote{This easily follows from the definitions $\ket{1,0}_{RL}=(1/\sqrt{2})(\ket{0,1}_{HV}-i\ket{1,0}_{HV})$ and $\ket{0,1}_{RL}=(1/\sqrt{2})(\ket{0,1}_{HV}+i\ket{1,0}_{HV})$.}. Similarly
\begin{align}\label{eq:Rdagger}
\begin{pmatrix}
R^{\dagger}a^{\dagger} R \\
R^{\dagger}b^{\dagger} R
\end{pmatrix}=r^{\dagger} 
\begin{pmatrix}
a^{\dagger} \\
b^{\dagger}
\end{pmatrix}=\begin{pmatrix}
\alpha'^* & -e^{-i\zeta}\beta' \\
\beta'^*&e^{-i\zeta}\alpha'
\end{pmatrix} \begin{pmatrix}
a^{\dagger} \\
b^{\dagger}
\end{pmatrix}.
\end{align}
Using Eqs.~\eqref{eq:R}, ~\eqref{eq:Rdagger}, ~\eqref{eq:aprime} and ~\eqref{eq:mdef} (by replacing the new Euler angles $\{\psi'_j\}$ with the old Euler angles $\{\psi_j\}$) to calculate $U' a^{\dagger}U'^{\dagger}$, we find
\begin{align} \label{eq:U11}
&\mathcal{M'}_{11}=\cos \left(\frac{\psi_2}{2} \right)\left(|\alpha'|^2e^{i\frac{(\psi_1+\psi_3)}{2}}+|\beta'|^2e^{-i\frac{(\psi_1+\psi_3)}{2}}\right) \nonumber \\
&+\sin\left(\frac{\psi_2}{2}\right)\left(\alpha'^* \beta'e^{-i\frac{(\psi_3-\psi_1)}{2}}-\alpha'\beta'^*e^{i\frac{(\psi_3-\psi_1)}{2}}\right) \nonumber \\
&\mathcal{M'}_{21}=-2i\alpha'\beta'e^{-i\zeta}\cos\left(\frac{\psi_2}{2}\right)\sin\left(\frac{\psi_3+\psi_1}{2}\right) \nonumber \\
&-e^{-i\zeta}\sin\left(\frac{\psi_2}{2}\right)\left(\alpha'^2e^{-i\frac{(\psi_3-\psi_1)}{2}}+\beta'^2e^{i\frac{(\psi_3-\psi_1)}{2}}\right),
\end{align}
where $\alpha$ and $\beta$ is defined by the new basis. Combining Eqs.~\eqref{eq:U11} and ~\eqref{eq:psiprimes},  we can now straightforwardly write the new Euler angles for the basis change $HV \rightarrow DA$ and $HV \rightarrow RL$ (see Appendix~\ref{sec:basischange}). 

In the most general case where the new probability distribution depends on all the new Euler angles $\psi'_1$, $\psi'_2$, $\psi'_3$, the Fisher information transforms under a change in measurement basis as
\begin{align} \label{eq:fprime23}
F'_{kl} =&\sum_{m'=-j=-\frac{N}{2}}^{m'=j=\frac{N}{2}}\frac{1}{p_{m'}(\psi'_1, \psi'_2, \psi'_3)}\frac{\partial p_{m'}(\psi'_1, \psi'_2, \psi'_3)}{\partial \psi_k}\frac{\partial p_{m'}(\psi'_1, \psi'_2, \psi'_3)}{\partial \psi_l} \nonumber \\
           =&\left(F_{11}(\psi'_1, \psi'_2, \psi'_3)W_{kl}^{(11)}+F_{22}(\psi'_1, \psi'_2, \psi'_3)W_{kl}^{(22)}+F_{33}(\psi'_1, \psi'_2, \psi'_3)W_{kl}^{(33)}\right) \nonumber \\
            &+F_{12}(\psi'_1, \psi'_2, \psi'_3)\left(W_{kl}^{(12)}+W_{lk}^{(12)}\right)+F_{13}(\psi'_1, \psi'_2, \psi'_3)\left(W_{kl}^{(13)}+W_{lk}^{(13)}\right) \nonumber \\
&+F_{23}(\psi'_1, \psi'_2, \psi'_3)\left(W_{kl}^{(23)}+W_{kl}^{(23)}\right),
\end{align}
where $p_{m'}$ are the probability distributions and 
\begin{align} \label{eq:wmatrix}
W_{kl}^{(ij)} \equiv \frac{\partial \psi'_i}{\partial \psi_k}\frac{\partial \psi'_j}{\partial \psi_l}.
\end{align} 
To obtain the total Fisher information in our protocol, we simply add the three Fisher information matrices derived with respect to each of the three bases. This comes from the additivity property of the Fisher information matrix (see Appendix~\ref{sec:FIadd}). 
\subsection{Parameterisation}
\label{sec:parameterisation}
So far we have been defining the Fisher information matrix with respect to Euler angle parameters, due to its usefulness in calculating probability distribution functions in terms of Wigner $D$-matrices and the convenience of the Euler angle decomposition. We now introduce a new parameterisation, called the \textit{locally independent parameters}, which is better suited for quantifying precision. The Euler angle representation is a very accessible starting point and we develop the formalism for converting the trace of the inverse Fisher information matrix (needed in the multi-parameter Cramer-Rao bound) with respect to Euler angles into our new parameterisation. 

Suppose we have the Fisher information matrix $F(\{\psi_m\})$ with respect to the set of Euler angles $\{\psi_m\}$. To find the Fisher information matrix $F(\{\alpha_m\})$ with respect to another set of parameters $\{\alpha_m\}$, we can write
\begin{equation} \label{eq:fchange}
F_{ij}(\{\alpha_m\})=\sum_{kl} J_{ki} F_{kl}(\{\psi_m\}) J_{lj},
\end{equation}
where $J_{ij} \equiv \partial \psi_i/\partial \alpha_j$. Thus the trace of the inverse Fisher information matrix transforms as
\begin{align} \label{eq:tracetransform}
\tr(F^{-1}_{ij}(\{\alpha_m\})=\sum_i \sum_{kl} (J^{-1}_{ki})^T J^{-1}_{il}F^{-1}_{lk}(\{\psi_m\})=\sum_{kl} V_{kl} F^{-1}_{lk}(\{\psi_m\}),
\end{align}
where
\begin{equation}
V_{kl} \equiv \sum_i (J^{-1}_{ki})^T J^{-1}_{il}.
\end{equation}
We set the parameters $\{\alpha_m\}$ be our \textit{locally independent parameters}, which can be motivated by the following scenario. 

Let us locally expand about a known $SU(2)$ matrix $\mathcal{M}_0$ using Taylor expansion to linear order which recovers our unknown $SU(2)$ matrix $\mathcal{M}$~\footnote{This is the matrix representation of unitary operator $U_s$ acting on the spin degrees of freedom, defined in section ~\ref{sec:mapping}.}. The higher precision to which this expansion is known, the better estimate we have of $\mathcal{M}$. Let $\mathcal{M}$ be dependent only on the locally independent parameters $\{\alpha_m\}$ and let $\mathcal{M}_0$ be defined in terms of special values of these parameters, denoted $\{\alpha_m^{(0)}\}$. Then, to linear order, $\mathcal{M}$ can be expanded in terms of the Taylor expansion about $\mathcal{M}_0$ as $\mathcal{M}(\{\alpha_m\}) \equiv \mathcal{M}_0(\{\alpha_m\})+\Delta \mathcal{M}(\{\alpha_m\},\{\alpha_m^{(0)}\})=\mathcal{M}_0(\{\alpha_m^{(0)}\})(\mathbf{1}-i\sum_m(i\mathcal{M}^{\dagger}\partial \mathcal{M}/\partial \alpha_m)\vert_{\alpha_m=\alpha_m^{(0)}}(\alpha_m-\alpha_m^{(0)}))$, where $t_{\alpha_m} \equiv i\mathcal{M}^{\dagger}\partial \mathcal{M}/\partial \alpha_m$ can be considered a generator to $SU(2)$. A good parameterisation to use to estimate $\mathcal{M}$ is one where $t_{\alpha_m}$ is independent of the value of $\{\alpha_m\}$ (and thus $t_{\alpha_m}\vert_{\alpha_m=\alpha_m^{(0)}}$ is independent of $\{\alpha_m^{(0)}\}$), which means that the precision of estimation should not depend on which $\mathcal{M}_0$ is used. One simple and natural choice is for $\{t_{\alpha_m}\}$ to be proportional to the Pauli spin matrices $\{\sigma_m\}$. For parameters $\{\alpha_m\}$ that satisfy $t_{\alpha_m}=\sigma_m/\sqrt{2}$ we call the \textit{locally independent parameters}~\footnote{These parameters also serve as coefficients of a basis of matrices for traceless Hermitian matrices.}. They are called \textit{locally independent} since in every local region about some $\mathcal{M}_0$, the generators $\{t_{\alpha_m}\}$ are independent of $\mathcal{M}_0$, whereas in the most general parameterisation, each $t_{\alpha_m}$ can depend on all $\{\alpha_m^{(0)}\}$ and hence $\mathcal{M}_0$. 

Having defined our new parameterisation, we can proceed to find our matrix $J^{-1}$ defined under Eq.~\eqref{eq:fchange} using
\begin{equation} \label{eq:sigmaj}
\frac{\sigma_m}{\sqrt{2}} \equiv t_{\alpha_m}=i\mathcal{M}^{\dagger}\sum_k \frac{\partial \mathcal{M}}{\partial \psi_k} J_{km}=\sum_k t_{\psi_k}J_{km},
\end{equation}
where $\{\psi_k\}$ are the Euler angles and $t_{\psi_k}$ can be computed directly from $\mathcal{M}$ using its Euler angle decomposition $\mathcal{M}=\exp(i \psi_1 \sigma_z/2)\exp(i \psi_2\sigma_y/2)\exp(i\psi_3\sigma_z/2)$. Therefore all the elements of $J^{-1}$ can be directly recovered from Eq.~\eqref{eq:sigmaj}. We can thus show the elements of $J^{-1}$ to be
\begin{align} \label{eq:inverseJ}
(J^{-1})_{ij}=-\frac{1}{\sqrt{2}}\begin{pmatrix}
\cos(\psi_3) \sin(\psi_2) & -\sin(\psi_3) &0 \\
\sin(\psi_3) \sin(\psi_2) & \cos(\psi_3) & 0 \\
\cos(\psi_2) & 0 & 1
\end{pmatrix}.
\end{align}
Using this equation for $J^{-1}$, we can compute the matrix $V$
\begin{align}\label{eq:V}
&V_{ij}=\sum_k (J^{-1})_{ik}^T J^{-1}_{kj} \nonumber \\
&=\frac{1}{2}\begin{pmatrix}
1 & 0 & \cos(\psi_2) \\
0 & 1 & 0 \\
\cos(\psi_2) & 0 & 1
\end{pmatrix}.
\end{align}
This can now be used in Eq.~\eqref{eq:tracetransform} to transform the inverse of any Fisher information matrix in the Euler angle parameterisation into our locally independent parameterisation.  
\section{Quantum fisher information matrix for unitary-estimation}
\label{sec:QFI}
We present in section ~\ref{sec:qfiderive} a derivation of the quantum Fisher information for the probe state used in our protocol described in section ~\ref{sec:protocol}. We demonstrate the necessary and sufficient conditions our probe state must satisfy to saturate the quantum Cramer-Rao bound in section ~\ref{sec:saturation}. We also show the conditions for reaching optimality (i.e. the lowest bound) in the quantum Cramer-Rao bound in section ~\ref{sec:optimality}. Our proof is an extension of \cite{ballester2005} that is adapted to our particular estimation scheme and also allows for a much wider class of optimal states. In addition, we utilise our mapping introduced in section ~\ref{sec:mapping} to interpret our results using spin states in terms of photonic states. We use this to identify photonic states that both saturate and reach optimality in the quantum Cramer-Rao bound.
\subsection{Quantum Fisher information matrix for multi-photon probe}
\label{sec:qfiderive}
We first derive the quantum Fisher information matrix for a general pure state before moving on to the protocol described in section ~\ref{sec:protocol}. We exploit the mathematical analogy between the dynamics of the photonic and spin states (explained in section ~\ref{sec:mapping}) to simplify our calculations. We start with the definition of the quantum Fisher information matrix $I_{\alpha \beta}$ for a pure $N$-particle initial spin state $\rho_0$ is 
\begin{gather} \label{eq:QFIdef}
I_{\alpha \beta}=\text{Re}\tr(\rho_f \lambda_{\alpha} \lambda_{\beta}),
\end{gather}
where the final state is given by $\rho_f=U_s^{\otimes N} \rho_0 (U_s^{\dagger})^{\otimes N}$ and $U_s$ is a unitary operator (satisfying $U_s^{\dagger}U_s=\mathbf{1}$) introduced earlier. If we started with the photonic state $\ket{M, N-M}$, the initial spin state is $\rho_0=\ket{\xi}_0 \bra{\xi}_0$ from the notation in our previous section. The symmetric logarithmic derivative $\lambda_{\alpha}$ for pure states is given by $\lambda_{\alpha}=2 \partial \rho_f/\partial \alpha$ like in the single-parameter estimation case.

We define $T_{\alpha} \equiv i(U_s^{\otimes N})^{\dagger} \partial U_s^{\otimes N}/\partial \alpha$ as the generator of $U_s^{\otimes N}$. Then using $T_{\alpha}=T_{\alpha}^{\dagger}$~\footnote{This follows from $0=\frac{\partial \mathbf{1}^{\otimes N}}{\partial \alpha}=\frac{\partial((U_s^{\otimes N})^{\dagger}U_s^{\otimes N})}{\partial \alpha}=(-iT_{\alpha})^{\dagger}+(-iT_{\alpha})$.}, the unitarity of $U_s^{\otimes N}$ and trace permutation invariance $\tr(AB)=\tr(BA)$ for matrices $A$, $B$ we find
\begin{gather}\label{eq:trrhof}
\tr(\rho_f \lambda_{\alpha} \lambda_{\beta})=4[\tr(\rho_0T_{\alpha}T_{\beta})-\tr(\rho_0 T_{\alpha})\tr(\rho_0 T_{\beta})].
\end{gather}
To express this in terms of the generators of $U_s$, which are $t_{\alpha} \equiv iU_s^{\dagger} \partial U_s/\partial \alpha$ (this we defined earlier in section ~\ref{sec:parameterisation}), we first need the easily derived relations
$T_{\alpha}=t_{\alpha}\otimes \mathbf{1}^{\otimes (N-1)}+\mathbf{1}\otimes t_{\alpha} \otimes \mathbf{1}^{\otimes (N-2)}+...+\mathbf{1}^{\otimes (N-1)} \otimes t_{\alpha}$ (which has $N$ terms altogether) and $T_{\alpha} T_{\beta}=t_{\alpha}t_{\beta}\otimes \mathbf{1}^{\otimes (N-1)}+t_{\alpha}\otimes t_{\beta}\otimes \mathbf{1}^{\otimes (N-2)}+t_{\alpha}\otimes\mathbf{1}^{\otimes (N-1)}\otimes t_{\beta}+...+\mathbf{1}^{\otimes (N-1)}\otimes t_{\alpha}t_{\beta}$ (which has $N^2$ terms altogether). We can also make use of the equality $\tr(\rho_0(\mathbf{1}_A \otimes B))=\tr(\tr_A(\rho)B)$. For example, suppose state $\rho_0=\rho_1 \otimes \rho_2$ is decomposed into two single-particle reduced states $\rho_1$ and $\rho_2$ where $\tr(\rho_1)=1=\tr(\rho_2)$. Then using $\tr(A \otimes B)=\tr(A) \tr(B)$ we can derive the total quantum Fisher information matrix as
\begin{align} \label{eq:totalqfisherinfo}
I_{\alpha \beta}=&4\text{Re}\sum_{i=1}^{N} \tr(\tr_{[i]}(\rho_0) t_{\alpha}t_{\beta})+4\text{Re}\sum_{i \neq j=1}^{N(N-1)}\tr(\tr_{[i,j]}(\rho_0) (t_{\alpha}\otimes t_{\beta})) \nonumber \\
&-4\text{Re}\left(\sum_{i=1}^{N} \tr(\tr_{[i]}(\rho_0) t_{\alpha})\right)\left(\sum_{i=1}^{N} \tr(\tr_{[i]}(\rho_0) t_{\beta})\right),
\end{align}
where the notation $\tr_{[i,j]}$ denotes a trace over all particles \textit{except} those particles labelled by positions $i,j$ (note that $\tr_{[i,j]}(\cdot)$ is equal to $\tr_{[j,i]]}(\cdot)$ after exchange of subsystems). 

In the case where $\rho_0$ is a symmetric state, all partial traces depends only on the number of subsystems traced out and not on \textit{which} subsystems are traced out. This implies that all single-particle reduced states are identical, and so we can define $\rho^{[1]} \equiv \tr_{[i]}(\rho_0)$ for any $i$. All two-particle reduced states are also identical and we define $\rho^{[2]} \equiv \tr_{[i,j]}(\rho_0)$ for any $i,j$ where $i \neq j$. The quantum Fisher information matrix then simplifies to the form
\begin{align} \label{eq:Ipart}
I_{\alpha \beta}=\text{Re}\{4N[\tr(\rho^{[1]}t_{\alpha}t_{\beta})+(N-1)\tr(\rho^{[2]}(t_{\alpha}\otimes t_{\beta}))-N\tr(\rho^{[1]}t_{\alpha}) \tr(\rho^{[1]}t_{\beta})]\}.
\end{align}
For the multi-photon probe protocol introduced in section ~\ref{sec:protocol}, we have access to three symmetric states $\rho_x$, $\rho_y$ and $\rho_z$, which are related to each other by a basis transformation. In the photonic version originally described, we used the HV, DA and RL polarisation bases, which map to the $z$, $x$ and $y$ basis in the spin case, respectively. The probe states are of the form $\rho_0=\rho_x \otimes \rho_y \otimes \rho_z$ and the quantum Fisher information matrix is just a sum of the quantum Fisher information matrix corresponding to the states $\rho_x$, $\rho_y$ and $\rho_z$ individually (see Appendix~\ref{sec:QFIadd}). A \textit{single run} of this protocol uses $\rho_x$, $\rho_y$, $\rho_z$ one after another, each with $N$ particles. Hence $\rho_0$ is a $3N$ particle state where each of $\rho_x$, $\rho_y$, $\rho_z$ are individually symmetric~\footnote{Note this does \textit{not} imply $\rho_0$ is  symmetric.}. The total quantum Fisher information for state $\rho_0$ over a single run is 
\begin{align} \label{eq:Itotal}
I_{\alpha \beta}(\rho_0)=\sum_{\xi=x, y, z}I_{\alpha \beta}(\rho_\xi).
\end{align}
Translating this into the photonic case just requires a mapping of the single-particle and two particle reduced spin states into their photonic counterparts. We will derive these in sections ~\ref{sec:saturation} and ~\ref{sec:optimality}. 
\subsection{Saturating the quantum Cramer-Rao bound}
\label{sec:saturation}
For the quantum Cramer-Rao bound to be saturated, it is a requirement that there exists a measurement for which the Fisher information matrix is the same as the quantum Fisher information matrix~\footnote{For single-parameter estimation where the Fisher and quantum Fisher information are scalars, it is known that there always exists a measurement that satisfies this condition \cite{braunsteincaves}. This is not always true for multi-parameter estimation.}. Finding the states and corresponding measurements which satisfy this condition is of important practical interest for experimentally realising the quantum advantage in ever more precise parameter estimation. It can be shown that a necessary and sufficient condition to attain this saturation is for the equality $\text{Im} \tr(\rho_f \lambda_{\alpha} \lambda_{\beta})=0$ to hold~\footnote{See Theorem 5 in \cite{matsumoto}}. Since $\lambda_{\alpha}$ are Hermitian, this condition can be written as 
\begin{align} \label{eq:matsu}
\tr(\rho_f [\lambda_{\alpha}, \lambda_{\beta}])=0.
\end{align}
For our protocol using $\rho_0=\rho_x \otimes \rho_y \otimes \rho_z$ we see that this can be further simplified as
\begin{align}
&\tr(\rho_f [\lambda_{\alpha}, \lambda_{\beta}]) \nonumber \\
&=4 \tr(\rho_0 (T_{\alpha}T_{\beta}-T_{\beta}T_{\alpha}) \nonumber \\
&=4 [N \sum_{\xi=x, y, z} \tr(\rho_{\xi}^{[1]} [t_{\alpha}, t_{\beta}]) \nonumber \\
&+\sum_{i \neq j}\tr(\tr_{\backslash \{i,j\}}(\rho_0)(t_{\alpha} \otimes t_{\beta}))-\tr(\tr_{\backslash \{i,j\}}(\rho_0)(t_{\beta} \otimes t_{\alpha}))] \nonumber \\
&=4 N\tr(\rho_{\text{tot}}^{[1]} [t_{\alpha}, t_{\beta}])=0=2 N\tr(\rho_{\text{tot}}^{[1]} [\sigma_{\alpha}, \sigma_{\beta}]),
\end{align}
where $\rho_{\text{tot}}^{[1]} \equiv \rho_x^{[1]}+\rho_{y}^{[1]}+\rho_z^{[1]}$ and $t_{\alpha} \equiv \sigma_{\alpha}/\sqrt{2}$, which defines the \textit{locally independent parameters} we introduced into section ~\ref{sec:parameterisation}. To see which states this condition permits, we can write a general single-particle reduced density matrix as $\rho_{\text{tot}}^{[1]}=a_0\mathbf{1}+a_x \sigma_x+a_y \sigma_y+a_z \sigma_z$. Since $[\sigma_{\alpha}, \sigma_{\beta}] \propto \sigma_{\gamma}$ for $\alpha \neq \beta$, Eq.~\eqref{eq:matsu} reduces to $\sum_{\xi=x, y, z}\tr(a_{\xi} \sigma_{\xi} \sigma_{\gamma})=0$, which is only satisfied if $a_x=0=a_y=a_z$. Thus the only possible state satisfying Eq.~\eqref{eq:matsu} is
\begin{align}
\rho^{[1]}_{\text{tot}} \propto \mathbf{1}.
\end{align}
To find out what state this implies for $\rho_x$, $\rho_y$, $\rho_z$ we first make the following observation on how $\rho_x$, $\rho_y$ and $\rho_z$ are related to one another. These states must give the same measurement statistics when measured with respect to bases $x,y$ and $z$ respectively. This means that the positive (negative) eigenstate $\ket{\uu}$ ($\ket{\dd}$) in the $\sigma_z$ basis correspond to the positive (negative) eigenstate $\ket{+}$ ($\ket{-}$) in the $\sigma_x$ basis and the positive (negative) eigenstate $\ket{0}_y$ ($\ket{1}_y$) in the $\sigma_y$ basis. Therefore, one can generate $\rho_x$ (and $\rho_y$) from $\rho_z$ by a cyclic permutation of the Pauli matrices, where $\sigma_x \rightarrow \sigma_z$ (and $\rightarrow \sigma_z$), $\sigma_y \rightarrow \sigma_y$ (and $\rightarrow -\sigma_x$), $\sigma_z \rightarrow \sigma_x$ (and $\rightarrow \sigma_y$). Thus if we write $\rho_z^{[1]}=\frac{\mathbf{1}}{2}+b_x \sigma_x+b_y \sigma_y+b_z \sigma_z$
then $\rho_x^{[1]}=\frac{\mathbf{1}}{2}+b_x \sigma_z+b_y \sigma_y+b_z \sigma_x$ and $\rho_y^{[1]}=\frac{\mathbf{1}}{2}+b_x \sigma_z-b_y \sigma_x+b_z \sigma_y$. Thus the requirement $\rho_{\text{tot}}^{[1]} =\frac{3}{2} \mathbf{1}+(b_x-b_y+b_z)\sigma_x+(2b_y+b_z) \sigma_y+(b_z+2b_x)\sigma_z \propto \mathbf{1}$ means $b_x=b_y=0=b_z$. Thus a sufficient and necessary condition for Eq.~\eqref{eq:matsu} to be satisfied is
\begin{align} \label{eq:rho1mixed}
\rho_{z}^{[1]}=\frac{\mathbf{1}}{2},
\end{align}
which equivalently ensures that the quantum Cramer-Rao bound is saturated. 

Following section ~\ref{sec:mapping}, we can now translate Eq.~\eqref{eq:rho1mixed} into an equivalent condition on photonic states. We only need to know the restriction on state $\rho_z$, which we know to be symmetric since all photonic states map onto symmetic spin states. The most general pure $N$-particle two-mode bosonic state is 
\begin{align} \label{eq:photonstate}
\ket{\Psi}=\sum_{M=0}^N c_M\ket{M, N-M}.
\end{align}
We transform the state in Eq.~\eqref{eq:photonstate} into its spin counterpart using the mapping in Eq.~\eqref{eq:mapping} and find its single-particle reduced state. We then transform the state back to this photonic form
\begin{align} \label{eq:rho1spin}
\rho^{[1]}=&\frac{1}{N}\sum_{M=0}^N|c_M|^2M\ket{1 0}\bra{1 0}+\frac{1}{N}\sum_{M=0}^N|c_M|^2(N-M)\ket{01}\bra{01} \nonumber \\
                &+\frac{1}{N}\sum_{M=0}^{N-1}c_M^*c_{M+1}\sqrt{(N-M)(M+1)}\ket{1 0}\bra{0 1} \nonumber \\
                 &+\frac{1}{N}\sum_{M=0}^{N-1}c_M c_{M+1}^*\sqrt{(N-M)(M+1)}\ket{0 1}\bra{1 0}.
\end{align}
Our state saturates the quantum Cramer-Rao bound if its corresponding single-particle reduced state is maximally mixed, which in the photonic form is 
\begin{align}
\rho^{[1]}=\frac{1}{2} \ket{01}\bra{01}+\frac{1}{2}\ket{10}\bra{10}.
\end{align}
Thus, we need only to satisfy the following easy-to-check conditions
\begin{align}
&\sum_{M=0}^N |c_M|^2M=\frac{N}{2} \nonumber \\
&\sum_{M=0}^{N-1}c_M^*c_{M+1}\sqrt{(N-M)(M+1)}=0=\sum_{M=0}^{N-1}c_M c_{M+1}^*\sqrt{(N-M)(M+1)}.
\end{align}
From this test we will see later that some commonly considered states in single-parameter estimation like $N00N$ states (except $N=1$), Holland-Burnett states and Yurke states ($\ket{\psi}=\frac{1}{\sqrt{2}}\left(\ket{\frac{N}{2}+1, \frac{N}{2}-1}+\ket{\frac{N}{2}-1, \frac{N}{2}+1}\right)$ all saturate the quantum Cramer-Rao bound. In fact, all states of the form $\ket{M,N-M}+\ket{N-M,M}$ \textit{except} when $M=(N-1)/2$ saturate the quantum Cramer-Rao bound. The Yurke states  $\ket{\psi}=\frac{1}{\sqrt{2}}\left(\ket{\frac{N}{2}+1, \frac{N}{2}-1}+\ket{\frac{N}{2}, \frac{N}{2}}\right)$ used in single-parameter estimation, however, do not satisfy these conditions.

To interpret these results in terms of quantities commonly measured in photonic experiments, let us rewrite Eq.~\eqref{eq:rho1spin} as
\begin{align}
\rho^{[1]}= &\frac{1}{N}\langle a^{\dagger}a \rangle \ket{10}\bra{10}+\frac{1}{N}\langle b^{\dagger}b \rangle \ket{01}\bra{01} \nonumber \\
                   &+\frac{1}{N} \langle a^{\dagger}b \rangle \ket{01}\bra{10}+\frac{1}{N} \langle b^{\dagger}a \rangle \ket{10}\bra{01}.
\end{align}
Thus the conditions for saturating the quantum Cramer-Rao bound (i.e. maximally-mixed one-particle density matrix) rewritten in the photonic form are
\begin{align} \label{eq:rho1conditions}
&\langle a^{\dagger} a \rangle =\langle b^{\dagger} b \rangle=\frac{N}{2}, \nonumber \\
&\langle a^{\dagger} b \rangle=0=\langle a b^{\dagger} \rangle.
\end{align}
The condition $\langle a^{\dagger} b \rangle=0$ corresponds to an absence of  first-order coherence for the state \cite{loudon}, which can be interpreted in the following way. Let the two modes of the photonic state be injected into an interferometer that takes $(a^{\dagger} \, \, \, \, \, b^{\dagger})^T$ through an arbitrary $SU(2)$ transformation. If the final measured intensities of the individual modes remain invariant with respect to the $SU(2)$ transformation, this is equivalent to $\langle a^{\dagger} b \rangle=0$.
\subsection{Reaching the optimal bound}
\label{sec:optimality}
To obtain the lowest bound in the quantum Cramer Rao inequality, we need to find a lower bound that the trace of the inverse quantum Fisher information matrix can attain. The purpose is to use this to identify the quantum states and measurements that together can achieve this lower bound. This lower bound we call the optimal bound that those states satisfying this bound we call an optimal state. To find this optimal bound, we make use of the Cauchy-Schwarz inequality on the total quantum Fisher information $I_{\alpha \beta}$
\begin{align}
9=\tr(\mathbf{1})^2=[\tr(I_{\alpha \beta}^{-\frac{1}{2}}I_{\alpha \beta}^{\frac{1}{2}})]^2\leq \tr(I_{\alpha \beta})\tr(I_{\alpha \beta}^{-1}).
\end{align}
The optimal bound (or minimum value) for $\tr(I_{\alpha \beta}^{-1})$ occurs when the Cauchy-Schwarz inequality is saturated, or 
\begin{align}
\tr(I^{-1}_{\alpha \beta})=\tr(I_{\alpha \beta}^{-1})\vert_{\text{min}}=\frac{9}{\tr(I_{\alpha \beta})}.
\end{align}
This is only obtained when $I_{\alpha \beta} \propto \delta_{\alpha \beta}$. The Fisher information matrix can only reach this limit if the condition for saturating the quantum Cramer-Rao bound is satisfied (i.e. $\rho_z^{[1]}=\mathbf{1}/2$ from Eq.~\eqref{eq:rho1mixed}). Inserting this into Eqs.~\eqref{eq:Ipart} and ~\eqref{eq:Itotal}, we find that the total quantum Fisher information takes the form
\begin{align} \label{eq:itotopt}
I_{\alpha \beta}=2[3N \delta_{\alpha \beta}+N(N-1)\tr[\rho_{\text{tot}}^{[2]}(\sigma_{\alpha}\otimes \sigma_{\beta})]],
\end{align}
where $\rho_{\text{tot}}^{[2]} \equiv \rho_{x}^{[2]}+\rho_{y}^{[2]}+\rho_{y}^{[2]}$ and we made the replacement $t_{\alpha}\equiv \sigma_{\alpha}/\sqrt{2}$, which we motivated in section ~\ref{sec:parameterisation} and defined in Eq.~\eqref{eq:sigmaj}. We require  $I_{\alpha \beta} \propto \delta_{\alpha \beta}$ for optimality. From Eq.~\eqref{eq:itotopt}, we see this is equivalent to requiring $\tr[\rho_{\text{tot}}^{[2]}(\sigma_{\alpha}\otimes \sigma_{\beta})] \propto \delta_{\alpha \beta}$, which is a restriction only on the two-particle reduced states. Thus we first write the general symmetric two particle reduced state in the $z$-basis $\rho_{z}^{[2]}$ as
\begin{align}
\rho_{z}^{[2]} =&\frac{\mathbf{1}\otimes\mathbf{1}}{4}+c_{xx}(\sigma_x \otimes \sigma_x)+c_{yy}(\sigma_y \otimes \sigma_y)+c_{zz}(\sigma_z \otimes \sigma_z) 
                        +c_{xy}(\sigma_x \otimes \sigma_y 
+\sigma_y \otimes \sigma_x) \nonumber \\
&+c_{xz}(\sigma_x \otimes \sigma_z+\sigma_z \otimes \sigma_x)+c_{yz}(\sigma_y \otimes \sigma_z+\sigma_z \otimes \sigma_y),
\end{align}
with corresponding $\rho_{x}^{[2]}$ found by replacing $\sigma_x \rightarrow \sigma_z$,  $\sigma_y \rightarrow \sigma_y$ and $\sigma_z \rightarrow \sigma_x$ in the expression above. Similarly for $\rho_{y}^{[2]}$ with the replacement $\sigma_x \rightarrow \sigma_z$,  $\sigma_y \rightarrow -\sigma_x$ and $\sigma_z \rightarrow \sigma_y$. Then it is possible to show $\tr[\rho_{\text{tot}}^{[2]}(\sigma_{\alpha}\otimes \sigma_{\beta})] \propto \delta_{\alpha \beta}$ if and only if $c_{xy}=0=c_{xz}=c_{yz}$ and $c_{xx}=c_{yy}$. Therefore, a necessary and sufficient condition for optimality is for
\begin{align} \label{eq:rhoz2}
\rho_{z}^{[2]}=\frac{\mathbf{1}\otimes \mathbf{1}}{4}+c_{xx}(\sigma_x \otimes \sigma_x+\sigma_y \otimes \sigma_y)+c_{zz}\sigma_z \otimes \sigma_z,
\end{align}
with corresponding expressions for $\rho_{x}^{[2]}$ and $\rho_{y}^{[2]}$. This means we can rewrite
\begin{align}
\rho_{\text{tot}}^{[2]}=\frac{3}{4}\mathbf{1}\otimes \mathbf{1}+K(\sigma_x \otimes \sigma_x+\sigma_y \otimes \sigma_y+\sigma_z \otimes \sigma_z),
\end{align}
where $K=(2c_{xx}+c_{zz})$. To find this constant $K$ we note that
\begin{align}
K=\frac{1}{12}\tr(\rho_{\text{tot}}^{[2]}(\sum_{\alpha=x,y,z} \sigma_{\alpha}\otimes \sigma_{\alpha})), 
\end{align}
where $\sum_{\alpha=x,y,z} \sigma_{\alpha}\otimes \sigma_{\alpha}$ may be rewritten in terms of the swap operator $S\equiv \sum_{ij}\ket{ji}\bra{ij}$ for $i,j=\uu, \dd$ as $\sum_{\alpha=x,y,z} \sigma_{\alpha} \otimes \sigma_{\alpha}=2S-\mathbf{1} \otimes \mathbf{1}/2$. We remark that since the swap operator just swaps the positions of the two modes, it leaves symmetric states invariant and $S \rho^{[2]}_{\text{tot}}=\rho^{[2]}_{\text{tot}}$. Using this feature we can show
\begin{align}
K&=\frac{1}{12}\tr(\rho_{\text{tot}}^{[2]} \sum_{\alpha} \sigma_{\alpha} \otimes \sigma_{\alpha}) =\frac{1}{12}\tr(\rho^{[2]}_{\text{tot}}(2S-\mathbf{1}\otimes \mathbf{1})) \nonumber \\
&=\frac{1}{12}\left(2 \tr(\rho^{[2]}_{\text{tot}})-\tr(\rho^{[1]}_{\text{tot}})\right)=\frac{1}{12}\left(2 \times 3-3\right)=\frac{1}{4},
\end{align}
where we use $\tr(\rho (\mathbf{1}_B \otimes C))=\tr(\tr_B(\rho)C)$ in the second last expression. Thus a sufficient two-particle reduced state in the $z$-basis takes the form
\begin{align}\label{eq:rhoz2new}
\rho_{z}^{[2]}=\frac{\mathbf{1}\otimes \mathbf{1}}{4}+\frac{1}{2}\left(\frac{1}{4}-c_{zz}\right)(\sigma_x \otimes \sigma_x+\sigma_y \otimes \sigma_y)+c_{zz}\sigma_z \otimes \sigma_z.
\end{align}
We can use $K=1/4$ to get $I_{\alpha \beta}=2N(N+2)\delta_{\alpha \beta}$, from which we find the minimum value of $\tr(I_{\alpha \beta}^{-1})$ to be
\begin{align} \label{eq:minI}
\tr(I_{\alpha \beta}^{-1})\vert_{\text{min}}=\frac{3}{2N(N+2)}.
\end{align}
We remark that this displays Heisenberg scaling $\tr(I_{\alpha \beta}^{-1}) \sim \mathcal{O}(N^{-2})$, signalling a quantum advantage in parameter estimation. This can be contrasted with the result using $\nu$ single-photon probes (i.e. $N=1$), where $\tr(I_{\alpha \beta}^{-1}) \sim \mathcal{O}(\nu^{-1})$, which shows shot-noise scaling. 

It is later shown that Eq.~\eqref{eq:minI} is satisfied for all Holland-Burnett states and states of the form $\ket{M,N-M}+\ket{N-M,M}$ (including $N00N$ states) \textit{except} when $M=(N-1)/2$ (which do not saturate the quantum Cramer-Rao bound) and when $M=N/2-1$ (Yurke states). 

We observe a very great advantage of the protocol set out in section ~\ref{sec:protocol} lies in the flexibility of states which are optimal compared to the case where there is only a single kind of projective measurement. In our present case, any state whose two-particle reduced state in the $z$-basis takes the form in Eq.~\eqref{eq:rhoz2} can give an optimal result in the estimation of any unitary applied to it. However, if we use a protocol that only uses one kind of input state (instead of three kinds of input states in our scheme), like the recent protocol in \cite{animesh} and an older protocol in \cite{ballester2005}, there are more strict conditions to satisfy. For example, suppose we only use a single kind of state in the $z$ basis. In this case, the optimal quantum Fisher information matrix must satisfy
\begin{align}
I^z_{\alpha \beta} \propto \delta_{\alpha \beta},
\end{align}
whereas in the multiple input state case we only require $I_{\alpha \beta} \propto  \delta_{\alpha \beta}$. Following the same derivation as before, now we find that the optimal result is only attained if the two-particle reduced state in the $z$ basis takes the form
\begin{align} \label{eq:rho2animesh}
\rho_{z}^{[2]} =\frac{\mathbf{1}\otimes\mathbf{1}}{4}+\frac{1}{12}(\sigma_x \otimes \sigma_x+\sigma_y \otimes \sigma_y+\sigma_z \otimes \sigma_z),
\end{align}
which is a much more restrictive special case of $\rho_{z}^{[2]}$ in Eq.~\eqref{eq:rhoz2}. Indeed, Eq.~\eqref{eq:rho2animesh} shows the special feature of the state used in \cite{animesh}, though without any proof to show this is the optimal case. We will see later that neither of the two well-known states in single-parameter estimation, the Holland-Burnett and the $N00N$ states, satisfy this condition. However, if we use the three-basis measurement protocol, we \textit{can} still achieve optimality with not only Holland-Burnett and $N00N$ states, but also a much larger class of other states. 

We can now go through a similar process to find the photonic form of the two-particle reduced state and find an intepretation for the optimal photonic states. Using the equation for a general $N$-particle pure state in Eq.~\eqref{eq:photonstate} we find the two-particle reduced state of its corresponding spin state. The photonic form of this two-particle reduced state is then
\begin{align}
\rho^{[2]}=& \frac{1}{N(N-1)}\sum_{M=0}^{N} c_M^*c_M(N-M)(N-M-1)\ket{02}\bra{ 02} \times \nonumber \\
                &+2c_{M+1}^* c_{M+1}(M+1)(N-M-1)\ket{11}\bra{11} \nonumber \\
                 &+c_{M+2}^*c_{M+2}(M+1)(M+2)\ket{20}\bra{20}+\sqrt{2}(M+1)\sqrt{(M+2)(N-M-1)} \times \nonumber \\
                &\left(c_{M+1}^*c_{M+2}\ket{02}\bra{11}+c_{M+2}^*c_{M+1}\ket{11}\bra{20}\right) \nonumber \\
                 &+\sqrt{2}(N-M-1)\sqrt{(M+1)(N-M)} \times \nonumber \\
                 &\left(c_{M}^*c_{M+1}\ket{11}\bra{02}+c_{M+1}^*c_{M}\ket{02}\bra{11}\right) \nonumber \\
                 &+\sqrt{(M+1)(M+2)(N-M)(N-M-1)} \times \nonumber \\
                 &\left(c_{M}^*c_{M+2}\ket{20}\bra{02}+c_{M+2}^*c_{M}\ket{02}\bra{20}\right).
\end{align}
Using photonic observables, we can in fact rewrite the two-particle reduced state as
\begin{align}\label{eq:rho2photon}
\rho^{[2]}=&\frac{1}{N(N-1)} \times \nonumber \\
                  &\langle (a^{\dagger}a)(a^{\dagger}a-1)\rangle \ket{20}\bra{20}+\langle (b^{\dagger}b)(b^{\dagger}b-1)\rangle \ket{02}\bra{02} \nonumber \\
                   &+2\langle (a^{\dagger}a)(b^{\dagger}b)\rangle \ket{11}\bra{11} \nonumber \\
                   &+\langle a^{\dagger 2}b^2\rangle \ket{02}\bra{20}+\langle (a^{\dagger 2}b^2)^{\dagger}\rangle \ket{20}\bra{02} \nonumber \\
                   &+\sqrt{2}\langle (a^{\dagger}b)(b^{\dagger}b-1) \rangle \ket{02}\bra{11}+\sqrt{2}\langle ((a^{\dagger}b)(b^{\dagger}b-1))^{\dagger} \rangle \ket{11} \bra{02} \nonumber \\
                  &+\sqrt{2}\langle (a^{\dagger}b)(a^{\dagger}a) \rangle \ket{11}\bra{20}+\sqrt{2}\langle ((a^{\dagger}b)(a^{\dagger}a))^{\dagger} \rangle \ket{20}\bra{11}.
\end{align}
We can convert the two-particle reduced state of optimal states in Eq.~\eqref{eq:rhoz2new} to its photonic counterpart as
\begin{align}
\rho^{[2]}=\left(\frac{1}{4}+c_{zz}\right)(\ket{20}\bra{20}+\ket{02}\bra{02})+2\left(\frac{1}{4}-c_{zz}\right)\ket{11}\bra{11}.
\end{align}
Comparing this to Eq.~\eqref{eq:rho2photon} and using the form of our photonic state in Eq.~\eqref{eq:photonstate}, we arrive at the following conditions
\begin{align} \label{eq:rho2conditions}
&\langle (a^{\dagger}a)^2 \rangle=\langle (b^{\dagger}b)^2 \rangle=\langle (a^{\dagger}a)^2 \rangle-2N\langle a^{\dagger}a \rangle+N^2 \nonumber \\
&\langle (a^{\dagger}a)(a^{\dagger}a-1)\rangle+\langle (a^{\dagger}a)(b^{\dagger}b)\rangle=\langle a^{\dagger}a\rangle (N-1)=\frac{N(N-1)}{2} \nonumber \\
&=\langle (b^{\dagger}b)(b^{\dagger}b-1)\rangle+\langle (a^{\dagger}a)(b^{\dagger}b)\rangle=\langle b^{\dagger}b\rangle (N-1) \nonumber \\
&\langle a^{\dagger}a^{\dagger}a b \rangle=0=\langle a^{\dagger}a^{\dagger}bb \rangle=\langle a^{\dagger}b^{\dagger}bb\rangle,
\end{align}
where the first condition automatically satisfies one requirement ($\langle a^{\dagger}a\rangle=N/2=\langle b^{\dagger} b \rangle$) for the saturation of the quantum Cramer-Rao bound in Eq.~\eqref{eq:rho1conditions} and it also contains the second condition in Eq.~\eqref{eq:rho2conditions}. All the conditions in Eq.~\eqref{eq:rho2conditions} and Eq.~\eqref{eq:rho1conditions} together are necessary and sufficient for a state to reach the optimal quantum Cramer-Rao bound. All these requirements can be written succinctly as
\begin{align}
&\langle a^{\dagger}a \rangle=\langle b^{\dagger} b\rangle=\frac{N}{2} \nonumber \\
&\langle a^{\dagger} b \rangle=0 =\langle a^{\dagger}a^{\dagger}a b \rangle=0=\langle a^{\dagger}a^{\dagger}bb \rangle=\langle a^{\dagger}b^{\dagger}bb\rangle.
\end{align}
We observe that while the saturation of the Cramer-Rao bound depends only on the first-order correlations, the optimality conditions depends only on the second-order correlations $\langle a^{\dagger}a^{\dagger}a b \rangle$, $\langle a^{\dagger}a^{\dagger}bb \rangle$ and $\langle a^{\dagger}b^{\dagger}bb\rangle$. This is not surprising since in the spin picture, the saturation condition is a constraint only on the one-particle reduced state which correspond only to the first-order photonic correlations. The optimality condition for the spin state, on the other hand, depends only on the two-particle reduced state, which can be written in terms of only second-order photonic correlations. 

\section{Applications}
\label{sec:applications}
Using the tools we have developed in this chapter, we examine in sections  ~\ref{sec:mn} and ~\ref{sec:mnplusnm} respectively the states $\ket{M,N-M}$ and $(\ket{M,N-M}+\ket{N-M, M})/\sqrt{2}$ in the context of multi-parameter estimation. These states are chosen because they contain two states useful in single-parameter estimation, namely $N00N$ and Holland-Burnett states. In particular, we show that both of these states satisfy the optimality conditions for the quantum Cramer-Rao bound and show that optimal bound using number-counting measurements is realised neither by Holland-Burnett states nor $N00N$ states for $N=2,3$.
\subsection{$\ket{M,N-M}$ states}
\label{sec:mn}
We first want to investigate which of these states saturate the quantum Cramer-Rao bound and in addition, if any of these are optimal states. It is a well-known result in single-parameter estimation that out of the $\ket{M,N-M}$ states, the Holland-Burnett
states (where $M=N/2$) saturate the Cramer-Rao bound and are not optimal. To see whether this is true in the multi-parameter estimation setting, we first compute the one-particle reduced state of its symmetric spin state counterpart in the $z$ basis as
\begin{align}
\rho_{z}^{[1]}=\frac{1}{N}\left(M\ket{\uu}\bra{\uu}+(N-M)\ket{\dd}\bra{\dd}\right).
\end{align}
We learnt in the last section that a sufficient and necessary condition for saturating the quantum Cramer-Rao bound is for $\rho_z^{[1]}=\mathbf{1}/2$. We easily see that this only occurs when $M=N/2$ (i.e. Holland-Burnett states), which we will now confine our attention to. 
\subsubsection{Holland-Burnett states}
We know that Holland-Burnett states saturate the quantum Cramer-Rao bound. Are these states also optimal? To find out we need to also compute the two-particle reduced state of its symmetric spin state counterpart in the $z$ basis, which is 
\begin{align}
\rho_{z}^{[2]}=\frac{\mathbf{1}\otimes \mathbf{1}}{4}-\frac{1}{4(N-1)} \sigma_z \otimes \sigma_z+\frac{N}{8(N-1)}(\sigma_x \otimes \sigma_x+\sigma_y \otimes \sigma_y).
\end{align}
We see that this satisfies the sufficient and necessary conditions for an optimal state since here $c_{xx}=c_{yy}=N/(4(N-1))$and $K=2c_{xx}+c_{zz}=1/4$. This means we expect the minimum trace of the inverse Fisher information matrix $F_{\alpha \beta, \text{tot}}$ to be 
\begin{align} \label{eq:hbclquant}
\tr(I^{-1}_{\alpha \beta})=\frac{3}{2N(N+2)}=\tr(F^{-1}_{\alpha \beta, \text{tot}}) \vert_{\text{min}}.
\end{align}
This result only tells us that Holland-Burnett states can reach this optimal value using some projective measurement, but it does not give any idea what kind of projective measurement is required. One of the more commonly-used projective measurements for discrete photonic states is the number-counting measurement, which counts how many photons belong to each incoming polarisation mode. The advantage of the Fisher information formalism is that it contains explicitly the projective measurement that is required. Below we compare the minimum trace of the inverse Fisher information matrix with respect to number-counting measurements to Eq.~\eqref{eq:hbclquant}. 

We know from previous results on $\ket{M,N-M}$ in section ~\ref{sec:prob} that the Fisher information matrix in Euler angles is only dependent on the second Euler angle $\psi_2$ in the Euler angle decomposition of $\mathcal{M}$. In terms of Euler angles, the total Fisher information matrix is 
\begin{align} \label{eq:ftot2}
F_{kl, \text{tot}}=F_{22}(\psi_2)W_{kl}^H+F_{22}(\psi'_2)W_{kl}^D+F_{22}(\psi''_2)W_{kl}^R,
\end{align}
where 
\begin{align}
W_{kl}^H=\begin{pmatrix}
0 & 0 & 0 \\
0 & 1 & 0 \\
0 & 0 & 0 
\end{pmatrix},
\end{align}
and $W_{kl}^D$, $W_{kl}^R$ are the transformation matrices from basis $HV$ to $DA$ and $RL$ respectively, defined in Eq.~\eqref{eq:wmatrix}. The Euler angles $\psi_2'$, $\psi_2''$ are found~\footnote{See Appendix~\ref{sec:basischange} for explicit expressions of these Euler angles.} using the procedure outlined in section ~\ref{sec:differentbases}. Since $F_{22}(\psi_2)$ is only dependent upon a single-parameter, we can make use of already known results in single-parameter estimation. Metrology results for $\ket{M, N-M}$ states give \cite{durkin2007}
\begin{align}
F_{22}(\psi_2)=N(1+2M)-2M^2,
\end{align}
which is independent of $\psi_2$. This means that we can rewrite Eq.~\eqref{eq:ftot2} for Holland-Burnett states as
\begin{align}
F_{kl, \text{tot}}=F_{22}W^{\text{tot}}_{kl}=\frac{N(N+2)}{2} W^{\text{tot}}_{kl},
\end{align}
where $W^{\text{tot}}_{kl}\equiv W^{H}_{kl}+W^{D}_{kl}+W^{R}_{kl}$. In terms of locally independent parameters, this reduces to
\begin{equation}
\tr(F^{-1}_{\text{tot}})=\frac{2}{N(N+2)}\tr(V W_{\text{tot}}^{-1}),
\end{equation}
where $V$ is the matrix defined in Eq.~\ref{eq:V}. Note that the factor $\tr(V W_{\text{tot}}^{-1})$ is dependent only on the measurement bases and the unitary $\mathcal{M}$ and not on the initial states themselves. We find the global minimum of $\tr(V W_{\text{tot}}^{-1})$ for any $\mathcal{M}$ to be $3/2$, which is worse by a factor of $2$ compared to the optimal value for any measurement as given by Eq.~\eqref{eq:hbclquant}. We plot our numerical simulation result of the minimum $\tr(F^{-1}_{\text{tot}})$ value for Holland-Burnett states when $N=2,4,6,8$ in Fig.~\ref{hb}. 

We see from these results that using number-counting on Holland-Burnett states in our protocol, the optimal $\tr(F^{-1}_{\text{tot}})$ is obtainable.
\begin{figure}[ht!]
\centering
\label{hb}
\includegraphics[scale=0.4]{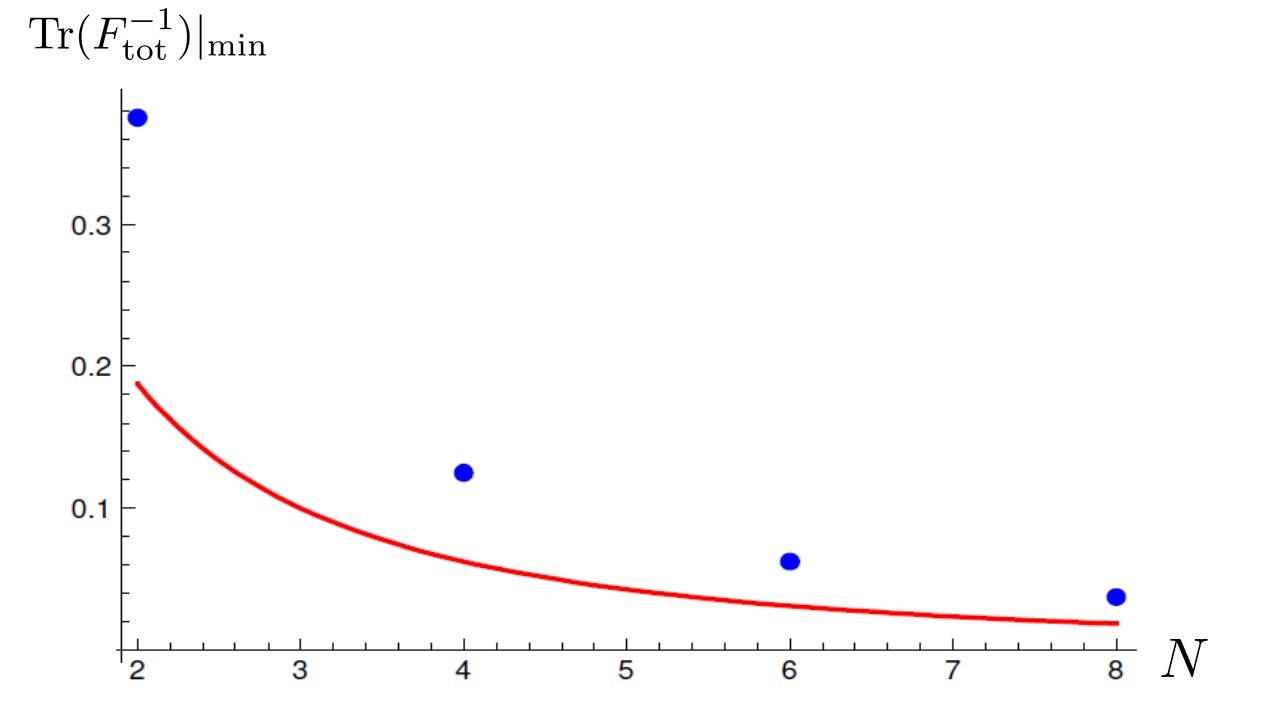}
\caption[ \textit{$\tr(F^{-1}_{\text{tot}})\vert_{\text{min}}$ for Holland-Burnett states}.] {\label{hb}\textit{$\tr(F^{-1}_{\text{tot}})\vert_{\text{min}}$ for Holland-Burnett states}. Plot of the minimum trace of the inverse Fisher information matrix $\tr(F^{-1}_{\text{tot}})\vert_{\text{min}}$ against $N$ for Holland-Burnett states with respect to number-counting measurements. The blue dots represent simulation results for $N=2,4,6,8$  Holland-Burnett states and the red line is theoretical minimum $\tr(F^{-1}_{\text{tot}})\vert_{\text{min}}$ for Holland-Burnett states.}
\end{figure}
We can also show that, unlike in the single-parameter estimation case where the Fisher information of Holland-Burnett states are independent of the unknown parameter, in fact in multi-parameter estimation $\tr(F^{-1}_{\text{tot}})$ \textit{is} dependent on the unknown parameters of the unitary. We can see this by plotting $\tr(V W_{\text{tot}}^{-1}) \propto \tr(F^{-1}_{\text{tot}})$ with respect to different unitaries, see Fig.~\ref{n3change}. The red line denotes the optimal lower bound $\tr(F^{-1}_{\text{tot}})$ for optimal measurements and the blue dots denote the minimmum $\tr(F^{-1}_{\text{tot}})$ at different values of $\psi_1$ for number-counting measurements. There is a clear cyclic pattern of the minimum value of $\tr(F^{-1}_{\text{tot}})$ for number-counting measurements as the Euler angle $\psi_1$ changes. From simulation results, the optimal lower bound $\tr(F^{-1}_{\text{tot}})$ occur at $\{\psi_1, \psi_2, \psi_3\}=\{0, \pi/2, \pi/2\}$, $\{\pi/2, \pi/2, 0\}$, $\{\pi, \pi/2, \pi/2\}$, $\{3\pi/2, \pi/2, 0\}$, which coincides with the results found in \cite{hugotomo} that uses process fidelity instead of Fisher information. The exact unitary-dependence deserves further investigation. 
\begin{figure}[ht!]
\centering
\includegraphics[scale=0.5]{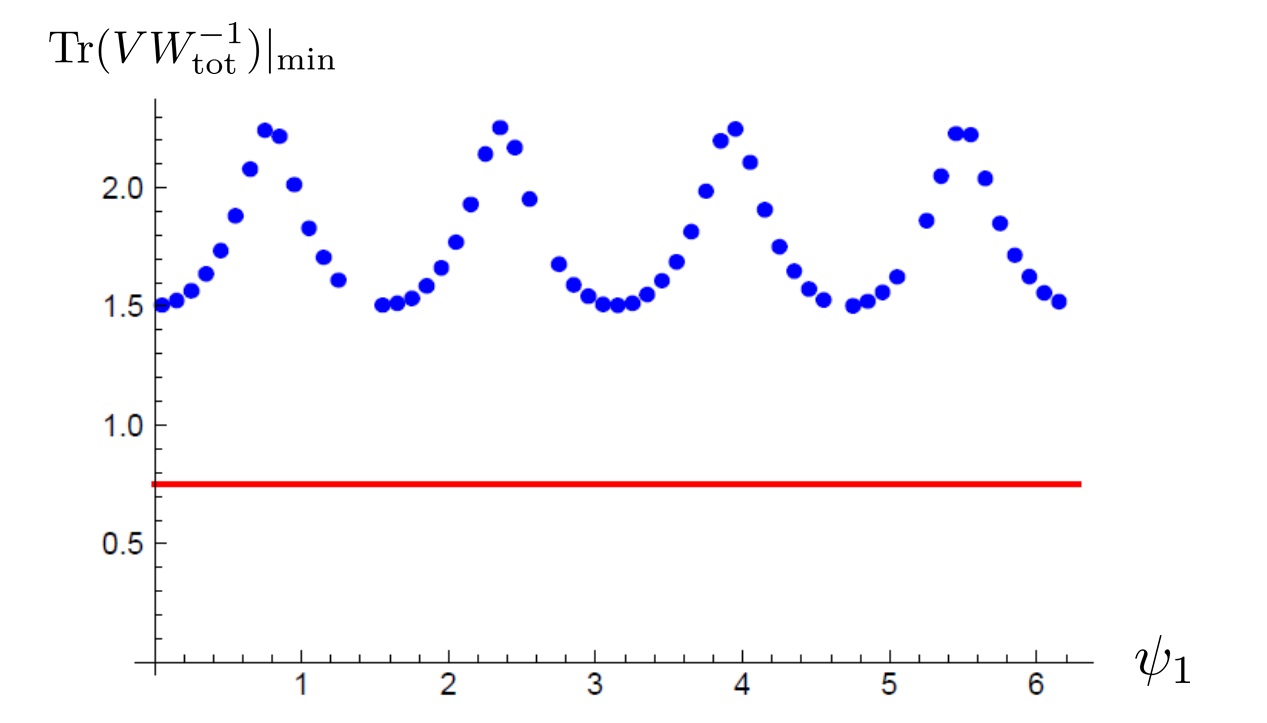}
\caption[\textit{$\tr(V W_{\text{tot}}^{-1})_{\text{min}}$ for Holland-Burnett states}.]{\label{n3change}\textit{$\tr(V W_{\text{tot}}^{-1})_{\text{min}}$ for Holland-Burnett states}. Plot of the local minimum of $\tr(V W_{\text{tot}}^{-1})$ (or  $\tr(V W_{\text{tot}}^{-1})_{\text{min}}$) along $\psi_1 \in[0, 2\pi]$. This is proportional to $\tr(F^{-1}_{\text{tot}})$ for Holland-Burnett states. The blue dots represent simulations results of $\tr(V W_{\text{tot}}^{-1})_{\text{min}}$ along $\psi_1$ and the underlying red line denotes the theoretical global minimum value at $\tr(V W_{\text{tot}}^{-1})=0.750$. The blue dots follow a distinct pattern, whose origin remains to be investigated.}
\end{figure}
This dependence on the unitary is in fact common in single-parameter estimation when not dealing with so-called path-symmetric states that include Holland-Burnett and $N00N$ states \cite{hofmann2009}. Adaptive schemes like in \cite{adaptive1, adaptive2} can be developed to deal with such dependencies. This is also a direction of future research. 
\subsection{$(\ket{M,N-M}+\ket{N-M, M})/\sqrt{2}$ states}
\label{sec:mnplusnm}
Now we will study another class of states of the form $(\ket{M,N-M}+\ket{N-M, M})/\sqrt{2}$. Which of these states saturate the quantum Cramer-Rao bound and in addition, which of these states are optimal? 

We know to satisfy the saturation condition, it is both sufficient and necessary that the one-particle reduced state of the symmetric spin state counterpart of $(\ket{M,N-M}+\ket{N-M, M})/\sqrt{2}$ is maximally mixed. Using our formalism earlier, these states are defined by $c_M=1/\sqrt{2}=c_{N-M}$ and all other coeffients vanish. Then it can be shown that $\rho^{[1]}_z=\frac{\mathbf{1}}{2}$ so long as $M \neq (N-1)/2$. This means that all $(\ket{M,N-M}+\ket{N-M, M})/\sqrt{2}$ states except $(\ket{(N-1)/2, (N+1)/2}+\ket{(N+1)/2, (N-1)/2})/\sqrt{2}$ saturate the quantum Cramer-Rao bound. 

Now we look at which states are optimal. If we neglect only two states, $M=(N-1)/2$ (which does not satisfy the saturation condition) and $M=N/2-1$, we find that the two-particle reduced states of the corresponding symmetric spin states takes the form
\begin{align}
\rho^{[2]}_z=\frac{\mathbf{1} \otimes \mathbf{1}}{4}+\frac{(N-2M)^2-N}{4N(N-1)} \sigma_z \otimes \sigma_z+\frac{M(N-M)}{2N(N-1)}(\sigma_x \otimes \sigma_x+\sigma_y \otimes \sigma_y).
\end{align}
These states all satisfy $c_{xx}=c_{yy}$ and $K=2c_{xx}+c_{yy}=1/4$. Therefore, \textit{all} states of the form $(\ket{M,N-M}+\ket{N-M, M})/\sqrt{2}$ are optimal \textit{except} $(\ket{(N-1)/2, (N+1)/2}+\ket{(N+1)/2, (N-1)/2})/\sqrt{2}$ and Yurke states $(\ket{N/2-1, N/2+1}+\ket{N/2+1, N/2-1})/\sqrt{2}$. When $M=0$, we have $N00N$ states, which are known to be optimal states for single-parameter estimation and is thus a very popular candidate, but have the disadvantage of being very hard to create and very sensitive to losses. We now turn our attention to $N00N$ states. 
\subsubsection{$N00N$ states}
We first want to identify the $N00N$ states which both saturate the quantum Cramer-Rao bound and are optimal. To saturate the quantum Cramer-Rao bound we only require $M \neq (N-1)/2$. Thus all $N00N$ states saturate the bound \textit{except} for $N=1$. The optimality condition requires only the saturation condition and $M \neq N/2-1$. Thus all $N00N$ states \textit{except} $N=1$ and $N=2$ satisfy both the saturation condition and are optimal. 

However, we do not know at this stage if $N \geq 3$ $N00N$ states are optimal under number-counting measurements. To find out, we need to find the minimum of the trace of the inverse Fisher information matrix with respect to number-counting measurements and compare this to the optimal bound. We begin with an $N$-particle $N00N$ state $(\ket{N0}+\ket{0N})/\sqrt{2}$. In the $\ket{j,m}$ notation this is equivalent to $\frac{1}{\sqrt{2}}\left(\ket{N/2, N/2}+\ket{N/2,-N/2}\right)$. The probability of being detected in state $\ket{M,N-M}$ (or equivalently $\ket{j=N/2, m'=M-N/2}$) after an arbitary linear process $U \equiv \exp(i\psi_1J_z)\exp(i\psi_2 J_y)\exp(i\psi_3 J_z)$ is
\begin{align}
&p_{m'}(\psi_2, \psi_3)  \nonumber \\
                                  &=\frac{1}{2}\ \left(d^{j=\frac{N}{2}}_{m',\frac{N}{2}}(\psi_2)\right)^2+\frac{1}{2}\ \left(d^{j=\frac{N}{2}}_{m',-\frac{N}{2}}(\psi_2)\right)^2+\cos(\psi_3 N)d^{j=\frac{N}{2}}_{m',\frac{N}{2}}(\psi_2)d^{j=\frac{N}{2}}_{m',-\frac{N}{2}}(\psi_2),
\end{align}
from which we can find the total inverse Fisher information matrix.

\begin{figure}[ht!]
\centering
\includegraphics[scale=0.5]{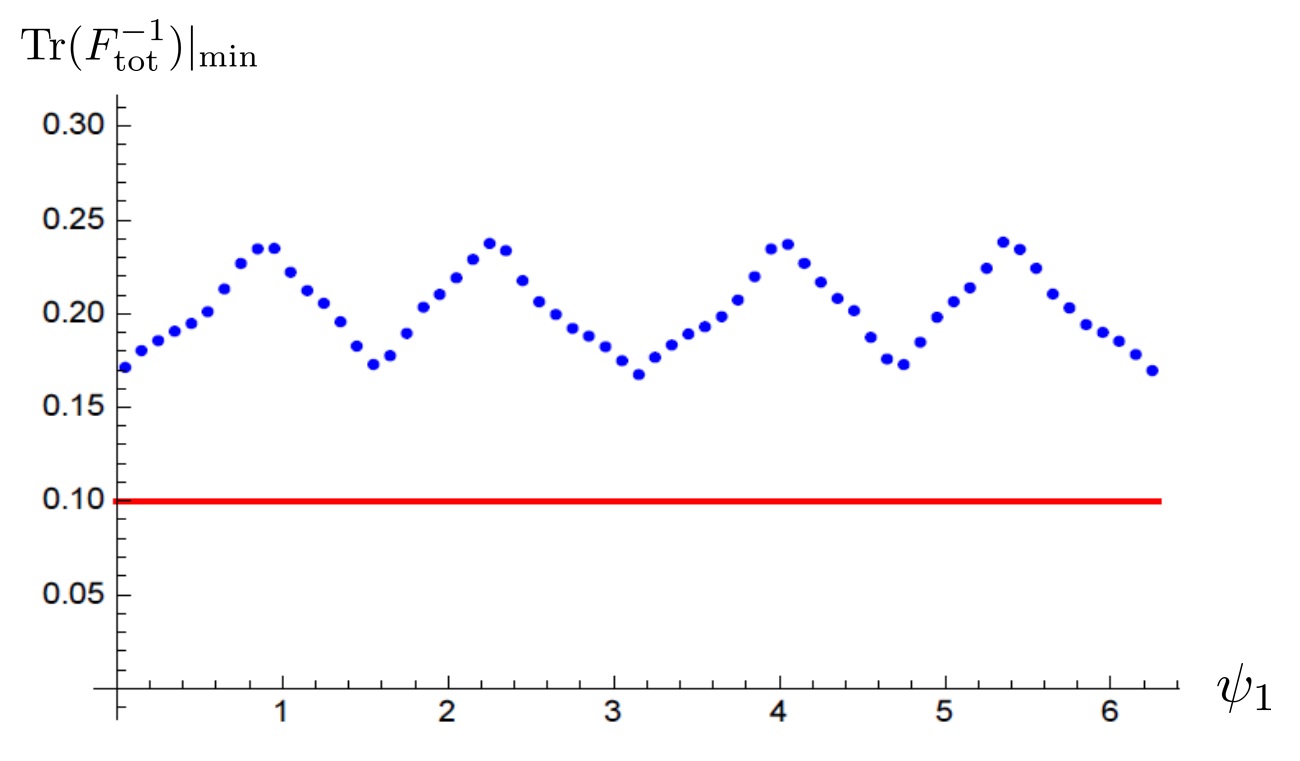}
\caption[\textit{$\tr(F^{-1}_{\text{tot}})\vert_{\text{min}}$ for $N=3$ $N00N$ state}.]{\label{n3new}\textit{$\tr(F^{-1}_{\text{tot}})\vert_{\text{min}}$ for $N=3$ $N00N$ state}. Plot of $\tr(F^{-1}_{\text{tot}})\vert_{\text{min}}$ against $\psi_1 \in [0, 2\pi]$. The blue dots represent simulation results for $N=3$ $N00N$ state with number-counting measurements and the red line represents the smallest  $\tr(F^{-1}_{\text{tot}})\vert_{\text{min}}$ given any measurement.}
\end{figure}

From our theoretical results, we learned that the $N=3$ $N00N$ state satisfies the optimality conditions. However, we demonstrate using simulation results presented in Fig.~\ref{n3new}, that for number-counting measurement the $N=3$ $N00N$ state does not achieve optimality. From our simulation we plot $\tr(F^{-1}_{\text{tot}})\vert_{\text{min}}$ at each value of the Euler angle $\psi_1 \in [0, 2\pi]$ for $N=3$ $N00N$ states under number-counting measurements (blue dots). The red line denotes the theoretical value of  $\tr(F^{-1}_{\text{tot}})\vert_{\text{min}}$ for an optimal measurement, where $\tr(I^{-1})=\tr(F^{-1}_{\text{tot}})\vert_{\text{min}}=3/(2N(N+2))\vert_{N=3}=0.1$. The minimum inverse Fisher information matrix for the $N=3$ $N00N$ state when using number-counting measurements by comparison is roughly $\tr(F^{-1})=0.167$. This gives a discrepancy of roughly $60 \%$ compared to the theoretical optimal measurement. Thus, unlike in single-parameter estimation, number-counting measurements are not optimal for $N00N$ states. 
\section{Discussion and further work}
\label{sec:discussion}
We have established a formalism for studying general $SU(2)$ estimation in the language of Fisher information for two-mode $N$-particle photonic states in the absence of photon loss and decoherence. With experimental feasibility as emphasis, we apply this theoretical framework to a new experimentally demonstrated protocol introduced in \cite{hugotomo}. We can show that this protocol can permit more optimal states than other schemes \cite{ballester2005,animesh}. The formalism we developed allow us to easily test for the best precision available for given photonic states, as well as testing which states are optimal (i.e. saturating the tightest bound for precision estimation). In addition, we present a mapping between photonic states and processes to their spin counterparts to establish a better connection between the results in both settings. This is the first concrete connection between photonic and spin states in the general multi-parameter estimation setting.

In particular, using our formalism we are able to find some key differences between multi-parameter and single-parameter estimation. For example, the Holland-Burnett state is sub-optimal in single-parameter estimation, but we have shown it be optimal for unitary estimation, although number-counting measurements are suboptimal (up to a constant factor in precision). Unlike Holland-Burnett states, photonic $N00N$ states are experimentally difficult to generate and extremely sensitive to photon loss. Furthermore, we have shown that $N00N$ states cannot reach their best precision under number-counting measurements, which is also contrary to the result in single-parameter estimation. We also confirm the findings in \cite{hugotomo} that, unlike in single-parameter estimation, Holland-Burnett states and $N00N$ states provide a precision in parameter estimation that is dependent on the unknown parameters of the unknown $SU(2)$ matrix itself. 

There are four main concrete directions to extend this work. 1) The first is using continuous variables input states, like squeezed states, which have not been studied in this context. 2) Another is the generalisation of our protocol to $SU(d)$ estimation for integer $d>2$, which amounts to multi-mode photonic inputs. This has obvious implications for quantum computing models like boson sampling. 3) An experimentally important direction is to focus on adaptive methods when the Fisher information is dependent on the unknown phases. Here it is also important to consider other measurement strategies. For example, taking four sets of measurements with respect to four different bases instead of three considered here. This also involves finding the origin of the exact dependency of the Fisher information on the unitary, such as was found in Fig.~\ref{n3change}. 4) In our present study, we have ignored all effects of photon loss and decoherence, which is present in any real experimental setting. It is thus important to explore general multi-parameter estimation in the presence of photon losses and decoherence, and comparing to the results attained in single-parameter estimation in \cite{escher2011general} and references therein. 

A more speculative and interesting direction is the exploration of similarities between multi-parameter estimation and phase estimation problems in quantum computing. It has sometimes been hinted, though often too vaguely, the connection between phase estimation and quantum computation problems. It is known already that linear optical quantum computing can benefit from advancements in parameter estimation and vice versa \cite{rosetta,dowlinglowdown}. Furthermore, one work suggests a more solid connection between $SU(2)$ estimation and phase estimation \cite{hayashi}. It would prove fruitful to study and compare the resources for both these problems.
\begin{subappendices}
\section{Euler angles under basis change}
\label{sec:basischange}
Here we derive the changes in Euler angles under the basis change $HV \rightarrow DA$ and $HV \rightarrow RL$, following the procedure in section ~\ref{sec:differentbases}. There we derived $\alpha'=\beta'=1/\sqrt{2}$ and $\zeta=0$ for $HV \rightarrow DA$. Inserting this into Eq.~\eqref{eq:U11}, we find
\begin{align} \label{eq:mda}
&\mathcal{M'}_{11}=\cos\left(\frac{\psi_2}{2}\right)\cos\left(\frac{\psi_1+\psi_3}{2}\right)-i\sin\left(\frac{\psi_2}{2}\right)\sin\left(\frac{\psi_3-\psi_1}{2}\right) \nonumber \\
&\mathcal{M'}_{21}=-\sin\left(\frac{\psi_2}{2}\right)\cos\left(\frac{\psi_3-\psi_1}{2}\right)+i\cos\left(\frac{\psi_2}{2}\right)\sin\left(\frac{\psi_1+\psi_3}{2}\right).
\end{align}
Then using Eqs. ~\eqref{eq:psiprimes} and ~\eqref{eq:mda} we have the new Euler angles $\{\psi'_i\}$ written as
\begin{align}
\psi'_2 &=2\cos^{-1}\sqrt{\cos\left(\frac{\psi_2}{2}\right)^2\cos\left(\frac{\psi_1+\psi_3}{2}\right)^2+\sin\left(\frac{\psi_2}{2}\right)^2\sin\left(\frac{\psi_3-\psi_1}{2}\right)^2} \nonumber \\
\psi'_1 &=\tan^{-1}\left(-\tan\left(\frac{\psi_2}{2}\right) \frac{\sin\left(\frac{\psi_3-\psi_1}{2}\right)}{\cos\left(\frac{\psi_1+\psi_3}{2}\right)}\right)
            -\tan^{-1}\left(-\frac{1}{\tan\left(\frac{\psi_2}{2}\right)}\frac{\sin\left(\frac{\psi_1+\psi_3}{2}\right)}{\cos\left(\frac{\psi_3-\psi_1}{2}\right)}\right) \nonumber \\
\psi'_3 &=\tan^{-1}\left(-\tan\left(\frac{\psi_2}{2}\right) \frac{\sin\left(\frac{\psi_3-\psi_1}{2}\right)}{\cos\left(\frac{\psi_1+\psi_3}{2}\right)}\right) 
            +\tan^{-1}\left(-\frac{1}{\tan\left(\frac{\psi_2}{2}\right)}\frac{\sin\left(\frac{\psi_1+\psi_3}{2}\right)}{\cos\left(\frac{\psi_3-\psi_1}{2}\right)}\right). 
\end{align}
For $HV \rightarrow RL$, $\alpha'=-i/\sqrt{2}$, $\beta'=1/\sqrt{2}$, $\zeta=-\pi/2$. Inserting this into Eq.~\eqref{eq:U11}, we arrive at 
\begin{align}\label{eq:u11rl}
&\mathcal{M'}_{11}=\cos\left(\frac{\psi_2}{2}\right)\cos\left(\frac{\psi_1+\psi_3}{2}\right)-i\sin\left(\frac{\psi_2}{2}\right)\cos\left(\frac{\psi_3-\psi_1}{2}\right) \nonumber \\
&\mathcal{M'}_{12}=\sin\left(\frac{\psi_2}{2}\right)\sin\left(\frac{\psi_3-\psi_1}{2}\right)-\cos\left(\frac{\psi_2}{2}\right)\sin\left(\frac{\psi_3+\psi_1}{2}\right).
\end{align}
Then using Eqs.~\eqref{eq:psiprimes} and ~\eqref{eq:u11rl} we arrive at the new Euler angles $\{\psi_i''\}$ written as
\begin{align}
\psi_2''&= \nonumber \\
& 2\cos^{-1}\left(\cos\left(\frac{\psi_2}{2}\right)^2\cos\left(\frac{\psi_1+\psi_3}{2}\right)^2+\sin\left(\frac{\psi_2}{2}\right)^2\cos\left(\frac{\psi_3-\psi_1}{2}\right)^2\right)^{\frac{1}{2}} \nonumber \\
\psi_1''&=\tan^{-1}\left(-\tan\left(\frac{\psi_2}{2}\right)\frac{\cos\left(\frac{\psi_3-\psi_1}{2}\right)}{\cos\left(\frac{\psi_1+\psi_3}{2}\right)}\right)
            -\tan^{-1}\left(-\frac{\sin\left(\frac{\psi_3+\psi_1}{2}\right)}{\tan\left(\frac{\psi_2}{2}\right)\sin\left(\frac{\psi_3-\psi_1}{2}\right)}\right) \nonumber \\
\psi_3''&=\tan^{-1}\left(-\tan\left(\frac{\psi_2}{2}\right)\frac{\cos\left(\frac{\psi_3-\psi_1}{2}\right)}{\cos\left(\frac{\psi_1+\psi_3}{2}\right)}\right)
            +\tan^{-1}\left(-\frac{\sin\left(\frac{\psi_3+\psi_1}{2}\right)}{\tan\left(\frac{\psi_2}{2}\right)\sin\left(\frac{\psi_3-\psi_1}{2}\right)}\right).
\end{align}
\section{Schwinger representation}
\label{sec:schwinger}
The Schwinger representation \cite{sakurai} refers to a mapping from a two-mode $N$-particle bosonic state to a single spin-$j$ state
\begin{equation}
\ket{\frac{N}{2}+m, \frac{N}{2}-m} \longrightarrow \ket{j=\frac{N}{2},m}, 
\end{equation}
where the spin $j$ is equal to $N/2$ and $m$ are the quantum numbers taking values $\{-j, -j+1,...,j-1,j\}$. Therefore, a linear optical process for two modes (represented by an $SU(2)$ matrix), which takes a photonic $N$-particle state to a superposition of other $N$-particle states, is mapped to a rotation in the spin-$j$ system, where $j$ is preserved. A general $SU(2)$ rotation $U$ on spin-$j$ has an Euler decomposition 
\begin{align}
U=e^{i\psi_1 J_z} e^{i\psi_2 J_y} e^{i\psi_3 J_z},
\end{align}
where $J_x$, $J_y$ and $J_z$ are the total angular momentum operators for spin-$j$ states along the $x$, $y,$, $z$ axis, obeying $[J_i,J_j]=i\epsilon_{ijk}J_k$~\footnote{$\epsilon_{ijk}$ is the Levi-Civita symbol.}. To compute the action of $U$ on the bosonic creation operators $a^{\dagger}$, $b^{\dagger}$ (i.e. the right hand side of Eq.~\eqref{eq:mdef}) we want to know the action of the angular momentum operators $J_y, J_z$ on the bosonic creation operators. The Schwinger representation defines the following mapping
\begin{align}
J_y &=\frac{1}{2i}(a^{\dagger}b-a b^{\dagger}) \\
J_z &=\frac{1}{2}(a^{\dagger}a-b^{\dagger}b).
\end{align}
These can be used to derive the commutation relations $[J_z,a^{\dagger}]=1/2 a^{\dagger}$, $[J_z,b^{\dagger}]=-1/2 a^{\dagger}$, $[J_y,a^{\dagger}]=-1/(2i)b^{\dagger}$ and $[J_y,b^{\dagger}]=1/(2i) a^{\dagger}$. Using these relations and the Campbell-Baker-Hausdorff formula for two operators $X$ and $Y$: $\exp(X)Y\exp(-X)=Y+[X,Y]+1/2![X,[X,Y]]+1/3![X,[X,[X,Y]]]+...$, we obtain
\begin{align}
&e^{i \psi J_z} a^{\dagger} e^{-i \psi J_z}=e^{i \frac{\psi}{2}} a^{\dagger} \nonumber \\
&e^{i \psi J_z} b^{\dagger} e^{-i \psi J_z}=e^{-i \frac{\psi}{2}} b^{\dagger} \nonumber \\ 
&e^{i \psi_2 J_y} a^{\dagger} e^{-i \psi_2 J_y}=\cos \left(\frac{\psi_2}{2} \right) a^{\dagger}-\sin \left(\frac{\psi_2}{2} \right) b^{\dagger} \nonumber \\
&e^{i \psi_2 J_y} b^{\dagger} e^{-i \psi_2 J_y}=\sin \left(\frac{\psi_2}{2} \right) a^{\dagger}+\cos \left(\frac{\psi_2}{2} \right) b^{\dagger}.
\end{align} 
Therefore, the action of $U$ on the bosonic creation operators $a^{\dagger}$ and $b^{\dagger}$ is given by
\begin{align}
\begin{pmatrix}
U a^{\dagger} U^{\dagger} \\
U b^{\dagger} U^{\dagger}
\end{pmatrix}=
\begin{pmatrix}
e^{i\frac{\psi_1+\psi_3}{2}} \cos \left(\frac{\psi_2}{2} \right) & -e^{i\frac{\psi_3-\psi_1}{2}} \sin \left(\frac{\psi_2}{2} \right) \\
e^{-i\frac{\psi_3-\psi_1}{2}} \sin \left(\frac{\psi_2}{2} \right) & e^{-i\frac{\psi_1+\psi_3}{2}} \cos \left(\frac{\psi_2}{2} \right) 
\end{pmatrix} \begin{pmatrix}
a^{\dagger} \\
b^{\dagger}
\end{pmatrix}=
\mathcal{M}^T
 \begin{pmatrix}
a^{\dagger} \\
b^{\dagger}
\end{pmatrix}.
\end{align}
\section{Additivity of Fisher information}
\label{sec:FIadd}
One property of the Fisher information matrix is that given two independent observables $x,y$, the joint Fisher information matrix  of $x,y$ is the sum of the Fisher information matrix for $x$ and $y$ individually. We begin with the joint probability distribution $P(x,y; \underline{\theta})=P(x; \underline{\theta})P(y; \underline{\theta})$ where we have normalisation $\sum_{x,y} P(x,y; \underline{\theta})=1=\sum_x P(x; \underline{\theta})=\sum_y P(y; \underline{\theta})$. Then we can show the joint Fisher information is
\begin{align}
F(x,y)_{ij} &=\sum_{x,y} \left(\frac{\partial \log P(x,y; \underline{\theta})}{\partial \theta_i} \right) \left(\frac{\partial \log P(x,y; \underline{\theta})}{\partial \theta_j} \right) P(x,y; \underline{\theta})=\sum_{x,y} P(x; \underline{\theta})P(y; \underline{\theta}) \times \nonumber \\
&\left(\frac{\partial \log P(x; \underline{\theta})}{\partial \theta_i}+\frac{\partial \log P(y; \underline{\theta})}{\partial \theta_i}\right) \left(\frac{\partial \log P(x; \underline{\theta})}{\partial \theta_j}+\frac{\partial \log P(y; \underline{\theta})}{\partial \theta_j} \right) \nonumber \\
&=F(x)_{ij}+F(y)_{ij},
\end{align}
where we used
\begin{equation}
\sum_x  P(x; \underline{\theta})\frac{\partial \log P(x; \underline{\theta})}{\partial \theta_i}=\frac{\partial}{\partial \theta_i} \sum_x P(x; \underline{\theta})=0.
\end{equation}
For example, if we are using conducting our protocol using three independent bases $X,Y,Z$ an equivalent number of times then the total Fisher information matrix $F$ is $F=F_X+F_Y+F_Z$. 
\section{Additivity of quantum Fisher information}
\label{sec:QFIadd}
Suppose we begin with two $N$-particle symmetric states $\rho_A$ and $\rho_B$. We want to show the quantum Fisher information matrix of the $2N$ particle state $\rho_0=\rho_A \otimes \rho_B$ is a sum of the quantum Fisher information matrices of $\rho_A$ and $\rho_B$ separately, i.e.
\begin{align}
I_{\alpha \beta}(\rho_0)=\sum_{\xi=x, y, z}I_{\alpha \beta}(\rho_\xi).
\end{align}
We know from Eq.~\eqref{eq:totalqfisherinfo} that there are only three main terms in the quantum Fisher information matrix which are $\sum_{i=1}^{2N} \tr(\tr_{\backslash \{i\}}(\rho_0) t_{\alpha}t_{\beta})$, $\sum_{i \neq j=1}^{2N(2N-1)}\tr(\tr_{\backslash \{i,j\}}(\rho_0) (t_{\alpha}\otimes t_{\beta}))$ and $\sum_{i=1}^{N} \tr(\tr_{\backslash \{i\}}(\rho_0) t_{\alpha})$. Since $\rho_A$ and $\rho_B$ are both symmetric with $N$-particles each, we can write
\begin{align}
&\sum_{i=1}^{2N} \tr(\tr_{\backslash \{i\}}(\rho_0) t_{\alpha})=N(\tr(\rho_{A}^{[1]} t_{\alpha})+\tr(\rho_{B}^{[1]}t_{\alpha})) \nonumber \\
&\sum_{i=1}^{2N} \tr(\tr_{\backslash \{i\}}(\rho_0) t_{\alpha}t_{\beta})=N(\tr(\rho_{A}^{[1]} t_{\alpha}t_{\beta})+\tr(\rho_{B}^{[1]} t_{\alpha}t_{\beta})) \nonumber \\
&\sum_{i \neq j=1}^{2N(2N-1)}\tr(\tr_{\backslash \{i,j\}}(\rho_0) (t_{\alpha}\otimes t_{\beta}))=N(N-1)
[\tr(\rho_{A}^{[2]} t_{\alpha} \otimes t_{\beta})+\tr(\rho_{B}^{[2]} t_{\alpha} \otimes t_{\beta})] \nonumber \\
&+N^2[\tr(\rho_{A}^{[1]} t_{\alpha})\tr(\rho_{B}^{[1]} t_{\beta})+\tr(\rho_{B}^{[1]} t_{\alpha})\tr(\rho_{A}^{[1]} t_{\beta})],
\end{align}
where the first two terms come from summing over  $\{i, j\} \in [1, N]$ and $\{i, j\} \in [N+1, 2N]$. The last two terms come from the summation over all $\{i, j\}$ that do not belong to any of those intervals. Then using
\begin{align}
I_{\alpha \beta}(\rho_A)=4 \text{Re}[N\tr(\rho_{A}^{[1]} t_{\alpha}t_{\beta})+N(N-1) \tr(\rho_{A}^{[2]} (t_{\alpha} \otimes t_{\beta}))-N^2\tr(\rho_{A}^{[1]} t_{\alpha})\tr(\rho_{A}^{[1]} t_{\beta})].
\end{align}
and similarly for $\rho_B$, we can show
\begin{align}
I_{\alpha \beta}(\rho_0)=I_{\alpha \beta}(\rho_A)+I_{\alpha \beta}(\rho_B).
\end{align}
\section{Useful relations with Pauli matrices}
\label{sec:useful}
Pauli matrices $\sigma_x$, $\sigma_y$, $\sigma_z$ satisfy the following properties:\\

1) $\tr(\sigma_i)=0$ \\

2) $\{\sigma_i, \sigma_j \}=2 \delta_{ij} \mathbf{1}$ \\

3) $[\sigma_i, \sigma_j]=2i\epsilon_{ijk} \sigma_k$
where $\epsilon_{ijk}$ is the Levi-Civita symbol. We can represent them in terms of $\sigma_z$ eigenstates $\ket{\uu}$ and $\ket{\dd}$ where
$\sigma_z \ket{\uu}=\ket{\uu}$ and $\sigma_z \ket{\dd}=-\ket{\dd}$. The Pauli matrices can be written as
\begin{align}
\sigma_x=&\ket{\uu}\bra{\dd}+\ket{\dd}\bra{\uu} \\
\sigma_y=&-i\ket{\uu}\bra{\dd}+i\ket{\dd}\bra{\uu} \\
\sigma_z=&=\ket{\uu}\bra{\uu}-\ket{\dd}\bra{\dd}.
\end{align}
We can now write
\begin{align}
\sigma_x \otimes \sigma_x=& \ket{\uu\uu}\bra{\dd\dd}+\ket{\dd\dd}\bra{\uu\uu}+\ket{\uu\dd}\bra{\dd\uu}+\ket{\dd\uu}\bra{\uu\dd} \\
\sigma_y \otimes \sigma_y=& \ket{\uu\dd}\bra{\dd\uu}+\ket{\dd\uu}\bra{\uu\dd}-\ket{\uu\uu}\bra{\dd\dd}-\ket{\dd\dd}\bra{\uu\uu} \\
\sigma_z \otimes \sigma_z=& \ket{\uu\uu}\bra{\uu\uu}+\ket{\dd\dd}\bra{\dd\dd}-\ket{\uu\dd}\bra{\uu\dd}-\ket{\dd\uu}\bra{\dd\uu} \\
\mathbf{1}\otimes \mathbf{1}=&\ket{\uu\uu}\bra{\uu\uu}+\ket{\dd\uu}\bra{\dd\uu}+\ket{\uu\dd}\bra{\uu\dd}+\ket{\dd\dd}\bra{\dd\dd}
\end{align}
and
\begin{align}
\sigma_x \otimes \sigma_y=&-i\ket{\uu\uu}\bra{\dd\dd}+i\ket{\dd\dd}\bra{\uu\uu}+i\ket{\uu\dd}\bra{\dd\uu}-i\ket{\dd\uu}\bra{\uu\dd} \\
\sigma_y \otimes \sigma_x=&-i\ket{\uu\uu}\bra{\dd\dd}+i\ket{\dd\dd}\bra{\uu\uu}+i\ket{\dd\uu}\bra{\uu\dd}-i\ket{\uu\dd}\bra{\dd\uu} \\
\sigma_x \otimes \sigma_z=&\ket{\uu\uu}\bra{\dd\uu}+\ket{\dd\uu}\bra{\uu\uu}-\ket{\uu\dd}\bra{\dd\dd}-\ket{\dd\dd}\bra{\uu\dd} \\
\sigma_z \otimes \sigma_x=&\ket{\uu\uu}\bra{\uu\dd}+\ket{\uu\dd}\bra{\uu\uu}-\ket{\dd\uu}\bra{\dd\dd}-\ket{\dd\dd}\bra{\dd\uu} \\
\sigma_y \otimes \sigma_z=&-i\ket{\uu\uu}\bra{\dd\uu}+i\ket{\dd\uu}\bra{\uu\uu}+i\ket{\uu\dd}\bra{\dd\dd}-i\ket{\dd\dd}\bra{\uu\dd} \\
\sigma_z \otimes \sigma_y=&-i\ket{\uu\uu}\bra{\uu\dd}+i\ket{\uu\dd}\bra{\uu\uu}+i\ket{\dd\uu}\bra{\dd\dd}-i\ket{\dd\dd}\bra{\dd\uu}.
\end{align} 
These can be used in rewriting the following useful combinations
\begin{align}
\ket{\uu\uu}\bra{\uu\uu}+\ket{\dd\dd}\bra{\dd\dd}=&\frac{1}{2}(\mathbf{1}\otimes \mathbf{1}+\sigma_z \otimes \sigma_z) \nonumber \\
(\ket{\dd\uu}+\ket{\uu\dd})(\bra{\dd\uu}+\bra{\uu\dd})=&\frac{1}{2}(\mathbf{1}\otimes \mathbf{1}-\sigma_z \otimes \sigma_z+\sigma_x \otimes \sigma_x+\sigma_y \otimes \sigma_y) \nonumber \\
\ket{\dd\dd}\bra{\uu\uu}+\ket{\uu\uu}\bra{\dd\dd}=&\frac{1}{2}(\sigma_x \otimes \sigma_x-\sigma_y \otimes \sigma_y).
\end{align}
Furthermore
\begin{align}
\tr(\ket{\uu}\bra{\uu} \sigma_x)&=0=\tr(\ket{\uu}\bra{\uu} \sigma_y)=\tr(\ket{\dd}\bra{\dd} \sigma_x)=\tr(\ket{\dd}\bra{\dd} \sigma_y) \nonumber \\
\tr(\ket{\uu}\bra{\uu} \sigma_z)&=1=-\tr(\ket{\dd}\bra{\dd} \sigma_z) \nonumber \\
\tr(\ket{\uu}\bra{\dd} \sigma_x) &=1=\tr(\ket{\dd}\bra{\uu} \sigma_x) \nonumber \\
\tr(\ket{\uu}\bra{\dd} \sigma_y) &=i=-\tr(\ket{\dd}\bra{\uu} \sigma_y) \nonumber \\
\tr(\ket{\uu}\bra{\dd} \sigma_z) &=0=\tr(\ket{\dd}\bra{\uu} \sigma_z).
\end{align}
Let $\ket{+}$ ($\ket{-}$) be the positive (negative) eigenstate of $\sigma_x$ and $\ket{0}_y$ ($\ket{1}_y$) be the positive (negative) eigenstate of $\sigma_y$. Then we can write
\begin{align}
\ket{+}&=\frac{1}{\sqrt{2}}(\ket{\dd}+\ket{\uu}) \nonumber \\
\ket{-}&=\frac{1}{\sqrt{2}}(\ket{\dd}-\ket{\uu}) \nonumber \\
\ket{0}_y&=\frac{1}{\sqrt{2}}(\ket{\dd}-i\ket{\uu}) \nonumber \\
\ket{1}_y&=\frac{1}{\sqrt{2}}(\ket{\dd}+i\ket{\uu}),
\end{align}
where
\begin{align}
\ket{+}\bra{+} &=\frac{1}{2}(\ket{\uu}\bra{\uu}+\ket{\uu}\bra{\dd}+\ket{\dd}\bra{\uu}+\ket{\dd}\bra{\dd}) \nonumber \\
\ket{-}\bra{-} &=\frac{1}{2}(\ket{\uu}\bra{\uu}-\ket{\uu}\bra{\dd}-\ket{\dd}\bra{\uu}+\ket{\dd}\bra{\dd}) \nonumber \\
\ket{+}\bra{-} &=\frac{1}{2}(\ket{\uu}\bra{\uu}+\ket{\uu}\bra{\dd}-\ket{\dd}\bra{\uu}-\ket{\dd}\bra{\dd}) \nonumber \\
\ket{-}\bra{+} &=\frac{1}{2}(\ket{\uu}\bra{\uu}-\ket{\uu}\bra{\dd}+\ket{\dd}\bra{\uu}-\ket{\dd}\bra{\dd}) 
\end{align}
and 
\begin{align}
\ket{0}\bra{0}_y &=\frac{1}{2}(\ket{\uu}\bra{\uu}-i\ket{\uu}\bra{\dd}+i\ket{\dd}\bra{\uu}+\ket{\dd}\bra{\dd}) \nonumber \\
\ket{1}\bra{1}_y &=\frac{1}{2}(\ket{\uu}\bra{\uu}+i\ket{\uu}\bra{\dd}-i\ket{\dd}\bra{\uu}+\ket{\dd}\bra{\dd}) \nonumber \\
\ket{0}\bra{1}_y &=\frac{1}{2}(\ket{\uu}\bra{\uu}-i\ket{\uu}\bra{\dd}-i\ket{\dd}\bra{\uu}-\ket{\dd}\bra{\dd}) \nonumber \\
\ket{1}\bra{0}_y &=\frac{1}{2}(\ket{\uu}\bra{\uu}+i\ket{\uu}\bra{\dd}+i\ket{\dd}\bra{\uu}-\ket{\dd}\bra{\dd}).
\end{align}
\end{subappendices}

\newpage\null\newpage

\startlist{toc}
\printlist{toc}{}{\section*{\textbf{Thermodynamics of a squeezed state in cosmology and other relativistic scenarios} \\
Chapter contents}}

\chapter{Thermodynamics of a squeezed state in cosmology and other relativistic scenarios}
\label{chap:cosmo}
\section{Introduction and motivation}
The three great pillars forming modern theoretical physics are the laws of quantum mechanics, thermodynamics and gravitation. An area of study that demands a merging of all three pillars is the thermodynamics of quantum processes in a gravitational field, which lie in the relativistic regime. Among the most interesting directions of modern research in these relativistic quantum systems is the origin of the current entropy content of the universe \cite{Guth:81, 0264-9381-26-14-145005,  0004-637X-710-2-1825}, accelerating reference frames \cite{unruh, unruhapplications, davies, fullingmovingmirror} and the thermodynamics of black holes \cite{PhysRevD.7.2333, wald, hawking}. To address these questions requires applying thermodynamics to these systems. Thermodynamics is one of the most exportable branches of physics and has successfully been applied to understand small and large systems, including cosmological models. It has provided important predictions for entropy-matter relations in expanding cosmological models and entropy bounds for black hole scenarios. Two-mode squeezed states, as familiar from quantum optics, also play an important role in these scenarios. 

One particularly interesting question in this area is the origin of the entropy content of the universe. Matter contributes to the large entropy content of the universe \cite{Guth:81, 0264-9381-26-14-145005,  0004-637X-710-2-1825}. The initial emergence of matter could be a consequence of the expansion of the spacetime \cite{PhysRevLett.21.562,  PhysRev.183.1057,  PhysRevD.3.346} and is a quantum mechanical phenomenon called cosmological particle creation. Therefore, it is often assumed, but without proof, that entropy production should be directly related to particle creation \cite{prigogine}. However, the laws of physics are fully reversible, which has led to the conclusion that entropy, in the form of von Neumann entropy, cannot be increased in processes governed by physical laws such as Einstein and Schr\"odinger equations \cite{prigogine}. This apparent contradiction is still not fully understood. 

Finding an entropic quantity with a corresponding thermodynamical interpretation for the rapid expansion of spacetime at the earliest stages of the universe (as predicted by the inflationary scenario \cite{Guth:81}~\footnote{The current estimate is that the universe expanded by a factor of at least $10^{26}$ within $10^{-4}$ to $10^{-3}$ seconds during the inflationary period \cite{liddle}.}) is also challenging, since this requires a thermodynamics suitable for out-of-equilbrium processes. The tools to study quantum processes for systems perturbed arbitrarily far from equilibrium \cite{tasaki} has only relatively recently been introduced and has not yet been applied, until this work, to the study of quantum systems in the relativistic and gravitational regime, which includes cosmological particle creation. Our aim is to provide a formalism in which these recently developed thermodynamical tools can be used to study these scenarios. In particular, we apply this formalism to the study of the rapid initial expansion of the universe and show that a new entropic quantity suitable for this scenario can allow entropy increase during the fully reversible (unitary) evolution of quantum states in an expanding spacetime. 

We also make use of the observation that cosmological particle creation (and also related phenomena like the Unruh effect and the radiating black hole), in the simplest models, occurs by particle pair creation via two-mode squeezing. Two-mode squeezed states thus play an important role and this was known early on by Parker \cite{PhysRevLett.21.562, PhysRev.183.1057, PhysRevD.3.346}. However, this was only later phrased in terms of two-mode squeezing, as a concept familiar to quantum opticians, by Hu \cite{husqueezed}. The entropy production in cosmological particle creation can then be considered to be the entropy production during the creation of two-mode squeezed states from vacuum or a thermal state. 

The question of the very large entropy content of the universe was originally addressed by Guth \cite{Guth:81}, which predicts entropy creation to have occurred mostly after the inflationary period when the universe has slowed its expansion~\footnote{This occurs in the `re-heating' phase after the inflationary period \cite{kolb}.}. However, any entropy production due to cosmological particle creation, which is quantum in origin, during the inflationary period was neglected. Furthermore, some models suggest the need for additional sources of entropy production \cite{kolb, kolb2, olive}, like that arising from cosmological particle creation during the inflationary period.  It is also an interesting question in its own right to investigate if there is any entropy contribution from the quantum mechanical process of particle creation arising from vacuum fluctuations. This question was first studied by Hu et al \cite{hukandrup, kandrup} by computing the change in the von Neumann entropy during particle creation, but only in a subsystem of the quantum field whose fluctuations induce particle creation. Since the whole system is not taken into account and the evolution is no longer unitary, the change of von Neumann entropy is nonzero and can be related to the number of particles created. 

Similar ideas were followed in later works by Hu and collaborators \cite{kokshu, hureview} that included more detail relevant to real cosmological models \cite{gasperini1, gasperini2, brandenberger} and further clarified the interpretation of this von Neumann entropy \cite{hulin} as `loss of information' when one neglects parts of the whole system. However, if one takes into account the whole system, which undergoes unitary evolution, von Neumann entropy does not change. To attempt to solve this conundrum, Hu et al \cite{hu1986} proposed a new measure whose definition is more mathematically motivated and is inspired by von Neumann entropy. Although this measure is also shown to be related to the number of particles created, it does not have a firm thermodynamical interpretation. Thus, a strongly motivated entropic measure from thermodynamics that can still explain the entropic increase in closed system evolution is lacking. 

We propose a new viewpoint to this old problem by using an entropy that has a different origin to von Neumann entropy and is inspired by the recent developments in out-of-equilibrium thermodynamics. The entropy inspired from this field has a strong thermodynamical motivation and is appropriate for out-of-equilibrium processes like the rapid initial expansion of the universe. We will use an entropy called inner friction, which also has a clear interpretation in terms of irreversibility in cosmological expansion, something which is unclear in interpretations based on von Neumann entropy. 

Now we describe a little more about one of our main tools: out-of-equilibrium thermodynamics. Out-of-equilibrium thermodynamics is a relatively recent development \cite{evans, jarzynski, crooks, sevick2007fluctuation} that describes the thermodynamics of systems perturbed arbitrarily far from equilibrium. This extends the regime of equilibrium thermodynamics, which applies for processes that take one equilibrium state to another equilibrium state. This extension is important for the study of processes like cosmological particle creation in the inflationary period, where spacetime undergoes very rapid expansion. It also has wider applicability to nonequilibrium thermodynamics, which only studies processes near-equilibrium \cite{noneqm0, groot, noneqm1, noneqm2} and, like equilibrium thermodynamics, is mostly applicable only in the thermodynamic limit (i.e. large particle numbers where volume grows in proportion to particle number). In the small particle limit for example, where large fluctuations about the average can dominate, the formalism of equilibrium thermodynamics breaks down as it allows violations of the second law of thermodynamics. 

Out-of-equilibrium thermodynamics extends the applicability of thermodynamics to the small particle limit by generalising the second law of thermodynamics to the small system regime and the results are encapsulated by the so-called the fluctuation relations. For small quantum systems (like a two-mode squeezed state), the validity of a thermodynamics outside the thermodynamic limit become essential. These fluctuation relations also introduce new concepts of entropy production where fluctuations about the average, instead of being an artefact of small numbers, actually play a central role \cite{jarzynski, crooks, campisireview}. In addition, these entropies have a direct interpretation in terms of irreversibility of a thermodynamics process where microscopic laws remain reversible \cite{crooks, campisi2011} and is different in origin to the traditional Gibbs interpretation of entropy (and also the Shannon and von Neumann entropies) as `loss of information' or 'lack of knowledge' about the system \cite{tolman, shannon2015mathematical, petz2001entropy}. As we will see in this chapter, the fluctuation relations also adds new insight to the connection between entropy and irreversibility in cosmological particle creation.

Originally proposed by Evans \cite{evans}, Jarzynski \cite{jarzynski} and Crooks \cite{crooks} for classical systems, out-of-equilibrium thermodynamics has also been extended to the quantum regime by Tasaki \cite{tasaki}. The main application of out-of-equilibrium thermodynamics is currently to the study of nano and quantum engines \cite{sevick2007fluctuation, morebangbuck} finding the free-energy landscape for out of equilibrium processes \cite{gupta2011experimental, frey2015reconstructing, luccioli2008free} and condensed matter systems \cite{sindona2014statistics, shchadilova2014quantum, dorner2012emergent, mascarenhas2014work, sotiriadis2013statistics}. Entropy production in the creation of two-mode squeezed states, however, have largely been ignored in the context of the fluctuation relations. One possibility is that, although the role of single-mode squeezed states have recently been mentioned in the context of quantum heat engines \cite{rossnagel2014nanoscale}, two-mode squeezed states have yet to find applications in this area~\footnote{Another possible application is to optico-mechanical oscillators, where two-mode squeezed states have recently been used \cite{li2016enhanced}.}. Another possibility is that two-mode squeezing does not traverse phase transitions, which has been the central interest for applications of the fluctuations relations to condensed matter systems. However, we show that the thermodynamics of generating two-mode squeezed states is useful in relativistic quantum scenarios like cosmological particle creation. 

The main contribution of this work is to jointly use tools from quantum field theory in curved spacetime \cite{BandD} and the recently developed concepts from out-of-equilibrium thermodynamics of quantum systems \cite{campisireview} to investigate a relationship between entropy production and particle creation in an expanding universe. We explore applications of this approach to a simple model of cosmological expansion. 

The main advantage of our formalism is that it provides a way of dealing with thermodynamics beyond the linear response regime in these quantum and relativistic settings. We give a thermodynamic meaning to particle creation in terms of a quantity called \emph{inner friction} \cite{kosloff, engine,plastina}. We show that inner friction arises due to the quantum fluctuations of the fields and has an entropic interpretation stemming from a quantum fluctuation relation \cite{campisireview} and is related to the irreversibility in cosmological expansion. This is an entropy different to von Neumann entropy previously considered in the literature and could provide a new contribution to entropy. Our main result can be considered a quantum version of the second law of thermodynamics for an expanding universe which accounts for the creation of matter. Our formalism also applies to the Unruh effect and the radiating black hole, which also relies on two-mode squeezing as the core underlying physical process. \\

\textbf{Chapter outline.} \\

We introduce the preliminaries of quantum field theory in curved spacetime in section ~\ref{sec:qftintro1}, including a basic model of cosmological particle creation and an introduction to the Unruh effect and the radiating black hole. We show how two-mode squeezing arises in these settings. In section ~\ref{sec:thermointro}, we introduce thermodynamics for classical and quantum systems that applies far beyond the linear response regime, which is the regime for our model. We present our model in section ~\ref{sec:resultscosmo} and derive our main results. We summarise our results and directions for further work in section ~\ref{sec:cosmodiscussion}. In this chapter, we assume familiarity with elementary general relativity and equilibrium thermodynamics.
\section{Quantum field theory in curved spacetime and applications}
\label{sec:qftintro1}
We begin by providing the preliminaries of quantum field theory in curved spacetime in section ~\ref{sec:qftintro2}, starting with Minkowski (flat) spacetime in section ~\ref{sec:mink}. We then move on to describe a simple example of cosmological particle creation in an expanding spacetime in section ~\ref{sec:squeezingcosmo}, showing the role of two-mode squeezing. This basic model we use later on in our analysis. In section ~\ref{sec:otherscenarios} we provide a brief description of two other scenarios that is characterised by the same mathematical description as the cosmological particle creation: namely the Unruh effect and the radiating black hole. Our results on the quantum thermodynamics of cosmological particle creation can thus also be applied to these two scenarios, with some modifications in interpretation.
\subsection{Introduction to quantum field theory in curved spacetime}
\label{sec:qftintro2}
\subsubsection{Minkowski spacetime}
\label{sec:mink}
We begin with a scalar field $\phi (x, t)$ in (1+1)-dimensional spacetime whose equation of motion is
\begin{equation} \label{eq:KG1}
(\square+m^2) \phi=0,
\end{equation}
where  $\square \equiv \eta^{\mu \nu} \partial_{\mu} \partial_{\nu}$ ($\eta^{\mu \nu}$ is the Minkowski metric with signature $(-,+)$)
and $m$ is the mass associated with the quanta of $\phi$ when the theory is quantized. Here the spacetime is treated purely classically~\footnote{This is called the semi-classical approximation, which is valid if one is far from the limit of requiring a full description of quantum gravity (i.e. at the Plank scale).}. This equation of motion can be obtained from the Lagrangian density $\mathcal{L}=\frac{1}{2} (\eta^{\mu \nu} \partial_{\mu} \phi \partial_{\nu} \phi-m^2 \phi^2)$. One can construct the full solution to Eq.~\eqref{eq:KG1} from the following set of solutions
\begin{align} \label{eq:ucomplete}
u_{k}=\frac{1}{\sqrt{2 \omega (2\pi)^3}} e^{i k x-i\omega t},
\end{align}
where $k$ is the momentum, $\omega$ is the frequency associated with the mode $u_k$. The 
dispersion relation $\omega^2= \sqrt{k^2+m^2}$ is obtained by substituting the $u_k$ solution into Eq.~\eqref{eq:KG1}. The modes $u_k(x,t)$ are called positive-frequency with respect to $t$ if they are eigenfunctions of the operator $i\partial/\partial t$ with eigenvalues $\omega$ (i.e. $i\partial/\partial t u_k(x,t)=-i \omega u_k(x, t)$). These eigenfunctions will be used as the basis with which to expand $\phi(x,t)$. In defining a basis, a definition of orthogonality is needed, where the inner product~\footnote{This can be easily check to obey the conditions an inner product needs to satisfy:  (i) conjugate symmetry $(x,y)=\bar{(x,y)}$, (ii) linearity with respect to the first argument
$(ax, y)=a(x,y)$, $(x+y, z)=(x,z)+(y,z)$ and (iii) postive-definiteness $(x,x) \leq 0$, $(x,x)=0 \Rightarrow x=0$.} between $\phi_1$ and $\phi_2$ is 
\begin{equation} \label{eq:innerproduct}
(\phi_1, \phi_2) =-i \int (\phi_1 \partial_t \phi^*_2-(\partial_t \phi_1) \phi^*_2) d^3 x.
\end{equation}
Modes ${u_k}$ and ${u_k'}$ are known as orthogonal if $(u_{k}, u_{k'})=0$, where $k \neq k'$. The modes also satisfy  $(u_{k}, u_{k'}) =\delta(k-k')=-(u^*_{k}, u^*_{k'})$ and $(u_{k}, u^*_{k'})=0$. To quantize the scalar field (in this canonical quantisation description of a quantum field, one is working in the \textbf{Heisenberg picture}), one imposes the following equal-time commutation relations
\begin{align} \label{eq:commutation}
[\phi (x,t), \phi(x',t)]& =0 \nonumber \\
[\phi (x,t), \pi(x',t)]&=0 \nonumber \\
[\phi (x,t), \pi(x',t)]&=i \delta^3(x-x'),
\end{align}
where $\pi(x,t)$ is the conjugate momentum to $\phi(x,t)$ defined by $\pi=\partial \mathcal{L}/\partial (\partial_t \phi) =\partial_t \phi$. 
The modes $u_{k}$ in Eq.~\eqref{eq:ucomplete} and their complex conjugates $u^*_{k}$ form a complete orthonormal basis with respect to the inner product in Eq.~\eqref{eq:innerproduct} so the full solution to the field equation in Eq.~\eqref{eq:KG1} may be expanded as
\begin{equation} \label{eq:phidecomp}
\phi (x,t)=\sum_{k} [a_{k} u_{k}(x,t)+a^{\dagger}_{k} u^*_{k}(x,t)],
\end{equation}
where $a_{k}$ and $a^{\dagger}_{k}$ are known respectively as the annihilation and creation operators for the mode of the field $\phi$ with momentum $k$. These operators satisfy the commutation relations which apply to bosons $[a_{k},a_{k'}]=0=[a^{\dagger}_{k},a^{\dagger}_{k'}]$ and
$[a_{k},a^{\dagger}_{k'}]=\delta_{k,k'}$, which follows from Eq.~\eqref{eq:commutation}.

One can choose the number (or Fock) representation of the state of the quantum field, which labels states by the particle numbers $n_{k_i}$ present in each mode $k_i$ of the field. The vacuum state is denoted by $\ket{0}$ and all the other number states ${\ket{n_{k_1}, n_{k_2}, n_{k_3}...}}$ can be generated from the vacuum state using the creation operators $a^{\dagger}_{k}$, where $a^{\dagger}_{k} \ket{n_{k}}=\sqrt{n+1}\ket{(n+1)_{k}}$. 
The vacuum state ${\ket{0}}$ is defined to be the state that is annihilated by the annihilation operator $a_{k} \ket{0}=0$.

The decomposition of the field in Eq.~\eqref{eq:phidecomp} for Minkowski space is unique for all inertial observers. The uniqueness of this decomposition also means that since $a_k$ is unique, so is the vacuum, which is defined by the annihilation operator. It is a well-established result that this decomposition is not unique if one is in curved spacetime, which contains non-inertial reference frames. It is this non-uniqueness that gives rise to different vacua, depending on which reference frame one uses. It is this feature of quantum field in curved spacetime that can be used to explain cosmological particle creation.
\subsubsection{Curved spacetime}
The key difference about quantum fields in curved spacetime is that while in flat spacetime a quantum field has a unique expansion in terms of plane wave solutions, in curved spacetime there is more than one complete set of mode solutions. In curved spacetime, to solve for the mode solutions $u_k$ , one must solve the modified Klein-Gordon second-order differential equation in curved spacetime, which is
\begin{equation} \label{eq:KG}
(\square+m^2) \phi=0,
\end{equation}
where $\square \phi \equiv \sqrt{-g} \partial_{\mu} [\sqrt{-g} g^{\mu \nu} \partial_{\nu} \phi]$ and $g$ is the determinant of our spacetime metric $g_{\mu \nu}$. In this work, we only be consider metrics conformal to the Minkowski metric, i.e. $g_{\mu \nu}=\Omega^2 \eta_{\mu \nu}$. The Klein-Gordon equation obeyed by the mode solutions $u_k$ is then $\Omega^{n}[\partial_{\eta} (\Omega^{n-2} \partial_{\eta} u_k) -\partial_{\zeta} (\Omega^{n-2} \partial_{\zeta} u_k)]+m^2 u_k^2=0$, where $\zeta$ and $\eta$ are the conformal space-like and time-like coordinates respectively. The Klein-Gordon equation in curved spacetime has in general two distinct sets of solutions, which we will call $u_k(\zeta, \eta)$ ( `in' modes) and $\tilde{u}_k(\zeta, \eta)$ (`out' modes). We may now expand the field in terms both these mode solutions
\begin{align} \label{eq:dualphi}
\phi (x,\eta) &=\sum_k [a_k u_k (x, \eta)+ a^{\dagger}_k  u^*_k(x,\eta)] \nonumber \\
                    &=\sum_k [\tilde{a}_k \tilde{u}_k(x, \eta)+\tilde{a}^{\dagger}_k\tilde{u}^*_k(x,\eta)], 
\end{align}
where $\tilde{a}_k$, $\tilde{a}^{\dagger}_k$ are the annihilation and creation operators corresponding to the `out-modes' $\tilde{u}_k$ such that the bosonic commutation relations still hold $[\tilde{a}^{\dagger}_{k'}, \tilde{a}_k]=\delta_{kk'}$ and $[\tilde{a}^{\dagger}_{k'}, \tilde{a}^{\dagger}_k]=0=[a_{k'}, a_k]$. 
We will show that in general the `out-mode' ladder operators differ from the `in-mode' ladder operators
and it is from this difference that the phenomenon of particle creation due to expanding spacetime arises. 

That the `out-mode' annihilation operator is different from the `in-mode' annihilation operator is a central result
of quantum field theory in curved spacetime. The crucial point is that the vacuum of the `out-modes', denoted $\ket{\tilde{0}}$), are different because they are defined by
\begin{equation}
\tilde{a}_k \ket{\tilde{0}}=0,
\end{equation}
whereas the vacuum annhilated by $a_k$  (denoted $\ket{0}$) is defined by $a_k \ket{0}=0$. We will show that the vacua are different since the annhilation operators $a_k$ and $\tilde{a}_k$ are different and
their relationship can be found through Eq.~\ref{eq:dualphi} and the relationship between the 
`in-mode' and `out-mode' solutions. This situation is actually a very familiar one in condensed matter and quantum optics, where one works with Bogoliubov transformations and we will define them accordingly for quantum field theory in curved spacetime. The solutions $\{u_k \}$ and $\{\tilde{u}_k\}$ form two complete orthonormal solution sets
with respect to the inner product 
\begin{equation} \label{eq:curvedinnerproduct}
(\phi_1, \phi_2) =-i \int (\phi_1 \partial_{\mu} \phi^*_2-(\partial_t \phi_1) \phi^*_2) \sqrt{-\text{det} g_{\mu \nu}} dS^{\mu},
\end{equation}
where $dS^{\mu}=n^{\mu}dS$, $n^{\mu}$ is the future-directed vector orthonormal to the Cauchy hypersurface $S$ and the inner product is independent of the choice of hypersurfaces. That the solutions $\{u_k \}$ and $\{\tilde{u}_k\}$ are complete means that they can be related
by a Bogoliubov transformation $\tilde{u}_k=\sum_l(\alpha_{kl}u_l+\beta_{kl}u^{*}_l)$ where $\alpha_{kl}$, $\beta_{kl}$ are the Bogoliubov coefficients. An equivalent expression is
\begin{equation} \label{eq:utildeu}
u_k=\sum_l(\alpha^*_{kl}\tilde{u}_l+\beta_{kl}\tilde{u}^{*}_l).
\end{equation}
Inserting Eq.~\eqref{eq:utildeu} into the decomposition of $\phi$ in Eq.~\eqref{eq:dualphi}, we obtain the following
relationship between the `out-mode' and `in-mode' ladder operators
\begin{equation} \label{eq:atildea}
\tilde{a}_k=\sum_m \alpha^*_{km}a_m-\beta^*_{km}a^{\dagger}_m.
\end{equation}
This is a central equation in quantum field theory in curved spacetime, where the Bogoliubov coefficients $\alpha_{km}$, $\beta_{km}$ contain all the physics of the scenarios we study. Eq.~\eqref{eq:atildea} applies to all the scenarios considered in quantum field theory in curved spacetime, like cosmological particle creation, the Unruh effect, the radiating black hole, the collapsing black hole and the moving mirror. We see later that the first three examples are even more intimately related since they do not have complicated `mixing' of modes but only mix between two modes in a specific way called two-mode squeezing, the terminology derived from quantum optics. Two-mode squeezing describes the scenario where the Bogoliubov coefficients $\alpha_{km}$ and $\beta_{km}$ simplify to having non-zero contributions only from $m=k,-k$. This means that only modes $k$ and $-k$ interact or `mix' with one another. 
\subsection{Two-mode squeezing and cosmology}
\label{sec:squeezingcosmo}
We now specialise to the Robertson-Walker spacetime in $1+1$ dimensions with coordinates $(t,x)$. Here the line element is $ds^2=-dt^2+a^2(t)dx^2=\Omega^2(\eta)(-d \eta^2+dx^2)$, where $a(t)$ is the scale factor and $\Omega^2(\eta)$ is the conformal scale factor. The conformal time $\eta$ is defined by $d \eta=dt/a(\eta)$. 

We notice that the (1+1)-dimensional Klein-Gordon equation, in the convenient coordinates $\eta,x$, reduce to a Klein-Gordon equation in Minkowski spacetime
\begin{align}
\left(-\frac{\partial^2}{\partial\eta^2}+\frac{\partial^2}{\partial x^2}+m^2\Omega^2(\eta)\right)\phi(x,\eta)=0.
\end{align}
\sloppy
There are two plane wave solutions to the field equation, the `in' modes 
$u_k=(1/\sqrt{4 \pi \omega})\exp(ikx-i\omega \eta)$ in the asymptotic past and the `out' modes $\tilde{u}_k=(1/\sqrt{4 \pi \tilde{\omega}})\exp(ikx-i\tilde{\omega}\eta)$ in the asymptotic future, with frequencies
\begin{align} 
\omega_k &=\sqrt{k^2+m^2 \Omega^2\vert_{\eta \rightarrow -\infty}}, \nonumber \\
 \tilde{\omega}_k &=\sqrt{k^2+m^2 \Omega^2\vert_{\eta \rightarrow +\infty}}.
\end{align}
Isotropy, homogeneity~\footnote{Assumptions of isotropy and homogeneity of spacetime expansion is currently a good approximation to explain observations of the large-scale structure of the universe \cite{maddox1}, based on galaxy surveys. However, with recent advances in astronomical imaging and the rise of the Planck telescope, there are proposals for detecting anisotropic expansion of the universe in future telescopes \cite{anisotropy1} and constraining the degree is anisotropy with current data \cite{anisotropy2}.} and the conservation of momentum and energy \cite{BandD} simplify the Bogoliubov transformation between the `in' and `out' bosonic operators to 
\begin{align} \label{eq:atildea}
\tilde{a}_k=\alpha_k a_k+\beta^*_k a^{\dagger}_{-k}.
\end{align}
The evolution between the `in' and `out' annihilation operators is known in quantum optics as two-mode squeezing~\footnote{The generation of two-mode squeezed states is well-established in quantum optics and further references can be found in \cite{gerryknight}.}. This generates an entangled state with strong correlations between modes $k$ and $-k$ of the field. Two-mode squeezing creates or annihilates particles pair-wise in the $(k,-k)$ mode pair, thus can be likened to the creation or annihilation of particle/anti-particle pairs from the vacuum. The connection between two-mode squeezing operations and quantum field theory has proved useful in analysing cosmological particle creation \cite{grishchuk,  hu1994, ivetteentanglement, Fuentes:Mann:10}, the Unruh effect \cite{PhysRevA.76.062112} and Hawking radiation \cite{adesso:correlation}. Two-mode squeezing is a generic feature in some of the key predictions of quantum field theory in curved spacetime. The Bogoliubov coefficients in Eq.~\eqref{eq:atildea} satisfy $|\alpha_k|=\cosh(r_k)$, $|\beta_k|=\sinh(r_k)$, where $r_k$ is known as the squeezing parameter~\footnote{The squeezing parameter we saw in the chapter~\ref{chap:qumode} refers to one-mode squeezed states and is different to the squeezing parameter here, which applies only to two-mode squeezed states.} \cite{barnett}. 

\begin{figure}[ht!]
\centering
\includegraphics[scale=0.5]{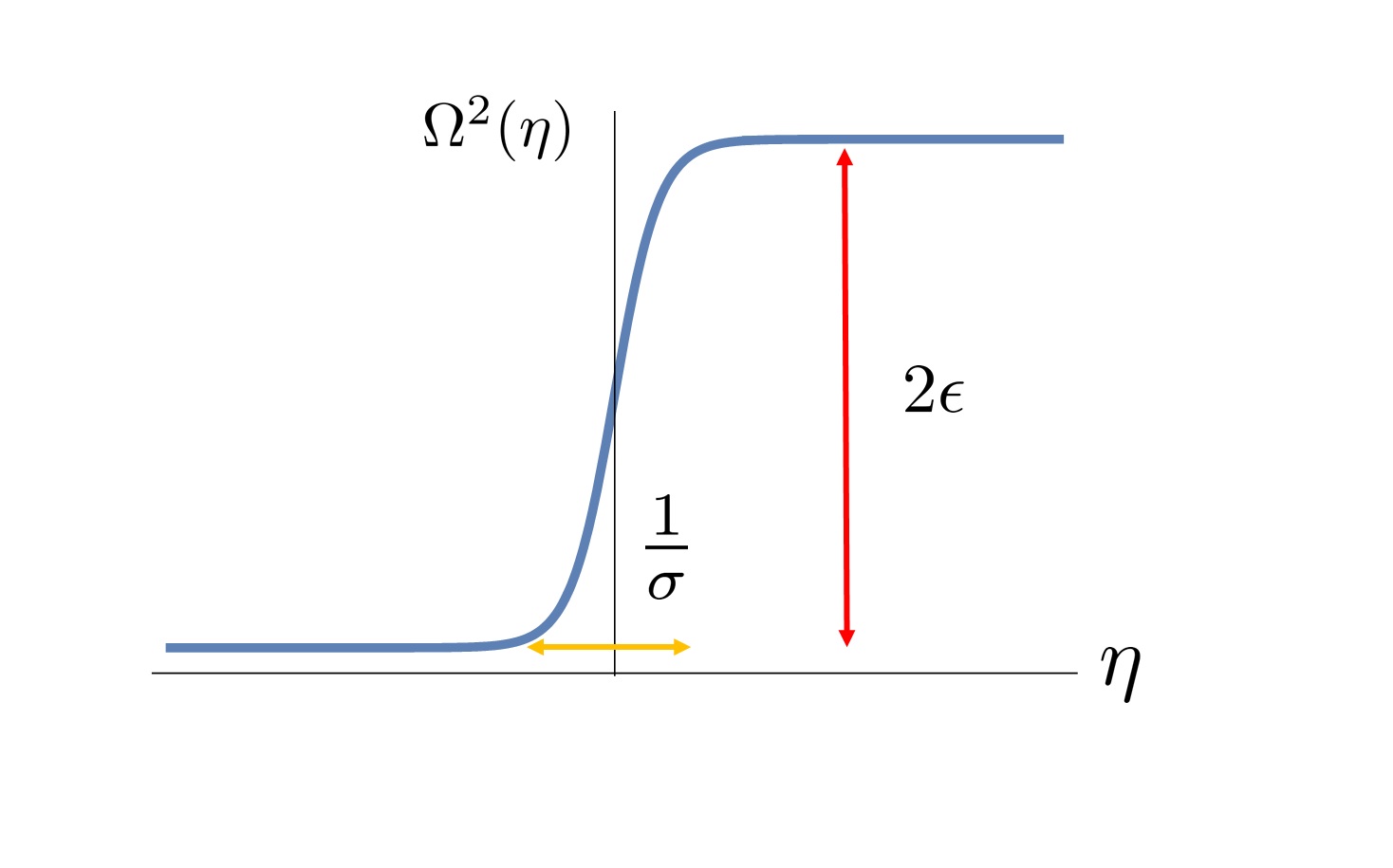}
\caption[\textit{Conformal factor}.]{\label{fig:conformal}\textit{Conformal factor}. This is the conformal factor in a simple model of cosmological expansion, where the spacetime expands isotropically and homogeneously. Using the chosen conformal factor $\Omega^2(\eta) = 1 + \epsilon ( 1 + \tanh(\sigma\,\eta))$, the spacetime expands to a size $2\epsilon$ within a conformal time interval $1/\sigma$.}
\end{figure}

If we choose the conformal form factor (see Fig.~\ref{fig:conformal})  $\Omega^2(\eta) = 1 + \epsilon ( 1 + \tanh(\sigma\,\eta))$, where $\epsilon,\sigma>0$ govern the total volume and rate of expansion respectively, it can be shown that the squeezing parameter obeys the equation $\tanh^2 (r_k)=\sinh^2 (\pi (\tilde{\omega}-\omega)/2 \sigma)/(\sinh^2 (\pi (\tilde{\omega}+\omega)/2 \sigma))$ \cite{BandD}.
The Bogoliubov coefffients, which relate to the squeezing parameter, are also related to expected number of particles one detects as a two-mode squeezed state is created from the vaccum. The expectation value of the `out-mode' number operator is $\bra{0} \tilde{n}_k \ket{0}=\bra{0} \tilde{a}^{\dagger}_k \tilde{a}_k \ket{0}= \sum_m |\beta_{km}|^2=\sinh^2(r_k)$. 

\begin{figure}[ht!]
\centering
\includegraphics[scale=0.5]{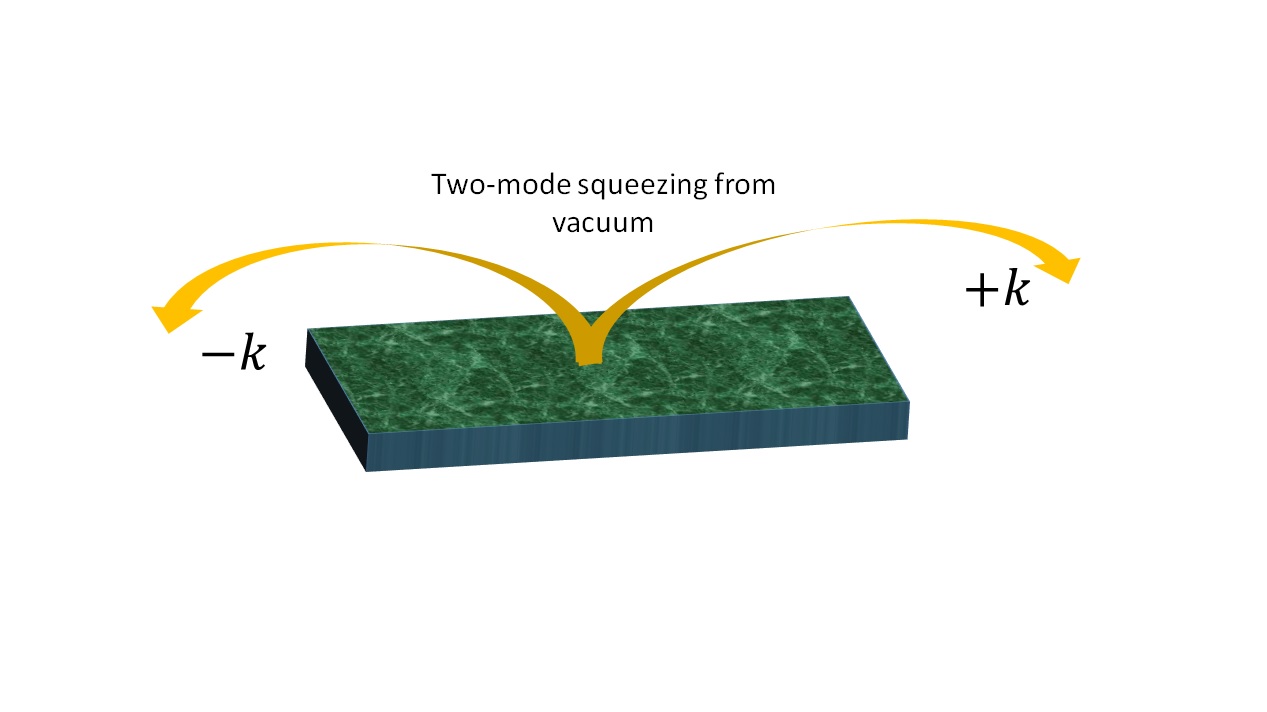}
\caption[\textit{Particle pair creation in cosmology}.]{\label{fig:cosmocreation}\textit{Particle pair creation in cosmology}. Cosmological particle creation from the vacuum can be modelled by two-mode squeezing (see text). Two-mode squeezing can be considered as particle creation by pairs of particles in modes $k$ and $-k$.}
\end{figure}

Now we can derive the initial (asymptotic past) and the final (asymptotic future) Hamiltonians (in the \textbf{Heisenberg picture}) corresponding to the creation of two-mode squeezed states from the vacuum (see Appendix~\ref{sec:DerivingHamiltonian} for a derivation)
\begin{align} \label{eq:hamneginfty}
\mathcal{H} (\eta \rightarrow -\infty) &=\sum_k \frac{\omega_k}{2}[a_k^{\dagger} a_k+a_k a_k^{\dagger}], \nonumber \\
 \mathcal{H} (\eta \rightarrow \infty) &=\sum_k \frac{\tilde{\omega}_k}{2}[\tilde{a}_k^{\dagger} \tilde{a}_k+\tilde{a}_k \tilde{a}_k^{\dagger}],
\end{align}
where each pair of modes $(k,-k)$ evolve unitarily. This means that there is interaction only between modes $k$ and $-k$ and no other modes of the field. Since each mode pair evolves independently under a change in spacetime, for simplicity we focus on one pair of modes $(k,-k)$ in order to illustrate our techniques (see Fig.~\ref{fig:cosmocreation}) . Thus we suppress all $k$ indices for the rest of this chapter. We define $a_k \equiv a$ and $a_{-k} \equiv b$ throughout the rest of the work. The initial and final Hamiltonian are
\begin{align} \label{eq:hamiltonian}
H &=\omega (a^{\dagger} a+b^{\dagger}b+1), \nonumber \\
\tilde{H} &= \tilde{\omega}(\tilde{a}^{\dagger} \tilde{a}+\tilde{b}^{\dagger} \tilde{b}+1).
\end{align}
The dynamics of the quantum field is therefore determined uniquely by the initial and final mode frequencies $\omega, \tilde{\omega}$ and the squeezing parameter $z$. Here we note that the energy spectrum is equally spaced for both the initial and final Hamiltonian. 
\subsection{The Unruh effect and the radiating black hole}
\label{sec:otherscenarios}
Here we provide a brief introduction to simple cases of two other scenarios in quantum field theory in curved spacetime that share the same mathematical desciption as cosmological particle creation: the Unruh effect and the radiating black hole. For a more detailed mathematical description of how two-mode squeezing plays a role in each case, please refer to Appendix~\ref{app:otherscenarios}.
\subsubsection{Unruh effect}
The Unruh effect describes a uniformally accelerating observer detecting a temperature $T$ in a quantum field which is in a vacuum state according to an inertial observer. This temperature is found to be proportional to the acceleration $a$ of the non-inertial observer. 

This result can be derived using quantum field theory in curved spacetime. A quantum field in a uniformly accelerating frame can be described by a quantum field in Rindler spacetime, which is related to Minkowski spacetime by a coordinate change that depends on the acceleration. This spacetime can be divided into two causally disconnected regions, called the left and right (which come from their location in the spacetime diagram representation). The right region is accessed by our uniformly accelerating observer and the left region cannot be accessed by this observer without faster-than-light travel. It turns out that the vacuum state in the original Minkowski spacetime (for the inertial observer) $\ket{0}_M$ is equivalent to a two-mode squeezed state in Rindler spacetime, where one mode $\ket{.}_L$ is in the left region and the second mode $\ket{.}_R$ resides in the right region. Choosing mode $k$ of a vacuum state of a quantum field with frequency $\omega_k$ in the inertial frame, a vacuum in Minkowski spacetime becomes 
\begin{align}
\ket{0}_M=\frac{1}{\cosh(r_k)} \sum_{n_k} \tanh^{n_k}(r_k) \ket{n_k}_L \ket{n_k}_R,
\end{align}
where $\tanh(r_k)=\exp(-2\pi \omega_k/a)$ and $\ket{n_k}_{L,R}$ is the number state in the left, right region of Rindler spacetime. However, this entire two-mode squeezed state is not accessible to our accelerating observer, who resides only within the right region of Rindler spacetime. Thus, to this observer, the state of mode $k$ of this quantum field $\rho_R$ is the single-mode reduced state of the two-mode squeezed state
\begin{align}
\rho_R=\tr_L(\ket{0}_M \bra{0}_M)=\frac{1}{\cosh^2(r_k)} \sum_{n_k}\tanh^{2n_k}(r_k)\ket{n_k}_R \bra{n_k}_R,
\end{align}
which is a thermal state with temperature $T=a/(2\pi)$. In fact, we can check its average particle excitation which shows a thermal spectrum of the above temperature
\begin{align}
\langle n_k \rangle \equiv \tr(a^{\dagger}_R a_R \rho_R)=\frac{1}{e^{-2\pi/(\omega_k a)}-1},
\end{align}
where $a^{\dagger}_R$, $a_R$ are creation and annihilation operators corresponding to the modes residing in the right region of Rindler spacetime. 

An initial and final Hamiltonian can also be derived in the same way as in the cosmological particle creation case and is of the same form as Eq.~\eqref{eq:hamneginfty} (with $\tilde{\omega_k}=\omega_k$), where the `initial' Hamiltonian describes the field in Minkowski spacetime and the `final' Hamiltonian applies in presence of acceleration. 
\subsubsection{Radiating black hole}
We look at an idealised case of a bosonic quantum field in the presence of a Schwarzschild black hole of mass $M$, which is spherically symmetric, stationary and only possess a mass parameter (i.e. no charge or angular momentum)~\footnote{The `no-hair' theorem for black holes \cite{nohair1, nohair2, nohair3} states that a black holes can, at most, be described by only three parameters: its mass, charge and angular momentum.}. 

The spacetime describing the black hole can also be divided into two causally disconnected regions, which is inside and outside the event horizon of a black hole. An observer in free-fall just outside the event horizon will fall through into the black hole, but will not be able to travel outside again without violating causality. This means for an observer to be maintained just outside the event horizon, the observer must accelerate uniformly in the opposite direction. The necessary acceleration is captured by the surface gravity of a black hole, which is $\kappa=1/4M$ for a Schwarzschild black hole. For this outside observer who is uniformaly accelerating, this is indistinguishable from Rindler spacetime. The inside and outside regions of this black hole become indistinguishable from the left and right regions of Rindler spacetime with acceleration $a=\kappa=1/4M$. Thus, all the mathematics from the Unruh case follows directly in this black hole example, including two-mode squeezing, now between the quantum field modes inside and outside the black hole. This squeezing $r_k$ satisfies $\tanh(r_k)=\exp(-2\pi \omega_k/\kappa)=\exp(-8\pi \omega_k M)$. Thus the presence of a black hole in the vacuum state of a quantum field appears thermal to an observer outside the black hole, so a black hole is said to radiate thermally. This temperature $T=\kappa/(2 \pi)=1/(8 \pi M)$ varies inversely with the mass of the black hole. The Hamiltonian for Minkowski spacetime (`initial' Hamiltonian) and for Schwarzschild spacetime (`final' Hamiltonian) can be equivalently derived as in the Unruh effect case by replacing the acceleration with the surface gravity of the black hole. 
\section{Thermodynamics in the out-of-equilibrium regime}
\label{sec:thermointro}
In this section, we introduce the preliminary concepts of out-of-equilibrium thermodynamics needed to understand how they can be applied to quantum systems perturbed arbitrarily far from equilibrium. The key results in out-of-equilibrium thermodynamics lie in the so-called fluctuation theorems which we define in sections ~\ref{sec:noneqmthermointro} and ~\ref{sec:noneqmthermointroquantum} for classical and quantum systems respectively. Here we introduce an entropic quantity called irreversible work, which is the simplest entry to thermodynamics in the out-of-equilibrium regime. We later use another entropic quantity in out-of-equilibrium thermodynamics regime called inner friction, that will play an important role in our analysis of cosmological expansion, which we briefly describe in section ~\ref{sec:innerfriction}. We later look into inner friction in more detail in section ~\ref{sec:resultscosmo}.

Equilibrium thermodynamical laws hold for processes where one system begins in equilibrium and undergoes a transformation where the state is always characterised by an equilibrium state. However, for many processes of physical interest, the perturbation may take one away from this regime. In particular, in the example in which we consider, as the universe expanded very rapidly during inflation, one's initial state is taken very far from equilibrium. 

Therefore, it is appropriate to consider thermodynamical laws that apply when states are perturbed arbitrarily far from equilibrium. The techniques of non-equilibrium thermodynamics \cite{groot, noneqm2} are insufficient for our purpose since it deals with transformations to states which are still near-equilibrium and hold only in the presence of large particle numbers, or the thermodynamic limit. It is not apparent that thermodynamical laws should generalise at all in this out-of-equilibrium limit. In fact, it is one of the very amazing discoveries in the past decade that in fact this is possible. In particular, the concept of entropy needs to be revisited for systems taken out-of-equilibrium since, as we soon see, traditional concepts of entropy appear to suggest violations of the second law of thermodynamics. 

\begin{figure}[ht!]
\centering
\includegraphics[scale=0.5]{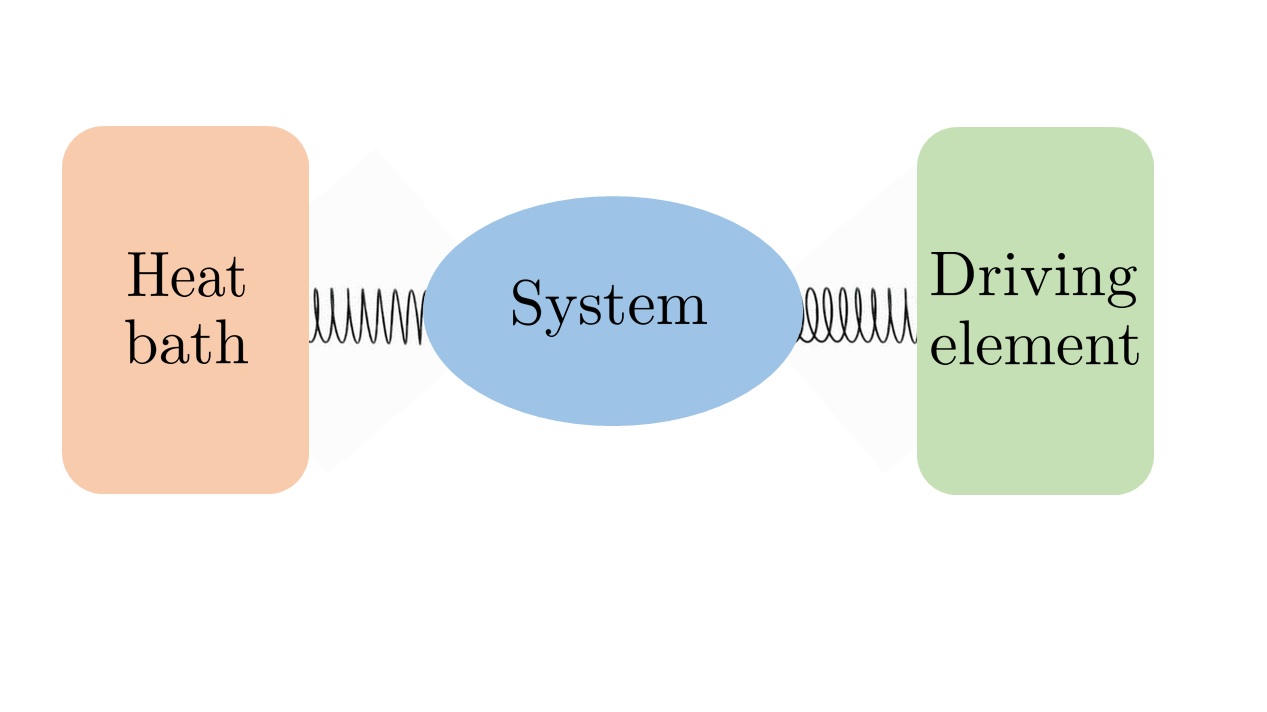}
\caption[\textit{A basic set-up in thermodynamics}.]{\label{setup}\textit{A basic set-up in thermodynamics}. It consists of a system, a heat bath to which the system is thermally coupled and a driving element that only couples to the system. Examples of the system/driving element are elastic band/ mechanical tweezer, atom/laser and quantum field/spacetime (see text).}
\end{figure}

Before we introduce the elements of out-of-equilibrium thermodynamics, we first note that one of the simplest set-ups in thermodynamics involves three main components (see Fig.~\ref{setup}). The first is the system under study. Another is a heat bath that the system is immersed in. This is a much larger system that is large enough to maintain a constant temperature. The third element is a driving element, which is mechanically, but not thermally, coupled to the system. It is capable of inputting energy to the system and transforming it. 

One example is that of an elastic band placed at room temperature with one end fixed and the other end being stretched by a mechanical tweezer. The elastic band is the system, the room the heat bath and the moving mechanical tweezer is the driving element. In a quantum mechanical set-up, one can also imagine a single multi-level atom, isolated in a cavity and held by an optical trap. Now imagine shining a laser from a small hole in the cavity onto the atom and changing its excitations. Here the system is the atom, the cavity is the heat bath and the laser is the driving element. We will see later that quantum fields in a changing spacetime can also be considered in this same framework, where the quantum field is the system and the background spacetime acts like a `mechanical' driving element.
\subsection{Classical thermodynamics in the out-of-equilibrium regime}
\label{sec:noneqmthermointro}
We can illustrate the difference between equilibrium and out-of-equilibrium thermodynamics using the example of an elastic band first undergoing an equilibrium process. Imagine an elastic band of original length $l_0$ in a room of temperature $T$. One end of the elastic band is fixed to the wall and the other is attached to a mechanical tweezer that can move set distances horizontally from the wall. Suppose the tweezer pulls the elastic band slowly until it reaches a final length of $l_f$. Let the total force exerted by the tweezer be $F$. Then the total work $W$ the tweezer has done on the elastic band is $W=\int_{l_0}^{l_f} F dl$. During this process, the elastic band also absorbs heat $Q$ from the room. The change in total internal energy $\Delta U$ of this elastic band can be expressed as the sum $\Delta U=W+Q$, which is a consequence of the first law of thermodynamics. 

One form of the second law of thermodynamics called the Clausius inequality \cite{adkins1983equilibrium}~\footnote{This is equivalent to the statement made earlier that entropy change is always positive. Here $\Delta S=\Delta S_{r}+\Delta S_{\text{irr}}$ denotes the total change in entropy where $\Delta S_r=\int_{\text{initial}}^{\text{final}} d Q/T $ is the entropy change associated with a reversible process and $\Delta S_{\text{irr}}$ is the entropy change associated with the irreversible part. Thus the Clausius inequality says $\Delta S_{\text{irr}}\geq 0$.} $\int_{\text{initial}}^{\text{final}} d Q/T \leq \Delta S$, where $dQ$ is an inexact differential and the integral is from the initial to the final state of the elastic band. The equality is only achieved for a fully reversible process, if the elastic band is pulled slowly enough. Combining the first and second law we have $W=\Delta U-Q=\Delta U-T \int_{\text{initial}}^{\text{final}} dQ/T \geq \Delta U-T \Delta S \equiv \Delta F$, where $F$ is known as the Helmholtz free energy. This equation can be rewritten as 
\begin{align}
W_{\text{irr}} \equiv W-\Delta F \geq 0,
\end{align}
where $W_{\text{irr}}$ is called the irreversible work. This is the difference between the total work done on a system and the work that would have been done on a system had the evolution being reversible and isothermal (captured by $\Delta F$). Thus, irreversible friction is zero only for a reversible isothermal process. The choice of a reversible isothermal process as a `benchmark' from which to compare the real evolution of the system is not a unique choice. One can also choose to compare the work done on a system and the work that would have been done had the evolution being reversible adiabatic. This latter case we will later examine in the case of inner friction. 

We can also understand the second law of thermodynamics by looking at cyclic processes (e.g. the elastic band starting and ending with the same length). The second law of thermodynamics for this process states that it is not possible to return an elastic band to its original length without heat being dissipated by the band to its surroundings, unless it is pulled in a fully reversible way. This means that there cannot be any process in which the only outcome is work being done. This statement forbids perpetual motion machines that can convert all of the heat produced in the process into work without heat being produced elsewhere. We can re-express this by writing the second law in terms of work performed in the `forward' process by $W_F$ and the work performed in the `reverse' process by $W_R$ as $W_F+\Delta F_F+W_R+\Delta F_R=W_F+W_R \equiv \Delta W \geq 0$, since $\Delta F_F=-\Delta F_R$. As the total internal energy is constant in a cyclic process, we have the inequality $\Delta Q=\Delta U-\Delta W=-\Delta W <0$. However, as we begin to access smaller and smaller systems and can realise ultra-fast external driving of a system, large fluctuations can dominate processes, where it may be possible that the observed outcome is \textit{not} the average value of the outcomes. Thus it is possible that during one experiment 
\begin{align}
W_F+W_R \leq 0.
\end{align}
This appears to be a momentary violation of the second law of thermodynamics. To maintain the second law of thermodynamics, it is important that we must interpret irreversibility as a statistical statement, in the sense that we only require entropy to increase for irreversible processes \textit{on average}, but not necessarily in any single process. Thus the central quantity to consider is no longer any single value for work or entropy,  but a \textit{distribution} of work $P(W)$ (or similarly distribution of entropy $P(s)$) of the system. It is thus more appropriate to have a second law where instead we only require a more general inequality to be obeyed
\begin{align}
\langle W_F+W_R \rangle \geq 0,
\end{align}
where $\langle . \rangle$ denotes an average respect to the number of trials. We can also rewrite the generalised second law as
\begin{align} \label{eq:workirraverage}
\langle W_{\text{irr}}  \rangle  \equiv \langle W \rangle -\Delta F\geq 0.
\end{align}
The inequality in Eq.~\eqref{eq:workirraverage} falls into the class of entropy production fluctuation theorems as originally introduced by Evans \cite{evans} and Crooks \cite{crooks} for classical systems. These relations, which relates entropy $s$ to the probability distributions of $s$ can be applied to systems perturbed arbitrarily far from equilibrium. This formulation is later extended to quantum systems \cite{tasaki, campisireview}. These relations take the general form
\begin{align}
e^s=P_F(s)/P_R(-s),
\end{align}
where $P_F(s)$ is the probability distribution of $s$ when the state begins in the initial thermal equilibrium state and is moved out of equilibrium. This is also called the `forward' distribution since it describes the $s$ distribution in the `forward' process (e.g. the elastic band being pulled from thermal equilibrium at length $l_0$ to $l_f$). $P_R(-s)$ is the 'reverse' probability distribution of $-s$ when a time-reversed protocol is performed on the systems when starting at the final thermal equilibrium state (e.g. the elastic band in thermal equilibrium with its surroundings at length $l_f$ and being contracted to length $l_0$ by the tweezer). Thus entropy $s$, defined in this context, quantifies the greater likelihood of the `forward' process occurring compared to the `reverse' process. 

The irreversible work is one example of an entropy obeying the fluctuation relation \cite{crooks} and it can be used to derive $\langle W_{\text{irr}} \rangle \geq 0$. It is possible to show that \cite{crooks} $P_F(W)/P_R(-W)=\exp((W-\Delta F)/T)=\exp(W_{\text{irr}}/T)$, where $\Delta F$ is the change of the Helmholtz free energy. This implies $\langle \exp(-W/T) \rangle =\exp(-\Delta F/T)$ which is known as the Jarzynski equality. Using the Jensen inequality $\langle \exp(X) \rangle \geq \exp(\langle X \rangle)$ it follows that $\exp(-\Delta F/T)=\langle \exp(-W/T) \rangle \geq \exp(\langle W \rangle/T)$ which recovers $\langle W_{\text{irr}} \rangle \geq 0$. This has been used to probe irreversibility of processes in a variety of systems like spin chains and ultracold gases \cite{dorner2012emergent, sindona2014statistics, silva2008statistics, apollaro2015work, yi2012work}. Another example of an entropy obeying a fluctuation relation is inner friction, which will be explored later in this chapter. 
\subsection{Quantum thermodynamics in the out-of-equilibrium regime}
\label{sec:noneqmthermointroquantum}
A quantum version of out-of-equilibrium thermodynamics have been recently introduced \cite{tasaki, campisireview}. Suppose a quantum system like an atom (in a cavity) begins at time $t=0$ in a thermal state $\rho(t=0)=\exp(-h_i/T)/Z_i$ described by some initial $h_i$ Hamiltonian (in the Schr\"odinger picture) and $Z_i$ is the initial partition function $Z_i=\text{Tr} \exp(-h_i/T)$. Let its energy eigenvalue and eigenstate be denoted by $E_n$ and $\ket{n}$ respectively. Let an external agent (like a laser shining onto the atom), which is treated classically, provide a perturbation to an initial Hamiltonian $h_i$ so the total Hamiltonian of the system evolves as $h(t)=h_i+v(t)h'$. Here $v(t)$ is a time-dependent parameter, $h'$ is a Hamiltonian and $h(t=0)=h_i$. At some final time $t=t_f$ let the external perturbation be turned off and we measure the energy of our time-evolved state $\rho_f \equiv \rho(t=t_f)$ to be $\bar{E}_{\bar{m}}$ and its corresponding energy eigenstate is $\ket{\bar{m}}$. The difference between these initial and final energies is
called work
\begin{equation}
W_{\bar{m}n}=\bar{E}_{\bar{m}}-E_n.
\end{equation}
This is analogous to the classical case with our elastic band, where the amount of work is also calculated by making two measurements: the length at the beginning and at the end after the external perturbation has been applied. 
The `forward' probability distribution of work in the \textbf{Schr\"odinger picture} is given by
\begin{align} \label{eq:probwork}
P_F(W) & =\sum_{\bar{m}n} P_{\bar{m}n} \delta(W_{\bar{m}n}-(\bar{E}_{\bar{m}}-E_n)) \nonumber \\
         &=\sum_{\bar{m}n}\bra{\bar{m}} U(t_f) \ket{n} \bra{n} \rho_i \ket{n} \bra{n} U^{\dagger}(t_f) \ket{\bar{m}}\delta (W_{\bar{m}n}-(\bar{E}_{\bar{m}}-E_n)),
\end{align}
where $P_{\bar{m} n}$ is the probability of obtaining the value of work $W_{\bar{m}n}$. The evolution operator is denoted $U(t)$ and is defined by the Schr\"odinger equation $i \hbar \partial U(t)/\partial t=h(t) U(t)$. 

\begin{figure}[ht!]
\centering
\includegraphics[scale=0.5]{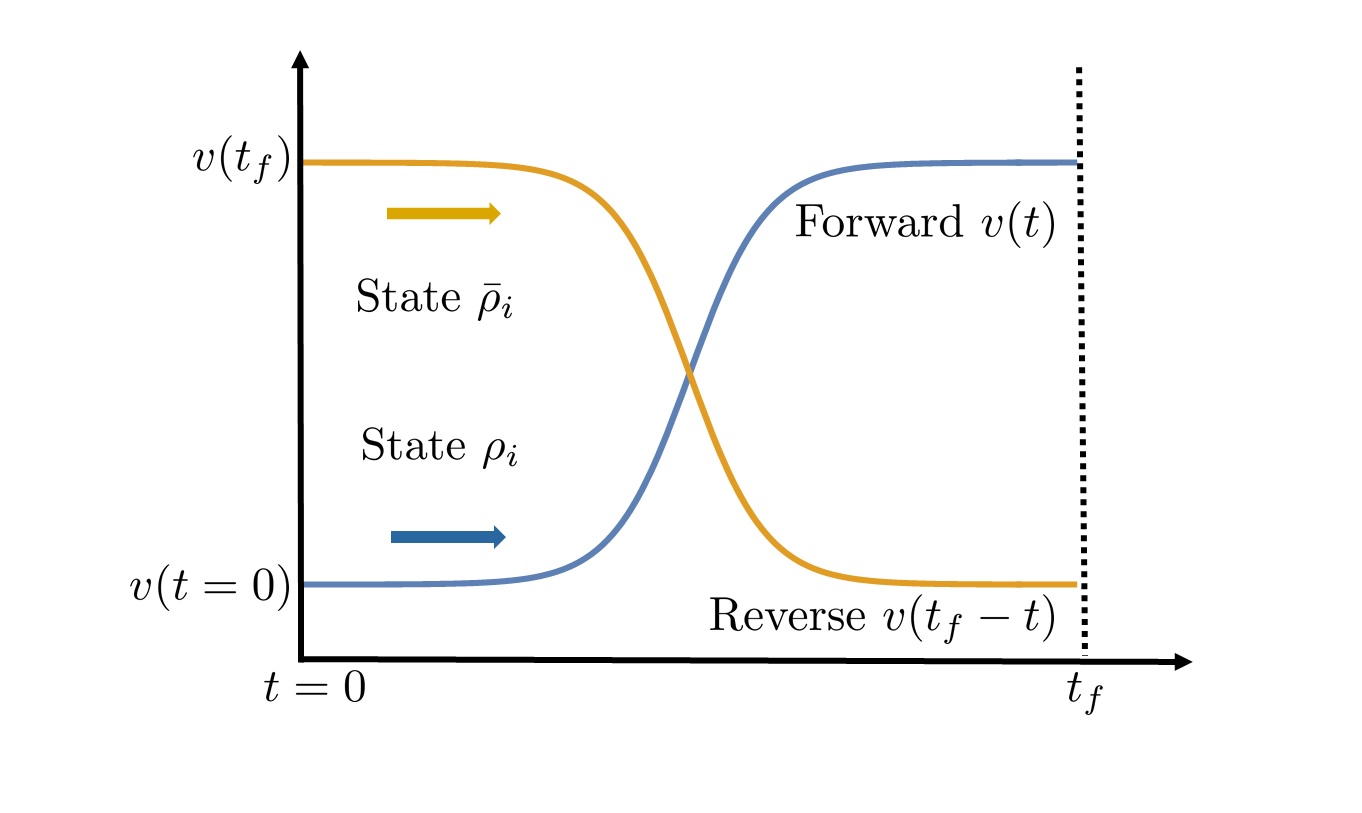}
\caption[\textit{`Foward' and `reverse' processes}.]{\label{fig:ForwardReverse}\textit{`Foward' and `reverse' processes}. Plot of the parameter $v(t)$ (blue line) that appear in the `forward' Hamiltonian $h(t)=h_i+v(t)h'$ and the parameter $v(t_f-t)$ (yellow line) that appear in the `reverse' Hamiltonian $h_R(t)=h_i+v(t_f-t)h'$ in the Schr\"odinger picture. The `forward' and `reverse' probability distributions in either work or entropy are defined with respect to $h(t)$ and $h_R(t)$ respectively.}
\end{figure}

The `reverse' probability distribution of work can likewise be defined as the process of beginning in the thermal state $\bar{\rho}_i=\exp(-h(t=t_f)/T)/Z_f$ of the \textit{final} Hamiltonian, where
$Z_f=\text{Tr} \exp(-h(t=t_f)/T)$ is the partition function corresponding to the final Hamiltonian. To perform a time-reversed protocol, one begins with the Hamiltonian $h(t=t_f)$ and end in the Hamiltonian $h_i$
while following the `reverse' Hamiltonian $h_R(t)=h_i+v(t_f-t)h'$ (see Fig.~\ref{fig:ForwardReverse}). We can thus write the `reverse' probability distribution of work as
\begin{align} \label{eq:probworkR}
P_R(-W)=\sum_{m\bar{n}}\bra{m} U_R(t_f) \ket{\bar{n}} \bra{\bar{n}} \bar{\rho}_i \ket{\bar{n}} \bra{\bar{n}} U_R^{\dagger}(t_f) \ket{m}\delta (W_{m\bar{n}}-(E_m-\bar{E}_{\bar{n}})),
\end{align}
where $U_R(t)$ is the evolution operator corresponding to the `reverse' Hamiltonian $h_R(t)$. 

Since the tools of quantum field theory in curved spacetime is developed using the Heisenberg picture, for our applications we need to define the distributions of work in the \textbf{Heisenberg picture}. In this case the state $\rho_i$ does not evolve. Let the `forward' Hamiltonian be denoted $H(t)$. The work distribution in Eq.~\eqref{eq:probwork} becomes
\begin{equation}
P_F(W)=\sum_{\tilde{m} n} \braket{\tilde{m}}{n}\bra{n} \rho_i \ket{n} \braket{n}{\tilde{m}}\delta(W-(\tilde{E}_{\tilde{m}}-E_n)),
\end{equation}
where $\{ \ket{\tilde{m}} \}$ are the energy eigenvectors of the final Hamiltonian in the Heisenberg picture and $\{ \tilde{E}_{\tilde{m}} \}$ are 
the corresponding energy eigenvalues. Likewise the `reverse' probability distribution of work is similarly defined with
\begin{align}
P_R(-W)=\sum_{m \tilde{n}} \braket{m}{\tilde{n}}\bra{\tilde{n}} \rho_f \ket{\tilde{n}} \braket{\tilde{n}}{m}\delta(W-(\tilde{E}_{\tilde{m}}-E_n)).
\end{align}
These distributions obey the normalisation $\int P_F(W) dW=1=\int P_R(-W)dW$.

Like in the classical scenario, irreversible work in the quantum case also obeys the fluctuation relation $P_F(W)/P_R(-W)=\exp((W-\Delta F)/T)=\exp(W_{\text{irr}}/T)$, where the `forward' and `reverse' work distributions are defined with respect to quantum evolution. Similarly to the classical case, irreversible work can be shown to obey $\langle W_{\text{irr}} \rangle \geq 0$.

Furthermore, irreversible work is also connected to relative entropy \cite{deffner1} and this interpretation is much more transparent in the Schr\"odinger picture. Let  $\rho_i=\sum_n p_n \ket{n}\bra{n}$ be the initial state where $p_n=\exp(-E_n/T)/Z_i$ and let $\rho_f$ be the final state. Irreversible work compares the difference between the actual process and a reversible isothermal path. So we define the final state after the hypothetical isothermal path as $\rho_{eq}=\sum_{\bar{m}} \bar{p}_{\bar{m}} \ket{\bar{m}}\bra{\bar{m}}$ where $\bar{p}_{\bar{m}}=\exp(-\bar{E}_{\bar{m}}/T)/Z_f$. Then the irreversible work is found to be equal to 
\begin{align}
\langle W_{\text{irr}} \rangle=T (\tr(\rho_f \ln \rho_f)-\tr(\rho_f \ln \rho_{eq})) \equiv T K(\rho_f || \rho_{eq}),
\end{align}
where $K(\tau|| \tau')$ is known as the quantum relative entropy \cite{kullbackleibler, sagawa} between states $\tau$ and $\tau'$ and it measures how far these two states can be distinguished from one another. For example $K(\tau||\tau)=0$ and Klein's inequality $K(\tau||\tau')>0$ for all states $\tau$, $\tau'$ always holds \cite{klein1931quantenmechanischen}. The positivity of this relative entropy term is an alternative way of showing $\langle W_{\text{irr}}\rangle>0$, which can be interpreted as a second law of thermodynamics in the out-of-equilibrium regime \cite{sagawa}. We will soon see that inner friction can also be expressed in terms of a quantum relative entropy between two states. 

We note here that unlike the entropy appearing in the fluctuation relations (which increase in irreversible and unitary processes),  the von Neumann entropy remains constant for unitary processes. Von Neumann entropy of a state $\rho_i$ is defined as
\begin{align}
S_{vN}(\rho_i)=-\tr(\rho_i \ln \rho_i),
\end{align}
which is a quantum mechanical analogue of the well-known Gibbs entropy in classical statistical mechanics~\footnote{Gibbs entropy $S_G$ of a classical state with a discrete energy spectrum is defined as $S_G=-\sum_i p_i \ln p_i$ where $p_i$ is the probability of finding the system at an energy $E_i$. Note that we have set Boltzsmann's constnat $k_B=1$.}. It depends only on the eigenvalues of the density matrix $\rho_i$, taking the value of zero for a pure state and its maximal value for a maximally mixed state. Suppose that the state undergoes a unitary transformation to state $\rho_f=U \rho_i U^{\dagger}=\sum_n p_n U\ket{n}\bra{n}U^{\dagger}=
\sum_n p_n \ket{n'}\bra{n'}$ where $\ket{n'}=U\ket{n}$ are not necessarily the energy eigenstates of the final Hamiltonian. Then $S_{vN}$ remains invariant under unitary processes since
\begin{align}
\tr(\rho_i \ln\rho_i) &=\sum_n \bra{n} \rho_i \sum_m \ln p_m \ketbra{m}{m}n\rangle=\sum_n \bra{n}\rho_i \ln p_n \ket{n}  \nonumber \\
                              &=\sum_n p_n \ln p_n=\sum_{n'} \bra{n'} \rho_f \sum_{n'} \ln p_n \ket{n'} \nonumber \\
                              &=\sum_{m'}\bra{m'} \rho_f \sum_{n'} \ln p_n \ket{n'} \braket{n'}{m'}=\tr(\rho_f \ln\rho_f).
\end{align}
\subsection{Inner friction}
\label{sec:innerfriction}
We have so far associated irreversible work with an entropic interpretation from the fluctuation relations and relating irreversible work to relative entropy. Although irreversible work provides the most accessible entry into the fluctuation relations (being a simple generalisation of a quantity familiar from traditional thermodynamics), it is not the only quantity that can be interpreted as an entropy through the fluctuation relations and the relative entropy. Another such quantity is called \textit{inner friction}. 

\begin{figure}[ht!]
\centering
\includegraphics[scale=0.5]{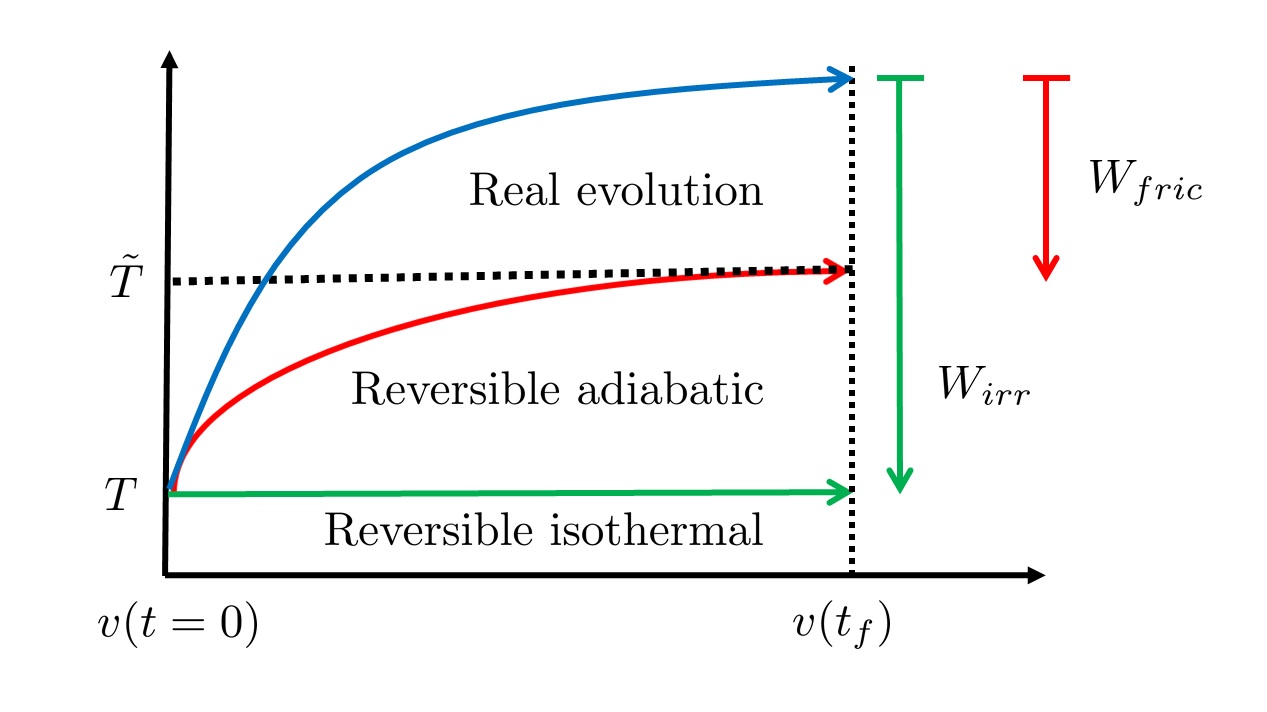}
\caption[\textit{Schematic diagram of the difference between inner friction and irreversible work}.]{\label{fig:innerfriction}\textit{Schematic diagram of the difference between inner friction $W_{fric}$ and irreversible work $W_{irr}$}. While irreversible work is the difference in work done during the real evolution and work done along a reversible isothermal path, inner friction is the difference in work done during the real evolution and work done along a reversible adiabatic (or quantum adiabatic) path. The quantum adiabatic path takes a thermal state at temperature $T$ to another thermal state at temperature $\tilde{T}$ (see text).}
\end{figure}
Recall that irreversible work is defined as the difference between the actual work done on a system and the work that would have been done on a system if the evolution of the system had been reversible and isothermal, where isothermal processes appear in reversible engine cycles like the Carnot cycle. Reversible adiabatic transformations also appear in these contexts (both Carnot and Otto engine cycles \cite{rossnagel2014nanoscale, abah2012single}). These transformations are also important in settings dominated by quantum mechanical finite-time processes, such as quantum mechanical systems undergoing a fast quench~\footnote{i.e. the evolution from the initial Hamiltonian to the final Hamiltonian happens almost instantaneously.}. 

Cosmological particle creation in a rapidly expanding spacetime is such a quantum mechanical quench, which makes reversible adiabatic transformations more interesting for our study. This motivates one to consider the work done for processes that depart from this ideal reversible adiabatic transformation. The `excess' work in not evolving along a quantum adiabatic path is captured by inner friction $W_{fric}$ (only recently introduced in \cite{plastina}) 
\begin{align}
W_{fric} \equiv W-W_{ad},
\end{align}
which is defined as the difference between the work done on a system and the work done in an ideal reversible adiabatic transformation $W_{ad}$, which we call the adiabatic work. This difference is called inner friction where $W_{ad}$ is the work done on the system supposing the system evolved in a reversible and adiabatic way (see Fig.~\ref{fig:innerfriction}). 

An adiabatic transformation in the quantum mechanical setting is known as a quantum adiabatic transformation, which does not generate any excitations, as we will see. This is easier to understand first in the \textbf{Schr\"odinger picture} (although the results are equivalent in the Heisenberg picture). Let us begin with a thermal state $\rho_i=\sum_n p_n \ket{n}\bra{n}$ with temperature $T$, where $p_n=\exp(-E_n/T)/Z_i$ is the occupation probability of the energy eigenstate $\ket{n}$ and $Z_i$ is the partition function of the initial Hamiltonian. A quantum adiabatic transformation does not change the occupation probabilities. This means an initial thermal state is transformed to another thermal state of a different temperature $\bar{T}$ when there are no level crossings. For example, if the final state is $\rho_f=\sum_{\bar{n}} \bar{p}_{\bar{n}} \ket{\bar{n}}\bra{\bar{n}}$, where $\{\bar{E}_{\bar{n}}\}$ are the energy eigenvalues of the final Hamiltonian and $Z_f$ is the partition function corresponding to the final Hamiltonian, then in an quantum adiabatic transformation
\begin{align}
\bar{p}_{\bar{n}}=\exp(-\bar{E}_{\bar{n}}/\bar{T})/Z_f=p_n=\exp(-E_n/T)/Z_i.
\end{align}
This constraint (which must hold for all $\bar{n},n$) means that the ratio between the initial and final energy eigenvalues is fixed by the ratio between the initial and final temperature $\bar{E}_{\bar{n}}/E_n=\bar{T}/T$. Thus the temperature $\bar{T}$ of the final state is $\bar{T}=\bar{E}_{\bar{n}}T/E_n$. The number of excitations is thus preserved since $\langle n\rangle_i =\sum_n n p_n=\sum_{\bar{n}}\bar{n}\bar{p}_{\bar{n}}=\langle n_f \rangle$, where $\langle n \rangle_i $ and $\langle n_f \rangle$ are the average initial and final number of particles. Thus departures from the quantum adiabatic limit can be associated with the particle creation, as we later see. 

Like irreversible work, inner friction can also be associated with a corresponding fluctuation relation and quantum relative entropy between two states (in the Schr\"odinger picture). We show in section ~\ref{sec:entprod} how inner friction is related to a fluctuation relation (in a different context and with a different derivation to \cite{plastina}).

We present here a derivation of the average inner friction (independently of \cite{plastina}) in terms of the quantum relative entropy between the actual final state and the state after a quantum adiabatic transformation. This formulation is more transparent in the Schr\"odinger picture  since the quantum relative entropy expresses a kind of distance between two states, whereas in the Heisenberg picture the state does not change. 

We note that the total average work is $\langle W \rangle=\tr(h_f \rho_f)-\tr(h_i \rho_i)$ where $h_f$ is the final Hamiltonian in the Schr\"odinger picture with $\{\bar{E}_{\bar{m}}\}$ and $\{\ket{\bar{M}}\}$ as its eigenvalues and eigenvectors. The final state in the Schr\"odinger picture is $\rho_f$ and the initial Hamltonian and state is the same as in the Heisenberg picture. Here $\rho_i=\sum_n p_n \ket{n}\bra{n}$ where $p_n=\exp(-E_n/T)/Z_i$. The total average work done in a quantum adiabatic transformation is $\langle W_{ad} \rangle=\tr(h_f \rho_{ad})$ where $\rho_{ad}=\sum_{\bar{m}}\bar{p}_{\bar{m}}\ket{\bar{m}}\bra{\bar{m}}$ is the final state after an adiabatic transformation, where $\bar{p}_{\bar{m}}=p_m$.  Therefore
\begin{align}
\langle W_{fric} \rangle &=\langle W \rangle -\langle W_{ad} \rangle =\tr(h_f \rho_f)-\tr(h_i \rho_i)-(\tr(h_f \rho_f)-\tr(h_f \rho_{ad})) \nonumber \\
&=\tr(h_f \rho_f)-\tr(h_f \rho_{ad})=\sum_{\bar{m}} \bar{E}_{\bar{m}}\bra{\bar{m}}\rho_f \ket{\bar{m}}-\sum_{\bar{m}}\bar{E}_{\bar{m}}\bar{p}_{\bar{m}}.
\end{align}
The first term $\tr(\rho_f \ln \rho_f)$ is minus the von Neumann entropy of the final state. Since we know that under unitary transformations the von Neumann entropy does not change and using $\ln(p_n)=-E_n/T-\ln(Z_i)=-\bar{E}_{\bar{n}}/\tilde{T}-\ln(Z_i)$ we can write
\begin{align}\label{eq:re1}
\tr(\rho_f \ln \rho_f)=\tr(\rho_i \ln \rho_i)=\sum_n p_n \ln p_n=-\frac{1}{\tilde{T}} \sum_n \bar{E}_{\bar{n}}p_n-\ln(Z_i).
\end{align}
The second term in the quantum relative entropy is equal to
\begin{align}\label{eq:re2}
\tr(\rho_f \ln \rho_{ad})&=\sum_{\bar{m}} \bra{\bar{m}} \rho_f \sum_{\bar{l}} \ln(p_l) \ket{\bar{l}} \braket{\bar{l}}{\bar{m}}     =\sum_{\bar{m}} \bra{\bar{m}} \rho_f \ket{\bar{m}} \ln(p_m) \nonumber \\
                                     &=-\frac{1}{\tilde{T}} \sum_{\bar{m}} \bar{E}_{\bar{m}} \bra{\bar{m}} \rho_f \ket{\bar{m}}-\ln(Z_i).
\end{align}
Taking the difference between terms in Eqs.~\eqref{eq:re1} and ~\eqref{eq:re2} we find
\begin{align}\label{eq:fricinf}
\avg{W}_{fric} =\tilde{T} (\tr(\rho_f \ln \rho_f)-\tr(\rho_f \ln \rho_{ad})) \equiv \tilde{T} \, K[\rho_f || \rho_{ad}],
\end{align}
where $\rho_f$ and $\rho_{ad}$ are the final states in the Schr\"odinger picture after the real transformation and a quantum adiabatic expansion respectively. From Klein's inequality $K(\tau||\tau')>0$ for any two states $\tau, \tau'$, thus $\avg{W}_{fric}>0$. Similarly to average irreversible work, the positivity of $\avg{W}_{fric}$ can also be interpreted as a second law of thermodynamics since $K(\tau||\tau')>0$ has been interpreted as a second law \cite{sagawa}. 
\section{Quantum thermodynamics for the expanding universe}
\label{sec:resultscosmo}
In this section we make use of the out-of-equilibrium thermodynamical formalism we have introduced and apply this to the scenario of cosmological particle creation.  In section ~\ref{sec:model} we describe the basic elements of our model. After specifying our key assumptions in section ~\ref{sec:keyassumptions}, we derive the average work done during cosmological particle creation in section ~\ref{sec:workspacetime}. We apply these results to relate inner friction directly to the expected number of particles created in an expanding spacetime.  In section ~\ref{sec:entprod} we show how inner friction can be interpreted in terms of an entropy through its corresponding fluctuation relation and well as an alternative representation in terms of relative entropy in the context of cosmological particle creation. Thus cosmological particle creation can be linked to entropy production. Lastly in section ~\ref{sec:unruhbhlastsummary} we interpret our mains results for cosmological particle creation in the context of the Unruh effect and the radiating black hole, since all three phenomena are similar in their mathematical descriptions. 
\subsection{Model}
\label{sec:model}
\begin{figure}[ht!]
\centering
\includegraphics[scale=0.5]{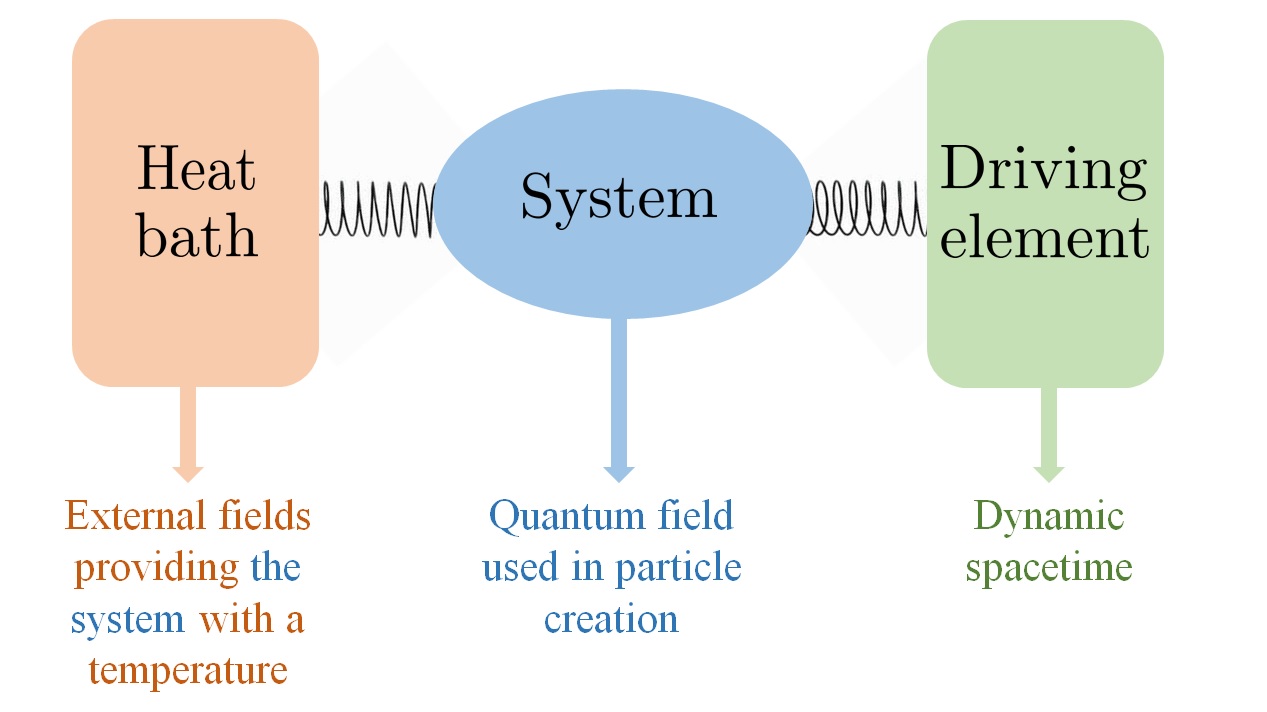}
\caption[\textit{Basic set-up of model}.]{\label{Cosmosetup2}\textit{Basic set-up of model}. The quantum field of interest (directly involved in particle creation) is the system. The external spacetime, being dynamical and coupled to the system, is the driving element and does work on the system. The heat bath is a collection of other fields that are present that can provide an initial temperature to the system.}
\end{figure}

The basic set-up of our model consist of three main elements: the system, the driving element and the heat bath (see Fig.~\ref{Cosmosetup2}). We choose our system to be a thermal state belonging to a massive quantum field. The spacetime couples to this field through its mass term and drives the dynamics of the quantum field like a parametric oscillator (i.e. by changing the effective mass of the field). The spacetime is classical and thus acts like a driving element that changes the Hamiltonian of our field through a classical coupling term. The heat bath can be considered like the surrounding `bath' of other quantum fields that provide our initial quantum field with a temperature. 
\subsubsection{Key assumptions}
\label{sec:keyassumptions}
Let us begin by specifying our main assumptions. (i) We confine our attention to two
modes of opposite momenta $(k, -k)$ of a quantum field for simplicity and define this to be our system. Since these two modes undergo two-mode squeezing with respect to each other as spacetime expands (in the model for cosmological particle creation in section ~\ref{sec:squeezingcosmo}), each mode pair evolves unitarily and independently from other $(k,-k)$ mode pairs. 

(ii) The ‘speed’ and ‘strength’ of interaction between
the spacetime and the quantum field is much greater than the ‘speed’ and ‘strength’ of
interaction between this field and any other external fields. In other words, we assume that
any interaction between our field and others that might be present in the Universe is negligible
during the time it takes for the mode pairs of our field of interest to be correlated via two-mode
squeezing. An example is in the inflationary scenario, where spacetime undergoes a
very rapid expansion. Therefore, each mode pair evolves unitarily and is described by the
Hamiltonian changing as $H \rightarrow \tilde{H}$. 

(iii) Our mode pair is approximated to start in a thermal
state $\rho$, which is a subsystem of the Universe. This comes from the result that any random
subsystem will be in a thermal state for almost all pure states of the Universe \cite{popescu}. This is due
to interaction between our state (the system) and other fields (the heat bath), before the spacetime begins expanding.
\subsubsection{Work done by expanding spacetime}
\label{sec:workspacetime}
Since we consider that spacetime is a classical external source, we can ask the question: \textit{how much work is done by the spacetime on the quantum fields as it expands}? In the case of a closed quantum system that (unitarily) evolves from the initial Hamiltonian $H$ to the final Hamiltonian $\tilde{H}$, the average work $\avg{W}$ is equivalent to the change in energy of the system \cite{campisireview}. We begin with a thermal state of a massive Klein-Gordon field. We measure its energy. Then we let the spacetime expand before making an energy measurement of the final state. The statistics of work is derived from making multiple such measurements in an ensemble of such expanding universes. We need two main ingredients. The first main ingredients
are the initial and final Hamiltonians in Eq.~\eqref{eq:hamiltonian}. The second ingredients are the Bogoliubov transformations between the ladder operators 
of the diagonalized initial and final Hamiltonians, whose origin we saw in the previous section. In the \textbf{Heisenberg picture}, when some initial Hamiltonian $H$ is changed to a final Hamiltonian $\tilde{H}$, the state $\rho$ remains the same. Let the initial Hamiltonian have eigenvalues and eigenvectors $\{E_n\}$ and $\{\ket{n}\}$. The final Hamiltonian
has corresponding eigenvalues and eigenvectors $\{\tilde{E}_{\tilde{m}}\}$ and $\{\ket{\tilde{m}}\}$. In this case, the `forward' work distribution can be written as
\begin{align}
P_F(W)=\sum_{\tilde{m}n}|\braket{\tilde{m}}{n}|^2\bra{n}\rho\ket{n} \delta(W_{\tilde{m}n}-(\tilde{E}_{\tilde{m}}-E_n)).
\end{align}
Then the average work done by the external agent responsible for the change in Hamiltonian is 
\begin{align}\label{eq:averageworkdef}
\langle W \rangle &=\int W P_F(W) dW \nonumber \\
                         &=\sum_{\tilde{m}n} |\braket{\tilde{m}}{n}|^2\bra{n}\rho_i\ket{n} (\tilde{E}_{\tilde{m}}-E_n) \nonumber \\
                         &=\sum_{n} \bra{n} \sum_{\tilde{m}}\tilde{E}_{\tilde{m}} \ketbra{\tilde{m}}{\tilde{m}}\rho\ket{n}-
\sum_{n} \bra{n}\sum_{\tilde{m}}\ketbra{\tilde{m}}{\tilde{m}}n \rangle E_n \bra{n}\rho\ket{n} \nonumber \\
                         &=\sum_n \bra{n}\tilde{H} \rho \ket{n}-\sum_n \bra{n}E_n \rho \ket{n} \nonumber \\
                         &=\tr(\tilde{H} \rho)-\tr(H\rho),
\end{align}
which holds when state is diagonal in the energy eigenstate ($\rho=\sum_n p_n \ket{n} \bra{n}$). 
Using Eqs.~\eqref{eq:averageworkdef} and ~\eqref{eq:hamiltonian} we can calculate average work to be
\begin{align} \label{eq:worknumber}
\langle W \rangle  &= \text{Tr} [\tilde{H} \rho]-\text{Tr}[H \rho] \nonumber \\
                           &= \text{Tr} [\rho \tilde{\omega} (\tilde{a}^{\dagger} \tilde{a}+\tilde{b}^{\dagger} \tilde{b}+1)]-\text{Tr}[\rho \omega (a^{\dagger}a+b^{\dagger}b+1)] \nonumber \\
                            &= \tilde{\omega} \langle n \rangle_{\text{tot}}-\omega \langle n \rangle_i+(\tilde{\omega}-\omega) \nonumber \\
                          &= \tilde{\omega} \langle n \rangle_{cr}+(\tilde{\omega}-\omega) \langle n \rangle_i+(\tilde{\omega}-\omega),
\end{align}
where $\avg{n}_{\text{tot}}$ is the total average number of excitations, $\avg{n}_i$ is the initial average number of excitations and $\avg{n}_{cr}$ is the average number of created particles due to the evolution of the system. This is essentially a statement of the conservation of energy and we can identify that there are three different contributions to work. The first term $\tilde{\omega} \avg{n}_{cr}$ is the work cost associated with the creation of new particles. The second term $(\tilde{\omega}-\omega) \avg{n}_i $ is the work cost in changing the frequencies of the particles already present in the initial thermal state. Finally, the cost $\tilde{\omega}-\omega$ of changing the ground state energy of the system. Note that the particle creation term $\tilde{\omega} \avg{n}_{cr}$ does not arise from particle interaction, like particle decay and collisions (entropies in these other regimes are treated elsewhere, see \cite{hucosmology}).

If the expansion occurs in a quantum adiabatic limit, then there are no transitions between different energy levels during the evolution. In this limit, the Bogoliubov coefficients $\beta_k$ vanish and the final adiabatic Hamiltonian is
\begin{align}\label{eq:adHam}
\tilde{H}_{ad} = \tilde{\omega} (a^{\dagger} a+b^{\dagger}b+1) = \frac{\tilde{\omega}}{\omega} H.
\end{align}
In this quantum adiabatic scenario, the average work done by spacetime onto the fields is defined as the adiabatic work $\avg{W}_{ad}$ which reads
\begin{align} \label{eq:adwork}
\avg{W}_{ad} \equiv \tr(\tilde{H}_{ad}\rho)-\tr(H \rho)=(\tilde{\omega}-\omega)(\avg{n}_i +1).
\end{align}
Note that no particles are created in an adiabatic evolution, as we saw in section ~\ref{sec:innerfriction}. This happens when either the rate of spacetime expansion is quasistatic (i.e. $\sigma \rightarrow 0$) or when the coupling between the field and spacetime disappears, which occurs for a massless scalar field. The difference between the average work $\avg{W}$ and the average adiabatic work $\avg{W}_{ad}$ defines the quantity $\avg{W}_{fric}$ called \textit{inner friction} that we introduced in section ~\ref{sec:innerfriction}. In our cosmological setting the inner friction is directly proportional to particle creation
\begin{align}\label{eq:workfriction}
\avg{W}_{fric} \equiv \avg{W}-\avg{W}_{ad}= \tilde{\omega}\, \avg{n}_{cr}.
\end{align}
This result fits one's intuition that the more particles are created, the farther one is from adiabatic evolution. In the absence of particle creation when the universe expands adiabatically, there is still a work cost in expanding without inner friction being produced, which is quantified by $\avg{W}_{ad}$. Our final step is to show how inner friction $\avg{W}_{fric}$ can be interpreted as an entropy production, to be defined below, in the cosmological context.
\subsection{Entropy production and cosmological particle creation}
\label{sec:entprod}
Inner friction can also be considered as quantifying entropy production during cosmological particle creation from the viewpoint of the entropy production fluctuation theorems. The fluctuation theorems in the context of entropy production were introduced first in classical systems by Evans \cite{evans} and Crooks \cite{crooks} to define entropy production $s$ for systems when perturbed arbitrarily away from equilibrium and were later extended to quantum systems \cite{campisireview,tasaki}. These relations take the general form $\exp(s)=P_F(s)/P_R(-s)$, where $P_F(s)$ is the probability distribution of $s$ when beginning from equilibrium. This is also called the `forward' distribution. $P_R(-s)$ is the (`reverse') probability distribution of $s$ when a time-reversed driving is applied to the system starting at equilibrium. Thus entropy production, defined in this sense, expresses the difference between the `forward' and `reverse' probability distributions. 

This way of viewing entropy production motivates us to define `forward' and `reverse' processes in cosmology and then to derive a corresponding fluctuation theorem. We define our `forward' process to be the expansion of spacetime beginning in an equilibriums state of $H$, where this state is $\rho=\sum_j \exp(-E_j/T) \ket{j}\bra{j}/Z$ and $Z=\sum_j \exp(-E_j/T)$ is the partition function. Here $\{E_j\}$, $\{\ket{j}\}$ are the energy eigenvalues and eigenstates of $H$ and $T$ can be considered as the temperature of the state. The `reverse' process is the contraction of this spacetime but beginning in the final adiabatic Hamiltonian $\tilde{H}_{ad}$. The state $\rho$ remains the same since we are working in the \textbf{Heisenberg picture}. Now let $p_{n}$ be the probability of $n$ particles being found initially in one run of spacetime expansion. Let $q_{m}$ be the probability that $m$ particles are initially found in the spacetime contraction process. We can associate the entropic quantities $-\log(p_{n})$ and $-\log(q_{m})$ to these probabilities. We can then define the difference of these entropic quantities as 
\begin{align}
s_{nm}=-\log(q_{m})+\log(p_{n}),
\end{align}
where $p_{n}=\bra{n}\rho \ket{n}= \exp(-E_{n}/T)/Z$ and $q_{m}=\bra{m}\rho \ket{m}= \exp(-E_{m}/T)/Z$. We can thus rewrite our new entropic random variable as
\begin{equation}\label{eq:snm}
s_{nm}=\frac{E_m}{T}-\frac{E_n}{T}=\frac{\tilde{E}_m}{\tilde{T}}-\frac{E_n}{T},
\end{equation}
where $\{\tilde{E}_j\}$ are the eigenvalues of $\tilde{H}$ and we define an effective temperature $\tilde{T} \equiv \tilde{E}_j T/E_j$.  In our model, $\tilde{T}$ is a constant since the energy spectrum of $H$ and $\tilde{H}$ are equally-spaced. This is also equivalent to $\tilde{T}=(\tilde{\omega}/\omega)T$ in our model.

We are now ready to define the probability distribution for an entropic quantity $s$ in the expansion process as
\begin{equation}\label{eq:pe}
P_E(s)=\sum_{n, m}\delta(s-s_{nm}) p_{m|n}p_{n},
\end{equation}
where $p_{m|n}=|\braket{\tilde{m}}{n}|^2$ and $\{\ket{\tilde{n}}\}$ are the eigenvectors of $\tilde{H}$. The term $p_{m|n}$ is the transition probability in going from $n$-particles in the beginning of expansion to $m$-particles at the end of expansion. Similarly, for the corresponding contraction process we can define 
\begin{align}
P_C(-s)=\sum_{n, m}\delta(s-s_{nm}) q_{n|m}q_{m}.
\end{align}
Here $q_{n|m}$ is the transition probability of going from $m$ particles to $n$ particles during spacetime contraction. Note that the normalisation conditions for both probability distributions are obeyed $\int P_E(s)ds=1=\int P_C(-s) ds$. Using Eqs.~\eqref{eq:snm}, ~\eqref{eq:pe} and the thermal state $\rho$ we find 
\begin{align}
\langle e^{-s} \rangle\equiv \int e^{-s}P_E(s)ds=\sum_{nm}e^{-s_{nm}}\braket{\tilde{m}}{n}\bra{n}\rho \ket{n}\braket{n}{\tilde{m}}=1.
\end{align} Combined with the normalisation conditions $\int \exp(-s)P_E(s)ds=1=\int P_C(-s)ds$, we find our entropic quantity $s$ satisfies the following fluctuation relation 
\begin{equation}
e^s=\frac{P_E(s)}{P_C(-s)}.
\end{equation}
This suggests the process in which $s$ is positive is exponentially more likely in the spacetime expansion case compared to the contraction process. Taking the logarithm on both sides and taking the average with $P_E(s)$ we have
\begin{align}
\avg{s} &\equiv \int s P_E(s)ds \nonumber \\
&=\int \ln\left(\frac{P_E(s)}{P_C(-s)}\right) P_E(s)=K[P_E(s) \| P_C(-s)] \geq 0, \label{entpos}
\end{align}
where $K[P_E(s) \| P_C(-s)]$ is the Kullback-Leibler divergence (or relative entropy) between probability distributions $P_E(s)$ and $P_C(-s)$ \cite{kullbackleibler}. This entropic quantity $\avg{s}$ is positive since the relative entropy $K[X\|Y]$ is always positive. It vanishes only when $P_E(s)=P_C(-s)$, i.e. for an adiabatic expansion of the spacetime, where no particles are created. This can be understood from the connection between this entropic quantity and particle creation. We show this relationship explicitly by demonstrating $\langle s \rangle$ is also proportional to our inner friction term. Using Eq.~\eqref{eq:pe} and $\rho=\sum_j p_j \ket{j}\bra{j}$ we have
\begin{align} \label{eq:ttildes}
\tilde{T} \langle s \rangle \equiv \tilde{T} \int sP_E(s)ds=\tilde{T} \sum_{mn} s_{nm}\braket{\tilde{m}}{n}p_n\braket{n}{\tilde{m}}.
\end{align}
Then inserting Eqs.~\eqref{eq:adHam}, ~\eqref{eq:workfriction} and ~\eqref{eq:snm} into Eq.~\eqref{eq:ttildes} we derive 
\begin{align}\label{eq:sandW}
\tilde{T} \langle s \rangle &=\sum_{nm} \left(\frac{\tilde{E}_m}{\tilde{T}}-\frac{E_n}{T}\right)\braket{\tilde{m}}{n}p_n\braket{n}{\tilde{m}} \nonumber \\
                                      &=\sum_{m} \tilde{E}_m\bra{\tilde{m}}\sum_n p_n \ket{n}\braket{n}{\tilde{m}}-\frac{\tilde{T}}{T} \sum_{n}E_n p_n \nonumber \\
                                      &=\sum_n\bra{n} \sum_m \tilde{E}_m \ketbra{\tilde{m}}{\tilde{m}} \rho \ket{n}-\frac{\tilde{T}}{T} \sum_n \bra{n} H \rho \ket{n} \nonumber \\
                                        &=\tr(\tilde{H}\rho)-\frac{\tilde{\omega}}{\omega} \tr(H \rho) \nonumber \\
                                        &=(\tr(\tilde{H}\rho)-\tr(H\rho))-(\tr(\tilde{H}_{ad}\rho)-\tr(H\rho)) \nonumber \\
&=\langle W \rangle -\langle W \rangle_{ad}=\langle W\rangle_{fric} .
\end{align}
From Eqs.~\eqref{eq:workfriction} and ~\eqref{eq:sandW} we now have an exact relationship between an entropy production and the number of particles created
\begin{equation}\label{eq:snumber}
\langle s \rangle =\frac{\langle W\rangle_{fric} }{\tilde{T}}=\frac{\tilde{\omega}}{\tilde{T}} \langle n \rangle_{cr}=\frac{\omega}{T} \langle n \rangle_{cr}.
\end{equation}
Since $\langle s \rangle \geq 0$, this implies that inner friction is also positive. The positivity of $\avg{W}_{fric}$ can be seen as a statement of the second law of thermodynamics \cite{jarzynski} in a statistical formulation~\footnote{It has been shown that inner friction is non-negative for any time-dependent Hamiltonian starting from a passive state \cite{alla} and non-positive for an active state \cite{campisi2011}. A thermal state, which is the assumption used in this chapter, is an example of a passive state.}. This is strong evidence that $\avg{s}$ should be considered a suitable entropic term (similar to the entropy production as originally defined by Crooks \cite{crooks}) to use in this cosmological context. This is the main result of this chapter. 

The intimate relationship between this particular measure of entropy production in spacetime expansion and particle creation is another main result in this chapter. We observe that if a state diagonal in the number basis (e.g. thermal state) undergoes two-mode squeezing we find $\avg{n}_{cr} \geq 0$. From Eq.~\eqref{eq:snumber} we see that this is not only consistent with the second law of thermodynamics $\avg{W}_{fric} \geq 0$ but it also provides an alternative interpretation for $\avg{n}_{cr} \geq 0$ in terms of the second law. 

We also saw in Eq.~\eqref{eq:fricinf} in section ~\ref{sec:innerfriction} that, in the context of cosmological particle creation, inner friction can also be related to the quantum relative entropy \cite{plastina} between the final state  $\rho_f$ after spacetime expansion and the state $\rho_{ad}$~\footnote{These are states in the corresponding the Schr\"odinger picture.} after a reversible adiabatic expansion of spacetime. Thus particle creation is also directly linked to a relative entropy, which has itself been related to a second law of thermodynamics \cite{sagawa}. 
\subsection{Entropy production in other scenarios}
\label{sec:unruhbhlastsummary}
The formalism developed in previous sections apply in a straightforward fashion to any other scenario described by two-mode squeezing, although with changes in interpretation. Scenarios of this type include the well-known Unruh effect \cite{unruh} and the radiating black hole scenario \cite{BandD}, which we saw in section ~\ref{sec:otherscenarios}.
In the Unruh effect there is one stationary observer and another observer uniformly accelerated with respect to the first observer with uniform acceleration $a$. In the radiating black hole scenario, the interesting parameter is the black hole mass $M$. In both cases, the `inertial' vacuum is perceived by a stationary observer as a state full of particles which are thermally distributed with a temperature $T_U\propto a$ in the Unruh case and $T_H\propto 1/M$ in the black hole case.
The squeezing parameter $r$ satisfies $\tanh r=\exp[-\omega/T_U]$ for the Unruh case and $\tanh r=\exp[-\omega/T_H]$ for the black hole case. The complete mathematical analogy between the cosmological model  and these two scenarios allow us to immediately export our results to these setups as well, where the only difference is how the squeezing $r$ depends on the relevant physical parameters.

For example, in the Unruh effect, the work done by a changing Rindler spacetime onto the quantum field can be considered as a work done by a driving element (e.g. the engine in a rocket) responsible for the acceleration of an observer in an initial inertial frame (e.g. observer inside a stationary rocket). The energy cost of the engine can then be considered to be `transferred' into the energy of the particles `created' by this change of reference frame. 

One can also loosely interpret the driving element in a black hole scenario to be the dense matter in a region of flat spacetime that induces the change of spacetime from Minkowski to black hole spacetime. The energy from the work done onto this spacetime by the dense matter can then be considered to be `transferred' to the thermal particles radiated  from the black hole. 

We finally note that in the Unruh effect and the radiating black hole scenario, event horizons are present. This suggests that in these cases, unlike in the cosmological scenario, observers will only be able to access one mode of the field (instead of both modes in the the $(k, -k)$ mode pair). We leave it to further work to apply techniques of open quantum systems to analyse the interpretation of these physical processes by localised observers.
\section{Discussion and further work}
\label{sec:cosmodiscussion}
We employed tools from quantum field theory in curved spacetime and thermodynamics in the out-of-equilibrium and quantum regimes to study the connection between entropy production and the creation of particles described by quantum field theory in curved spacetime. In particular, we examine the case of cosmological particle creation during the earliest stages of cosmological expansion, which can be characterised by two-mode squeezing. 

There are two main contributions of this work. Firstly, we introduce a framework in which to study scenarios described by quantum field theory in curved spacetime using recently developed out-of-equilibrium thermodynamics. Exploring this out-of-equilibrium regime in these contexts is currently lacking in the literature, although it is necessary in cases like cosmological particle creation during rapid initial expansion of spacetime (like in the inflationary scenario). 

Secondly, by using concepts from out-of-equilibrium thermodynamics, we identify an entropy production term, called inner friction, that can be used to quantify the entropy associated with the creation of particles due to the expansion of spacetime. This presents a new way of thinking about contributions to entropy in cosmological particle creation that is different from von Neumann entropy, which has so far dominated research in this area. We show that inner friction produced during cosmological particle creation is positive and is directly associated with particle creation (unlike the von Neumann entropy which does not change under unitary evolution). Furthermore, we can interpret inner friction in terms of the `irreversibility' of particle creation in spacetime expansion by presenting a fluctuation relation corresponding to inner friction. 

There are three main directions for future research: 1) The first direction is considering the theoretical extensions to the simple cosmological model we have considered. The next challenge is to work with more realistic cosmological models and to compute the contribution of inner friction to the total entropy production during spacetime expansion. In these more realistic models, it is possible that the evolution of the initial quantum state is no longer described by two-mode squeezing and is replaced by combinations of two-mode squeezing, one-mode squeezing and beam-splitting. 2) Extension of our results to the Unruh effect and the radiating black hole is less physical than our application to cosmological particle creation. This is largely due to the presence of physical event horizons present in both the Unruh effect and the radiating black hole. Taking into account the presence of the event horizons require out-of-equilbrium thermodynamics for open systems. Working towards such a thermodynamics will be useful not only for these relativistic scenarios, but also the thermodynamics of quantum computing in the presence of noise. 3) The rapidly developing field of analogue gravity \cite{barcelo2005analogue} provides very exciting opportunities to test key predictions of quantum field theory in curved spacetime (like black hole radiation, which is currently beyond direct experimental probing) in the laboratory. Analogue gravity describes a collection of models (of systems accessible in table-top experiments like Bose-Einstein condensates \cite{barcelo2001analogue}) that are near-exact mathematical analogues to quantum fields in curved spacetime. 
With experimental protocol development also underway in the measurement of work statistics in real systems like Bose-Einstein condensates \cite{ngo2012demonstration}, the field of analogue gravity provides an excellent opportunity to experimentally test our results of out-of-equilibrium thermodynamics in the relativistic regime. 
\begin{subappendices}
\section{The Unruh effect and the radiating black hole}
\label{app:otherscenarios}
\subsection{The Unruh effect}
The Unruh effect is a very interesting example in which one may also apply the techniques developed earlier in this chapter, but which is 
conceptually quite different to the cosmological case. We begin with a uniformly accelerating observer in Minkowski spacetime, which is equivalent to
a stationary observer in Rindler spacetime. We work in (1+1)-dimensions all throughout for simplicity. The time $t$ and space $x$ coordinates in Minkowski spacetime
is transformed into the equivalent time $\eta$ and space $\xi$ coordinates in Rindler spacetime (Rindler coordinates) by
\begin{align}
x&=\xi \cosh(a \eta) \nonumber \\
t&=\xi \sinh(a \eta)
\end{align}
in the region  $x>0$, $x>|t|$ and $a$ is the acceleration of the observer in the frame of the inertial observer. This is called the right (R) region since they reside on the right-hand-side in the Minkowski spacetime diagram. This is also the region where we choose our accelerating observer to reside. Since these coordinates do not cover all of Minkowski spacetime, we need another wedge in the region  $x<0$, $x>|t|$, called the left (L) region, which is causally disconnected from the right region. The Rindler coordinates here relate to $t$ and $x$ by 
\begin{align}
x&=-\xi \cosh(a \eta) \nonumber \\
t&=\xi \sinh(a \eta).
\end{align}
The Rindler spacetime metric is equivalent to the Minkowski metric and can be written as
\begin{align}\label{eq:rindlermetric}
ds^2=-dt^2-dx^2=(a\xi)^2 d\eta^2-d\xi^2.
\end{align}
Let us place a (1+1)-dimensional massless Klein-Gordon field $\phi$ in this spacetime. In the right region, the Klein-Gordon equation is
\begin{align}
\square_R \phi \equiv \xi^2\left(\frac{1}{a^2} \frac{\partial^2}{\partial \eta^2}-\frac{\partial^2}{\partial \ln(a\xi)^2}\right)\phi=0.
\end{align}
The positive and negative-frequency solutions are
\begin{align}
u^R_k&=\frac{1}{\sqrt{4\pi\omega}}e^{i(k\ln(a\xi)/a-\omega \eta)} \nonumber \\
(u^R_k)^*&=\frac{1}{\sqrt{4\pi\omega}}e^{-i(k\ln(a\xi)/a-\omega \eta)},
\end{align}
with the dispersion relation $\omega=|k|$. These modes do not form a complete basis without the positive and negative-frequency solutions to the Klein-Gordon equation in the
left region 
\begin{align}
\square_L \phi \equiv \xi^2\left(\frac{1}{a^2} \frac{\partial^2}{\partial (-\eta)^2}-\frac{\partial^2}{\partial \ln(-a\xi)^2}\right)\phi=0.
\end{align}
These solutions in the left region are
\begin{align}
u^L_k&=\frac{1}{\sqrt{4\pi\omega}}e^{i(-k\ln(-a\xi)/a+\omega \eta)} \nonumber \\
(u^L_k)^*&=\frac{1}{\sqrt{4\pi\omega}}e^{-i(-k\ln(-a\xi)/a+\omega \eta)}.
\end{align}
These two sets of solutions together form a complete basis with respect to the inner product defined in Eq.~\eqref{eq:innerproduct} Thus a Klein-Gordon field can be expanded 
\begin{align}
\phi=\sum_{k=-\infty}^{\infty} a_k^L u_k^L+a_k^{L \dagger }u_k^{L*}+a_k^R u_k^R+a_k^{ R \dagger }u_k^{R \dagger},
\end{align}
where $a^L_k$, $(a^L_k)^{\dagger}$ and $a^R_k$, $(a^R_k)^{\dagger}$ are respectively the annihilation and creation operators corresponding to the modes
residing in the left and right regions. They satisfy the commutation relations $[a^L_k, (a^L_{k'})^{\dagger}]=\delta(k-k')=[a^R_k, (a^R_{k'})^{\dagger}]$ with all other commutators between $L,R$ operators vanishing
since the left and right regions are causally disconnected. Thus the Rindler spacetime vacuum can be written as
\begin{align}
\ket{0}_{\text{Rind}}=\ket{0}_L \ket{0}_R,
\end{align}
with $a^L_k\ket{0}_L=0=a^R_k\ket{0}_R$. 

It turns out there set of two modes $\{u^I_k, (u^I_k)^*\}$, $\{(u^{II}_k), (u^{II}_k)^*\}$ called Unruh modes, which are related to the the left region and right region Rindler modes by two-mode squeezing. These are chosen as an alternative to Minkowski modes since they have a simpler relationship to Rindler modes while having an equivalent vacuum to Minkowski vacuum. We can define these Unruh modes by the following Bogoliubov transformation
\begin{align}
u^I_k&=\cosh(r_k)u^R_k+\sinh(r_k)(u^L_k)^* \nonumber \\
(u^{II}_k)^*&=\sinh(r_k)u^R_k+\cosh(r_k)(u^L_k)^*.
\end{align}
The equation for the annihilation and creation operators $\{a^I_k, (a^I_k)^{\dagger}\}$, $\{a^{II}_k, (a^{II}_k)^{\dagger}\}$, corresponding to these Unruh modes are then also similarly related. This means our field can be equivalently expanded in terms of these Unruh modes and operators by
\begin{align}
\phi=\sum_{k} a_k^I u_k^I+a_k^{I \dagger} u_k^{I *}+a_k^{II}u_k^{II}+a_k^{II \dagger} u_k^{II *}. 
\end{align}
The Unruh vacuum can then be defined as
\begin{equation}
\ket{0}_U=\ket{0}_{I} \ket{0}_{II},
\end{equation}
where $a_k^I \ket{0}_U=a_k^{II} \ket{0}_U=0$. The Unruh vacuum can be shown to be equivalent to the Minkowski vacuum since the Minkowski annihilation operator $a^M$ is a linear combination of annihilation operators of Unruh modes
\begin{align}
a^M_k=c^I_k a^I_k+c^{II}_k a^{II}_k,
\end{align}
where $c^I_k$, $c^{II}_k$ are constants. This shows the Minkowski vacuum is also annihilated by annihilation operators of Unruh modes (which annihilates the Unruh vacuum), so the Unruh vacuum coincides with Minkowski vacuum. Since the Unruh vacuum corresponds to a two-mode squeezed state in Rindler spacetime by the Bogoliubov transformation in Eq.~\eqref{eq:atildea}, the Minkowski vacuum can also be rewritten as the following two-mode squeezed state \cite{gerryknight}
\begin{align}
\ket{0}_M=\frac{1}{\cosh(r_k)} \sum_{n_k} \tanh^{n_k}(r_k) \ket{n_k}_L \ket{n_k}_R,
\end{align}
where $\tanh(r_k)=\exp(-2\pi \omega_k/a)$ and $\ket{n_k}_{L,R}$ is the number state in the left, right region of Rindler spacetime. However, this entire two-mode squeezed state is not accessible to our accelerating observer, who resides only within the right region of Rindler spacetime. This means to this observer, the state of mode $k$ of this quantum field $\rho_R$ is the single-mode reduced state of the two-mode squeezed state
\begin{align}
\rho_R=\tr_L(\ket{0}_M \bra{0}_M)=\frac{1}{\cosh^2(r_k)} \sum_{n_k} \sum_{n_k}\tanh^{2n_k}(r_k)\ket{n_k}_R \bra{n_k}_R,
\end{align}
which is a thermal state with temperature $T=a/(2\pi)$. In fact, we can check its average particle excitation which shows a thermal spectrum of the above temperature
\begin{align}
\langle n_k \rangle \equiv \tr(a^{\dagger}_R a_R \rho_R)=\frac{1}{e^{-2\pi/(\omega_k a)}-1},
\end{align}
where $a^{\dagger}_R$, $a_R$ are creation and annihilation operators corresponding to the modes residing in the right region of Rindler spacetime. 

We can derive the Hamiltonian corresponding to Minkowski and Rindler spacetime in the same way as in Appendix~\ref{sec:DerivingHamiltonian}. It can be shown
\begin{equation}
H_{\text{Rind}}=\sum_k \frac{\omega_k}{2} (a_k^{R \dagger} a_k^R+a_k^R a_k^{R \dagger}+a_k^{L \dagger} a_k^L+a_k^L a_k^{L \dagger}),
\end{equation}
while the Minkowski space Hamiltonian may be written as 
\begin{equation}
H_M=\sum_k \frac{\omega_k}{2}(a_k^{M \dagger}a_k^M+a_k^M a_k^{M \dagger}).
\end{equation}
These Hamiltonians can be loosely considered as the initial and final Hamiltonian (before and after acceleration). These is a difference between the two Hamiltonians since there is expected to be an energetic cost to changing reference frames, as one goes from stationary to an accelerated frame (e.g. fuel in a rocket). 
\subsection{Radiating black hole}
The case of an evaporating Schwarzschild black hole works in a very similar fashion to the Unruh effect. A Schwarzschild black hole is a stationary, spherically symmetric black hole of mass $M$ whose spacetime is described by the Schwarzschild metric
\begin{align}
ds^2=-\left(1-\frac{2M}{r}\right) dt^2+1/\left(1-\frac{2M}{r}\right)dr^2
\end{align}
in (1+1)-dimensions. In the presence of the black hole, spacetime can be separated into two causually disconnected regions, namely inside and outside a black hole. The separation between the inside and outside is the event horizon, which is situated at $r=2M$. Note that this is where the Schwarszchild metric becomes singular. An observer in free-fall just outside the event horizon will fall through into the black hole, but will not be able to travel outside again without violating causality. This means for an observer to be maintained just outside the event horizon, the observer must accelerate uniformly in the opposite direction. The necessary acceleration is captured by the surface gravity of a black hole, which is $\kappa=1/4M$ for a Schwarzschild black hole. This suggests that since the outside observer is uniformly accelerating with acceleration $\kappa$, its spacetime is indistinguishable from Rindler spacetime with acceleration equal to $\kappa$. To see this explicitly, let us use a transformation of coordinates $r=x^2/(8M)+2M$. Then we can see that near the surface of a black hole at the event horizon $r=2M$
\begin{align}
0 \approx \left(1-\frac{2M}{r}\right)=\frac{(\kappa x)^2}{1+(\kappa x)^2} \approx (\kappa x)^2.
\end{align}
So we can rewrite the Schwarzschild metric as
\begin{align}
ds^2=-(\kappa x)^2 dt^2+dx^2,
\end{align}
since $dr^2=(\kappa x)^2 dx^2$. This is equivalent to the Rindler metric in Eq.~\eqref{eq:rindlermetric} (in the right region) with the replacement $\kappa=a$, like we anticipated.  Furthermore, it can be shown that the inside region of this black hole becomes indistinguishable from the left region of Rindler spacetime with acceleration $a=\kappa=1/4M$. Thus all the mathematics from the Unruh case follows directly in this black hole example with the substitution $a=\kappa=1/(4M)$. For example, the Minkowski vacuum is equivalent to a two-mode squeezed state in black hole spacetime, where the squeezing is between the quantum field modes inside and outside the black hole. This squeezing $r_k$ satisfies $\tanh(r_k)=\exp(-2\pi \omega_k/\kappa)=\exp(-8\pi \omega_k M)$. Thus the presence of a black hole in the vacuum state of a quantum field appears thermal to an observer outside the black hole, so a black hole is said to evaporate through this thermal radiation. This temperature $T=\kappa/(2 \pi)=1/(8 \pi M)$ varies inversely with the mass of the black hole. An initial and final Hamiltonian can be equivalent derived as in the Unruh effect case, except with the replacement of acceleration with surface gravity of the black hole. 
\section{The Hamiltonian in quantum field theory in curved spacetime}
\label{sec:DerivingHamiltonian}
Throughout this work, we will be using the simplest case of a massive Klein-Gordon field in curved
spacetime to illustrate our ideas. The Klein-Gordon Lagrangian density in curved space is
\begin{equation}\label{eq:KGlag}
\mathcal{L}=\frac{1}{2} \sqrt{-g} [g^{\mu \nu} \partial_{\mu} \phi \partial_{\nu} \phi-m^2 \phi^2].
\end{equation}
We can construct a Hamiltonian density from $\mathcal{H}_{dens}=\pi \partial_0 \phi-\mathcal{L}$, where $\pi$ is the conjugate momentum $\pi=\partial \mathcal{L}/\partial (\partial_0 \phi)$. Then the Hamiltonian is 
\begin{equation} \label{eq:Ham}
\mathcal{H}=\int dx^{n-1} \mathcal{H}_{dens},
\end{equation}
where $(n-1)$ is the number of spatial dimensions in our scenario. Now, let our metric be conformal to the Minkowski metric
\begin{equation}
g_{\mu \nu}=\Omega^2 \eta_{\mu \nu},
\end{equation}
where we use the metric signature $\eta_{\mu \nu}=\text{diag} (-1, 1, 1,...)$. Then our Lagrangian density becomes
$\mathcal{L}=(1/2)( \Omega^{n-2}(\eta^{\mu \nu} \partial_{\mu} \phi \partial_{\nu} \phi)-\Omega^{n} m^2 \phi^2)$
Then we can find our Hamiltonian density to be 
\begin{equation}
\mathcal{H}_{dens}=\frac{1}{2}(\Omega^{n-2} \delta^{\mu \nu} \partial_{\mu} \phi \partial_{\nu} \phi+\Omega^{n} m^2 \phi^2),
\end{equation}
where $\delta_{\mu \nu}=\text{diag}(1, 1, 1, ...)$. To find the Hamiltonian, we must first quantize the field by making an expansion in terms of the classical mode solutions and its associated annihilation and creation
operators. Quantization of the field operator means that it can be written as an expansion in terms of the classical mode solutions $u_k$ of the classical field equations and
the associated annihilation and creation operators, like
\begin{equation} \label{eq:phiexpansion}
\phi=\sum_k a_k u_k+a_k^{\dagger} u_k^*.
\end{equation}
We note here that we consider a finite sum over $k$. This is because, in a finite-sized universe, just like for harmonic modes in a finite box, $k$ takes on discrete values. The larger this box,
the better we can approximate this sum by an integral. Let us now restrict to $(1+1)$-dimensions. In Minkowski space, we have plane wave solutions $u_k =(1/\sqrt{4 \pi \omega}) \exp(i(kx-\omega t))$ In curved spacetimes with metrics conformal to the Minkowski metric, let the coordinates be $(\zeta, \eta)$. There one can also find plane wave solutions
\begin{equation} \label{eq:planewave}
u_k=\sum_k \frac{1}{\sqrt{4 \pi \omega}} e^{i(k \zeta-\omega \eta)}.
\end{equation}
We note that the dispersion relation for Klein-Gordon fields in curved spacetime can be used to simplify the Hamiltonian. The equation of motion for a massive Klein-Gordon field is
\begin{equation} \label{eq:KG}
(\square+m^2) \phi=0,
\end{equation}
where $\square \phi \equiv \sqrt{-g} \partial_{\mu} [\sqrt{-g} g^{\mu \nu} \partial_{\nu} \phi]= \Omega^{n} \partial_{\mu} (\Omega^{n-2} \eta^{\mu \nu} \partial_{\nu} \phi)$
for metrics conformal to the Minkowski metric. Inserting the plane wave solution in Eq.~\eqref{eq:planewave} into the Klein-Gordon equation ~\eqref{eq:KG} we obtain the dispersion relation
\begin{equation} \label{eq:dispersion}
\omega^2=k^2+m^2 \Omega^2.
\end{equation}
Now we can insert the plane wave expansion from Eqs.~\eqref{eq:planewave} and ~\eqref{eq:phiexpansion} into the Hamiltonian in Eq.~\eqref{eq:Ham} to obtain
\begin{align} \label{eq:hamexpansion1}
\mathcal{H} & =\frac{1}{2} \int_{-\infty}^{\infty} d \zeta (\partial_{\eta} \phi \partial_{\eta} \phi+\partial_{\zeta} \phi \partial_{\zeta} \phi+\Omega^2 m^2 \phi^2) \nonumber \\
    &= \frac{1}{2}\sum_{k, k'} \int_{-\infty}^{\infty} d \zeta (-\omega \omega'-k k'+\Omega^2 m^2)a_k a_{k'} u_k u_{k'} +
                                                                                                (- \omega \omega'-k k'+\Omega^2 m^2)a_k^{\dagger}a_{k'}^{\dagger} u_k^* u_{k'}^*  \nonumber \\
& + (\omega \omega'+k k'+\Omega^2 m^2)(a_{k}a_{k'}^{\dagger} u_k u_{k'}^*+a_{k}^{\dagger}a_{k'} u_{k}^* u_{k'}) \nonumber \\
&=\sum_{k} \frac{\omega_{k}}{2} (a_{k} a_{k}^{\dagger}+a_{k}^{\dagger}a_{k}),
\end{align}
where we used the dispersion relation in Eq.~\eqref{eq:dispersion} and the normalisation relations $\int_{-\infty}^{\infty} d \zeta u_k u_{k'}= (1/2 \sqrt{\omega \omega'}) \exp(-i(\omega+\omega') \eta) \delta (k+k')$, $\int_{-\infty}^{\infty} d \zeta u_k^* u_{k'}^*=(1/2 \sqrt{\omega \omega'})\exp(i(\omega+\omega') \eta) \delta(k+k')$, $\int_{-\infty}^{\infty} d \zeta u_k u_{k'}^*=(1/2 \sqrt{\omega \omega'})\exp(-i(\omega-\omega') \eta) \delta(k-k')$
and $\int_{-\infty}^{\infty} d \zeta u_k^* u_{k'}=(1/2 \sqrt{\omega \omega'}) \exp(i(\omega-\omega') \eta)\delta(k-k')$. 

Thus the initial Hamiltonian $\mathcal{H}(\eta \rightarrow -\infty)$ can be derived using $\omega^2_k \equiv k^2+m^2 \Omega(\eta)\vert_{\eta\rightarrow -\infty}$ and the annihilation and creation operators corresponding to modes $k$ of the field on the flat spacetime in the asymptotic past are $a_k, a^{\dagger}_k$. Therefore
\begin{align}
\mathcal{H}(\eta \rightarrow -\infty)=\sum_k \frac{\omega_k}{2} (a^{\dagger}_k a_k+a_k a^{\dagger}_k).
\end{align}
In the asymptotic future, the plane wave solutions have frequencies $\tilde{\omega}^2_k \equiv k^2+m^2 \Omega(\eta)\vert_{\eta\rightarrow \infty}$, 
where the annihilation and creation operators are now $\tilde{a}_k$, $\tilde{a}^{\dagger}_k$. Thus the final total Hamiltonian $\mathcal{H}(\eta \rightarrow \infty)$ takes the form
\begin{align}
\mathcal{H}(\eta \rightarrow \infty)=\sum_k \frac{\tilde{\omega}_k}{2} (\tilde{a}^{\dagger}_k \tilde{a}_k+\tilde{a}_k \tilde{a}^{\dagger}_k).
\end{align}
\end{subappendices}
\bibliographystyle{unsrt}
\bibliography{TotalRef}

\begin{thebibliography}{100}

\bibitem{nanaqumode}
Nana Liu, Jayne Thompson, Christian Weedbrook, Seth Lloyd, Vlatko Vedral, Mile
  Gu, and Kavan Modi.
\newblock Power of one qumode for quantum computation.
\newblock {\em Phys. Rev. A}, 93(5):052304, 2016.

\bibitem{liu2016quantum}
Nana Liu and Hugo Cable.
\newblock Quantum-enhanced multi-parameter estimation for unitary photonic
  systems.
\newblock {\em arXiv:1612.03621}, 2016.

\bibitem{nanacosmo}
Nana Liu, John Goold, Ivette Fuentes, Vlatko Vedral, Kavan Modi, and
  David~Edward Bruschi.
\newblock Quantum thermodynamics for a model of an expanding universe.
\newblock {\em Class. Quantum Grav}, 33(3):035003, 2016.

\bibitem{dj}
David Deutsch and Richard Jozsa.
\newblock Rapid solution of problems by quantum computation.
\newblock {\em Proc. R. Soc. London A}, 439(1907):553--558, 1992.

\bibitem{shor}
Peter~W Shor.
\newblock Polynomial-time algorithms for prime factorization and discrete
  logarithms on a quantum computer.
\newblock {\em SIAM J. Comput}, 26(5):1484--1509, 1997.

\bibitem{grover}
Lov~K Grover.
\newblock A fast quantum mechanical algorithm for database search.
\newblock In {\em Proceedings of the twenty-eighth annual ACM symposium on
  Theory of computing}, pages 212--219. ACM, 1996.

\bibitem{harrowlloyd}
Aram~W Harrow, Avinatan Hassidim, and Seth Lloyd.
\newblock Quantum algorithm for linear systems of equations.
\newblock {\em Phys. Rev. Lett}, 103(15):150502, 2009.

\bibitem{jozsalinden}
Richard Jozsa and Noah Linden.
\newblock On the role of entanglement in quantum-computational speed-up.
\newblock {\em Proc. R. Soc. London A}, 459(2036):2011--2032, 2003.

\bibitem{ekertjozsa}
Artur Ekert and Richard Jozsa.
\newblock Quantum algorithms: entanglement-enhanced information processing.
\newblock {\em Philos. Trans. R. Soc. London A}, pages 1769--1781, 1998.

\bibitem{meyer}
David~A Meyer.
\newblock Sophisticated quantum search without entanglement.
\newblock {\em Phys. Rev. Lett}, 85(9):2014, 2000.

\bibitem{lloyd1999}
Seth Lloyd.
\newblock Quantum search without entanglement.
\newblock {\em Phys. Rev. A}, 61(1):010301, 1999.

\bibitem{jozsa1997}
Richard Jozsa.
\newblock Entanglement and quantum computation.
\newblock {\em arXiv: quant-ph/9707034}, 1997.

\bibitem{knill1998power}
Emanuel Knill and Raymond Laflamme.
\newblock Power of one bit of quantum information.
\newblock {\em Phys. Rev. Lett}, 81(25):5672, 1998.

\bibitem{white}
BP~Lanyon, M~Barbieri, MP~Almeida, and AG~White.
\newblock Experimental quantum computing without entanglement.
\newblock {\em Phys. Rev. Lett}, 101(20):200501, 2008.

\bibitem{datta2007}
Animesh Datta and Guifre Vidal.
\newblock Role of entanglement and correlations in mixed-state quantum
  computation.
\newblock {\em Phys. Rev. A}, 75(4):042310, 2007.

\bibitem{parkerplenio}
S~Parker and MB~Plenio.
\newblock Efficient factorization with a single pure qubit and log n mixed
  qubits.
\newblock {\em Phys. Rev. Lett}, 85(14):3049, 2000.

\bibitem{GKP}
Daniel Gottesman, Alexei Kitaev, and John Preskill.
\newblock Encoding a qubit in an oscillator.
\newblock {\em Phys. Rev. A}, 64(1):012310, 2001.

\bibitem{terhal}
BM~Terhal and D~Weigand.
\newblock Encoding a qubit into a cavity mode in circuit qed using phase
  estimation.
\newblock {\em Phys. Rev. A}, 93(1):012315, 2016.

\bibitem{schoelkopf}
Brian Vlastakis, Gerhard Kirchmair, Zaki Leghtas, Simon~E Nigg, Luigi Frunzio,
  Steven~M Girvin, Mazyar Mirrahimi, Michel~H Devoret, and Robert~J Schoelkopf.
\newblock Deterministically encoding quantum information using 100-photon
  schr{\"o}dinger cat states.
\newblock {\em Science}, 342(6158):607--610, 2013.

\bibitem{caves}
Carlton~M Caves.
\newblock Quantum-mechanical noise in an interferometer.
\newblock {\em Phys. Rev. D}, 23(8):1693, 1981.

\bibitem{monras}
Alex Monras.
\newblock Optimal phase measurements with pure gaussian states.
\newblock {\em Phys. Rev. A}, 73(3):033821, 2006.

\bibitem{pinel}
Olivier Pinel, Pu~Jian, N~Treps, C~Fabre, and Daniel Braun.
\newblock Quantum parameter estimation using general single-mode gaussian
  states.
\newblock {\em Phys. Rev. A}, 88(4):040102, 2013.

\bibitem{lloydCV}
Seth Lloyd and Samuel~L Braunstein.
\newblock Quantum computation over continuous variables.
\newblock {\em Phys. Rev. Lett}, 82(8):1784, 1999.

\bibitem{mile2009}
Mile Gu, Christian Weedbrook, Nicolas~C Menicucci, Timothy~C Ralph, and Peter
  van Loock.
\newblock Quantum computing with continuous-variable clusters.
\newblock {\em Phys. Rev. A}, 79(6):062318, 2009.

\bibitem{algorithms}
Mikhail~J Atallah and Marina Blanton.
\newblock {\em Algorithms and theory of computation handbook, volume 2: special
  topics and techniques}.
\newblock CRC press, 2009.

\bibitem{lloydhybrid}
Seth Lloyd.
\newblock Hybrid quantum computing.
\newblock In {\em Quantum information with continuous variables}, pages 37--45.
  Springer, 2003.

\bibitem{hybridbook}
Akira Furusawa and Peter Van~Loock.
\newblock {\em Quantum teleportation and entanglement: a hybrid approach to
  optical quantum information processing}.
\newblock John Wiley \& Sons, 2011.

\bibitem{cook71}
Stephen~A Cook.
\newblock The complexity of theorem-proving procedures.
\newblock In {\em Proceedings of the third annual ACM symposium on Theory of
  computing}, pages 151--158. ACM, 1971.

\bibitem{papbook}
Christos~H Papadimitriou.
\newblock Computational complexity.
\newblock {\em Addison Weseley Publishing Co.}, 1994.

\bibitem{landauer}
Rolf Landauer.
\newblock The physical nature of information.
\newblock {\em Phys. Lett. A}, 217(4):188--193, 1996.

\bibitem{bernvaz}
Ethan Bernstein and Umesh Vazirani.
\newblock Quantum complexity theory.
\newblock {\em SIAM J. Comp}, 26(5):1411--1473, 1997.

\bibitem{bennett}
Charles~H Bennett.
\newblock The thermodynamics of computation—a review.
\newblock {\em Int. J. Theor. Phys}, 21(12):905--940, 1982.

\bibitem{vlatkodiscord}
Leah Henderson and Vlatko Vedral.
\newblock Classical, quantum and total correlations.
\newblock {\em J. Phys. A}, 34(35):6899, 2001.

\bibitem{zurekdiscord}
Harold Ollivier and Wojciech~H Zurek.
\newblock Quantum discord: a measure of the quantumness of correlations.
\newblock {\em Phys. Rev. Lett}, 88(1):017901, 2001.

\bibitem{coherenceref}
Iman Marvian and Robert~W Spekkens.
\newblock How to quantify coherence: distinguishing speakable and unspeakable
  notions.
\newblock {\em arXiv:1602.08049}, 2016.

\bibitem{hakop}
Hakop Pashayan, Joel~J Wallman, and Stephen~D Bartlett.
\newblock Estimating outcome probabilities of quantum circuits using
  quasiprobabilities.
\newblock {\em Phys. Rev. Lett}, 115(7):070501, 2015.

\bibitem{turing}
Alan~Mathison Turing.
\newblock On computable numbers, with an application to the
  entscheidungsproblem.
\newblock {\em J. Math}, 58(345-363):5, 1936.

\bibitem{church}
Alonzo Church.
\newblock An unsolvable problem of elementary number theory.
\newblock {\em Am. J. Math}, 58(2):345--363, 1936.

\bibitem{PNP}
William~I Gasarch.
\newblock Guest column: The second p=? np poll.
\newblock {\em ACM SIGACT News}, 43(2):53--77, 2012.

\bibitem{pap}
Christos~H Papadimitriou.
\newblock The euclidean travelling salesman problem is np-complete.
\newblock {\em Theor. Comp. Sci}, 4(3):237--244, 1977.

\bibitem{garey}
Michael~R Garey and David~S Johnson.
\newblock A guide to the theory of np-completeness.
\newblock {\em WH Freemann, New York}, 1979.

\bibitem{lund2014boson}
AP~Lund, A~Laing, S~Rahimi-Keshari, T~Rudolph, JL~O’Brien, and TC~Ralph.
\newblock Boson sampling from a gaussian state.
\newblock {\em Phys. Rev. Lett}, 113(10):100502, 2014.

\bibitem{mon}
Ashley Montanaro.
\newblock Quantum algorithms: an overview.
\newblock {\em NPJ Quantum Information}, 2:15023, 2016.

\bibitem{shi}
Yaoyun Shi.
\newblock Both toffoli and controlled-not need little help to do universal
  quantum computing.
\newblock {\em Quantum Information \& Computation}, 3(1):84--92, 2003.

\bibitem{shorjordan}
Peter~W Shor and Stephen~P Jordan.
\newblock Estimating jones polynomials is a complete problem for one clean
  qubit.
\newblock {\em Quantum Inf. Comput}, 8(8):681--714, 2008.

\bibitem{shepherd}
Dan Shepherd.
\newblock Computation with unitaries and one pure qubit.
\newblock {\em arXiv: quant-ph/0608132}, 2006.

\bibitem{hugoDQC1}
Hugo Cable, Mile Gu, and Kavan Modi.
\newblock Power of one bit of quantum information in quantum metrology.
\newblock {\em arXiv:1504.02460}, 2015.

\bibitem{fidelitydecay}
David Poulin, Robin Blume-Kohout, Raymond Laflamme, and Harold Ollivier.
\newblock Exponential speedup with a single bit of quantum information:
  Measuring the average fidelity decay.
\newblock {\em Phys. Rev. Lett}, 92(17):177906, 2004.

\bibitem{knill2001}
Emanuel Knill and Raymond Laflamme.
\newblock Quantum computing and quadratically signed weight enumerators.
\newblock {\em Inform. Process. Lett}, 79(4):173--179, 2001.

\bibitem{datta2005}
Animesh Datta, Steven~T Flammia, and Carlton~M Caves.
\newblock Entanglement and the power of one qubit.
\newblock {\em Phys. Rev. A}, 72(4):042316, 2005.

\bibitem{animeshphd}
Animesh Datta.
\newblock {\em Studies on the role of entanglement in mixed-state quantum
  computation}.
\newblock PhD thesis, The University of New Mexico, 2008.

\bibitem{cleve}
Richard Cleve, Artur Ekert, Chiara Macchiavello, and Michele Mosca.
\newblock Quantum algorithms revisited.
\newblock {\em Proc. R. Soc. London A}, 454(1969):339--354, 1998.

\bibitem{nandc}
Michael~A Nielsen and Isaac~L Chuang.
\newblock {\em Quantum computation and quantum information}.
\newblock Cambridge university press, 2010.

\bibitem{freeth2006decoding}
Tony Freeth, Y~Bitsakis, X~Moussas, JH~Seiradakis, A~Tselikas, H~Mangou,
  M~Zafeiropoulou, R~Hadland, D~Bate, A~Ramsey, et~al.
\newblock Decoding the ancient greek astronomical calculator known as the
  antikythera mechanism.
\newblock {\em Nature}, 444(7119):587--591, 2006.

\bibitem{gerardoreview}
Gerardo Adesso, Sammy Ragy, and Antony~R Lee.
\newblock Continuous variable quantum information: Gaussian states and beyond.
\newblock {\em Open Systems \& Information Dynamics}, 21(01n02):1440001, 2014.

\bibitem{knightgerry}
Christopher Gerry and Peter Knight.
\newblock {\em Introductory quantum optics}.
\newblock Cambridge university press, 2005.

\bibitem{localhamiltonian}
Julia Kempe, Alexei Kitaev, and Oded Regev.
\newblock The complexity of the local hamiltonian problem.
\newblock {\em SIAM J. Comput}, 35(5):1070--1097, 2006.

\bibitem{abramslloyd}
Daniel~S Abrams and Seth Lloyd.
\newblock Quantum algorithm providing exponential speed increase for finding
  eigenvalues and eigenvectors.
\newblock {\em Phys. Rev. Lett}, 83(24):5162, 1999.

\bibitem{andersen201530}
Ulrik~L Andersen, Tobias Gehring, Christoph Marquardt, and Gerd Leuchs.
\newblock 30 years of squeezed light generation.
\newblock {\em arXiv:1511.03250}, 2015.

\bibitem{lan}
Rolf Landauer.
\newblock Irreversibility and heat generation in the computing process.
\newblock {\em IBM journal of research and development}, 5(3):183--191, 1961.

\bibitem{braunstein2005}
Samuel~L Braunstein and Peter Van~Loock.
\newblock Quantum information with continuous variables.
\newblock {\em Rev. Mod. Phys}, 77(2):513, 2005.

\bibitem{christian2012}
Christian Weedbrook, Stefano Pirandola, Raul Garcia-Patron, Nicolas~J Cerf,
  Timothy~C Ralph, Jeffrey~H Shapiro, and Seth Lloyd.
\newblock Gaussian quantum information.
\newblock {\em Rev. Mod. Phys}, 84(2):621, 2012.

\bibitem{andersen}
Ulrik~L Andersen, Jonas~S Neergaard-Nielsen, Peter van Loock, and Akira
  Furusawa.
\newblock Hybrid discrete-and continuous-variable quantum information.
\newblock {\em Nat. Phys}, 11(9):713--719, 2015.

\bibitem{mandelstam}
L~Mandelstam and Igor Tamm.
\newblock The uncertainty relation between energy and time in nonrelativistic
  quantum mechanics.
\newblock {\em J. Phys.(USSR)}, 9(249):1, 1945.

\bibitem{bioimaging}
Michael Taylor.
\newblock {\em Quantum Microscopy of Biological Systems}.
\newblock Springer, 2015.

\bibitem{magneto}
Jonatan~Bohr Brask, Rafael Chaves, and J~Ko{\l}ody{\'n}ski.
\newblock Improved quantum magnetometry beyond the standard quantum limit.
\newblock {\em Phys. Rev. X}, 5(3):031010, 2015.

\bibitem{quantumclock}
Thomas Udem, Ronald Holzwarth, and Theodor~W H{\"a}nsch.
\newblock Optical frequency metrology.
\newblock {\em Nature}, 416(6877):233--237, 2002.

\bibitem{bloom2014optical}
BJ~Bloom, TL~Nicholson, JR~Williams, SL~Campbell, M~Bishof, X~Zhang, W~Zhang,
  SL~Bromley, and J~Ye.
\newblock An optical lattice clock with accuracy and stability at the 10-18
  level.
\newblock {\em Nature}, 506(7486):71--75, 2014.

\bibitem{quantumlitho1}
Zeyang Liao, Mohammad Al-Amri, and M~Suhail Zubairy.
\newblock Quantum lithography beyond the diffraction limit via rabi
  oscillations.
\newblock {\em Physi. Rev. Lett}, 105(18):183601, 2010.

\bibitem{quantumlitho2}
Agedi~N Boto, Pieter Kok, Daniel~S Abrams, Samuel~L Braunstein, Colin~P
  Williams, and Jonathan~P Dowling.
\newblock Quantum interferometric optical lithography: exploiting entanglement
  to beat the diffraction limit.
\newblock {\em Phys. Rev. Lett}, 85(13):2733, 2000.

\bibitem{luis}
Luis~A Correa, Mohammad Mehboudi, Gerardo Adesso, and Anna Sanpera.
\newblock Individual quantum probes for optimal thermometry.
\newblock {\em Physi. Rev. Lett}, 114(22):220405, 2015.

\bibitem{robthermo}
Robert~B Mann and Eduardo Mart{\'\i}n-Mart{\'\i}nez.
\newblock Quantum thermometry.
\newblock {\em Foundations of Physics}, 44(5):492--511, 2014.

\bibitem{LIGO}
BP~Abbott, R~Abbott, TD~Abbott, MR~Abernathy, F~Acernese, K~Ackley, C~Adams,
  T~Adams, P~Addesso, RX~Adhikari, et~al.
\newblock Observation of gravitational waves from a binary black hole merger.
\newblock {\em Phys. Rev. Lett}, 116(6):061102, 2016.

\bibitem{cable2007efficient}
Hugo Cable and Jonathan~P Dowling.
\newblock Efficient generation of large number-path entanglement using only
  linear optics and feed-forward.
\newblock {\em Phys. Rev. Lett}, 99(16):163604, 2007.

\bibitem{carlos}
Marcin Zwierz, Carlos~A P{\'e}rez-Delgado, and Pieter Kok.
\newblock General optimality of the heisenberg limit for quantum metrology.
\newblock {\em Phys. Rev. Lett}, 105(18):180402, 2010.

\bibitem{hofmann2009}
Holger~F Hofmann.
\newblock All path-symmetric pure states achieve their maximal phase
  sensitivity in conventional two-path interferometry.
\newblock {\em Phys. Rev. A}, 79(3):033822, 2009.

\bibitem{heitler}
Walter Heitler.
\newblock {\em The quantum theory of radiation}.
\newblock Courier Corporation, 1954.

\bibitem{hugotomo}
Xiao-Qi Zhou, Hugo Cable, Rebecca Whittaker, Peter Shadbolt, Jeremy~L
  O’Brien, and Jonathan~CF Matthews.
\newblock Quantum-enhanced tomography of unitary processes.
\newblock {\em Optica}, 2(6):510--516, 2015.

\bibitem{peterh}
Peter~C Humphreys, Marco Barbieri, Animesh Datta, and Ian~A Walmsley.
\newblock Quantum enhanced multiple phase estimation.
\newblock {\em Phys. Rev. Lett}, 111(7):070403, 2013.

\bibitem{yue}
Jie-Dong Yue, Yu-Ran Zhang, and Heng Fan.
\newblock Quantum-enhanced metrology for multiple phase estimation with noise.
\newblock {\em Sci. Rep}, 4, 2014.

\bibitem{mihai}
Mihai~D Vidrighin, Gaia Donati, Marco~G Genoni, Xian-Min Jin, W~Steven
  Kolthammer, MS~Kim, Animesh Datta, Marco Barbieri, and Ian~A Walmsley.
\newblock Joint estimation of phase and phase diffusion for quantum metrology.
\newblock {\em Nat. Commun}, 5, 2014.

\bibitem{philip}
Philip~JD Crowley, Animesh Datta, Marco Barbieri, and Ian~A Walmsley.
\newblock Tradeoff in simultaneous quantum-limited phase and loss estimation in
  interferometry.
\newblock {\em Phys. Rev. A}, 89(2):023845, 2014.

\bibitem{childspreskill}
Andrew~M Childs, John Preskill, and Joseph Renes.
\newblock Quantum information and precision measurement.
\newblock {\em J. Mod. Opt}, 47(2-3):155--176, 2000.

\bibitem{peres}
Asher Peres and Petra~F Scudo.
\newblock Entangled quantum states as direction indicators.
\newblock {\em Phys. Rev. Lett}, 86(18):4160, 2001.

\bibitem{bagan1}
E~Bagan, M~Baig, and R~Munoz-Tapia.
\newblock Entanglement-assisted alignment of reference frames using a dense
  covariant coding.
\newblock {\em Phys. Rev. A}, 69(5):050303, 2004.

\bibitem{bagan2}
E~Bagan, M~Baig, and R~Munoz-Tapia.
\newblock Quantum reverse engineering and reference-frame alignment without
  nonlocal correlations.
\newblock {\em Phys. Rev. A}, 70(3):030301, 2004.

\bibitem{chiri}
G~Chiribella, GM~D’Ariano, P~Perinotti, and MF~Sacchi.
\newblock Efficient use of quantum resources for the transmission of a
  reference frame.
\newblock {\em Phys. Rev. Lett}, 93(18):180503, 2004.

\bibitem{hayashi}
Masahito Hayashi.
\newblock Parallel treatment of estimation of su (2) and phase estimation.
\newblock {\em Phys. Lett. A}, 354(3):183--189, 2006.

\bibitem{fujiwara}
Akio Fujiwara.
\newblock Estimation of su (2) operation and dense coding: An information
  geometric approach.
\newblock {\em Phys. Rev. A}, 65(1):012316, 2001.

\bibitem{matsumoto}
Keiji Matsumoto.
\newblock A new approach to the cram{\'e}r-rao-type bound of the pure-state
  model.
\newblock {\em J. Phys. A}, 35(13):3111, 2002.

\bibitem{fujiwaralate}
Hiroshi Imai and Akio Fujiwara.
\newblock Geometry of optimal estimation scheme for su (d) channels.
\newblock {\em J. Phys. A}, 40(16):4391, 2007.

\bibitem{kahn}
Jonas Kahn.
\newblock Fast rate estimation of a unitary operation in su (d).
\newblock {\em Phys. Rev. A}, 75(2):022326, 2007.

\bibitem{ballester2004}
Manuel~A Ballester.
\newblock Estimation of unitary quantum operations.
\newblock {\em Phys. Rev. A}, 69(2):022303, 2004.

\bibitem{ballester2005}
Manuel~A Ballester.
\newblock Optimal estimation of su (d) using exact and approximate 2-designs.
\newblock {\em arXiv: quant-ph/0507073}, 2005.

\bibitem{animesh}
Tillmann Baumgratz and Animesh Datta.
\newblock Quantum enhanced estimation of a multidimensional field.
\newblock {\em Phys. Rev. Lett}, 116(3):030801, 2016.

\bibitem{shajicaves}
Carlton~M Caves and Anil Shaji.
\newblock Quantum-circuit guide to optical and atomic interferometry.
\newblock {\em Opt. Commun}, 283(5):695--712, 2010.

\bibitem{HB1993}
MJ~Holland and K~Burnett.
\newblock Interferometric detection of optical phase shifts at the heisenberg
  limit.
\newblock {\em Phys. Rev. Lett}, 71(9):1355, 1993.

\bibitem{datta2011quantum}
Animesh Datta, Lijian Zhang, Nicholas Thomas-Peter, Uwe Dorner, Brian~J Smith,
  and Ian~A Walmsley.
\newblock Quantum metrology with imperfect states and detectors.
\newblock {\em Phys. Rev. A}, 83(6):063836, 2011.

\bibitem{Bollinger1996}
JJ~Bollinger, WM~Itano, DJ~Wineland, and DJ~Heinzen.
\newblock Optimal frequency measurements with maximally correlated states.
\newblock {\em Phys. Rev. A}, 54(6):R4649, 1996.

\bibitem{durkin2007}
Gabriel~A Durkin and Jonathan~P Dowling.
\newblock Local and global distinguishability in quantum interferometry.
\newblock {\em Phys. Rev. Lett}, 99(7):070801, 2007.

\bibitem{andersen1970asymptotic}
Erling~Bernhard Andersen.
\newblock Asymptotic properties of conditional maximum-likelihood estimators.
\newblock {\em J. R. Stat. Soc. Series B}, pages 283--301, 1970.

\bibitem{le1990maximum}
Lucien Le~Cam.
\newblock Maximum likelihood: an introduction.
\newblock {\em Int. Stat. Rev}, pages 153--171, 1990.

\bibitem{parisreview}
Matteo~GA Paris.
\newblock Quantum estimation for quantum technology.
\newblock {\em Int. J. Quantum Inf}, 7(supp01):125--137, 2009.

\bibitem{braunsteincaves}
Samuel~L Braunstein and Carlton~M Caves.
\newblock Statistical distance and the geometry of quantum states.
\newblock {\em Phys. Rev. Lett}, 72(22):3439, 1994.

\bibitem{dorner2009optimal}
U~Dorner, R~Demkowicz-Dobrzanski, BJ~Smith, JS~Lundeen, W~Wasilewski,
  K~Banaszek, and IA~Walmsley.
\newblock Optimal quantum phase estimation.
\newblock {\em Phys. Rev. Lett}, 102(4):040403, 2009.

\bibitem{knysh2011scaling}
Sergey Knysh, Vadim~N Smelyanskiy, and Gabriel~A Durkin.
\newblock Scaling laws for precision in quantum interferometry and the
  bifurcation landscape of the optimal state.
\newblock {\em Phys. Rev. A}, 83(2):021804, 2011.

\bibitem{lee2009optimization}
Tae-Woo Lee, Sean~D Huver, Hwang Lee, Lev Kaplan, Steven~B McCracken, Changjun
  Min, Dmitry~B Uskov, Christoph~F Wildfeuer, Georgios Veronis, and Jonathan~P
  Dowling.
\newblock Optimization of quantum interferometric metrological sensors in the
  presence of photon loss.
\newblock {\em Phys. Rev. A}, 80(6):063803, 2009.

\bibitem{highnoon}
Itai Afek, Oron Ambar, and Yaron Silberberg.
\newblock High-noon states by mixing quantum and classical light.
\newblock {\em Science}, 328(5980):879--881, 2010.

\bibitem{thomas2011}
Nicholas Thomas-Peter, Brian~J Smith, Animesh Datta, Lijian Zhang, Uwe Dorner,
  and Ian~A Walmsley.
\newblock Real-world quantum sensors: evaluating resources for precision
  measurement.
\newblock {\em Phys. Rev. Lett}, 107(11):113603, 2011.

\bibitem{dicke}
Robert~H Dicke.
\newblock Coherence in spontaneous radiation processes.
\newblock {\em Phys. Rev}, 93(1):99, 1954.

\bibitem{toth}
G{\'e}za T{\'o}th and Iagoba Apellaniz.
\newblock Quantum metrology from a quantum information science perspective.
\newblock {\em J. Phys. A}, 47(42):424006, 2014.

\bibitem{numbercount}
Jonathan~CF Matthews, Xiao-Qi Zhou, Hugo Cable, Peter~J Shadbolt, Dylan~J
  Saunders, Gabriel~A Durkin, Geoff~J Pryde, and Jeremy~L O'Brien.
\newblock Practical quantum metrology.
\newblock {\em arXiv:1307.4673}, 2013.

\bibitem{sakurai}
Jun~John Sakurai and San~Fu Tuan.
\newblock {\em Modern quantum mechanics}, volume~1.
\newblock Addison-Wesley Reading, Massachusetts, 1985.

\bibitem{loudon}
Rodney Loudon.
\newblock {\em The quantum theory of light}.
\newblock OUP Oxford, 2000.

\bibitem{adaptive1}
Takanori Sugiyama, Peter~S Turner, and Mio Murao.
\newblock Adaptive experimental design for one-qubit state estimation with
  finite data based on a statistical update criterion.
\newblock {\em Physi. Rev. A}, 85(5):052107, 2012.

\bibitem{adaptive2}
DH~Mahler, LA~Rozema, A~Darabi, C~Ferrie, R~Blume-Kohout, and AM~Steinberg.
\newblock Adaptive quantum state tomography improves accuracy quadratically.
\newblock {\em Phys. Rev. Lett}, 111(18):183601, 2013.

\bibitem{escher2011general}
BM~Escher, RL~de~Matos~Filho, and L~Davidovich.
\newblock General framework for estimating the ultimate precision limit in
  noisy quantum-enhanced metrology.
\newblock {\em Nat. Phys}, 7(5):406--411, 2011.

\bibitem{rosetta}
Hwang Lee, Pieter Kok, and Jonathan~P Dowling.
\newblock A quantum rosetta stone for interferometry.
\newblock {\em J. Mod. Opt}, 49(14-15):2325--2338, 2002.

\bibitem{dowlinglowdown}
Jonathan~P Dowling.
\newblock Quantum optical metrology--the lowdown on high-n00n states.
\newblock {\em Contemp. Phys}, 49(2):125--143, 2008.

\bibitem{Guth:81}
Alan Guth.
\newblock Inflationary universe: A possible solution to the horizon and
  flatness problems.
\newblock {\em Phys. Rev. D}, 23:347--356, Jan 1981.

\bibitem{0264-9381-26-14-145005}
Paul~H Frampton, Stephen~DH Hsu, Thomas~W Kephart, and David Reeb.
\newblock What is the entropy of the universe?
\newblock {\em Class. Quantum Grav}, 26(14):145005, 2009.

\bibitem{0004-637X-710-2-1825}
Chas~A. Egan and Charles~H. Lineweaver.
\newblock A larger estimate of the entropy of the universe.
\newblock {\em Astrophys. J}, 710(2):1825, 2010.

\bibitem{unruh}
William~G Unruh.
\newblock Notes on black-hole evaporation.
\newblock {\em Phys. Rev. D}, 14(4):870, 1976.

\bibitem{unruhapplications}
Lu{\'\i}s~CB Crispino, Atsushi Higuchi, and George~EA Matsas.
\newblock The unruh effect and its applications.
\newblock {\em Rev. Mod. Phys}, 80(3):787, 2008.

\bibitem{davies}
Paul~CW Davies.
\newblock Scalar production in schwarzschild and rindler metrics.
\newblock {\em J. Phys. A}, 8(4):609, 1975.

\bibitem{fullingmovingmirror}
Stephen~A Fulling and Paul~CW Davies.
\newblock Radiation from a moving mirror in two dimensional space-time:
  conformal anomaly.
\newblock In {\em Proc. R. Soc. A}, volume 348, pages 393--414, 1976.

\bibitem{PhysRevD.7.2333}
Jacob~D. Bekenstein.
\newblock Black holes and entropy.
\newblock {\em Phys. Rev. D}, 7:2333--2346, Apr 1973.

\bibitem{wald}
Robert~M Wald.
\newblock {\em Quantum field theory in curved spacetime and black hole
  thermodynamics}.
\newblock University of Chicago Press, 1994.

\bibitem{hawking}
James~M Bardeen, Brandon Carter, and Stephen~W Hawking.
\newblock The four laws of black hole mechanics.
\newblock {\em Comm. Math. Phys}, 31(2):161--170, 1973.

\bibitem{PhysRevLett.21.562}
Leonard Parker.
\newblock Particle creation in expanding universes.
\newblock {\em Phys. Rev. Lett}, 21:562--564, Aug 1968.

\bibitem{PhysRev.183.1057}
Leonard Parker.
\newblock Quantized fields and particle creation in expanding universes. i.
\newblock {\em Phys. Rev}, 183:1057--1068, Jul 1969.

\bibitem{PhysRevD.3.346}
Leonard Parker.
\newblock Quantized fields and particle creation in expanding universes. ii.
\newblock {\em Phys. Rev. D}, 3:346--356, Jan 1971.

\bibitem{prigogine}
I~Prigogine, J~Geheniau, E~Gunzig, and P~Nardone.
\newblock Thermodynamics and cosmology.
\newblock {\em Gen. Rel. Grav}, 21(8):767--776, 1989.

\bibitem{liddle}
Andrew~R Liddle and David~H Lyth.
\newblock {\em Cosmological inflation and large-scale structure}.
\newblock Cambridge University Press, 2000.

\bibitem{tasaki}
Hal Tasaki.
\newblock Jarzynski relations for quantum systems and some applications.
\newblock {\em arXiv:cond-mat/0009244}, 2000.

\bibitem{husqueezed}
Bei~Lok Hu, GW~Kang, and Andrew Matacz.
\newblock Squeezed vacua and the quantum statistics of cosmological particle
  creation.
\newblock {\em Int. J. Mod. Phys. A}, 9(07):991--1007, 1994.

\bibitem{kolb}
Edward~W Kolb and Michael~Stanley Turner.
\newblock The early universe.
\newblock {\em Front. Phys., Vol. 69,}, 69, 1990.

\bibitem{kolb2}
Edward~W Kolb.
\newblock First-order inflation.
\newblock {\em Phys. Scripta}, 1991(T36):199, 1991.

\bibitem{olive}
Keith~A Olive.
\newblock Inflation.
\newblock {\em Phys. Rep}, 190(6):307--403, 1990.

\bibitem{hukandrup}
Bei~Lok Hu and Henry~E Kandrup.
\newblock Entropy generation in cosmological particle creation and
  interactions: A statistical subdynamics analysis.
\newblock {\em Phys. Rev. D}, 35(6):1776, 1987.

\bibitem{kandrup}
Henry~E Kandrup.
\newblock Entropy generation, particle creation, and quantum field theory in a
  cosmological spacetime: When do number and entropy increase?
\newblock {\em Phys. Rev. D}, 37(12):3505, 1988.

\bibitem{kokshu}
Don Koks, Andrew Matacz, and Bei~Lok Hu.
\newblock Entropy and uncertainty of squeezed quantum open systems.
\newblock {\em Phys. Rev. D}, 55(10):5917, 1997.

\bibitem{hureview}
Bei~Lok Hu.
\newblock Nonequilibrium quantum fields in cosmology: Comments on selected
  current topics.
\newblock {\em arXiv:gr-qc/9409053}, 1994.

\bibitem{gasperini1}
M~Gasperini and Massimo Giovannini.
\newblock Quantum squeezing and cosmological entropy production.
\newblock {\em Class. Quantum Grav}, 10(9):L133, 1993.

\bibitem{gasperini2}
M~Gasperini and Massimo Giovannini.
\newblock Entropy production in the cosmological amplification of the vacuum
  fluctuations.
\newblock {\em Phys. Lett. B}, 301(4):334--338, 1993.

\bibitem{brandenberger}
R~Brandenberger, V~Mukhanov, and T~Prokopec.
\newblock Entropy of a classical stochastic field and cosmological
  perturbations.
\newblock {\em Phys. Rev. Lett}, 69(25):3606, 1992.

\bibitem{hulin}
Shih-Yuin Lin, Chung-Hsien Chou, and Bei~Lok Hu.
\newblock Quantum entanglement and entropy in particle creation.
\newblock {\em Phys. Rev. D}, 81(8):084018, 2010.

\bibitem{hu1986}
Bei~Lok Hu and D~Pavon.
\newblock Intrinsic measures of field entropy in cosmological particle
  creation.
\newblock {\em Phys. Lett. B}, 180(4):329--334, 1986.

\bibitem{evans}
Denis~J Evans and Debra~J Searles.
\newblock The fluctuation theorem.
\newblock {\em Adv. in Phys}, 51(7):1529--1585, 2002.

\bibitem{jarzynski}
Christopher Jarzynski.
\newblock Nonequilibrium equality for free energy differences.
\newblock {\em Phys. Rev. Lett}, 78(14):2690, 1997.

\bibitem{crooks}
Gavin~E Crooks.
\newblock Entropy production fluctuation theorem and the nonequilibrium work
  relation for free energy differences.
\newblock {\em Phys. Rev. E}, 60(3):2721, 1999.

\bibitem{sevick2007fluctuation}
Edith~M Sevick, R~Prabhakar, Stephen~R Williams, and Debra~J Searles.
\newblock Fluctuation theorems.
\newblock {\em arXiv:0709.3888}, 2007.

\bibitem{noneqm0}
Lars Onsager.
\newblock Reciprocal relations in irreversible processes. i.
\newblock {\em Phys. Rev}, 37(4):405, 1931.

\bibitem{groot}
SR~De~Groot, P~Mazur, and J~Th~G Overbeek.
\newblock Nonequilibrium thermodynamics of the sedimentation potential and
  electrophoresis.
\newblock {\em J. Chem. Phys}, 20(12):1825--1829, 1952.

\bibitem{noneqm1}
P~Glansdorff and I~Prigogine.
\newblock Structure, stability and fluctuations.
\newblock {\em New York, NY: Interscience}, 1971.

\bibitem{noneqm2}
Georgy Lebon, David Jou, and Jos{\'e} Casas-V{\'a}zquez.
\newblock {\em Understanding non-equilibrium thermodynamics}.
\newblock Springer, 2008.

\bibitem{campisireview}
Michele Campisi, Peter H{\"a}nggi, and Peter Talkner.
\newblock Colloquium: Quantum fluctuation relations: Foundations and
  applications.
\newblock {\em Rev. Mod. Phys}, 83(3):771, 2011.

\bibitem{campisi2011}
Michele Campisi and Peter H{\"a}nggi.
\newblock Fluctuation, dissipation and the arrow of time.
\newblock {\em Entropy}, 13(12):2024--2035, 2011.

\bibitem{tolman}
Richard~Chace Tolman.
\newblock {\em The principles of statistical mechanics}.
\newblock Courier Corporation, 1938.

\bibitem{shannon2015mathematical}
Claude~E Shannon and Warren Weaver.
\newblock {\em The mathematical theory of communication}.
\newblock University of Illinois press, 2015.

\bibitem{petz2001entropy}
D{\'e}nes Petz.
\newblock Entropy, von neumann and the von neumann entropy.
\newblock pages 83--96, 2001.

\bibitem{morebangbuck}
Adolfo Del~Campo, J~Goold, and M~Paternostro.
\newblock More bang for your buck: Super-adiabatic quantum engines.
\newblock {\em Sci. Rep}, 4, 2014.

\bibitem{gupta2011experimental}
Amar~Nath Gupta, Abhilash Vincent, Krishna Neupane, Hao Yu, Feng Wang, and
  Michael~T Woodside.
\newblock Experimental validation of free-energy-landscape reconstruction from
  non-equilibrium single-molecule force spectroscopy measurements.
\newblock {\em Nat. Phys}, 7(8):631--634, 2011.

\bibitem{frey2015reconstructing}
Eric~W Frey, Jingqiang Li, Sithara~S Wijeratne, and Ching-Hwa Kiang.
\newblock Reconstructing multiple free energy pathways of dna stretching from
  single molecule experiments.
\newblock {\em J. Phys. Chem. A}, 119(16):5132--5135, 2015.

\bibitem{luccioli2008free}
Stefano Luccioli, Alberto Imparato, and Alessandro Torcini.
\newblock Free-energy landscape of mechanically unfolded model proteins:
  Extended jarzinsky versus inherent structure reconstruction.
\newblock {\em Phys. Rev. E}, 78(3):031907, 2008.

\bibitem{sindona2014statistics}
A~Sindona, J~Goold, N~Lo Gullo, and F~Plastina.
\newblock Statistics of the work distribution for a quenched fermi gas.
\newblock {\em New J. Phys}, 16(4):045013, 2014.

\bibitem{shchadilova2014quantum}
Yulia~E Shchadilova, Pedro Ribeiro, and Masudul Haque.
\newblock Quantum quenches and work distributions in ultralow-density systems.
\newblock {\em Phys. Rev. Lett}, 112(7):070601, 2014.

\bibitem{dorner2012emergent}
Ross Dorner, John Goold, Cecilia Cormick, Mauro Paternostro, and Vlatko Vedral.
\newblock Emergent thermodynamics in a quenched quantum many-body system.
\newblock {\em Phys. Rev. Lett}, 109(16):160601, 2012.

\bibitem{mascarenhas2014work}
Eduardo Mascarenhas, Helena Bragan{\c{c}}a, Ross Dorner, M~Fran{\c{c}}a Santos,
  Vlatko Vedral, Kavan Modi, and John Goold.
\newblock Work and quantum phase transitions: Quantum latency.
\newblock {\em Phys. Rev. E}, 89(6):062103, 2014.

\bibitem{sotiriadis2013statistics}
Spyros Sotiriadis, Andrea Gambassi, and Alessandro Silva.
\newblock Statistics of the work done by splitting a one-dimensional
  quasicondensate.
\newblock {\em Phys. Rev. E}, 87(5):052129, 2013.

\bibitem{rossnagel2014nanoscale}
Johannes Ro{\ss}nagel, Obinna Abah, Ferdinand Schmidt-Kaler, Kilian Singer, and
  Eric Lutz.
\newblock Nanoscale heat engine beyond the carnot limit.
\newblock {\em Phys. Rev. Lett}, 112(3):030602, 2014.

\bibitem{li2016enhanced}
Peng-Bo Li, Hong-Rong Li, and Fu-Li Li.
\newblock Enhanced electromechanical coupling of a nanomechanical resonator to
  coupled superconducting cavities.
\newblock {\em Sci. Rep}, 6, 2016.

\bibitem{BandD}
Nicholas~David Birrell and Paul Charles~William Davies.
\newblock {\em Quantum fields in curved space}.
\newblock Cambridge University press, 1984.

\bibitem{kosloff}
Tova Feldmann and Ronnie Kosloff.
\newblock Performance of discrete heat engines and heat pumps in finite time.
\newblock {\em Phys. Rev. E}, 61:4774--4790, May 2000.

\bibitem{engine}
Adolfo del Campo, J.~Goold, and M.~Paternostro.
\newblock More bang for your buck: Super-adiabatic quantum engines.
\newblock {\em Sci. Rep}, 4:6208, 2014.

\bibitem{plastina}
F~Plastina, A~Alecce, TJG Apollaro, G~Falcone, G~Francica, F~Galve, N~Lo Gullo,
  and R~Zambrini.
\newblock Irreversible work and inner friction in quantum thermodynamic
  processes.
\newblock {\em Phys. Rev. Lett}, 113(26):260601, 2014.

\bibitem{maddox1}
SJ~Maddox, G~Efstathiou, WJ~Sutherland, and J~Loveday.
\newblock The apm galaxy survey: Pt. 1.
\newblock 1990.

\bibitem{anisotropy1}
Claudia Quercellini, Miguel Quartin, and Luca Amendola.
\newblock Possibility of detecting anisotropic expansion of the universe by
  very accurate astrometry measurements.
\newblock {\em Phys. Rev. Lett}, 102(15):151302, 2009.

\bibitem{anisotropy2}
Rong-Gen Cai, Yin-Zhe Ma, Bo~Tang, and Zhong-Liang Tuo.
\newblock Constraining the anisotropic expansion of the universe.
\newblock {\em Phys. Rev. D}, 87(12):123522, 2013.

\bibitem{gerryknight}
Christopher Gerry and Peter Knight.
\newblock {\em Introductory quantum optics}.
\newblock Cambridge University Press, 2005.

\bibitem{grishchuk}
L~P Grishchuk and YV~Sidorov.
\newblock Squeezed quantum states of relic gravitons and primordial density
  fluctuations.
\newblock {\em Phys. Rev. D}, 42(10):3413, 1990.

\bibitem{hu1994}
Bei~Lok Hu, GW~Kang, and Andrew Matacz.
\newblock Squeezed vacua and the quantum statistics of cosmological particle
  creation.
\newblock {\em Int. J. Mod. Phys. A}, 9(07):991--1007, 1994.

\bibitem{ivetteentanglement}
Jonathan~L Ball, Ivette Fuentes-Schuller, and Frederic~P Schuller.
\newblock Entanglement in an expanding spacetime.
\newblock {\em Phys. Lett. A}, 359(6):550--554, 2006.

\bibitem{Fuentes:Mann:10}
Ivette Fuentes, Robert Mann, Eduardo Mart\'in-Mart\'inez, and Shahpoor Moradi.
\newblock Entanglement of dirac fields in an expanding spacetime.
\newblock {\em Phys. Rev. D}, 82:045030, 2010.

\bibitem{PhysRevA.76.062112}
Gerardo Adesso, Ivette Fuentes-Schuller, and Marie Ericsson.
\newblock Continuous-variable entanglement sharing in noninertial frames.
\newblock {\em Phys. Rev. A}, 76:062112, 2007.

\bibitem{adesso:correlation}
Gerardo Adesso and Ivette Fuentes-Schuller.
\newblock Correlation loss and multipartite entanglement across a black hole
  horizon.
\newblock {\em Quantum Information \& Computation}, 9:657--665, 2009.

\bibitem{barnett}
Stephen Barnett and Paul~M. Radmore.
\newblock {\em Methods in theoretical quantum optics}.
\newblock Oxford University Press, 2002.

\bibitem{nohair1}
Werner Israel.
\newblock Event horizons in static vacuum space-times.
\newblock {\em Phys. Rev.}, 164(5):1776, 1967.

\bibitem{nohair2}
Werner Israel.
\newblock Event horizons in static electrovac space-times.
\newblock {\em Comm. Math. Phys}, 8(3):245--260, 1968.

\bibitem{nohair3}
Brandon Carter.
\newblock Axisymmetric black hole has only two degrees of freedom.
\newblock {\em Phys. Rev. Lett}, 26(6):331, 1971.

\bibitem{adkins1983equilibrium}
Clement~John Adkins.
\newblock {\em Equilibrium thermodynamics}.
\newblock Cambridge University Press, 1983.

\bibitem{silva2008statistics}
Alessandro Silva.
\newblock Statistics of the work done on a quantum critical system by quenching
  a control parameter.
\newblock {\em Phys. Rev. Lett}, 101(12):120603, 2008.

\bibitem{apollaro2015work}
Tony~JG Apollaro, Gianluca Francica, Mauro Paternostro, and Michele Campisi.
\newblock Work statistics, irreversible heat and correlations build-up in
  joining two spin chains.
\newblock {\em Phys. Scripta}, 2015(T165):014023, 2015.

\bibitem{yi2012work}
Juyeon Yi, Yong~Woon Kim, and Peter Talkner.
\newblock Work fluctuations for bose particles in grand canonical initial
  states.
\newblock {\em Phys. Rev. E}, 85(5):051107, 2012.

\bibitem{deffner1}
Sebastian Deffner and Eric Lutz.
\newblock {\em Phys. Rev. Lett}, 105:170402, 2010.

\bibitem{kullbackleibler}
Solomon Kullback and Richard~A Leibler.
\newblock On information and sufficiency.
\newblock {\em Ann. Math. Statist}, 22:79--86, 1951.

\bibitem{sagawa}
Takahiro Sagawa.
\newblock In {\em Lectures on Quantum Computing, Thermodynamics and Statistical
  Physics}, volume~8, page 127. World Scientific, 2013.

\bibitem{klein1931quantenmechanischen}
Otto Klein.
\newblock Zur quantenmechanischen begr{\"u}ndung des zweiten hauptsatzes der
  w{\"a}rmelehre.
\newblock {\em Zeitschrift f{\"u}r Physik}, 72(11-12):767--775, 1931.

\bibitem{abah2012single}
Obinna Abah, Johannes Rossnagel, Georg Jacob, Sebastian Deffner, Ferdinand
  Schmidt-Kaler, Kilian Singer, and Eric Lutz.
\newblock Single-ion heat engine at maximum power.
\newblock {\em Phys. Rev. Lett}, 109(20):203006, 2012.

\bibitem{popescu}
Sandu Popescu, Anthony~J Short, and Andreas Winter.
\newblock Entanglement and the foundations of statistical mechanics.
\newblock {\em Nat. Phys}, 2(11):754--758, 2006.

\bibitem{hucosmology}
Bei~Lok Hu.
\newblock {\em Cosmology of the early universe}.
\newblock World Scientific, Singapore, 1984.

\bibitem{alla}
AE~Allahverdyan and Th~M Nieuwenhuizen.
\newblock A mathematical theorem as the basis for the second law: Thomson's
  formulation applied to equilibrium.
\newblock {\em Phys. Rev. A}, 305(3):542--552, 2002.

\bibitem{barcelo2005analogue}
Carlos Barcel{\'o}, Stefano Liberati, Matt Visser, et~al.
\newblock Analogue gravity.
\newblock {\em Living Rev. Rel}, 8(12):214, 2005.

\bibitem{barcelo2001analogue}
Carlos Barcelo, Stefano Liberati, and Matt Visser.
\newblock Analogue gravity from bose-einstein condensates.
\newblock {\em Class. Quantum Grav}, 18(6):1137, 2001.

\bibitem{ngo2012demonstration}
Van~A Ngo and Stephan Haas.
\newblock Demonstration of jarzynski's equality in open quantum systems using a
  stepwise pulling protocol.
\newblock {\em Phys. Rev. E}, 86(3):031127, 2012.

\end{thebibliography}

\end{document}